\documentclass[final,3p,11pt,pdflatex]{elsarticle}
\usepackage[utf8x]{inputenc}      
\usepackage{amsmath,amssymb,mathtools}
\usepackage[T1]{fontenc}          
\usepackage{booktabs,tabularx}
\usepackage{graphicx,subfig}
\usepackage{xspace}
\usepackage[usenames]{xcolor}\definecolor{fscolor}{RGB}{44,118,255}
\usepackage{tikz,tikz-uml}
\usetikzlibrary{arrows,calc,matrix,positioning,shapes}
\usepackage{listings}
\usepackage[absolute]{textpos}
\usepackage[many]{tcolorbox}
\usepackage{xparse}
\usepackage[font=small,labelfont=bf,format=plain,margin=0.05\textwidth]{caption}
\usepackage{bbm}
\bibstyle{elsarticle-num}
\usepackage[pdftitle={FlexibleSUSY 2.0: Extensions to investigate the phenomenology of
  SUSY and non-SUSY models},
  pdfauthor={Peter Athron, Markus Bach, Dylan Harries, Thomas Kwasnitza, Jae-hyeon Park,
    Dominik Stockinger, Alexander Voigt, Jobst Ziebell},
  pdfkeywords={FlexibleSUSY,supersymmetry,spectrum,generator,MSSM,NMSSM,E6SSM},
  bookmarks=true, linktocpage, colorlinks=true, allbordercolors=white,
  allcolors=fscolor]{hyperref}
\biboptions{sort&compress}
\allowdisplaybreaks

\newcommand{\modelname}[1]{\texttt{#1}\@\xspace}
\newcommand{\sarah}{\texttt{SARAH}\@\xspace}
\newcommand{\spheno}{\texttt{SPheno}\@\xspace}
\newcommand{\suspect}{\texttt{SuSpect}\@\xspace}
\newcommand{\isasusy}{\texttt{ISASUSY}\@\xspace}
\newcommand{\suseflav}{\texttt{SuSeFLAV}\@\xspace}
\newcommand{\nmspec}{\texttt{NMSPEC}\@\xspace}
\newcommand{\fs}{\texttt{FlexibleSUSY}\@\xspace}
\newcommand{\fsbreak}{\texttt{Flex\-ib\-le\-SUSY}\@\xspace}
\newcommand{\HSSUSY}{\modelname{HSSUSY}}
\newcommand{\susyhd}{\texttt{SusyHD}\@\xspace}
\newcommand{\MhEFT}{\texttt{MhEFT}\@\xspace}
\newcommand{\softsusy}{\texttt{SOFTSUSY}\@\xspace}
\newcommand{\micromegas}{\texttt{micrOMEGAS}\@\xspace}

\newcommand{\calchep}{\texttt{CALCHEP}\@\xspace}
\newcommand{\lanhep}{\texttt{LANHEP}\@\xspace}
\newcommand{\madgraph}{\texttt{MadGraph}\@\xspace}
\newcommand{\whizard}{\texttt{WHIZARD}\@\xspace}
\newcommand{\sherpa}{\texttt{SHERPA}\@\xspace}
\newcommand{\FAFC}{\texttt{FeynArts}\@\xspace}
\newcommand{\helac}{\texttt{HELAC}\@\xspace}
\newcommand{\herwig}{\texttt{HERWIG}\@\xspace}
\newcommand{\pythia}{\texttt{PYTHIA}\@\xspace}
\newcommand{\checkmate}{\texttt{CheckMate}\@\xspace}
\newcommand{\madanalysis}{\texttt{MadAnalysis}\@\xspace}
\newcommand{\smodels}{\texttt{SModelS}\@\xspace}
\newcommand{\fastlim}{\texttt{Fastlim}\@\xspace}
\newcommand{\colliderbit}{\texttt{ColliderBit}\@\xspace}
\newcommand{\higgsbounds}{\texttt{HiggsBounds}\@\xspace}
\newcommand{\higgssignals}{\texttt{HiggsSignals}\@\xspace}
\newcommand{\htm}{\texttt{H3m}\@\xspace}
\newcommand{\Himalaya}{\texttt{Himalaya}\@\xspace}
\newcommand{\FeynHiggs}{\texttt{FeynHiggs}\@\xspace}
\newcommand{\FeynRules}{\texttt{FeynRules}\@\xspace}
\newcommand{\NMSSMCalc}{\texttt{NMSSMCalc}\@\xspace}
\newcommand{\multinest}{\texttt{MultiNest}\@\xspace}
\newcommand{\CPsuperH}{\texttt{CPsuperH}\@\xspace}
\newcommand{\GAMBIT}{\texttt{GAMBIT}\@\xspace}
\newcommand{\GMTCalc}{\texttt{GM2Calc}\@\xspace}
\newcommand{\fsone}{\fs 1.0\@\xspace}
\newcommand{\fstwo}{\fs 2.0\@\xspace}
\newcommand{\fsh}{\texttt{FlexibleSUSY+}\Himalaya\xspace}
\newcommand{\fbsm}{\texttt{FlexibleBSM}\@\xspace}
\newcommand{\fcpv}{\texttt{FlexibleCPV}\@\xspace}
\newcommand{\fmw}{\texttt{FlexibleMW}\@\xspace}
\newcommand{\famu}{\texttt{FlexibleAMU}\@\xspace}
\newcommand{\fsas}{\texttt{FlexibleSAS}\@\xspace}
\newcommand{\feft}{\texttt{Flex\-ib\-le\-EFT\-Higgs}\@\xspace}
\newcommand{\mathematica}{\texttt{Ma\-the\-ma\-ti\-ca}\xspace}

\newcommand{\ESSM}{E$_6$SSM\@\xspace}
\newcommand{\code}[1]{\lstinline|#1|}  
\newcommand{\ol}[1]{\overline{#1}}
\newcommand{\MSbar}{\ensuremath{\ol{\text{MS}}}\xspace}
\newcommand{\DRbar}{\ensuremath{\ol{\text{DR}}}\xspace}
\newcommand{\MDRbar}{\ensuremath{\ol{\text{MDR}}}\xspace}
\newcommand{\unit}[1]{\,\text{#1}}      

\newcommand{\SM}{\ensuremath{\text{SM}}\xspace}
\newcommand{\BSM}{\ensuremath{\text{BSM}}\xspace}
\newcommand{\THDM}{\ensuremath{\text{THDM}}\xspace}
\newcommand{\THDMII}{\ensuremath{\text{THDM-II}}\xspace}
\newcommand{\MSSM}{\ensuremath{\text{MSSM}}\xspace}
\newcommand{\MS}{\ensuremath{M_S}\xspace}
\newcommand{\MLCP}{\ensuremath{M_\text{LCP}}\xspace}
\newcommand{\MEWSB}{\ensuremath{M_\text{EWSB}}\xspace}
\newcommand{\Qmatch}{\ensuremath{Q_\text{match}}}
\newcommand{\QpoleBSM}{\ensuremath{Q_\text{pole,BSM}}}
\newcommand{\QpoleSM}{\ensuremath{Q_\text{pole,SM}}}
\newcommand{\QED}{\ensuremath{\text{QED}}}

\newcommand{\amu}{\ensuremath{a_\mu}\xspace}
\newcommand{\amuBSM}{\ensuremath{\amu^{\BSM}}\xspace}
\newcommand{\amuMSSM}{\ensuremath{\amu^{\MSSM}}\xspace}
\newcommand{\edm}[1]{\ensuremath{d_{#1}}\xspace}
\newcommand{\edmBSM}[1]{\ensuremath{\edm{#1}^{\BSM}}\xspace}
\newcommand{\deltaVB}{\ensuremath{\delta_{\text{VB}}}\xspace}
\newcommand{\figref}[1]{\figurename~\ref{#1}}
\newcommand{\secref}[1]{Section~\ref{#1}}
\newcommand{\appref}[1]{Appendix~\ref{#1}}
\newcommand{\tabref}[1]{\tablename~\ref{#1}}
\newcommand{\exref}[1]{Example~\ref{#1}}
\newcommand{\ptitle}[1]{\emph{#1}}
\renewcommand{\ptitle}[1]{}

\newcommand{\scoeff}[2]{[ c^{#1}_{#2}(Q) ]}
\newcommand{\barlog}{\overline{\log}}
\newcommand{\GSM}{\ensuremath{G_{\SM}}\xspace}
\newcommand{\CP}{\ensuremath{CP}\xspace}
\newcommand{\azero}{\ensuremath{A_0}\xspace}
\newcommand{\mhalf}{\ensuremath{M_{1/2}}\xspace}
\newcommand{\mzero}{\ensuremath{m_0}\xspace}
\newcommand{\SQCD}{SUSY-QCD\xspace}

\newcommand{\multilinecell}[2][c]{%
  \begin{tabular}[#1]{@{}c@{}}#2\end{tabular}}

\newtcolorbox[
  auto counter
]{example}[1][]{%
  breakable,
  before skip=1em,
  after skip=1em,
  enhanced,
  colback=white,
  colbacktitle=white,
  arc=0pt,
  leftrule=1.5pt,
  rightrule=0.5pt,
  toprule=0.5pt,
  bottomrule=0.5pt,
  titlerule=0.5pt,
  colframe=fscolor,
  fonttitle=\bfseries\normalcolor,
  overlay={},
  parbox=false,
  title=Example~\thetcbcounter,
  #1
}

\NewDocumentEnvironment{OptionTable}{m m O{llXX}}{%
\table[tbh!]
  \tabularx{\textwidth}{#3}%
    \toprule
    Symbol & Default value & Allowed values & Description \\
    \midrule
}{\endtabularx\caption{#1}\label{#2}\endtable}

\makeatletter
\lstnewenvironment{numlstlisting}[1][]{%
  \lstset{%
    #1,
    numbers=left,
    firstnumber=auto,
    numberstyle=\tiny\sffamily}%
  \csname\@lst @SetFirstNumber\endcsname
}{%
  \csname \@lst @SaveFirstNumber\endcsname
}
\makeatother
\DeclareMathOperator{\diag}{diag}
\DeclareMathOperator{\sign}{sign}
\DeclareMathOperator{\re}{Re}
\DeclareMathOperator{\im}{Im}
\def\at{\alpha_t}
\def\ab{\alpha_b}
\def\as{\alpha_s}
\def\atau{\alpha_{\tau}}
\def\aem{\alpha_{\text{em}}}

\journal{Computer Physics Communications}
\begin{document}
\begin{frontmatter}
 \vspace*{0.5cm}
 \title{\Large\bf FlexibleSUSY 2.0: Extensions to investigate the phenomenology of SUSY and non-SUSY models}

\author[Monash]{Peter Athron}
\author[dresden]{Markus Bach}
\author[adelaide,prague]{Dylan Harries}
\author[dresden]{Thomas Kwasnitza}
\author[kiasquc]{Jae-hyeon Park}
\author[dresden]{Dominik St\"ockinger}
\author[rwth]{Alexander Voigt\corref{cor1}}
\ead{Alexander.Voigt@physik.rwth-aachen.de}
\cortext[cor1]{Corresponding author}
\author[dresden]{Jobst Ziebell}
\address[Monash]{ARC Centre of Excellence for Particle Physics at
  the Terascale, School of Physics, Monash University, Melbourne,
  Victoria 3800, Australia}
\address[dresden]{Institut f\"ur Kern- und Teilchenphysik,
TU Dresden, Zellescher Weg 19, 01069 Dresden, Germany}
\address[adelaide]{ARC Centre of Excellence for Particle Physics at
the Terascale, Department of Physics, The University of Adelaide,
Adelaide, South Australia 5005, Australia}
\address[prague]{Institute of Particle and Nuclear Physics, Faculty of
  Mathematics and Physics, Charles University in Prague, V
  Hole\v{s}ovi\v{c}k\'{a}ch 2, 180 00 Praha 8, Czech Republic}
\address[kiasquc]{Quantum Universe Center,
Korea Institute for Advanced Study,
85 Hoegiro Dongdaemungu,
Seoul 02455, Republic of Korea}
\address[rwth]{Institute for Theoretical Particle Physics and Cosmology, RWTH Aachen University, 52074 Aachen, Germany}

  \begin{abstract}
    We document major new features and improvements of \fs, a
    \mathematica and C++ package with a dependency on the external
    package \sarah, that generates fast and precise spectrum
    generators. The extensions presented here significantly increase
    the generality and capabilities of the \fs package, which already
    works with a wide class of models, while
    maintaining an elegant structure and easy to use interfaces. The
    \fbsm extension makes it possible to also create spectrum
    generators for non-supersymmetric extensions of the Standard
    Model. The \fcpv extension adds the option of complex parameters
    to the spectrum generators, allowing the study of many interesting
    models with new sources of \CP violation. \fmw computes the decay
    of the muon for the generated model and thereby allows \fs to
    predict the mass of the $W$ boson from the input parameters by
    using the more precise electroweak input of $\{ G_F, M_Z, \aem \}$
    instead of $\{ M_W, M_Z, \aem \}$. The \famu extension provides a
    calculator of the anomalous magnetic moment of the muon in any
    model \fs can generate a spectrum for.  \fsas introduces a new
    solver for the boundary value problem which makes use of
    semi-analytic expressions for dimensionful parameters to find
    solutions in models where the classic two-scale solver will not
    work such as the constrained E$_6$SSM\@. \feft is a hybrid
    calculation of the Higgs mass which combines the virtues of both
    effective field theory calculations and fixed-order calculations.
    All of these extensions are included in \fstwo, which is released
    simultaneously with this manual.
  \end{abstract}

\end{frontmatter}

\begin{textblock*}{10em}(\textwidth,1.5cm)
\raggedleft\noindent\footnotesize
KIAS--Q17043 \\
TTK--17--31 \\
CoEPP--MN--17--16
\end{textblock*}

\clearpage
\newgeometry{top=2.5cm,left=3cm,right=3cm,bottom=4cm,footskip=3em}
\section*{New version program summary}
\noindent
{\em Program title:} \fs\\[0.5em]
{\em Licensing provisions:} GPLv3\\[0.5em]
{\em Programming language:} C++, Wolfram/Mathematica, FORTRAN, Bourne shell\\[0.5em]
{\em Journal reference of previous version:} Comput.Phys.Commun.\ 190 (2015) 139-172\\[0.5em]
{\em Does the new version supersede the previous version?:} yes \\[0.5em]
{\em Reasons for the new version:} Program extension including new models, observables and algorithms\\[0.5em]
{\em Summary of revisions:} Extension to non-supersymmetric models
(\fbsm), complex parameters (\fcpv), prediction of $W$ boson mass from
muon decay (\fmw), calculation of anomalous magnetic moment of the
muon (\famu), semi-analytic boundary value problem solver (\fsas),
improved hybrid Higgs mass calculation (\feft).\\[0.5em]
{\em Nature of problem:}
Determining the mass spectrum, mixings and further observables for an arbitrary extension
of the Standard Model, input by the user. The generated code must find
simultaneous solutions to constraints that are specified at two or
more different renormalization scales, which are connected by
renormalization group equations forming a large set of coupled
first-order differential
equations. \\[0.5em]
{\em Solution method:}
Nested iterative algorithm and numerical
minimization of the Higgs potential.\\[0.5em]
{\em Restrictions:}
The couplings must remain perturbative at all scales between the
highest and lowest boundary condition.  Tensor-like Lagrangian
parameters of rank 3 are currently not supported.  The automatic
determination of the Standard Model-like gauge and Yukawa couplings is
only supported for models that have the Standard Model gauge group
$SU(3)_C\times SU(2)_L\times U(1)_Y$ as a gauge symmetry group factor.
However, due to the modular nature of the generated code, adapting and
extending it to overcome restrictions in scope is quite straightforward.

\clearpage
\tableofcontents

\newpage
\section{Introduction}

Popular and well studied new physics extensions of the Standard Model
(SM) are coming under increasing pressure from the searches at the
Large Hadron Collider (LHC) as well as other current experiments.  At
the same time, there remain many outstanding physics problems that are
not solved by the SM and require new physics.  For example, the origin
of dark matter is still unexplained and it requires some new particle
to fit the relic density.  In addition, the gauge structure and
fractional charges of the SM have no explanation, but some
observations hint at the possibility of a grand unified theory (GUT)
from which these can be derived after the breakdown of the GUT gauge
group.  Finally, the stability of the weak scale, which is $17$ orders
of magnitude smaller than the Planck scale, has no explanation within
the SM \cite{Weinberg:1975gm, Weinberg:1979bn,
  Gildener:1976ai,Susskind:1978ms, tHooft:1980xss} and has been the
driving motivation for the construction and study of concepts that go
beyond the Standard Model (BSM), most notably supersymmetry (SUSY) \cite{Miyazawa:1966mfa,Miyazawa:1968zz,Ramond:1971gb,Golfand:1971iw,Gervais:1971ji,Neveu:1971rx,Neveu:1971iv,Volkov:1973ix,Wess:1973kz,Wess:1974tw,Salam:1974yz} which ensures that the quadratic corrections from fermions and bosons in the loop diagrams cancel at all orders in perturbation theory \cite{Dimopoulos:1981au,Witten:1981nf,Dine:1981za,Dimopoulos:1981zb,Kaul:1981hi}.

These issues strongly motivate the development of new ideas and new
models.  However, in principle there are a huge number of models that
can solve some, or all, of these problems and many more that may be
motivated by other principles that have not yet gained widespread
interest in high energy physics (HEP)\@. Previously, most work has
been done on the simplest variants, on scenarios that are easiest to
test and on those that individual researchers consider to be the very
best motivated models. Examples include the Minimal Supersymmetric
Standard Model\footnote{For a review see Ref.\ \cite{Chung:2003fi}.}
(MSSM), scalar singlet dark matter models (SSDM)
\cite{Silveira:1985rk,McDonald:1993ex,Burgess:2000yq}, type II two
Higgs doublet models (THDM-II) \cite{Lee:1973iz,Glashow:1976nt,
  Donoghue:1978cj}, and universal extra dimensions
\cite{Appelquist:2000nn}.  Now, it is becoming increasingly well
motivated to also examine more complicated model variants, explore
scenarios that are calculationally difficult or hard to observe, or
motivated from an entirely new perspective.  Such advanced models can
avoid phenomenological or conceptual difficulties of simpler models,
and might provide explanations for the lack of experimental evidence
for new physics at the LHC.

Faced with this challenge, it is important to reduce
the calculational hurdle required to explore a new model and
look at its phenomenology as much as possible. This makes it easier for
new ideas to get developed and for many more models to
be studied together for much more general conclusions.

\fs \cite{Athron:2014yba,Athron:2016fuq} already made a significant
push in this direction, allowing the automatic creation of a spectrum
generator for a very wide range of supersymmetric models.  \fs uses
\sarah \cite{Staub:2010ty,Staub:2009bi,Staub:2010jh,Staub:2012pb,
  Staub:2013tta} to obtain \mathematica expressions for model dependent
components, the 2-loop renormalization group equations (RGEs),
1-loop self energies, 1-loop tadpoles, mass matrices and
electroweak symmetry breaking (EWSB) conditions. \fs then translates these
expressions into C++ routines and embeds them inside a code structure
for solving the boundary value problem (BVP)\@.  It also uses some numerical
routines from \softsusy \cite{Allanach:2001kg,Allanach:2013kza} and is
heavily unit-tested against the MSSM and Next-to-MSSM versions of
\softsusy every night to ensure bugs are avoided in updates.

Spectrum generators determine the pole masses and couplings of a
particular model from assumptions about the parameters at some
high scale, such as the grand unification scale, or directly from
on-shell parameters or running parameters\footnote{Usually defined in
  the \MSbar or \DRbar scheme.} given at the new physics scale.
This requires computing self energies, tadpole corrections to the EWSB
conditions and threshold corrections to SM-like gauge and Yukawa
couplings.  In the case of high scale assumptions a BVP must be
solved, which additionally requires integrating the RGEs for the model.
These codes are essential for testing hypotheses about the parameters of
the model, and determining if they lead to phenomenologically viable masses.

Spectrum generators are widely used in studies of supersymmetry, and for the
MSSM and the Next-to-MSSM\footnote{For a review of the model see
  Refs.\ \cite{Ellwanger:2009dp, Maniatis:2009re}.} (NMSSM) there are
a number of them that solve the BVP for particular choices of breaking
mechanism inspired boundary conditions (\softsusy
\cite{Allanach:2001kg,Allanach:2009bv, Allanach:2011de,
  Allanach:2013kza, Allanach:2014nba,
  Allanach:2016rxd,Allanach:2017hcf}, \suspect \cite{Djouadi:2002ze},
\spheno \cite{Porod:2003um,Porod:2011nf}, \isasusy \cite{Baer:1993ae},
\nmspec \cite{Ellwanger:2004xm,Ellwanger:2005dv, Ellwanger:2006rn} and
\suseflav \cite{Chowdhury:2011zr}) and several more that start from \DRbar
parameters at the SUSY scale or on-shell inputs (\FeynHiggs
\cite{Heinemeyer:1998yj, Heinemeyer:1998np, Degrassi:2002fi,
  Frank:2006yh, Hahn:2013ria, Drechsel:2016jdg, Bahl:2016brp,
  Bahl:2017aev}, \NMSSMCalc \cite{Baglio:2013iia, King:2015oxa} and
\CPsuperH \cite{Lee:2003nta, Lee:2007gn, Lee:2012wa}). This means that
the parameter space of the MSSM (and to a lesser extent the NMSSM) has
been extensively explored in a fast and reliable way.

However, for other models few, if any, public software packages exist.
Until recently this meant that one would have to spend a very long
time writing and testing code in order to explore the parameter space
of the model and do phenomenological investigations, and even then
bugs are more likely if one uses a private tool rather than a well
tested public one.  With the recent development of \fs and generated
modules for \spheno from \sarah \cite{Porod:2003um,
  Porod:2011nf,Staub:2017jnp,Staub:2009bi,Staub:2010jh,Staub:2012pb,
  Staub:2013tta,Athron:2014yba} it is now possible to obtain spectrum
generators in a much wider range of models.  \fs can generate spectrum
generators of various kinds, with boundary conditions at some high
scale or purely with low-energy input.  This push towards calculators
that are not model specific has also happened for other major types of
analyses.  For example, one may create new models in \micromegas
\cite{Belanger:2001fz,
  Belanger:2004yn,Belanger:2006is,Belanger:2008sj,Belanger:2010gh,
  Belanger:2013oya,Belanger:2014vza} using \calchep \cite{Boos:1994xb,
  Belyaev:2012qa} and \lanhep \cite{Semenov:1996es,Semenov:1998eb,
  Semenov:2002jw,Semenov:2008jy, Semenov:2010qt} to study the relic
density and direct and indirect detection of dark matter. \FeynRules
\cite{Christensen:2008py,Alloul:2013bka} can
also be used to generate the Feynman rules after inputting the
Lagrangian, in a similar manner to \sarah and \lanhep. The output of
these codes can be used in matrix element generators, such as \calchep,
\madgraph
\cite{Stelzer:1994ta,Maltoni:2002qb,Alwall:2007st,Alwall:2011uj,
  Alwall:2014hca}, \whizard \cite{Moretti:2001zz,Kilian:2007gr},
\sherpa \cite{Gleisberg:2003xi,Gleisberg:2008ta,Hoeche:2012yf}, \FAFC
\cite{Hahn:1998yk,Hahn:2000kx,Hahn:2006zy,Hahn:2010zi,Hahn:2016ebn}
and \helac \cite{Kanaki:2000ey,Cafarella:2007pc}, with showering and
hadronization handled by the event generators \herwig
\cite{Bahr:2008pv, Bellm:2015jjp}, \pythia
\cite{Bengtsson:1987kr,Sjostrand:2014zea, Sjostrand:2006za} and
\sherpa. Results from collider experiments can be applied to different
BSM models using re-interpretation tools, with some examples being \higgsbounds
\cite{Bechtle:2008jh,Bechtle:2011sb,Bechtle:2013wla, Bechtle:2015pma},
\higgssignals \cite{HiggsSignals}, \checkmate
\cite{Drees:2013wra,Dercks:2016npn}, \smodels \cite{Kraml:2013mwa,
  Kraml:2014sna,Ambrogi:2017neo}, \fastlim \cite{Papucci:2014rja},
\madanalysis\ \cite{Conte:2012fm,Conte:2014zja,Dumont:2014tja}, and
the native re-interpretation of \colliderbit \cite{Balazs:2017moi}.
Finally, very recently \GAMBIT \cite{Athron:2017ard} has been released
which uses auto-generated code from \fs \cite{Workgroup:2017bkh} and
\micromegas \cite{Workgroup:2017lvb}, along with many other packages,
to perform global fits of a user implemented BSM model.

Indeed \fs has been used extensively to study new physics models, on
its own or in combination with some of the codes mentioned above, in a
large number of physics studies, see, e.g.,
Refs.\ \cite{Diessner:2014ksa,
  Athron:2014pua,Athron:2015tsa,Diessner:2015yna,Staub:2015aea,Athron:2015vxg,
  Diessner:2015iln,Bagnaschi:2015pwa,Staub:2016dxq,Athron:2016fuq,
  deFlorian:2016spz,Athron:2016gor,Athron:2016qqb,Drechsel:2016htw,
  Bagnaschi:2017xid,Athron:2017yua,Athron:2017kgt,Athron:2017qdc,
  Workgroup:2017bkh,Chakravarty:2017hcy,Harlander:2017kuc,Ellis:2017hdw}.
Nonetheless, in the first release of \fs there were still a number of
limitations that restricted the models and phenomenology that could be studied
and the precision of the calculations.

In this paper, we document extensions to \fs, all now available in
version 2.0, that make a substantial push further in expanding the
number of models that can be explored and the observables that can be
calculated within them.
Now the list of models that can be investigated with
\fstwo also includes non-supersymmetric models (\fbsm), those with
complex parameters (\fcpv) and, with the new BVP solver
\fsas, constrained versions of certain non-minimal SUSY models like the NMSSM
\cite{Ellwanger:2009dp, Maniatis:2009re,Ellwanger:1993xa,Elliott:1994ht,
  King:1995vk,Ellwanger:1995ru,Ellwanger:1996gw} and the Exceptional
Supersymmetric Standard Model (\ESSM)
\cite{King:2005jy,King:2005my,Athron:2010zz, Athron:2009ue,Athron:2009bs} that
cannot be solved using the two-scale BVP solver
\cite{Barger:1993gh} approach implemented in public spectrum
generators.
In addition, \fstwo extends its repertoire of calculations.
The user may now calculate the mass of the $W$ boson as a
prediction of the model (\fmw) with the Fermi
constant treated as an input, as well as the anomalous magnetic
moment of the muon (\famu) and some electric dipole moments as
an application of \fcpv. Furthermore, we document and update our \feft
calculation of the Higgs pole mass, the physics of which has been discussed
in Ref.~\cite{Athron:2016fuq}.  Besides these new physics features, we
also document the new \mathematica interface.

In \secref{sec:download} we give a quick start guide explaining how to
download and compile the code using basic options.  In
\secref{sec:Mathematica_interface} we present the \mathematica
interface of \fs's spectrum generators and in
\secref{sec:model_file_extensions} we describe the new model file
extensions.  The
following Sections \ref{sec:fbsm}--\ref{sec:feft} document
each of the major extensions to \fs in a
modular fashion so that readers
can easily skip to a particular section.  These extensions are
as follows.  \fbsm, documented in \secref{sec:fbsm}, allows \fs to work in
non-SUSY models as well as SUSY models; the calculation of the
anomalous magnetic moment of the muon is
given in \secref{sec:famu}; \fcpv, described in
\secref{sec:fcpv}, introduces complex parameters to \fs so that \CP
violating effects may be considered; \fmw calculates the
muon decay, allowing \fs to predict $M_W$, and
is documented in \secref{sec:fmw}; a new BVP solver \fsas is
introduced in \secref{sec:fsas} and finally the hybrid Higgs mass
calculation, \feft, which combines the benefits of both effective field theory
and fixed-order calculations is documented in \secref{sec:feft}.
After this, we briefly summarize remaining limitations of the code in
\secref{sec:limitations} before adding concluding remarks in
\secref{sec:conclusions}.
In the appendices we provide a reference of all input parameters and
configuration options of \fstwo as well as new models.

\section{Quick start}
\label{sec:download}

\subsection{Requirements}

The build process of a custom spectrum generator using \fs requires
the following:
\begin{itemize}
\item \mathematica, version 7 or higher
\item \sarah, version 4.11.0 or higher [\url{http://sarah.hepforge.org}]
\item C++11 compatible compiler (\texttt{g++} 4.8.5 or higher,
  \texttt{clang++} 3.8 or higher, \texttt{icpc} 15.0 or higher)
\item FORTRAN compiler (\texttt{gfortran}, \texttt{ifort} etc.)
\item Eigen library, version 3.1 or higher
  [\url{http://eigen.tuxfamily.org}]
\item Boost library, version 1.37.0 or higher
  [\url{http://www.boost.org}]
\item GNU scientific library [\url{http://www.gnu.org/software/gsl}]
\end{itemize}
Optional:
\begin{itemize}
\item an implementation of LAPACK [\url{http://www.netlib.org/lapack}]
  such as ATLAS [\url{http://math-atlas.sourceforge.net}] or
  Intel Math Kernel Library\\\ [\url{http://software.intel.com/intel-mkl}]
\item LoopTools, version 2.8 or higher
  [\url{http://www.feynarts.de/looptools}]
\end{itemize}

\subsection{Downloading \fs and generating a first spectrum generator}
\label{sec:quick-start-cmssm}
\fs is available as gzipped tarball on
[\url{http://flexiblesusy.hepforge.org}] or
[\url{https://github.com/FlexibleSUSY/FlexibleSUSY}] under the version
control system \texttt{git}.
To download and install \fstwo as a gzipped tarball run at the command
line:
\begin{lstlisting}[language=bash]
$ wget https://www.hepforge.org/archive/flexiblesusy/FlexibleSUSY-2.0.1.tar.gz
$ tar -xf FlexibleSUSY-2.0.1.tar.gz
$ cd FlexibleSUSY-2.0.1
\end{lstlisting}
\fstwo is distributed with a huge selection of predefined ``models'', 
including several MSSM and NMSSM scenarios such as the CMSSM (called
\modelname{CMSSM}), high-scale SUSY and split-SUSY (called \HSSUSY and
\modelname{SplitMSSM}; these  models have been created for
Ref.~\cite{Bagnaschi:2015pwa}), the semi-constrained and
fully constrained NMSSM (\modelname{NMSSM}, \modelname{CNMSSM}). The
distribution also contains BSM models such as the
$R$-symmetric MSSM (\modelname{MRSSM}; for a definition of the model
see Ref.~\cite{Kribs:2007ac}), the NUHM \ESSM
(\modelname{E6SSM}, see \cite{King:2005jy,King:2005my,Athron:2010zz,
  Athron:2009ue,Athron:2009bs,Athron:2007en}) and the two-Higgs 
doublet model type II
(\modelname{THDMII}).  See the 
contents of \code{model_files/} for all predefined model files.

A spectrum generator for any of these models can be built with just
three commands.  For example, the CMSSM spectrum generator can be
created by running the following shell commands:
\begin{lstlisting}[language=bash]
$ ./createmodel --name=CMSSM
$ ./configure --with-models=CMSSM
$ make
\end{lstlisting}
The \code{createmodel} command creates the model directory
\code{models/CMSSM/} where the code will be generated and adds the
CMSSM model file and an example SLHA input file to it.  The \code{configure}
script sets up the \code{Makefile}, checking the system requirements
and dependencies.  For more options see \code{./configure
  --help}.  The last command creates the code for the spectrum
generator of this model and compiles it.
This generated spectrum generator can then be run from the command
line as:
\begin{lstlisting}[language=bash]
$ cd models/CMSSM
$ ./run_CMSSM.x --slha-input-file=LesHouches.in.CMSSM \
                --slha-output-file=LesHouches.out.CMSSM
\end{lstlisting}
The spectrum generator reads the CMSSM input parameters from the SLHA input file
\code{LesHouches.in.CMSSM} and first solves a BVP to find a
set of running parameters at the SUSY scale that are consistent with all
boundary conditions specified in the model file, and then calculates the pole
masses, mixing matrices and potentially further observables.
The mass spectrum etc.\ obtained in this way is output
in SLHA format \cite{Skands:2003cj,Allanach:2008qq} as
\code{LesHouches.out.CMSSM}.  See \code{./run_CMSSM.x --help} for more
options.

\fs also provides a \mathematica interface, introduced in version 1.7.0,
to call the generated spectrum generators.  For each spectrum
generator, an example \mathematica script named
\code{models/<model>/run_<model>.m} is created for illustration.  For
example, the \modelname{CMSSM} spectrum generator can be called from within
\mathematica like this:
\begin{lstlisting}
Get["models/CMSSM/CMSSM_librarylink.m"];

handle = FSCMSSMOpenHandle[
    fsModelParameters -> {
        m0 -> 125,
        m12 -> 500,
        TanBeta -> 10,
        SignMu -> 1,
        Azero -> 0
    }
];

spectrum = FSCMSSMCalculateSpectrum[handle];
FSCMSSMCloseHandle[handle];

Print[spectrum];
\end{lstlisting}
Execute \code{?FSCMSSMOpenHandle} for a list of all allowed options and
input parameters.  In the example above, the \code{spectrum} variable
contains the mass spectrum and the running parameters in the form of a
list of replacement rules:
\begin{lstlisting}
{CMSSM -> {
   Pole[M[Glu]] -> 1147.35,
   Pole[M[Sd]] -> {957.993, 997.56, 1000.49, 1000.5, 1045.93, 1045.94},
   Pole[M[Sv]] -> {350.753, 351.913, 351.917},
   Pole[M[Su]] -> {796.653, 1002.67, 1003.96, 1005.06, 1043.07, 1043.07},
   Pole[M[Se]] -> {222.916, 229.983, 230.008, 360.842, 360.846, 361.978},
   Pole[M[hh]] -> {114.836, 713.119},
   Pole[M[Ah]] -> {88.5864, 712.848},
   Pole[M[Hpm]] -> {77.2642, 717.628},
   Pole[M[Chi]] -> {204.054, 385.012, 629.649, 643.612},
   Pole[M[Cha]] -> {385.017, 643.924},
   Pole[M[VWm]] -> 80.3935, ...}
}
\end{lstlisting}
More details about the \mathematica interface as well as a neat
example of running \HSSUSY through it can be found in Sections
\ref{sec:Mathematica_interface} and \ref{sec:application_HSSUSY},
respectively.

\subsection{Spectrum generators for alternative models}
\label{sec:quick-start-alternative-models}
If the user instead wants to create a spectrum generator for a model
for which there is no pre-existing model file distributed in
\fs, then the model file can be written. Before a \fs
model file can be written for the spectrum generator, there must exist
\sarah model files, which \fs uses to obtain model dependent
information. \sarah also comes with many pre-defined models, but if an
appropriate model is not available, the users may create their own
\sarah model files and add them to the directory
\code{sarah/<modelname>/} in \fs.  For the writing of a \sarah model
file we refer the reader to the extensive \sarah documentation, for
example Refs.\
\cite{Staub:2017jnp,Staub:2009bi,Staub:2010jh,Staub:2012pb,
  Staub:2013tta,Athron:2014yba}.

Creating a new \fs model file is straightforward. Full details are
given in the original \fs manual Ref.~\cite{Athron:2014yba}.
Here we just repeat a basic example
in the context of the NMSSM that illustrates the main points:
The semi-constrained NMSSM (\modelname{NMSSM}) distributed in \fs has
all the soft-breaking trilinear scalar couplings set to a unified
$\azero$ at the GUT scale.  However, often $A_\lambda$ and $A_\kappa$ are
taken to be non-universal in semi-constrained variants of the NMSSM
since the non-universality of the soft singlet mass already violates
the standard universality assumptions of constrained
models.\footnote{See \secref{sec:fsas} for a new approach that
  allows the fully constrained NMSSM to be solved.} To allow separate
values for $A_\lambda$ and $A_\kappa$ at the GUT scale the \fs model
file \code{model_files/NMSSM/FlexibleSUSY.m.in} should be changed from
\begin{lstlisting}[language=Mathematica]
EXTPAR = { {61, LambdaInput} };

HighScaleInput = {
  ...
  {T[\[Kappa]], Azero \[Kappa]},
  {T[\[Lambda]], Azero LambdaInput}
   ...
};
\end{lstlisting}
into
\begin{lstlisting}[language=Mathematica]
EXTPAR = { {61, LambdaInput},
           {63, ALambdaInput},
           {64, AKappaInput} };

HighScaleInput = {
  ...
  {T[\[Kappa]], AKappaInput \[Kappa]},
  {T[\[Lambda]], ALambdaInput LambdaInput},
  ...
};
\end{lstlisting}
The GUT scale values of $A_\lambda$ and $A_\kappa$ can then be specified in the SLHA input file in the \code{EXTPAR} block by entries $63$ and $64$,
\begin{lstlisting}
Block EXTPAR
   61   0.1                  # LambdaInput
   63   -100                 # ALambdaInput
   64   -300                 # AKappaInput

\end{lstlisting}

\section{\mathematica interface}
\label{sec:Mathematica_interface}

The spectrum generators created with \fs can be called from within
\mathematica.  To do that, first the spectrum generator must be built,
as described in \secref{sec:quick-start-cmssm}.  Afterwards, the
provided \mathematica interface functions for the model must be
loaded.  For a given model \code{<model>} this is done by including
the following file in the \mathematica session:
\begin{lstlisting}
Get["models/<model>/<model>_librarylink.m"];
\end{lstlisting}
This script loads the library
\code{models/<model>/<model>_librarylink.so} into the \mathematica
session (assuming the user is in the \code{FlexibleSUSY/} directory).
Afterwards, the \mathematica interface functions listed in
\tabref{tab:mma_ifce_functions} are available.
\begin{table}[tbh]
  \centering
  \begin{tabularx}{\textwidth}{lX}
    \toprule
    Function & Description \\

    \midrule \code{FS<model>OpenHandle[...]} & Takes all model input
    parameters as argument and returns a ``handle'' (a reference) to
    the given parameter point, the associated mass spectrum and observables.\\

    \code{FS<model>CloseHandle[handle]} & Releases the resources
    associated to a given handle.\\

    \code{FS<model>CalculateSpectrum[handle]} & Calculates the mass
    spectrum for a given handle.\\

    \code{FS<model>CalculateObservables[handle]} & Calculates the
    observables for a given handle.\\

    \code{FS<model>ToSLHA[handle]} & Returns a string containing the
    mass spectrum and observables associated to a given handle in SLHA format.\\

    \code{FS<model>Set[handle, ...]} & Changes the input parameters
    associated to a given handle.\\

    \code{FS<model>GetSettings[handle]} & Returns the spectrum generator
    settings (precision goal, loop orders, etc.) associated to
    a given handle.\\

    \code{FS<model>GetSMInputParameters[handle]} & Returns the SM
    input parameters associated to a given handle.\\

    \code{FS<model>GetInputParameters[handle]} & Returns the
    model-specific input parameters associated to a given handle.\\

    \code{FS<model>GetProblems[handle]} & Returns a list of problems
    that occurred when calculating the spectrum.\\

    \code{FS<model>GetWarnings[handle]} & Returns a list of warnings
    that occurred when calculating the spectrum.\\
    \bottomrule
  \end{tabularx}
  \caption{\mathematica interface functions provided for a \fs model with the name \code{<model>}.}
  \label{tab:mma_ifce_functions}
\end{table}

To run the spectrum generator for a given parameter point, a handle to that point must be
created first, using the \code{FS<model>OpenHandle[...]} function.
The returned handle represents a reference to the given parameter
point, the associated mass spectrum and observables.  The concept of
handles allows the user to calculate mass spectra for different
parameter points in parallel using multiple \mathematica kernels: Each
kernel can open a handle to a separate parameter point, calculate the
mass spectrum and finally close the handle.  In this way there
is no ambiguity in the parameter point used by each kernel.
\exref{ex:HSSUSY_parallel} in \secref{sec:application_HSSUSY} illustrates the usage of handles by
performing a parallel scan over the MSSM parameter space with \HSSUSY.
In the most general form, the \code{FS<model>OpenHandle[...]} function
can take the following three arguments:
\begin{lstlisting}
handle = FS<model>OpenHandle[
   fsSettings -> {...},
   fsSMParameters -> {...},
   fsModelParameters -> {...}
]
\end{lstlisting}
The \code{fsSettings} symbol can be used to set the spectrum generator
options.  All possible options are listed in
\tabref{tab:sg_mma_options} in \appref{app:fs_config_options}.
The \code{fsSMParameters} symbol can be used to set the SM
input parameters.  The possible SM input parameters are
listed in \tabref{tab:sm_mma_options} in \appref{app:sm_input_parameters}.
The \code{fsModelParameters} symbol can be used to set the \BSM
model-specific input parameters.  Unspecified model input parameters
are set to zero by default.  The names of the model input parameters
are identical to the ones specified in the \code{MINPAR},
\code{IMMINPAR}, \code{EXTPAR}, \code{IMEXTPAR} and
\code{FSAuxiliaryParameterInfo} variables in the \fs model file.  The
settings, the SM and the \BSM input parameters
associated to a given handle can be obtained using the
\code{FS<model>GetSettings[]}, \code{FS<model>GetSMInputParameters[]}
and \code{FS<model>GetInputParameters[]} functions, respectively.  The
opened handle can then be used to calculate the mass spectrum and the
observables:
\begin{lstlisting}
spectrum    = FS<model>CalculateSpectrum[handle];
observables = FS<model>CalculateObservables[handle];
FS<model>CloseHandle[handle];
\end{lstlisting}
Finally, the handle should be closed to release the associated
resources by calling the function \code{FS<model>CloseHandle[handle]}.
\begin{example}
  In the CMSSM, the \BSM model-specific input parameters are named as
  \code{m0}, \code{m12}, \code{TanBeta}, \code{SignMu} and
  \code{Azero}, see the \modelname{CMSSM} model file provided with \fstwo.
  Thus, an example \mathematica session for the CMSSM could look like:
\begin{lstlisting}
Get["models/CMSSM/CMSSM_librarylink.m"];

handle = FSCMSSMOpenHandle[
    fsSettings -> {
        poleMassLoopOrder -> 2,
        ewsbLoopOrder -> 2,
        thresholdCorrectionsLoopOrder -> 2,
        betaFunctionLoopOrder -> 3
    },
    fsSMParameters -> {
        Mt -> 173.34,
        alphaSMZ -> 0.1184
    },
    fsModelParameters -> {
        m0 -> 125,
        m12 -> 500,
        TanBeta -> 10,
        SignMu -> 1,
        Azero -> 0
    }
];

spectrum    = FSCMSSMCalculateSpectrum[handle];
observables = FSCMSSMCalculateObservables[handle];

FSCMSSMCloseHandle[handle];
\end{lstlisting}
\end{example}
The output of \code{FS<model>CalculateSpectrum[handle]} is a list
that contains the pole mass spectrum as well as the running masses
and parameters at the chosen output scale.
The running parameters are named as defined in the \sarah model.  For
example, \code{g1}, \code{g2}, \code{g3} usually denote the running
gauge couplings and \code{Yu}, \code{Yd}, \code{Ye} the running Yukawa
couplings.  The running masses are denoted as \code{M[<p>]}, where
\code{<p>} is the name of the particle as defined in the \sarah model
file.  All running parameters and masses are given at the parameter
output scale, \code{SCALE}.  The pole masses and mixing matrices carry
the additional \code{Pole[]} head.  For example, \code{Pole[M[hh]]}
usually denotes the pole mass(es) of the Higgs boson(s).

In the \mathematica output, the running parameters, the masses and
mixing matrices are defined in the \sarah convention, \emph{not} in the
SLHA convention.  This means in particular that the Yukawa matrices,
the soft-breaking squark mass matrices and the soft-breaking trilinear
couplings are \emph{not} defined in the (super)-CKM basis.  In the
\mathematica output, the particle masses are always non-negative and mixing
matrices are in general complex.
\begin{example}
  In the CMSSM, the output of \code{FSCMSSMCalculateSpectrum[handle]}
  may look like (skipping some entries for brevity):
\begin{lstlisting}
{
  CMSSM -> {
    M[Glu] -> 1116.4857717819132,
    M[Sd] -> {929.5770939936384, 963.6803089181217,
              965.7750791635142, 965.7786645820956,
              1010.5301444258299, 1010.5317308607175},
    M[Su] -> {770.2929836944288, 969.5389940603632,
              969.5446720936367, 975.4073015489867,
              1007.5435846398124, 1007.5443071514123},
    M[Se] -> {219.54939719808144, 226.45230058860656,
              226.4768407746941, 356.2452631868304,
              356.2501661663376, 357.5576772510361},
    M[hh] -> {88.16467333922309, 726.2603417238729},
    M[Ah] -> {90.09835220027803, 726.0229889725828},
    M[Hpm] -> {78.48914789145176, 730.2533306006959},
    M[Chi] -> {207.1963755793879, 375.7416364302936,
               627.5178023483583, 641.6676783271736},
    M[Cha] -> {375.56991892585705, 641.3578484531205},
    ...
    ZH -> {{0.10592570722508611, 0.9943740465985952},
           {0.9943740465985952, -0.10592570722508611}},
    ...
    Pole[M[Glu]] -> 1147.3536227374905,
    Pole[M[Sd]] -> {957.9934299811302, 997.5603867095314,
                    1000.4932601265115, 1000.4969819618583,
                    1045.9354429433467, 1045.9372472457565},
    Pole[M[Su]] -> {796.653619369782, 1002.6690741336473,
                    1003.9614916607435, 1005.0642702137084,
                    1043.0672831732345, 1043.067920812505},
    Pole[M[Se]] -> {222.90126096766593, 229.9832415178622,
                    230.00840279144913, 360.84198174065307,
                    360.8462569384804, 361.9798562942742},
    Pole[M[hh]] -> {114.83583179574276, 713.1187313487922},
    Pole[M[Ah]] -> {88.58641341930426, 712.8473602456997},
    Pole[M[Hpm]] -> {77.26414997655887, 717.6270868215212},
    Pole[M[Chi]] ->  {204.05370940499517, 385.0116889026496,
                      629.6500252267041, 643.6127224060953},
    Pole[M[Cha]] -> {385.0164604772902, 643.924798526633},
    Pole[ZH] -> {{0.1066307364997843, 0.9942986905520461},
                 {0.9942986905520461, -0.1066307364997843}},
    ...
    Yd -> {{0.00013999141660755535, 0., 0.},
           {0., 0.0030650771019348657, 0.},
           {0., 0., 0.1317656023934078}},
    Ye -> {{0.00002895462310608183, 0., 0.},
           {0., 0.005986898727723582, 0.},
           {0., 0., 0.1006931798596906}},
    Yu -> {{7.267255094462218*^-6, 0., 0.},
           {0., 0.0033082824973678956, 0.},
           {0., 0., 0.8606532901364391}},
    \[Mu] -> 624.160899893032,
    g1 -> 0.4679063156949638,
    g2 -> 0.6430285180350706,
    g3 -> 1.0655340318624051,
    vd -> 25.099612589273388,
    vu -> 242.8296409176676,
    T[Yd] -> {{-0.19442921534444055, 0., 0.},
              {0., -4.256965123994014, 0.},
              {0., 0., -171.13078241755716}},
    T[Ye] -> {{-0.008660431147847696, 0., 0.},
              {0., -1.790668166947902, 0.},
              {0., 0., -29.953008563262035}},
    T[Yu] -> {{-0.008251796354121698, 0., 0.},
              {0., -3.7564603713866354, 0.},
              {0., 0., -755.7309107649228}},
    B[\[Mu]] -> 53907.68839928095,
    mq2 -> {{1.0178399270038805*^6, 0., 0.},
            {0., 1.0178348758707164*^6, 0.},
            {0., 0., 865711.3590482193}},
    ml2 -> {{124853.43144557778, 0., 0.},
            {0., 124850.91737364854, 0.},
            {0., 0., 124143.31667980803}},
    mHd2 -> 109509.1756551005,
    mHu2 -> -377534.5544643501,
    md2 -> {{932089.8366766714, 0., 0.},
            {0., 932084.7501918572, 0.},
            {0., 0., 923097.970074961}},
    mu2 -> {{941294.0634724408, 0., 0.},
            {0., 941288.914981822, 0.},
            {0., 0., 639354.6597906639}},
    me2 -> {{49375.97115879859, 0., 0.},
            {0., 49370.8410356076, 0.},
            {0., 0., 47926.68179837005}},
    MassB -> 209.15138268684439,
    MassWB -> 387.9365053016521,
    MassG -> 1116.4857717819132,
    SCALE -> 866.8060753250803
  }
}
\end{lstlisting}
\end{example}
The function \code{FS<model>CalculateObservables[handle]} returns the
observables for a given handle in the form of a list of replacement rules.
\begin{example}
  In the CMSSM, the output of \code{FSCMSSMCalculateObservables[handle]}
  might look like
\begin{lstlisting}
{
  CMSSM -> {
    FlexibleSUSYObservable`aMuon -> 8.46316731749956*^-10,
    FlexibleSUSYObservable`CpHiggsPhotonPhoton ->
      {0.000029645107712580034 - 2.1094207517495443*^-7*I,
       7.984875731996049*^-7 + 9.125974009524432*^-7*I},
     FlexibleSUSYObservable`CpHiggsGluonGluon ->
       {-0.00006704582692080136 - 2.68821584064419*^-6*I,
        2.8335074218510527*^-6 + 4.966199588265877*^-6*I},
     FlexibleSUSYObservable`CpPseudoScalarPhotonPhoton ->
       1.066762880186648*^-6 - 8.27198259619208*^-7*I,
     FlexibleSUSYObservable`CpPseudoScalarGluonGluon ->
       6.825816917497379*^-6 + 8.151893730825134*^-7*I
  }
}
\end{lstlisting}
\end{example}
The symbol \code{aMuon} represents the anomalous magnetic moment of
the muon, $\amu$, calculated as described in \secref{sec:famu}.  The
symbols \code{CpHiggsPhotonPhoton}, \code{CpHiggsGluonGluon},
\code{CpPseudoScalarPhotonPhoton} and \code{CpPseudoScalarGluonGluon}
denote the effective couplings of the physical \CP-even and \CP-odd Higgs
boson(s) to two photons and gluons, respectively, as described in
Ref.~\cite{Staub:2016dxq}.  Note that if the \CP-even or \CP-odd Higgs states
are multiplets, as is the case for the \CP-even Higgs in the MSSM, for
example, the relevant couplings are calculated for all members of the
multiplet and the result is returned as a list.

The calculated spectrum can be printed in an SLHA-compatible format
using the \code{FS<model>ToSLHA[]} function.
\begin{example}
  For the CMSSM, an example output of \code{FSCMSSMToSLHA[handle]}
  could look like:
\begin{lstlisting}
Block SPINFO
     1   FlexibleSUSY
     2   2.0.1
     5   CMSSM
     9   4.11.0
Block MASS
   1000021     1.14735362E+03   # Glu
        24     8.03935152E+01   # VWm
   1000024     3.85016460E+02   # Cha(1)
   1000037     6.43924799E+02   # Cha(2)
        25     1.14835832E+02   # hh(1)
        35     7.13118731E+02   # hh(2)
...
\end{lstlisting}
\end{example}
In \secref{sec:application_HSSUSY} several examples can be found that
illustrate how to perform parameter scans and uncertainty
estimates using the \mathematica interface of \fs's spectrum
generators.

\section{\fs model file extensions}
\label{sec:model_file_extensions}

\subsection{Model-specific higher-order contributions}
\label{sec:model_specific_contributions}
To improve the accuracy in some specific models, \fs provides a few
model file switches to enable further higher-order contributions in
the calculation of the running parameters, the $\beta$ functions or
the Higgs pole mass.  \tabref{tab:fbsm_options} lists all switches
available in \fstwo, which are explained in the following subsections.
These switches can usually be enabled in many models, providing that
the corresponding requirements are fulfilled.  However, the user
should be aware that contributions may be missing if the switches are
enabled in models beyond their scope of application.

\begin{OptionTable}{\fs model file switches to enable/disable
    model-specific higher-order contributions in the SM,
    MSSM, NMSSM and split-MSSM.}{tab:fbsm_options}[lllX]
  \multicolumn{4}{c}{\textit{SM}}\\
  \midrule
  \code{UseHiggs2LoopSM} & \code{False} & \code{True} or \code{False}
  & 2-loop contributions $O(\at^2 + \at \as)$ to $M_h$ in the SM \\
  \code{UseHiggs3LoopSM} & \code{False} & \code{True} or \code{False} &
  3-loop contributions $O(\at^3 + \at^2 \as \allowbreak + \at \as^2)$ to $M_h$ in
  the SM \\
  \code{UseSM3LoopRGEs} & \code{False} & \code{True} or \code{False} &
  3-loop RGEs in the SM \\
  \code{UseYukawa3LoopQCD} & \code{Automatic} & \multilinecell[t]{\code{True} or \code{False}\\ or \code{Automatic}} &
  2-loop and 3-loop QCD contributions $O(\as^2 + \as^3)$ to the \MSbar $y_t$ in the SM \\
  \code{UseSMAlphaS3Loop} & \code{False} & \code{True} or \code{False}
  & 2-loop and 3-loop QCD threshold corrections
  $O(\as^2 + \as^3)$
  to the \MSbar $\as$ in the SM \\
  \midrule
  \multicolumn{4}{c}{\textit{MSSM}}\\
  \midrule
  \code{UseHiggs2LoopMSSM} & \code{False} & \code{True} or \code{False} &
  2-loop contributions $O((\at + \ab)\as \allowbreak + (\at +
  \ab)^2 \allowbreak + \atau^2)$ to $M_h$, $M_H$ and $M_A$ in
  the MSSM \\
  \code{UseHiggs3LoopMSSM} & \code{False} & \code{True} or
  \code{False} & 3-loop contributions
  $O(\at\as^2 \allowbreak + \ab\as^2)$ to $M_h$ in
  the MSSM (requires \Himalaya) \\
  \code{UseMSSM3LoopRGEs} & \code{False} & \code{True} or \code{False} &
  3-loop RGEs in the MSSM \\
  \code{UseMSSMYukawa2Loop} & \code{False} & \code{True} or \code{False} &
  2-loop \SQCD contribution $O(\as^2)$ to the \DRbar $y_b$ and $y_t$ in the MSSM \\
  \code{UseMSSMAlphaS2Loop} & \code{False} & \code{True} or
  \code{False} & 2-loop \SQCD contribution
  $O(\as^2 + \as\at + \as\ab)$
  to the \DRbar $\as$ in the MSSM \\
  \midrule
  \multicolumn{4}{c}{\textit{NMSSM}}\\
  \midrule
  \code{UseHiggs2LoopNMSSM} & \code{False} & \code{True} or \code{False} &
  2-loop contributions $O((\at + \ab)\as \allowbreak + (\at +
  \ab)^2 \allowbreak + \atau^2)$ to $M_{h_i}$ and $M_{A_i}$ in
  the NMSSM \\
  \midrule
  \multicolumn{4}{c}{\textit{split-MSSM}}\\
  \midrule
  \code{UseHiggs3LoopSplit} & \code{False} & \code{True} or \code{False} &
  3-loop contributions $O(\at\as^2)$ to $M_h$ in the split-MSSM \\
  \bottomrule
\end{OptionTable}

\subsubsection{SM-specific higher-order contributions}

\paragraph{2-loop and 3-loop contributions to the SM Higgs mass}

In the SM, 2-loop and leading 3-loop contributions to the Higgs pole
mass are known in the \MSbar scheme
\cite{Degrassi:2012ry,Vega:2015fna,Martin:2014cxa}.  In \fs the 2-loop
contributions of $O(\at^2 + \at \as)$
\cite{Vega:2015fna,Martin:2014cxa} and the 3-loop
$O(\at^3 + \at^2 \as + \at \as^2)$ contributions \cite{Martin:2014cxa}
can be taken into account to calculate the SM-like Higgs pole mass for
non-SUSY models with only one Higgs. In order to enable these loop
contributions, the following switches must be set in the model file:
\begin{lstlisting}
UseHiggs2LoopSM = True;
UseHiggs3LoopSM = True;
\end{lstlisting}
The 2-loop and 3-loop
contributions enter the mass of the Higgs boson as
\begin{align} 
 \begin{split}
   M_h^2 &= m_h^2 + (\Delta m_h^2)_{1L}(p^2)
   + (\Delta m_h^2)_{2L}(p^2) +(\Delta m_h^2)_{3L} \,,
 \end{split} \label{eq:Mh_polp^2}
\end{align}
where $m_h^2$ and $(\Delta m_h)_{1L}(p^2)$ correspond to the tree-level 
and a 1-loop expression, respectively.
The enabled 2-loop SM contributions of
$O(\at^2 + \at \as)$ to the Higgs mass read
\begin{align}
  (\Delta m_h^2)_{2L}(p^2) &= (\Delta m_h^2)_{2L}^{(\at^2)} + 
     (\Delta m_h^2)_{2L}^{(\at\as)}(p^2) \,, \\
  (\Delta m_h^2)_{2L}^{(\at^2)} &=
  \frac{2 t}{(4\pi)^4} \left[
    -3 y_t^4 \left(3 \;\barlog^2(t) - 7 \;\barlog(t) + 2 + \frac{\pi^2}{3}\right)
  \right] , \label{eq:2L_Mh_atat}\\
  (\Delta m_h^2)_{2L}^{(\at\as)} (p^2)&=
      \frac{g_3^2 y_t^2}{(4\pi)^4} \;\bigg[
        \frac{37}{3} p^2 - \frac{122 p^4}{135 t} - 4 (5 p^2 - 8 t) \;\barlog(t) -
        12 (p^2 - 8 t) \;\barlog^2(t) \bigg] \label{eq:2L_Mh_atas} \\
  &\stackrel{\mathmakebox[\widthof{=}]{p^2=0}}{=} \
  \frac{2 t}{(4\pi)^4} \left[
    16 g_3^2 y_t^2 \left(3 \;\barlog^2(t) + \barlog(t)\right)
  \right] \,.
  \label{eq:2L_Mh_atas p2=0}
\end{align}
In Eqs.~\eqref{eq:2L_Mh_atat} ff.\ the abbreviations $t \equiv m_t^2$
and $\barlog(t) \equiv \log(t/Q^2)$ have been used for brevity, 
analogous to the notation of Ref.~\cite{Martin:2014cxa}, where $m_t$ 
is the \MSbar top mass in the SM\@. 
By performing a momentum iteration in the computation of the Higgs 
pole mass, the momentum entering the 1-loop self energy in 
Eq.~\eqref{eq:Mh_polp^2} consists of a tree-level part and a loop 
correction.
In combination with Eq.~\eqref{eq:2L_Mh_atat}, the latter yields the 
complete Higgs mass contribution at $O(\at^2 t)$ which is identical 
to the corresponding correction given in Ref.~\cite{Degrassi:2012ry}.
The 2-loop self energy at $O(\at\as)$ \cite{Martin:2014cxa} together
with the occurring integral functions from Ref.~\cite{Martin:2003qz}
have been evaluated for a small external momentum argument. Neglecting
higher orders, the expansion in powers of the momentum over the \MSbar
top mass up to $O(p^4/t)$ results in the Eq.~\eqref{eq:2L_Mh_atas}. 
Note that the momentum dependence is included in the 2-loop expression 
in order to generate implicit 3-loop $O(\at^2\as)$ terms.
The explicit 3-loop effective potential contributions of
$O(\at^3 + \at^2 \as + \at \as^2)$ to the
Higgs mass included by \fs read
\begin{align}
  (\Delta m_h^2)_{3L} &=
  \frac{m_t^2}{(4\pi)^6} \Bigg[
     g_3^4 y_t^2 \left(248.122 + 839.197 \;\barlog(t) + 160 \;\barlog^2(t) - 736 \;\barlog^3(t)\right) \nonumber \\
     &\phantom{=\frac{m_t^2}{(4\pi)^6} \Bigg[}
     + g_3^2 y_t^4 \left(2764.365 + 1283.716 \;\barlog(t) - 360 \;\barlog^2(t) + 240 \;\barlog^3(t)\right) \nonumber \\
     &\phantom{=\frac{m_t^2}{(4\pi)^6} \Bigg[}
     + y_t^6 \Big(-3199.017 + 36 \;\barlog(h) - 2653.511 \;\barlog(t) + 756 \;\barlog(h) \;\barlog(t) \nonumber \\
     &\phantom{=\frac{m_t^2}{(4\pi)^6} \Bigg[}\qquad~~
     + \frac{27}{2} \;\barlog^2(t) + 324 \;\barlog(h) \;\barlog^2(t) - 225 \;\barlog^3(t)\Big)
  \Bigg] \,,
\end{align}
where $h\equiv m_h^2$ is the squared \MSbar Higgs mass in the SM\@.
Note that the 3-loop Higgs mass calculation in
the \MSbar scheme in \fsbreak is only complete at $O(\at\as^2)$.  The
3-loop contributions of $O(\at^3 + \at^2 \as)$ are currently incomplete
because they would require the \MSbar top Yukawa coupling to be determined
from the top pole mass at the 2-loop $O(\at^2 + \at\as)$.
However, these 2-loop contributions to the top Yukawa coupling are
currently not available in \fs. Furthermore, the proper inclusion of
corrections to the Higgs pole mass at $O(\at^3)$
would require the extension of Eq.~\eqref{eq:2L_Mh_atat} by momentum
dependent $O(\at^2)$ terms. Alternatively, the evaluation of the Higgs mass at
the renormalization scale $Q^2 = t$ implies that the neglected
3-loop contributions at $O(\at^3 t)$ vanish
\cite{Martin:2014cxa}.\footnote{Note that terms of $O(\at^3 h)$ are
  neglected here and in Ref.\ \cite{Martin:2014cxa}.}
Likewise, Eq.~\eqref{eq:2L_Mh_atat} neglects contributions of $O(\at^2h)$ which are
subdominant in comparison to the implemented $O(\at^2t)$ corrections.
In contrast, the QCD corrections to the pole mass at 2-loop level, Eq.~\eqref{eq:2L_Mh_atas}, involve
terms proportional to the quartic Higgs coupling $\lambda$ up to $O(\as\at h^2/t)$, which are not
neglected here.

\begin{figure}[tbh]
  \centering
  \includegraphics[width=0.5\textwidth]{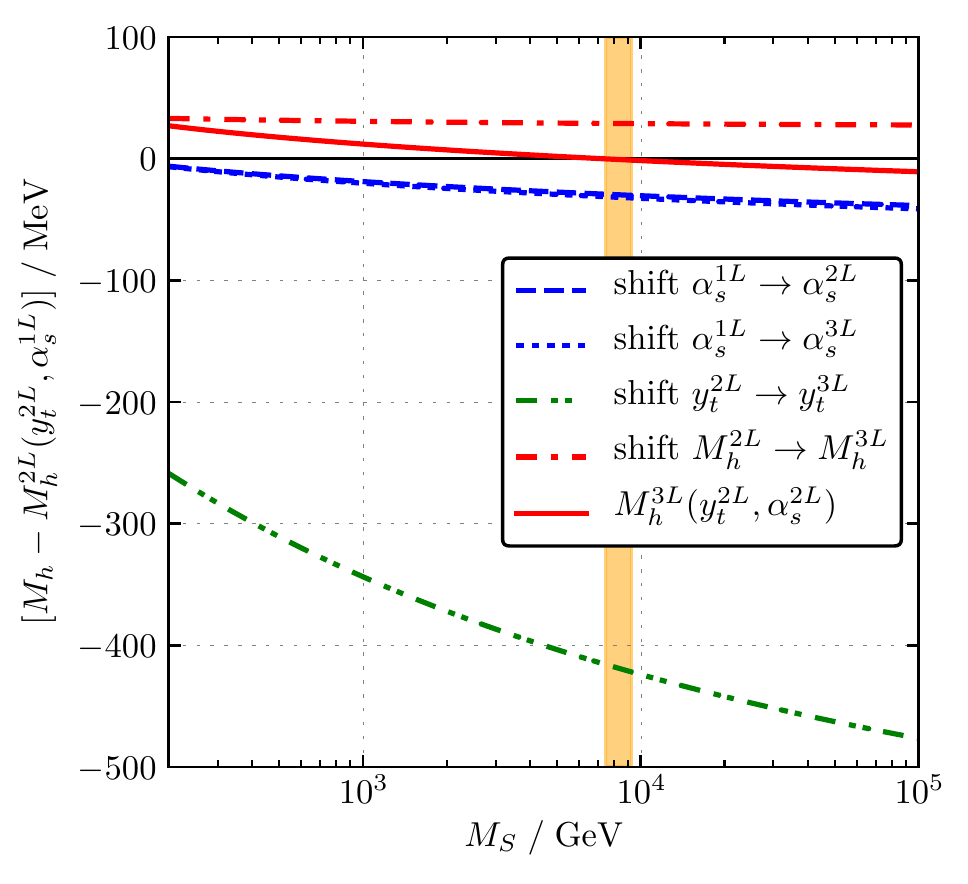}
  \caption{Effect of different SM corrections on the Higgs
    pole mass in the pure EFT calculation of \HSSUSY for
    $\tan\beta = 5$ and $X_t = \sqrt{6}\MS$ as a function of the SUSY
    scale \MS.  The orange band marks the region where \HSSUSY
    predicts a Higgs mass compatible with the experimental value of
    $(125.09 \pm 0.32)\unit{GeV}$ \cite{Aad:2015zhl}.}
  \label{fig:SM_thresholds}
\end{figure}

The 2-loop and 3-loop Higgs mass contributions from above can be
enabled at runtime by setting the following flags in the SLHA input
file:
\begin{lstlisting}
Block FlexibleSUSY
    4   3         # pole mass loop order
    5   3         # EWSB loop order
    7   2         # threshold corrections loop order
    8   1         # Higgs 2-loop corrections O(alpha_t alpha_s)
   10   1         # Higgs 2-loop corrections O(alpha_t^2)
   24   122111221 # individual threshold correction loop orders
   26   1         # Higgs 3-loop corrections O(alpha_t alpha_s^2)
   28   1         # Higgs 3-loop corrections O(alpha_t^2 alpha_s)
   29   1         # Higgs 3-loop corrections O(alpha_t^3)
\end{lstlisting}
In \fs's \mathematica interface, the above SLHA configuration options
correspond to
\begin{lstlisting}[language=Mathematica]
handle = FS<model>OpenHandle[
    fsSettings -> {poleMassLoopOrder -> 3,
                   ewsbLoopOrder -> 3,
                   thresholdCorrectionsLoopOrder -> 2,
                   higgs2loopCorrectionAtAs -> 1,
                   higgs2loopCorrectionAtAt -> 1,
                   thresholdCorrections -> 122111221,
                   higgs3loopCorrectionAtAsAs -> 1,
                   higgs3loopCorrectionAtAtAs -> 1,
                   higgs3loopCorrectionAtAtAt -> 1}
    ...];
\end{lstlisting}
See \appref{app:fs_config_options} for a list and description of
all of \fs's configuration options.
In \figref{fig:SM_thresholds} the impact of the leading 3-loop
Standard Model corrections of $O(\at^3 + \at^2 \as + \at \as^2)$ on
the light \CP-even Higgs pole mass in the MSSM in the pure effective
field theory (EFT) calculation of \HSSUSY is shown with the red
dashed-dotted line.  The line shows the predicted 3-loop Higgs mass
relative to the one calculated at the 2-loop level as a function of
the SUSY scale \MS, by taking into account the 1-loop threshold
correction for $\as^{\SM}(M_Z)$ and the 2-loop QCD correction to
$y_t^{\SM}(M_Z)$.  We find that the explicit 3-loop Standard Model
contributions to the Higgs mass lead to a small positive shift by
around $30\unit{MeV}$.

\paragraph{3-loop renormalization group running in the SM}

In the SM, the 3-loop $\beta$ functions are known
\cite{Mihaila:2012fm,Bednyakov:2012rb,Bednyakov:2012en,Chetyrkin:2012rz,Bednyakov:2013eba}.
\fs allows the user to take these 3-loop $\beta$ functions into
account in the running of the \MSbar SM parameters by
setting the following switch in the model file:
\begin{lstlisting}
UseSM3LoopRGEs = True;
\end{lstlisting}
The expression for the 3-loop $\beta$ function of the $\mu^2$ parameter in
the SM has been extracted from the file \code{smh3l.m} available on
the arXiv page of Ref.~\cite{Bednyakov:2013eba}.  The remaining 3-loop
$\beta$ functions have been kindly provided by the authors of \susyhd
1.0.1 \cite{Vega:2015fna}.  To use 3-loop running at runtime, the
following flag should also be set in the SLHA input file:
\begin{lstlisting}
Block FlexibleSUSY
    6   3         # beta-functions loop order
\end{lstlisting}
In \fs's \mathematica interface, the above SLHA configuration
corresponds to
\begin{lstlisting}[language=Mathematica]
handle = FS<model>OpenHandle[
    fsSettings -> {betaFunctionLoopOrder -> 3}
    ...];
\end{lstlisting}

\paragraph{3-loop QCD corrections to the \MSbar top Yukawa coupling in the SM}

The pure SM QCD contribution to the \MSbar top Yukawa
coupling is known up to the 3-loop level of $O(\as^3)$
\cite{Chetyrkin:1999qi,Melnikov:2000qh}.  This 3-loop QCD expression
can be taken into account in \fs to extract the \MSbar top Yukawa
coupling in the SM from the top pole mass $M_t$ by setting
the following switch in the model file:
\begin{lstlisting}
UseYukawa3LoopQCD = True;
\end{lstlisting}
By default, \code{UseYukawa3LoopQCD} is set to \code{Automatic}, which
means that the 3-loop QCD contribution to the top Yukawa coupling is
taken into account only if the model is \MSbar renormalized.  To take
this 3-loop correction into account at runtime, the following flags
should be set in the SLHA input file:
\\\begin{minipage}{\linewidth}
\begin{lstlisting}
Block FlexibleSUSY
    7   3         # threshold corrections loop order
   24   123111221 # individual threshold correction loop orders
\end{lstlisting}
\end{minipage}
In \fs's \mathematica interface, the above SLHA configuration options
correspond to
\begin{lstlisting}[language=Mathematica]
handle = FS<model>OpenHandle[
    fsSettings -> {thresholdCorrectionsLoopOrder -> 3,
                   thresholdCorrections -> 123111221}
    ...];
\end{lstlisting}
The 3-loop QCD contribution is combined with the full 1-loop and QCD
2-loop contribution as
\begin{align}
\begin{split}
  m_t(Q) &= M_t + \re\Sigma_{t}^S(p^2=M_t^2,Q) \\
  &\phantom{={}} + M_t \Big[ \re\Sigma_{t}^L(p^2=M_t^2,Q) +
    \re\Sigma_{t}^R(p^2=M_t^2,Q) \\
  &\phantom{={} + M_t \Big[}
    + \Delta m_t^{(1),\text{qcd}}(Q) + \Delta m_t^{(2),\text{qcd}}(Q) + \Delta m_t^{(3),\text{qcd}}(Q) \Big]
  \,,
\end{split} \label{eq:mt_MSbar}
\end{align}
where $\Sigma_t^{S,L,R}(p^2,Q)$ denote the scalar, left- and
right-handed parts of the \MSbar-renormalized 1-loop top quark self
energy, $M_t$ is the top pole mass and
\begin{align}
  \label{eq:top-selfenergy-qcd-1L}
  \Delta m_t^{(1),\text{qcd}}(Q) &=
  -\frac{g_3^2}{12 \pi^2} \left[4 - 3 \;\barlog(t)\right], \\
  \label{eq:top-selfenergy-qcd-2L}
  \begin{split}
    \Delta m_t^{(2),\text{qcd}}(Q) &= \left(\Delta
      m_t^{(1),\text{qcd}}(Q)\right)^2 - \frac{g_3^4}{4608 \pi^4}
    \Bigg[396 \;\barlog^2(t) - 1452 \;\barlog(t) \\
    &\phantom{={}}
    - 48 \zeta(3) + 2053 + 16 \pi^2 (1+\log 4)\Bigg] ,
  \end{split} \\
\begin{split}
  \Delta m_t^{(3),\text{qcd}}(Q) &= \frac{g_3^6}{2430\, (4\pi)^6}
  \Bigg\{
  48600 \;\barlog^3(t) - 208980 \;\barlog^2(t) \\
  &\phantom{=\;} +540 \Big[-1560\,\zeta (3)+2993+40 \pi ^2 (1+\log 4)\Big] \barlog(t)\\
  &\phantom{=\;} +15 \left[69120\,\text{Li}_4\left(\frac{1}{2}\right)+113040\,\zeta(3)-
    94800 \,\zeta(5)-280853+2880 \,\log^4 2\right]\\
  &\phantom{=\;} +4 \pi^2 \Big[129510 \,\zeta(3)-388781 + 240 (733+24 \log 2) \,\log 2\Big] - 10500 \pi^4
  \Bigg\}
\end{split}\\
  &\approx
  \frac{g_3^6}{(4\pi)^6}\, 20 \left[\barlog^3(t)
    - \frac{43}{10} \;\barlog^2(t)
    + 22.8874 \;\barlog(t)
    -172.937 \right] \,.
\end{align}
In \figref{fig:SM_thresholds} the impact of the 3-loop correction to
the \MSbar top Yukawa coupling in the SM $y_t^{\SM}(M_Z)$
on the prediction of the light \CP-even Higgs pole mass in the MSSM in
the pure EFT calculation of \HSSUSY is shown as the green
dashed-double-dotted line.  As already discussed in
Ref.~\cite{Vega:2015fna}, we find that the inclusion of the 3-loop
correction to $y_t^{\SM}(M_Z)$ reduces the Higgs mass by up to
$500\unit{MeV}$.  Note that this is formally a (partial) 4-loop
effect on the Higgs mass, which is beyond the current accuracy of
\HSSUSY.

\paragraph{3-loop QCD corrections to the \MSbar strong coupling in the SM}

The pure SM QCD contribution to the \MSbar strong coupling is known up
to the 3-loop level of $O(\as^3)$
\cite{Rodrigo:1993hc,Rodrigo:1997zd,Chetyrkin:1997un,Chetyrkin:2000yt}.
This 3-loop QCD expression can be taken into account in \fs to extract
the \MSbar strong coupling in the SM from the input value
$\as^{\SM(5)}(M_Z)$ by setting the following switch in the
model file:
\begin{lstlisting}
UseSMAlphaS3Loop = True;
\end{lstlisting}
To take this 3-loop threshold correction into account at runtime, the
following flags should also be set in the SLHA input file:
\begin{lstlisting}
Block FlexibleSUSY
    7   3         # threshold corrections loop order
   24   123111321 # individual threshold correction loop orders
\end{lstlisting}
In \fs's \mathematica interface, the above SLHA configuration options
correspond to
\begin{lstlisting}[language=Mathematica]
handle = FS<model>OpenHandle[
    fsSettings -> {thresholdCorrectionsLoopOrder -> 3,
                   thresholdCorrections -> 123111321}
    ...];
\end{lstlisting}
The 3-loop QCD contributions are combined as
\begin{align}
  \as^{\SM}(Q) &=
  \frac{\as^{\SM(5)}(Q)}{1 - \Delta\as^{1L}(Q)
    - \Delta\as^{2L}(Q) - \Delta\as^{3L}(Q)} \,,
\end{align}
where $\Delta\as^{nL}(Q)$ are the $n$-loop threshold corrections,
which read \cite{Chetyrkin:2000yt}
\begin{align}
  \Delta\as^{1L}(Q) &= \Delta_1 \,,\\
  \Delta\as^{2L}(Q) &= \Delta_2 - \Delta_1^2 \,,\\
  \Delta\as^{3L}(Q) &= \Delta_3 + \Delta_1^3 - 2\Delta_1\Delta_2 \,,\\
  \Delta_1 &= \left(\frac{\as^{\SM(5)}(Q)}{\pi}\right) \frac{L}{6} \,, \\
  \Delta_2 &= \left(\frac{\as^{\SM(5)}(Q)}{\pi}\right)^2 \Bigg[
  -\frac{11}{72} + \frac{11}{24} L + \frac{1}{36} L^2
  \Bigg] \,,\\
  \Delta_3 &=
  \left(\frac{\as^{\SM(5)}(Q)}{\pi}\right)^3 \Bigg[
  \frac{1}{216} L^3
  + \frac{167}{576} L^2
  + \frac{2645}{1728}L
  + n_f \left(\frac{1}{36} L^2
  - \frac{67}{576} L
  + \frac{2633}{31104}\right) \nonumber \\
  &\phantom{={}}
  + \frac{82043}{27648} \zeta_3
  - \frac{564731}{124416}
  \Bigg] \,, \\
  &\approx
  \left(\frac{\as^{\SM(5)}(Q)}{\pi}\right)^3 \Bigg[
  0.00462963 L^3
  + 0.428819 L^2
  + 0.94907 L
  - 0.54880 \Bigg] \,,
\end{align}
with $n_f = 5$, $L = \log(Q^2/(m_t^{\SM}(Q))^2)$ and $m_t^{\SM}(Q)$ being the \MSbar
top quark mass in the SM at the scale $Q$.

\figref{fig:SM_thresholds} shows the impact of the 2- and 3-loop QCD
threshold corrections for $\as^{\SM}(M_Z)$ on the light \CP-even Higgs
pole mass in the MSSM, as predicted by the pure EFT calculation of
\HSSUSY as a function of the SUSY scale \MS.  We find that the
inclusion of $\Delta\as^{2L}$ (blue dashed line) reduces the Higgs
mass by up to $40\unit{MeV}$, depending on the SUSY scale.  Taking
into account $\Delta\as^{3L}$ (blue dotted line) reduces the Higgs
mass further by around $3\unit{MeV}$.

\subsubsection{(N)MSSM-specific higher-order contributions}
\label{sec:MSSM-specific_ho_corrections}
\paragraph{2-loop contributions to the (N)MSSM Higgs masses}

As already described in Ref.~\cite{Athron:2014yba}, the known dominant 2-loop
effective potential contributions to the Higgs masses in the (N)MSSM
of $O((\at + \ab)\as + (\at +
\ab)^2 + \atau^2)$
\cite{Degrassi:2001yf,Brignole:2001jy,Dedes:2002dy,Brignole:2002bz,Dedes:2003km,Degrassi:2009yq}
can be taken into account by \fs.
Note, however, that the implemented NMSSM 2-loop contributions of
$O((\at + \ab)^2 + \atau^2)$ are currently available only in the
MSSM-limit.  Furthermore, NMSSM-specific 2-loop corrections to the
running vacuum expectation value beyond the MSSM limit, which would be
required for a consistent treatment of the $O((\at + \ab)\as)$
contributions, are currently not implemented.
To take the implemented 2-loop corrections to the Higgs masses in the
(N)MSSM into account, the following switches must be set in the model
file:
\begin{lstlisting}
UseHiggs2LoopMSSM = True;
UseHiggs2LoopNMSSM = True;
\end{lstlisting}
These flags can be enabled in all real (N)MSSM-like models with 2(3)
\CP-even Higgs bosons, 1(2) \CP-odd Higgs boson(s) and 1 electrically neutral
Goldstone boson.  In addition to these flags, the (effective) $\mu$
parameter in the convention of Ref.~\cite{Haber:1984rc} and the effective
squared \CP-odd Higgs tree-level mass $m_A^2$ of the model must be identified.
In the \modelname{NMSSM} model file of \sarah they would read for
example:
\begin{lstlisting}
EffectiveMu = \[Lambda] vS / Sqrt[2];
EffectiveMASqr = (T[\[Lambda]] vS / Sqrt[2] + 0.5 \[Lambda] \[Kappa] vS^2) (vu^2 + vd^2) / (vu vd);
\end{lstlisting}
These 2-loop contributions can then be enabled at runtime by setting
the following flags in the SLHA input file:
\begin{lstlisting}
Block FlexibleSUSY
    4   2         # pole mass loop order
    5   2         # EWSB loop order
    7   2         # threshold corrections loop order
    8   1         # Higgs 2-loop corrections O(alpha_t alpha_s)
    9   1         # Higgs 2-loop corrections O(alpha_b alpha_s)
   10   1         # Higgs 2-loop corrections O((alpha_t + alpha_b)^2)
   11   1         # Higgs 2-loop corrections O(alpha_tau^2)
   24   122111221 # individual threshold correction loop orders
\end{lstlisting}
In \fs's \mathematica interface, the above SLHA configuration options
correspond to
\begin{lstlisting}[language=Mathematica]
handle = FS<model>OpenHandle[
    fsSettings -> {poleMassLoopOrder -> 2,
                   ewsbLoopOrder -> 2,
                   thresholdCorrectionsLoopOrder -> 2,
                   higgs2loopCorrectionAtAs -> 1,
                   higgs2loopCorrectionAbAs -> 1,
                   higgs2loopCorrectionAtAt -> 1,
                   higgs2loopCorrectionAtauAtau -> 1,
                   thresholdCorrections -> 122111221}
    ...];
\end{lstlisting}

\paragraph{3-loop contributions to the light \CP-even MSSM Higgs mass}

The 3-loop contributions to the light \CP-even Higgs mass in the MSSM
have been calculated to $O(\at\as^2)$ in the \DRbar and
\MDRbar scheme \cite{Harlander:2008ju,Kant:2010tf,Kunz:2014gya}.  The
expressions are available in the public spectrum generator \htm
\cite{h3murl}, where they are added to the 2-loop on-shell result of
\FeynHiggs
\cite{Heinemeyer:1998yj,Heinemeyer:1998np,Degrassi:2002fi,Frank:2006yh,Hahn:2013ria,Drechsel:2016jdg,Bahl:2016brp,Bahl:2017aev}.
In Ref.~\cite{Harlander:2017kuc} the 3-loop contributions of
$O(\at\as^2 + \ab\as^2)$ have been studied for the first time in a
pure \DRbar MSSM spectrum generator and have been made available in
the public C++ library \Himalaya \cite{himalaya}.

The explicit 3-loop contributions of $O(\at\as^2 + \ab\as^2)$ to the
light \CP-even Higgs mass from \Himalaya can be taken into account in
\fs by setting the following flag in the \fs model file:
\begin{lstlisting}[language=Mathematica]
UseHiggs3LoopMSSM = True;
\end{lstlisting}
The model is required to be real MSSM-like with two \CP-even Higgs bosons and
one electrically neutral Goldstone boson.
However, in a pure \DRbar calculation, another source of such 3-loop
contributions originates from the 2-loop \SQCD contribution to the MSSM
\DRbar top Yukawa coupling, which must be included.  Then, however, in
order to be consistent with respect to the loop orders of the running and
decoupling, also 3-loop renormalization group running should be
performed and the 2-loop threshold correction $\Delta\as^{2L}$ of
the strong coupling should be included, see below.  Therefore, we
strongly recommend setting the following flags in addition in the \fs
model file:
\begin{lstlisting}[language=Mathematica]
UseHiggs2LoopMSSM = True;  (* 2-loop contribution to Higgs mass *)
EffectiveMu = \[Mu];       (* specify mu parameter, see above *)
UseMSSM3LoopRGEs = True;   (* 3-loop running *)
UseMSSMYukawa2Loop = True; (* 2-loop SUSY-QCD correction to yt *)
UseMSSMAlphaS2Loop = True; (* 2-loop threshold correction to alpha_s *)
\end{lstlisting}
When building \fs, the path to the \Himalaya headers and to the
\Himalaya library must be specified:
\begin{lstlisting}[language=bash]
$ ./configure --with-models=[...] \
   --enable-himalaya \
   --with-himalaya-incdir=$HIMALAYA_PATH/source/include \
   --with-himalaya-libdir=$HIMALAYA_PATH/build
$ make
\end{lstlisting}
where \code{$HIMALAYA_PATH} is the \Himalaya directory. 
To calculate the light \CP-even Higgs mass in the MSSM at the 3-loop
level with \fs, the following flags must be set at runtime: In the
SLHA input file we recommend setting at least
\begin{lstlisting}
Block FlexibleSUSY
    4   3         # pole mass loop order
    5   3         # EWSB loop order
    6   3         # beta-functions loop order
    7   2         # threshold corrections loop order
   24   122111221 # individual threshold correction loop orders
   25   0         # ren. scheme for 3L corrections (0 = DR, 1 = MDR)
   26   1         # Higgs 3-loop corrections O(alpha_t alpha_s^2)
   27   1         # Higgs 3-loop corrections O(alpha_b alpha_s^2)
\end{lstlisting}
In \fs's \mathematica interface, the above SLHA configuration options
correspond to
\begin{lstlisting}[language=Mathematica]
handle = FS<model>OpenHandle[
    fsSettings -> {poleMassLoopOrder -> 3,
                   ewsbLoopOrder -> 3,
                   betaFunctionLoopOrder -> 3,
                   thresholdCorrectionsLoopOrder -> 2,
                   thresholdCorrections -> 122111221,
                   higgs3loopCorrectionRenScheme -> 0,
                   higgs3loopCorrectionAtAsAs -> 1,
                   higgs3loopCorrectionAbAsAs -> 1}
    ...];
\end{lstlisting}

\begin{figure}[h!]
  \centering
  \includegraphics[width=0.49\textwidth]{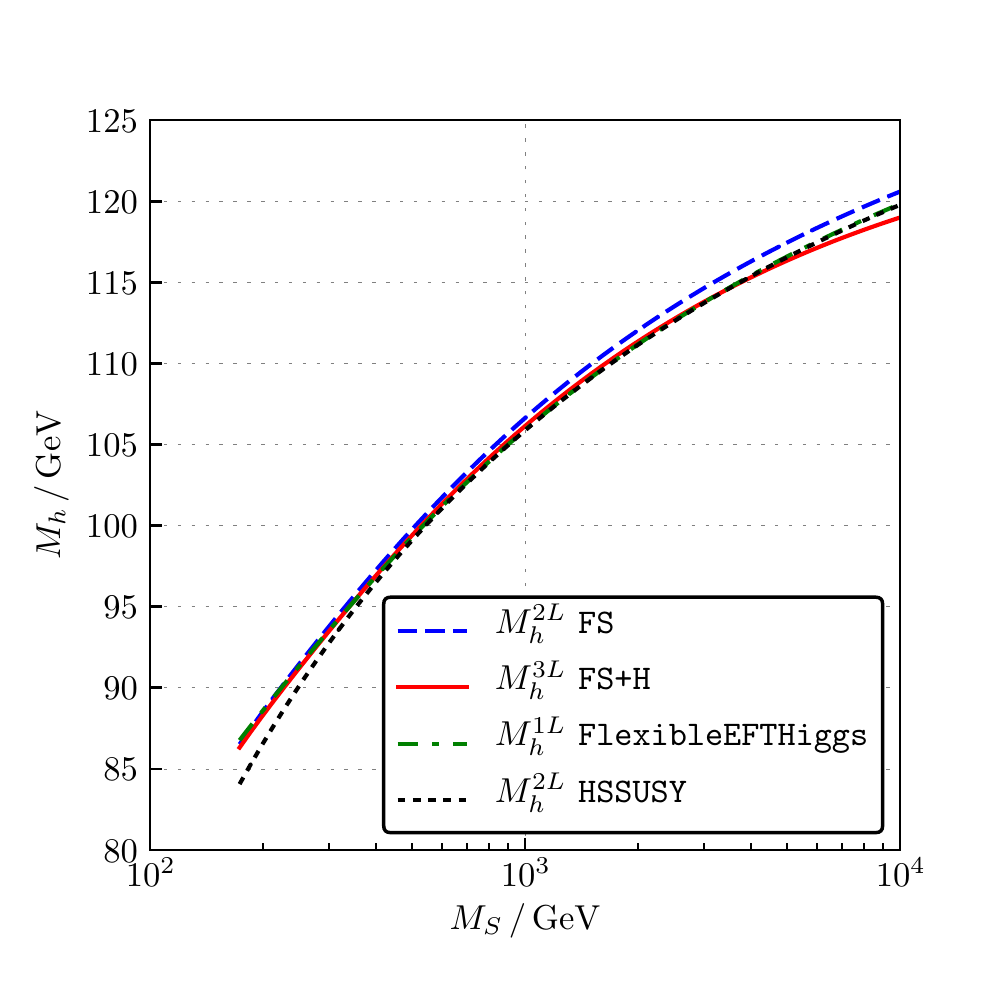}\hfill
  \includegraphics[width=0.49\textwidth]{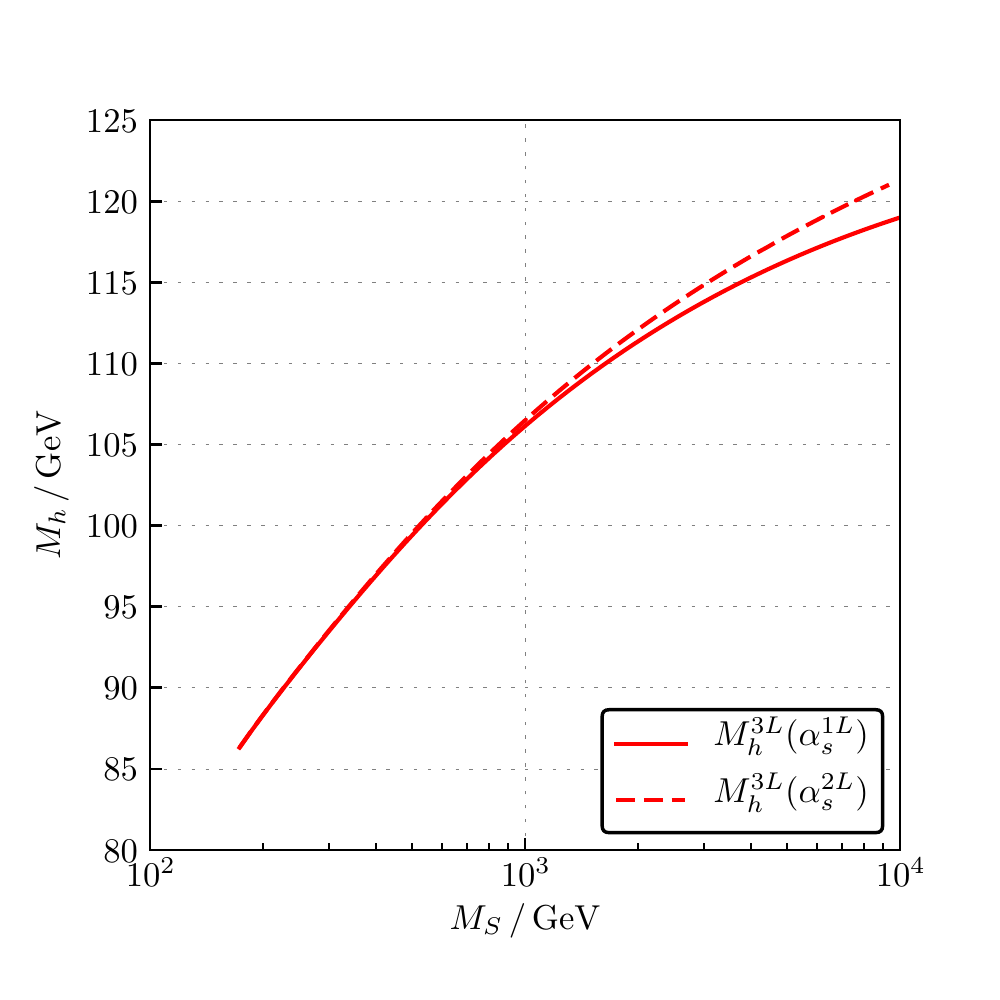}\\
  \includegraphics[width=0.49\textwidth]{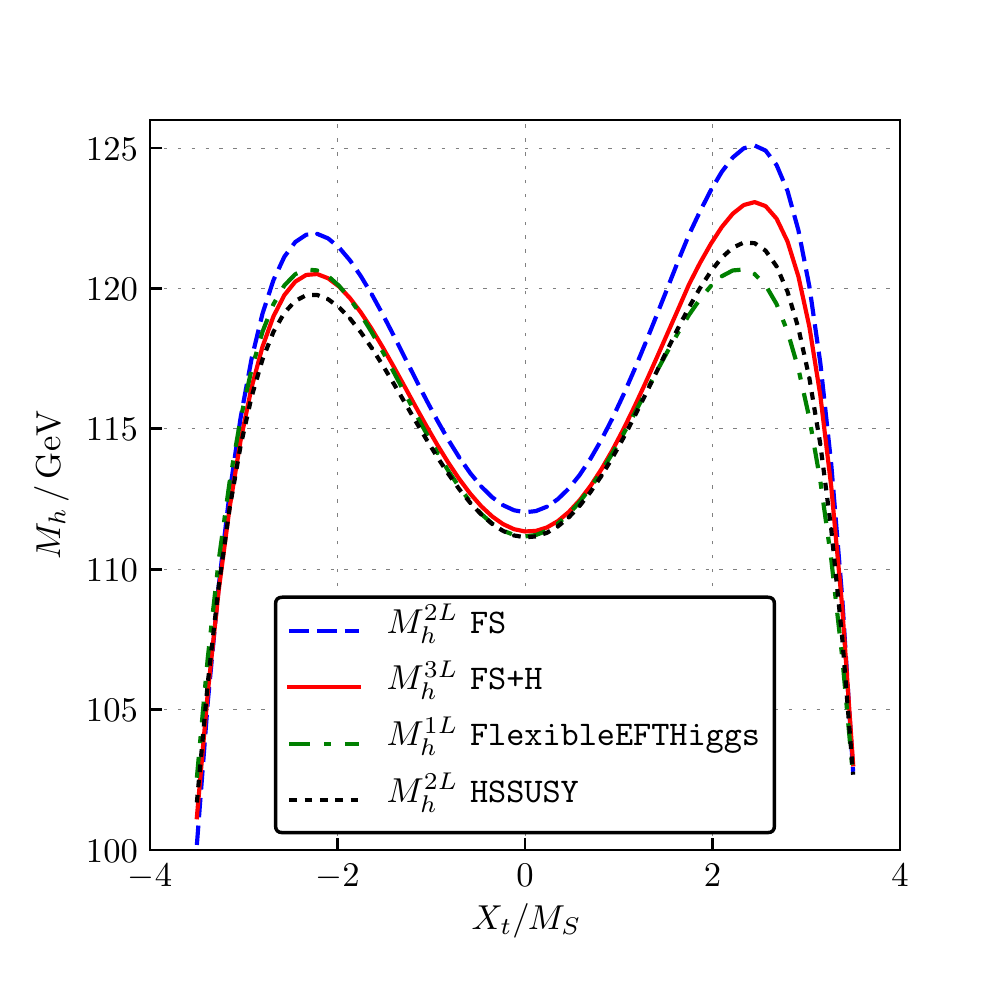}\hfill
  \includegraphics[width=0.49\textwidth]{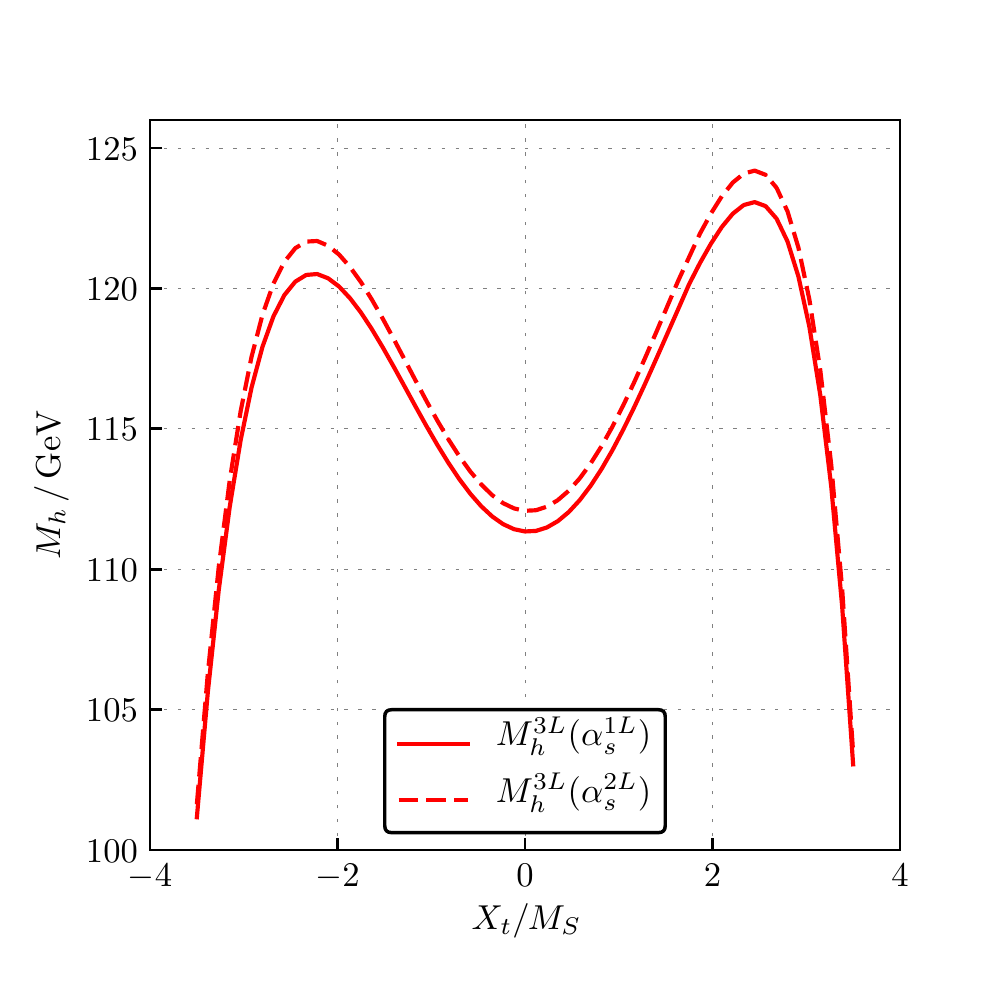}
  \caption{Comparison of different 2-loop and 3-loop calculations of
    the lightest \CP-even Higgs mass in the MSSM with different \fs
    spectrum generators for $\tan\beta = 5$.  In the top row we set
    $X_t = 0$ and in the bottom row we use $\MS = 2\unit{TeV}$.}
  \label{fig:Mh-himalaya}
\end{figure}
\figref{fig:Mh-himalaya} shows a comparison of the different light
\CP-even Higgs mass calculations available in \fs.  The scenario is
chosen such that all soft-breaking \DRbar mass parameters, the
superpotential
$\mu$ parameter and the \DRbar \CP-odd Higgs mass are equal to the SUSY
scale $\MS$ at the scale $Q =
\MS$ and all \DRbar sfermion mixing parameters
$X_f$ are set to zero, except for the stop mixing parameter
$X_t$.  Furthermore, we set $M_t =
173.34\unit{GeV}$, $\as^{\SM(5)}(M_Z) =
0.1184$ and $\aem^{\SM(5)}(M_Z) =
1/127.944$.  The top row of \figref{fig:Mh-himalaya} shows the Higgs
mass as a function of $\MS$ for $X_t = 0$ and $\tan\beta =
5$.  In the left panel of the top row the classic 2-loop fixed-order
calculation with \fs \cite{Athron:2014yba} is shown as the blue dashed
line and the pure EFT calculation with \HSSUSY is shown as the black
dotted line.  The reason why these two curves do not deviate from each
other logarithmically when
$\MS$ increases is an accidental cancellation of large
logarithmic contributions in the fixed-order calculation
\cite{Athron:2016fuq}.  The green dashed-dotted line shows the
improved version of the \feft calculation \cite{Athron:2016fuq}, which
interpolates
between the fixed-order calculation at low \MS and the pure EFT
calculation at large \MS.  See \secref{sec:feft} for a discussion of
the improvements of \feft.  The red solid line shows the 3-loop
calculation up to $O(\at\as^2 +
\ab\as^2)$ with \fsh \cite{Harlander:2017kuc}.  In this calculation, the
full 2-loop \SQCD contributions to the \DRbar top and bottom Yukawa
couplings $y_{t,b}(M_Z)$ \cite{Bednyakov:2007vm,Bednyakov:2002sf,Bednyakov:2005kt}
are taken into account.  The sum of these contributions
and the explicit 3-loop contribution to the \CP-even
Higgs mass matrix leads to a downward shift of the predicted Higgs
mass by $1$--$2\unit{GeV}$, depending on \MS, compared to the classic 2-loop
calculation with \fs (blue dashed line).  As also discussed in
Ref.~\cite{Harlander:2017kuc}, this brings the prediction closer to the pure
EFT calculation, which is expected to lead to a more precise result above
the TeV scale.  In the left panel of the bottom row of
\figref{fig:Mh-himalaya}, the calculated Higgs mass is shown as a
function of the stop mixing parameter $X_t$ for fixed
$\MS = 2\unit{TeV}$.  Also for non-zero
$X_t$, one finds a reduction of the Higgs mass by
$1$--$3\unit{GeV}$, which brings the prediction closer to the one of
the pure EFT calculation.

In the right panels of \figref{fig:Mh-himalaya}, the effect of the 2-loop
\SQCD threshold correction $\Delta\as^{2L}$
\cite{Harlander:2005wm,Bauer:2008bj,Bednyakov:2010ni} to the strong
coupling $\as$ (see below) on the 3-loop Higgs mass calculation is shown for
the same scenario as in the corresponding left panels.
The threshold correction $\Delta\as^{2L}$ is included at the scale
$Q = M_Z$, the same scale at which all dimensionless MSSM \DRbar
parameters are determined from the Standard Model input parameters.
The inclusion of $\Delta\as^{2L}$ is formally a 4-loop effect on the light
\CP-even Higgs mass in the MSSM\@.  However, $\Delta\as^{2L}$ should
be taken into account for a consistent running and decoupling
procedure with 3-loop renormalization group running.  The red solid
lines in the right panels correspond to the 3-loop calculation of
Ref.~\cite{Harlander:2017kuc}, which uses only the 1-loop threshold correction
$\Delta\as^{1L}$.  These red solid lines are the same as in the
corresponding left panels.  The effect of including
$\Delta\as^{2L}$ is shown as the red dashed line.  We find that the
inclusion of this 2-loop threshold correction leads to an upwards
shift of the Higgs mass by up to $2\unit{GeV}$, depending on \MS and
$X_t$.  Note, that large 4-loop contributions of multiple GeV hint at
a large theoretical uncertainty of the fixed-order calculation of the
light \CP-even Higgs pole mass in parameter regions with multi-TeV
stop masses.

\paragraph{3-loop renormalization group running in the MSSM}

\begin{figure}[tbh]
  \centering
  \includegraphics[width=0.49\textwidth]{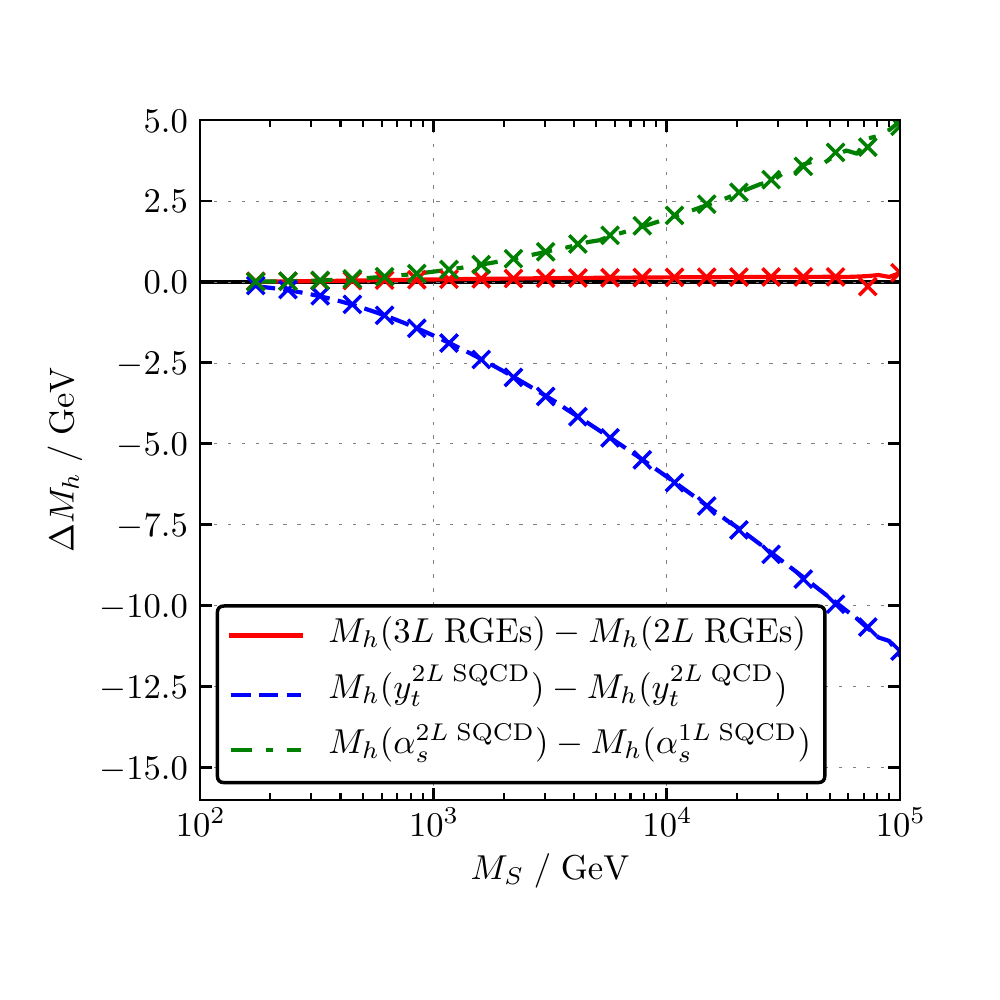}\hfill
  \includegraphics[width=0.49\textwidth]{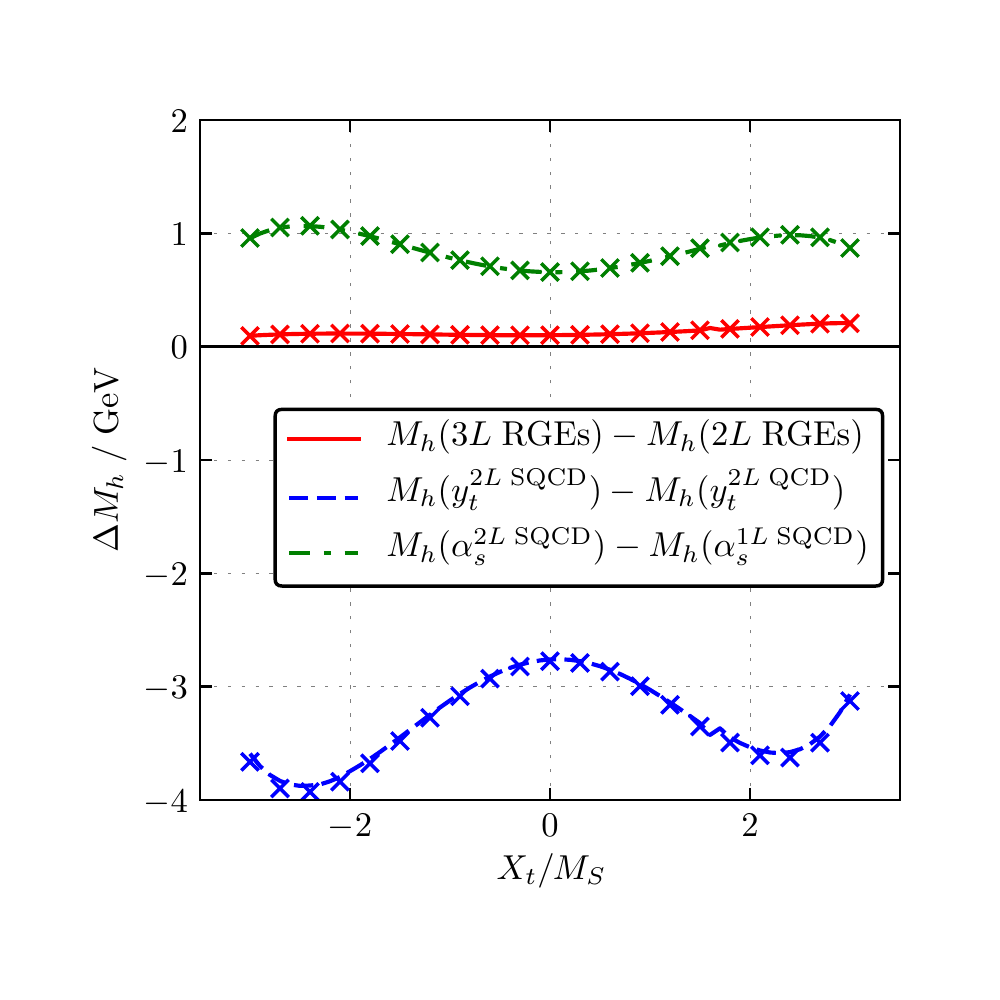}
  \caption{Effect of the 3-loop RGEs for all MSSM parameters (red),
    2-loop \SQCD contributions to $y_b$ and $y_t$ (blue) and 2-loop
    \SQCD contributions to $\as$ (green) on the lightest \CP-even
    Higgs pole mass in the MSSM\@.  The lines show the shift in the
    Higgs pole mass in \fstwo compared to the one obtained with 2-loop
    RGEs, full 1-loop \SQCD + 2-loop SM-QCD contributions to $y_t$ and
    1-loop \SQCD contributions to $\as$.  The crosses show the
    corresponding shift obtained with \softsusy~4.0.1.  In the left
    panel we use $\tan\beta = 5$ and $X_t = 0$ and in the right panel
    we fix $\tan\beta = 5$ and $\MS = 2\unit{TeV}$.}
  \label{fig:MSSMMuBMu}
\end{figure}

In the MSSM, the 3-loop $\beta$ functions are also known
\cite{Jack:2003sx,Jack:2004ch}.  \fs allows the user to take these 3-loop
$\beta$ functions into account in the running of the \DRbar MSSM parameters by
setting the following switch in the model file:
\begin{lstlisting}
UseMSSM3LoopRGEs = True;
\end{lstlisting}
To use 3-loop running at runtime, the following flag should also be set in
the SLHA input file:
\\\begin{minipage}{\linewidth}
\begin{lstlisting}
Block FlexibleSUSY
    6   3         # beta-functions loop order
\end{lstlisting}
\end{minipage}
In \fs's \mathematica interface, the above SLHA configuration
corresponds to
\begin{lstlisting}[language=Mathematica]
handle = FS<model>OpenHandle[
    fsSettings -> {betaFunctionLoopOrder -> 3}
    ...];
\end{lstlisting}
The expressions for the 3-loop $\beta$ functions have been extracted
from the official FORM file provided by the authors of
Refs.~\cite{Jack:2003sx,Jack:2004ch}.\footnote{\url{https://www.liverpool.ac.uk/~dij/betas/allgennb.log}}
We have numerically compared the expressions with the ones implemented
in \softsusy 3.7.0 \cite{Allanach:2014nba} and found exact agreement.
The effect of the 3-loop RGEs on the Higgs pole mass in the MSSM is of
the order of a few $100\unit{MeV}$ as discussed in
Ref.~\cite{Allanach:2014nba} and is shown in \figref{fig:MSSMMuBMu}.
Due to the complexity of the 3-loop $\beta$ functions, the runtime of
the MSSM spectrum generators is increased by a factor of $4$--$5$ if
3-loop running is enabled in the MSSM\@.

\paragraph{2-loop \SQCD corrections to the \DRbar top and bottom Yukawa couplings in the MSSM}
In the MSSM, the full 2-loop \SQCD corrections of $O(\as^2)$
to the \DRbar top Yukawa coupling
\cite{Bednyakov:2007vm,Bednyakov:2002sf,Bednyakov:2005kt} as well as
the 2-loop \SQCD $O(\as^2)$ and partial electroweak contributions
to the \DRbar bottom Yukawa coupling \cite{Bednyakov:2009wt} are
known.  These 2-loop corrections have already been made available in
\softsusy~3.7.0 \cite{Allanach:2014nba}.  The 2-loop \SQCD
corrections of $O(\as^2)$ are now also incorporated in \fstwo and can be
used by setting the following switch in the model file:\footnote{We
  kindly thank Alexander Bednyakov for providing the 2-loop \SQCD
  expressions.}
\begin{lstlisting}
UseMSSMYukawa2Loop = True;
\end{lstlisting}
To take these 2-loop contributions into account at runtime, the
following flags should also be set in the SLHA input file:
\begin{lstlisting}
Block FlexibleSUSY
    7   2         # threshold corrections loop order
   24   122111221 # individual threshold correction loop orders
\end{lstlisting}
In \fs's \mathematica interface, the above SLHA configuration options
correspond to
\begin{lstlisting}[language=Mathematica]
handle = FS<model>OpenHandle[
    fsSettings -> {thresholdCorrectionsLoopOrder -> 2,
                   thresholdCorrections -> 122111221}
    ...];
\end{lstlisting}
In \figref{fig:MSSMMuBMu} we show in blue the effect of the full
2-loop \SQCD corrections on the lightest \CP-even Higgs pole mass in
the MSSM\@.  The blue dashed line shows the effect in \fstwo and the
crosses in \softsusy~4.0.1.
As can be seen from the figure, the full 2-loop \SQCD corrections are
negative and can affect the Higgs mass by several GeV, as has been
observed in Ref.~\cite{Allanach:2014nba}.  In the left panel of
\figref{fig:MSSMMuBMu}, the shift in the Higgs mass is shown as a
function of the SUSY scale.  For scales above $\approx 2\unit{TeV}$ we
find a logarithmic shift in $M_h$, which is caused by new large
logarithms originating from the 2-loop contribution of the SUSY
particles to the \DRbar top Yukawa coupling.  We also find that these
new logarithms alone would spoil the accidental cancellation of large
logarithms described in Ref.~\cite{Athron:2016fuq}.
Note that in the MSSM the effect of the 2-loop \SQCD corrections to
the top and bottom Yukawa couplings is a partial 3-loop contribution
to the light \CP-even Higgs pole mass.  Thus, these 2-loop \SQCD corrections
must be taken into account if the explicit 3-loop Higgs mass
contributions of $O(\at \as^2 + \ab \as^2)$ from
\Himalaya are used, see above.

\paragraph{2-loop \SQCD corrections to the \DRbar strong gauge coupling in the MSSM}

In the MSSM, the full 2-loop \SQCD corrections of
$O(\as^2 + \as\at + \as\ab)$ to the \DRbar strong gauge coupling are known
\cite{Harlander:2005wm,Bauer:2008bj,Bednyakov:2010ni} and have been
made available in \softsusy~3.7.0 \cite{Allanach:2014nba}.  In \fstwo,
these corrections can be taken into account by setting in the model
file:\footnote{We kindly thank Ben Allanach and Alexander Bednyakov
  for providing the 2-loop \SQCD expressions.}
\begin{lstlisting}
UseMSSMAlphaS2Loop = True;
\end{lstlisting}
To take these 2-loop threshold corrections into account at runtime, the
following flags should also be set in the SLHA input file:
\begin{lstlisting}
Block FlexibleSUSY
    7   2         # threshold corrections loop order
   24   122111221 # individual threshold correction loop orders
\end{lstlisting}
In \fs's \mathematica interface, the above SLHA configuration options
correspond to
\begin{lstlisting}[language=Mathematica]
handle = FS<model>OpenHandle[
    fsSettings -> {thresholdCorrectionsLoopOrder -> 2,
                   thresholdCorrections -> 122111221}
    ...];
\end{lstlisting}
In \figref{fig:MSSMMuBMu} we show in green the effect of the 2-loop
\SQCD corrections to $\as$ on the lightest \CP-even Higgs pole
mass in the MSSM\@.  The green dashed-dotted line shows the effect in
\fstwo and the crosses in \softsusy~4.0.1.  Both implementations agree
exactly.  We furthermore find that the inclusion of the 2-loop
threshold corrections to $\as$ leads to a logarithmic enhancement
of the Higgs mass as a function of the SUSY scale.  The enhancement is
around $+1\unit{GeV}$ for $\MS\approx 2\unit{TeV}$ and maximal stop
mixing.
Note that in the MSSM the effect of the 2-loop \SQCD corrections to
the strong coupling is formally a partial 4-loop contribution to the
Higgs pole mass.

\subsubsection{Split-MSSM-specific higher-order contributions}
\paragraph{3-loop contribution to the Higgs mass in the split-MSSM}

In the split-MSSM \cite{Bagnaschi:2014rsa}, part of the 3-loop
contribution of $O(\at \as^2)$
to the SM-like Higgs pole mass is known in the \MSbar
scheme \cite{Benakli:2013msa}.  \fs allows the user to take this
3-loop contribution into account by setting the following switch in
the model file:
\begin{lstlisting}
UseHiggs3LoopSplit = True;
\end{lstlisting}
This is used in the distributed \modelname{SplitMSSM} model file. For use in
other models, it is a requirement that the model contains a single
Higgs boson and a gluino.  The enabled 3-loop Higgs pole mass
contribution reads
\begin{align}
  (\Delta m_h^2)_{3L} &= \frac{64 g_3^4 y_t^2}{(4\pi)^6} m_t^2\;\barlog^3(g) \,,
\end{align}
where $g = m_{\tilde{g}}^2$ and $m_{\tilde{g}}$ is the \MSbar gluino
mass.  Furthermore, for consistency the 2-loop gluino contribution in
the calculation of the \MSbar top mass from Ref.\
\cite{Benakli:2013msa} is taken into account by adding the following
term to Eq.\ \eqref{eq:mt_MSbar}:
\begin{align}
  \Delta m_t^{(2),\text{split-qcd}}(Q) =
  -\frac{g_3^4}{(4\pi)^4} M_t \left\{
     \frac{89}{9}
     + 4\;\barlog(g) \left[
        \frac{13}{3}
        + \barlog(g)
        - 2\;\barlog(t)
     \right]
  \right\} \,.
\end{align}
To take these contributions into account at runtime, the following flags
should also be set in the SLHA input file:
\begin{lstlisting}
Block FlexibleSUSY
    4   3         # pole mass loop order
    5   3         # EWSB loop order
    7   2         # threshold corrections loop order
   24   122111221 # individual threshold correction loop orders
   26   1         # Higgs 3-loop corrections O(alpha_t alpha_s^2)
\end{lstlisting}
In \fs's \mathematica interface, the above SLHA configuration options
correspond to
\\\begin{minipage}{\linewidth}
\begin{lstlisting}[language=Mathematica]
handle = FS<model>OpenHandle[
    fsSettings -> {poleMassLoopOrder -> 3,
                   ewsbLoopOrder -> 3,
                   thresholdCorrectionsLoopOrder -> 2,
                   thresholdCorrections -> 122111221,
                   higgs3loopCorrectionAtAsAs -> 1}
    ...];
\end{lstlisting}
\end{minipage}

\subsection{New features for definition of boundary conditions}
In \fstwo, the expressions to define boundary conditions are allowed
to be more complicated and to involve trigonometric functions,
dilogarithms, branches and more.  This is particularly useful for defining
high-scale boundary conditions that match a model to its UV-completion.  An
example is \fs's \HSSUSY model file, which implements the known
1- and 2-loop high-scale matching condition on the quartic Higgs
coupling of the SM against the MSSM at the SUSY scale
\cite{Bagnaschi:2014rsa,Vega:2015fna,Bagnaschi:2017xid}.  The list of
special functions and symbols to be used in boundary conditions can be
found in \tabref{tab:bcs_functions}.
\begin{table}[h!]
  \centering
 \resizebox{\textwidth}{!}{%
  \begin{tabularx}{1.05\textwidth}{lX}
    \toprule
    Function & Description \\
    \midrule

    \code{Abs[a_]} &
    Returns the magnitude of a real or complex number \code{a}, $|a|$.
    If \code{a} is a vector, a vector is returned with \code{Abs} applied to each element. \\

    \code{AbsSqr[a_]} &
    Returns the squared magnitude of a real or complex number \code{a}, $|a|^2$. \\

    \code{AbsSqrt[a_]} &
    Returns the square root of the magnitude of \code{a}, $\sqrt{|a|}$. \\

    \code{ArcSin[a_]}, \code{ArcCos[a_]} &
    Returns $\arcsin a$ and $\arccos a$, respectively. \\

    \code{ArcTan[a_]} & Returns $\arctan a$. \\

    \code{Arg[z_]} &
    Returns the phase angle of a complex number, $\arg z$. \\

    \code{Cbrt[a_]} &
    Returns the cubic root of \code{a}, $\sqrt[3]{a}$. \\

    \code{CKM}, \code{PMNS} &
    CKM and PMNS matrices, respectively, as defined in \cite{Allanach:2008qq}. \\

    \code{Conjugate[a_]} &
    Returns the complex conjugate of \code{a}. \\

    \code{Exp[a_]} & Returns $e^a$ for real or complex \code{a}. \\

    \code{FiniteLog[a_]} &
    Returns $\log a$ if $\log a$ is well-defined, otherwise returns $0$. \\

    \code{FSThrow[msg_]} & Throws an exception of type
    \code{PhysicalError} with the message \code{msg}. \\

    \code{I} & Imaginary unit. \\

    \code{If[cond_, a_, b_]} &
    If \code{cond} is true, \code{a} is returned, otherwise \code{b}. \\

    \code{Im[a_]} &
    Returns the imaginary part of \code{a}. \\

    \code{IsClose[a_, b_, eps_]} &
    Returns \code{True} if \code{Abs[a - b]} $<$ \code{eps}, otherwise \code{False}. \\

    \code{IsCloseRel[a_, b_, eps_]} &
    Returns \code{True} if \code{Abs[(a - b)/a]} $<$ \code{eps}, otherwise \code{False}. \\

    \code{IsFinite[a_]} &
    Returns \code{True} if \code{a} is neither \code{nan} nor \code{inf}. \\

    \code{KroneckerDelta[i_, j_]} &
    Returns the Kronecker $\delta_{ij}$. \\

    \code{Log[a_], ComplexLog[a_]} &
    Returns the natural logarithm for real and complex arguments, respectively. \\

    \code{Max[a_, ...]} &
    Returns the maximum of all given arguments. \\

    \code{Min[a_, ...]} &
    Returns the minimum of all given arguments. \\

    \code{Not[cond_]} &
    Returns the logical negation of \code{cond}. \\

    \code{PolyLog[2, z_]} &
    Returns the dilogarithm of the real or complex number \code{z}. \\

    \code{Re[a_]} &
    Returns the real part of \code{a}. \\

    \code{Round[a_]} &
    Returns \code{Floor[a + 0.5]} if \code{a} $\geq 0$, otherwise
    \code{Floor[a - 0.5]}. \\

    \code{Print<type>[msg_]} & Prints a debug, info, error, warning or
    fatal message, depending on the \code{<type>}, and returns zero.
    \code{<type>} can be \code{DEBUG}, \code{INFO}, \code{WARNING},
    \code{ERROR} or \code{FATAL}.  \code{PrintFATAL[msg]}
    throws an exception after \code{msg} has been printed.\\

    \code{SCALE} & Returns the renormalization scale at which
    the boundary condition is imposed. \\

    \code{Sign[a_]} &
    Returns $1$ if \code{a} $\geq 0$, otherwise $-1$. \\

    \code{SignedAbsSqrt[a_]} &
    Returns \code{Sign[a]*Sqrt[Abs[a]]}. \\

    \code{Sin[a_]}, \code{Cos[a_]}, \code{Tan[a_]} &
    Returns $\sin a$, $\cos a$ and $\tan a$, respectively. \\

    \code{Total[vec_]} &
    Returns the sum of all elements of \code{vec}, $\sum_i v_i$. \\

    \code{UnitStep[value_]} &
    Returns $0$ if \code{value} $<0$, $1$ otherwise. \\

    \multilinecell[t]{\code{Which[test1_, value1_,}\\\code{\ test2_, value2_, ...]}} &
    If \code{test1} is true, \code{value1} is returned, otherwise
    if \code{test2} is true, \code{value2} is returned, etc. \\

    \code{ZeroSqrt[a_]} &
    Returns $\sqrt{a}$ if $a>0$, otherwise returns $0$. \\

    \bottomrule
  \end{tabularx}}
  \caption{Available special functions and symbols in the boundary conditions.}
  \label{tab:bcs_functions}
\end{table}

\section{\fbsm extension}
\label{sec:fbsm}

Since version 1.1.0, \fs can generate spectrum generators not
only for SUSY models, but also for non-SUSY models.  We document this
feature here for the first time.  In Subsections
\ref{sec:application_THDM} and \ref{sec:application_HSSUSY}, we will
describe important applications of this feature to the two-Higgs
doublet model and to an effective low-energy theory of the MSSM
(\HSSUSY).

The generated non-SUSY spectrum generators have the same features as the SUSY
spectrum generators:
\begin{itemize}
\item The running gauge and Yukawa couplings of the non-SUSY model are
  calculated automatically at the 1-loop level from the known
  low-energy SM parameters
  $\aem^{\SM(5)}(M_Z)$, $\as^{\SM(5)}(M_Z)$
  and from the known quark and lepton masses as well as $M_Z$ and either $G_F$
  or $M_W$.  2-loop and 3-loop QCD corrections can be taken into
  account to determine the running top Yukawa coupling of the model,
  see \secref{sec:model_specific_contributions}.
\item Up to three boundary conditions can be specified to fix the
  running parameters of the model at different user-defined scales.
\item 2-loop renormalization group running is used between the scales
  at which the boundary conditions are imposed.\footnote{Note that
    the $\beta$ functions of scalar tadpole terms in
    non-supersymmetric models \cite{Goodsell:2012fm} are currently not
    generated by \sarah.  For this reason, such tadpole terms do not
    run in \sarah/\spheno or \fs.} In the SM and in the MSSM, also
  3-loop running is available, see
  \secref{sec:model_specific_contributions}.
\item The pole mass spectrum is calculated at the full 1-loop level,
  taking into account all BSM contributions.  Some 2-loop and 3-loop
  corrections can be added in specific non-SUSY models, see
  \secref{sec:model_specific_contributions}.
\end{itemize}

\subsection{Setting up a \fbsm model}

The \fs user interface for creating spectrum generators for non-SUSY
models is exactly the same as in the case of SUSY models, except that
all non-SUSY parameters are defined in the \MSbar scheme.  In
particular, at the low-energy scale \fs automatically determines
the gauge and Yukawa couplings of the non-SUSY model in the \MSbar
scheme.  For gauge-dependent quantities like running masses and VEVs,
\fs adopts the Feynman gauge, where all gauge fixing parameters
$\xi_i$ are set to unity.

\subsection{Determination of the \MSbar gauge and Yukawa couplings}
\label{sec:determination_of_g_Y}

If the considered BSM model has a gauge symmetry with the SM gauge group $SU(3)_C\times SU(2)_L\times U(1)_Y$ as a factor,
then \fs automatically fixes the three corresponding normalized running gauge
couplings $g_1$, $g_2$ and $g_3$ at the low-energy boundary condition
from the given input parameters $M_Z$, $\aem^{\SM(5)}(M_Z)$,
$\as^{\SM(5)}(M_Z)$ and $G_F$ or $M_W$ as
\begin{align}
  g_1(Q) &= N_{g_Y} \, g_Y(Q) \,,
  & g_Y(Q) &= \frac{\sqrt{4\pi\aem(Q)}}{\cos\theta_W(Q)} \,,\\
  g_2(Q) &= N_{g_L} \, g_L(Q) \,,
  & g_L(Q) &= \frac{\sqrt{4\pi\aem(Q)}}{\sin\theta_W(Q)} \,,\\
  g_3(Q) &= N_{g_{\text{s}}} \, g_{\text{s}}(Q) \,,
  & g_{\text{s}}(Q) &= \sqrt{4\pi\as(Q)} \,.
\end{align}
Here, $\aem$ and $\as$
denote the \MSbar electromagnetic and strong coupling constants of the
non-SUSY model, respectively, and $\theta_W$ is the \MSbar\ weak
mixing angle.  The coefficients $N_{g_i}$ denote the potential
normalization factors defined in the \sarah model file.  The
renormalization scale $Q$, at which the gauge couplings are calculated,
can be specified using the \code{LowScale} variable in the model file.
The coupling constants of the model are related to the corresponding
ones of the SM with five active quark flavors,
$\aem^{\SM(5)}(Q)$ and $\as^{\SM(5)}(Q)$, which are input,
via the relations
\begin{align}
  \aem(Q) &=
  \frac{\aem^{\SM(5)}(Q)}{1 - \Delta\aem(Q)} \,,\\
  \as(Q) &=
  \frac{\as^{\SM(5)}(Q)}{1 - \Delta\as(Q)} \,.
\end{align}
The threshold corrections $\Delta\alpha_i(Q)$ have the form
\begin{align}
  \Delta\aem(Q) &=
  \frac{\aem}{2\pi} \sum_i C_i^{\text{em}} \log\frac{m_i}{Q}
  \,,\label{eq:alpha_em_general_threshold} \\
  \Delta\as(Q) &=
  \frac{\as}{2\pi} \sum_i C_i^{\text{s}} \log\frac{m_i}{Q} \,,
\end{align}
where the sum runs over all non-SM particles plus the top quark
with running \MSbar masses $m_i(Q)$.
The constants $C_i^{\text{em}}$ and $C_i^{\text{s}}$ depend on the
representation of the particle $i$ with respect to the Lorentz and gauge group.
The \MSbar\ weak mixing angle $\theta_W$ in the non-SUSY model is
determined either
\begin{itemize}
\item from the Fermi constant $G_F$ and $M_Z$ using the iterative
  approach described in Ref.~\cite{Degrassi:1990tu} taking into account the
  full 1-loop corrections and leading 2-loop SM
  corrections to $\Delta\hat{\rho}$ and $\Delta\hat{r}$, see
  \secref{sec:fmw}.
\item or from the running $W$ and $Z$ masses, which
  are obtained from the corresponding pole masses via a 1-loop
  calculation.  See \secref{sec:fmw} for more details.
\end{itemize}
If the considered BSM model does not contain the SM gauge
group as a factor, it is of course still possible to fix the gauge
couplings at the low-energy boundary condition by defining them to be
input parameters.
\begin{example}
  In a left-right-symmetric model with the gauge group $SU(3)_C\times
  SU(2)_L\times SU(2)_R\times U(1)_L\times U(1)_R$, one could for
  example fix the running BSM gauge couplings $g_3$, $g_L$, $g_R$, $g_{1L}$
  and $g_{1R}$ by the running SM-like gauge couplings $g_3^\SM$,
  $g_2^\SM$ and $g_1^\SM$ which are given as input via the
  \code{EXTPAR} block:
\begin{lstlisting}
EXTPAR = {
  {100, g1SMInput},
  {101, g2SMInput},
  {102, g3SMInput}
};

LowScaleInput = {
   {g3, g3SMInput},
   {gL, g2SMInput},
   {gR, g2SMInput},
   {g1L, g1SMInput g2SMInput / Sqrt[2 (-g1SMInput^2 + g2SMInput^2)]},
   {g1R, g1SMInput g2SMInput / Sqrt[2 (-g1SMInput^2 + g2SMInput^2)]},
   {g1L1R, 0},
   {g1R1L, 0}
};
\end{lstlisting}
Note that, in addition to fixing the BSM gauge couplings, in this example
the off-diagonal gauge couplings $g_{1L1R}$ and $g_{1R1L}$ that arise due to
$U(1)$ mixing are being set to zero at the low-energy scale.
\end{example}

The \MSbar\ Yukawa couplings $Y_f(Q)$ of the SM-like
fermions $f$ in the non-SUSY model are determined from the
corresponding \MSbar\ masses $m_f(Q)$ using the tree-level
relation. For example, in the SM this relation reads
\begin{align}
  y_f^{\SM}(Q) = \frac{\sqrt{2} m_f^{\SM}(Q)}{v^{\SM}(Q)} \,,
\end{align}
with $f = u,d,c,s,t,b,e,\mu,\tau$ and the \MSbar vacuum expectation value $v^{\SM}(Q)$.
The running top quark \MSbar\ mass in the non-SUSY model, $m_t(Q)$, is
calculated from the top pole mass $M_t$
using the full 1-loop self energy plus 2-loop SM QCD
corrections as shown in Eq.\ \eqref{eq:mt_MSbar}.
In the SM, 3-loop QCD contributions can be taken into
account as well, see \secref{sec:model_specific_contributions}.
The bottom quark \MSbar\ mass in the non-SUSY model, $m_b(Q)$, is
obtained from the \MSbar\ mass $m_b^{\SM(5)}(m_b)$ in the SM with $5$ active quark flavors by first evolving
$m_b^{\SM(5)}(m_b)$ to the scale $Q$ using the 1-loop QED and 3-loop
QCD RGEs.  Afterwards, $m_b^{\SM(5)}(Q)$ is converted to $m_b(Q)$ as
\begin{align}
  m_b(Q) &= \frac{m_b^{\SM(5)}(Q)}{1 - \Delta m_b} \,, \\
  \Delta m_b &= \re\Sigma_{b}^S(p^2=(m_b^{\SM(5)})^2,Q)/m_b \notag\\
  &\phantom{={}}+ \re\Sigma_{b}^L(p^2=(m_b^{\SM(5)})^2,Q)
  + \re\Sigma_{b}^R(p^2=(m_b^{\SM(5)})^2,Q) \,,
  \label{eq:bottom-conversion}
\end{align}
where $\Sigma_{b}^{S,L,R}$ are the scalar, left- and right-handed
parts of the 1-loop bottom quark self energy in the \MSbar\ scheme, in
which all loops that contain only SM(5) particles are omitted.  Finally, the
\MSbar\ mass of the $\tau$ lepton, $m_\tau(Q)$, is calculated by first
identifying the $\tau$ pole mass, $M_\tau$, with the \MSbar\ mass in
the SM with 5 active quark flavors at the scale $M_\tau$,
\begin{align}
  m_\tau^{\SM(5)}(M_\tau) &=  M_\tau \,.
\end{align}
In this identification, the 1-loop SM electroweak
corrections to $m_\tau^{\SM(5)}(M_\tau)$ are neglected.  Afterwards,
$m_\tau^{\SM(5)}(M_\tau)$ is evolved to the scale $Q$ using the 1-loop QED RGE
and $m_\tau^{\SM(5)}(Q)$ is converted to $m_\tau(Q)$ as
\begin{align}
  m_\tau(Q) &= \frac{m_\tau^{\SM(5)}(Q)}{1 - \Delta m_\tau} \,, \\
  \Delta m_\tau &=
  \re\Sigma_{\tau}^S(p^2=(m_\tau^{\SM(5)})^2,Q) / m_\tau^{\SM(5)}(Q) \notag \\
  &\phantom{={}}
  + \re\Sigma_{\tau}^L(p^2=(m_\tau^{\SM(5)})^2,Q)
    + \re\Sigma_{\tau}^R(p^2=(m_\tau^{\SM(5)})^2,Q) \,,
\end{align}
where $\Sigma_{\tau}^{S,L,R}$ are the scalar, left- and right-handed
parts of the 1-loop $\tau$ self energy in the \MSbar\ scheme,
from which all loops that contain only SM(5) particles are omitted.

In most models, it is necessary to also fix the running SM-like vacuum
expectation value (VEV), $v$, at the low-energy scale.  For this purpose \fs
provides the symbols \code{MZMSbar} and \code{MWMSbar} in the model
file to access the \MSbar $W$ and $Z$ masses $m_W$ and $m_Z$ in the
non-SUSY model at the low-energy scale.  These running masses can be
used to calculate the \MSbar vacuum expectation value $v$, as for
example in the SM,
\begin{align}
  v(Q) &= \frac{2 m_Z(Q)}{\sqrt{3 g_1^2(Q) / 5+ g_2^2(Q)}} \,.
\end{align}

\subsection{Structure of the generated code}

In analogy to SUSY models, the parameters of a non-supersymmetric
model are distributed among two classes in the model class
hierarchy, see \figref{fig:parameter-classes}:
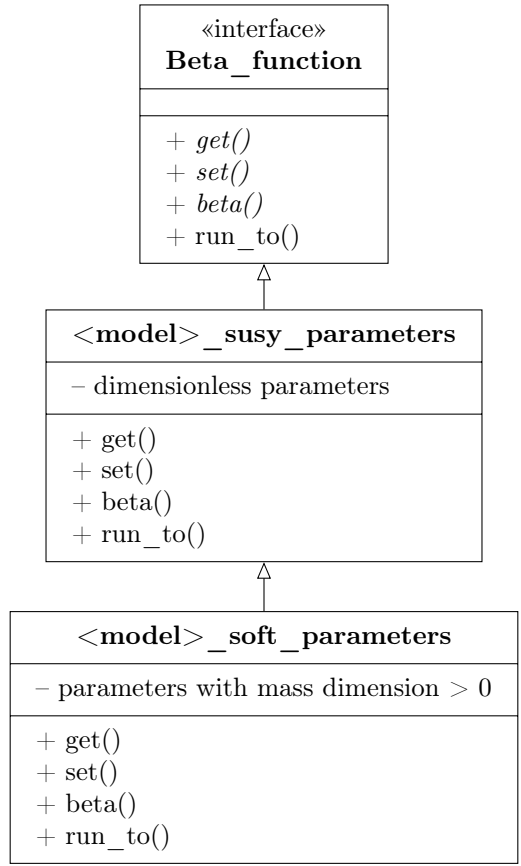
\begin{figure}[tbh]
  \centering
  \tikzumlset{fill class=white}
  \begin{tikzpicture}
    \umlclass[x=0, y=8, type=interface]{Beta\_function}{
    }{
      + \umlvirt{get()}\\
      + \umlvirt{set()}\\
      + \umlvirt{beta()}\\
      + run\_to()
    }
    \umlclass[x=0, y=4]{<model>\_susy\_parameters}{
      -- dimensionless parameters
    }{
      + get()\\
      + set()\\
      + beta()\\
      + run\_to()
    }
    \umlclass[x=0, y=0]{<model>\_soft\_parameters}{
      -- parameters with mass dimension > 0
    }{
      + get()\\
      + set()\\
      + beta()\\
      + run\_to()
    }
    \umlinherit{<model>\_susy\_parameters}{Beta\_function}
    \umlinherit{<model>\_soft\_parameters}{<model>\_susy\_parameters}
  \end{tikzpicture}
  \caption{Model class hierarchy.}
  \label{fig:parameter-classes}
\end{figure}
At the top of the model class hierarchy stands the
\code{Beta_function} interface class, which defines the interface for
the RGE integrator of the model parameters.  It provides the interface
function \code{run_to()}, which integrates the RGEs up to a given
scale using an adaptive Runge-Kutta algorithm.  The Runge-Kutta
algorithm makes use of the virtual functions \code{get()},
\code{set()} and \code{beta()} to obtain the model parameters at the
current renormalization scale, set the model parameters to new values
or calculate the $\beta$ functions.  These virtual functions are
implemented by the derived classes, \code{<model>_susy_parameters} and
\code{<model>_soft_parameters}.  Dimensionless parameters, like
gauge, Yukawa or quartic scalar couplings, are contained in the
\code{<model>_susy_parameters} class.  Parameters with mass dimension
greater than zero are contained in the \code{<model>_soft_parameters}
class.  The distribution of the model parameters between these two
classes reflects the dependency of the
$\beta$ functions of the dimensionful parameters upon the
dimensionless parameters.  Furthermore, it allows
the RGEs of the dimensionless parameters to be integrated independently of the
dimensionful parameters.

\subsection{Application: High-scale MSSM with light Higgs sector}
\label{sec:application_THDM}

As an application of \fbsm we consider the Higgs pole mass prediction
in an MSSM scenario with very heavy sfermions, Higgsinos and gauginos
at the SUSY scale $\MS$, but a light Higgs sector.  If $\MS$ is larger
than a few TeV, an EFT approach should be considered and the heavy SUSY
particles should be integrated out at $\MS$.  The resulting EFT below $\MS$ is
the Two-Higgs-Doublet-Model (\THDM).  Our aim is to calculate the Higgs pole
masses in this effective \THDM, where the quartic Higgs couplings are fixed by the
MSSM at $\MS$.\footnote{This \fs setup was also used in
  Ref.~\cite{Bagnaschi:2015pwa} to study the vacuum stability at very
  high SUSY scales in different \THDM variants with the MSSM as a
  supersymmetric UV completion.}  We use a \THDM of type II here, for
which the full 1- and leading 2-loop threshold corrections at the SUSY
scale are known \cite{Haber:1993an,Gorbahn:2009pp,Lee:2015uza}.

\begin{figure}[tbh]
  \centering
  \includegraphics[width=0.49\textwidth]{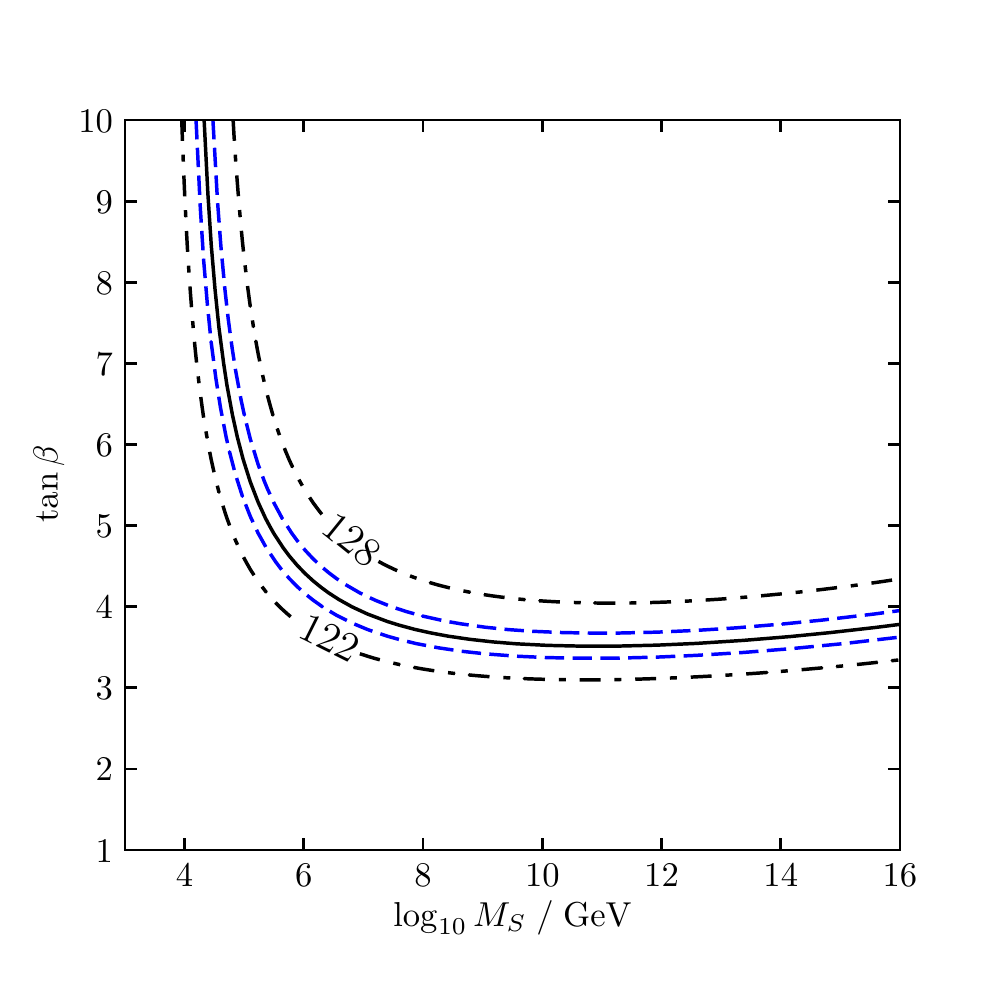}\hfill
  \includegraphics[width=0.49\textwidth]{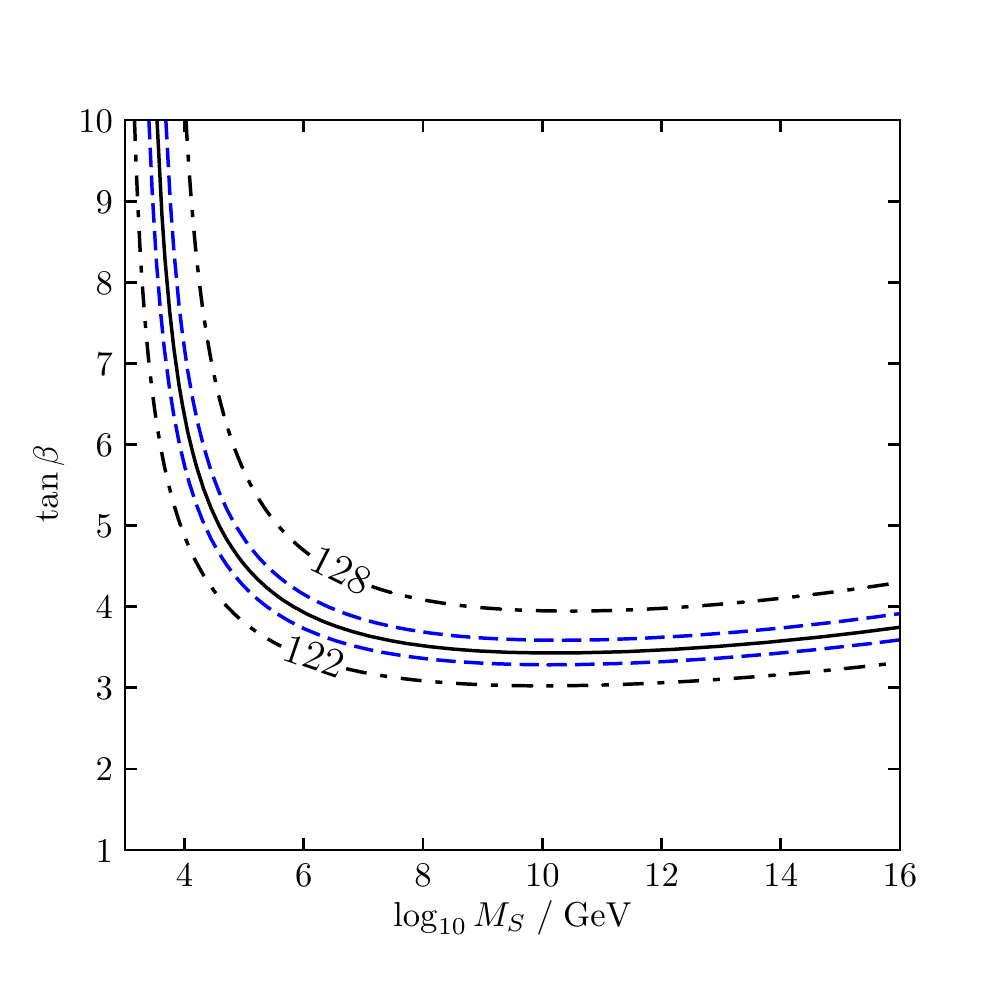}
  \caption{Lightest \CP-even Higgs pole mass calculated at the 1-loop
    level in the effective \THDM setup as a function of $\tan\beta$ and
    $\MS$ for $m_A(M_t) = 200\unit{GeV}$, $A_t = \mu/\tan\beta +
    X_t$, $A_b = A_\tau = A_t$, $\mu = \MS$, $M_t = 173.34\unit{GeV}$
    and $\as^{\SM(5)}(M_Z) = 0.1184$.  The left panel shows the results for $X_t
    = 0$ and the right panel for $X_t = \sqrt{6}\MS$.  The solid line
    corresponds to a Higgs pole mass of $125\unit{GeV}$ and the dashed
    lines to $124\unit{GeV}$ and $126\unit{GeV}$, respectively.}
  \label{fig:THDMIIMSSMBC_Mh_MS}
\end{figure}

In order to construct such an EFT setup we have to build the \THDM-II
with \sarah.  We start by specifying the gauge group, the field
content and the Lagrangian:
\begin{lstlisting}
(* gauge groups *)
Gauge[[1]]={B,   U[1], hypercharge, g1,False};
Gauge[[2]]={WB, SU[2], left,        g2,True};
Gauge[[3]]={G,  SU[3], color,       g3,False};

(* field content *)
FermionFields[[1]] = {q , 3, {uL,dL}   , 1/6, 2,  3};
FermionFields[[2]] = {l , 3, {vL,eL}   ,-1/2, 2,  1};
FermionFields[[3]] = {d , 3, conj[dR]  , 1/3, 1, -3};
FermionFields[[4]] = {u , 3, conj[uR]  ,-2/3, 1, -3};
FermionFields[[5]] = {e , 3, conj[eR]  ,   1, 1,  1};
ScalarFields[[1]]  = {H1, 1, {H1p, H10}, 1/2, 2,  1};
ScalarFields[[2]]  = {H2, 1, {H2p, H20}, 1/2, 2,  1};

DEFINITION[GaugeES][Additional] = {
    {LagHC  , { AddHC->True  }},
    {LagNoHC, { AddHC->False }}
};

LagNoHC = -(M112 conj[H1].H1 + M222 conj[H2].H2 \
            + Lambda1 conj[H1].H1.conj[H1].H1 \
            + Lambda2 conj[H2].H2.conj[H2].H2 \
            + Lambda3 conj[H2].H2.conj[H1].H1 \
            + Lambda4 conj[H2].H1.conj[H1].H2 );

LagHC = -(-M122 conj[H1].H2
          + Lambda5/2 conj[H2].H1.conj[H2].H1
          + Lambda6 conj[H1].H1.conj[H1].H2
          + Lambda7 conj[H2].H2.conj[H1].H2
          + Yd conj[H1].d.q + Ye conj[H1].e.l + Yu H2.u.q);
\end{lstlisting}
The neutral components of the two Higgs doublets acquire vacuum
expectation values $v_1$ and $v_2$:
\begin{lstlisting}
DEFINITION[EWSB][VEVs] = {
    {H10, {v1, 1/Sqrt[2]},
          {sigma1, \[ImaginaryI]/Sqrt[2]}, {phi1, 1/Sqrt[2]}},
    {H20, {v2, 1/Sqrt[2]},
          {sigma2, \[ImaginaryI]/Sqrt[2]}, {phi2, 1/Sqrt[2]}}
};
\end{lstlisting}
and the Higgs field components mix to \CP-even, \CP-odd and charged
Higgs mass eigenstates \code{hh}, \code{Ah} and \code{Hm},
respectively:
\begin{lstlisting}
DEFINITION[EWSB][MatterSector] = {
    {{phi1, phi2}          , {hh, ZH}},
    {{sigma1, sigma2}      , {Ah, ZA}},
    {{conj[H1p], conj[H2p]}, {Hm, ZP}}
};
\end{lstlisting}
Now we need a \fs model file in which we specify the boundary
conditions for all \THDM-II parameters.  As input we use the \MSbar
parameter
$\tan\beta^{\THDMII}(M_t)$, the MSSM \DRbar parameters
$\mu(\MS)$ and $A_f(\MS)$
($f=t,b,\tau$) at the SUSY scale, and the \MSbar \CP-odd Higgs mass
$m_A^{\THDMII}(\MEWSB)$, where \MEWSB is the scale of the electroweak
symmetry breaking, at which we calculate the light \CP-even Higgs pole mass in the
end:
\begin{lstlisting}
MINPAR = {
    {3, TanBeta}
};

EXTPAR = {
    {0, MSUSY},
    {1, MEWSB},
    {2, MuInput},
    {6, MAInput},
    {7, AtInput},
    {8, AbInput},
    {9, AtauInput}
};
\end{lstlisting}
At the low-energy scale
$M_t$, we let \fs calculate the gauge and Yukawa couplings of the
\THDMII and we fix the two Higgs VEVs using
$m_Z^{\THDMII}(M_t)$ and $\tan\beta^{\THDMII}(M_t)$:
\begin{lstlisting}
LowScale = LowEnergyConstant[MT];

LowScaleInput = {
   {Yu, Automatic},
   {Yd, Automatic},
   {Ye, Automatic},
   {v1, 2 MZMSbar / Sqrt[GUTNormalization[g1]^2 g1^2 + g2^2] \
        Cos[ArcTan[TanBeta]]},
   {v2, 2 MZMSbar / Sqrt[GUTNormalization[g1]^2 g1^2 + g2^2] \
        Sin[ArcTan[TanBeta]]}
};
\end{lstlisting}
At the scale $\MEWSB$, we fix the \code{M122} parameter using the input
value of $m_A^{\THDMII}(\MEWSB)$ and we impose the EWSB conditions by
fixing \code{M112} and \code{M222}:
\begin{lstlisting}
EWSBOutputParameters = { M112, M222 };

SUSYScale = MEWSB;

SUSYScaleInput = {
    {M122, MAInput^2 Sin[ArcTan[v2/v1]] Cos[ArcTan[v2/v1]]},
    FSSolveEWSBFor[EWSBOutputParameters]
};
\end{lstlisting}
Finally, we need to fix the quartic Higgs couplings of the \THDMII at
the scale $\MS$.  The necessary relations between the MSSM parameters
and the quartic Higgs couplings of the \THDMII are known at the full
1-loop and leading 2-loop level
\cite{Haber:1993an,Gorbahn:2009pp,Lee:2015uza}.
We can use expressions from these references to write the boundary conditions
on the quartic Higgs couplings at the SUSY scale,
shown in lines 23--57 of \appref{app:THDMIIMSSMBC},
in terms of the 1- and 2-loop threshold corrections,
shown in lines 110--398 of the same listing
which displays the complete
\fs model file.
\figref{fig:THDMIIMSSMBC_Mh_MS} shows the lightest \CP-even
Higgs pole mass calculated at the 1-loop level with \fs in this EFT
setup as a function of $\tan\beta^{\THDMII}(M_t)$ and $\MS$ for
$m_A^{\THDMII}(\MEWSB) = 200\unit{GeV}$, $\MEWSB = M_t$ and maximal
stop mixing.  The figure shows that using this setup, \fs can reproduce
the results presented in the left panels of Figure~2 of
Ref.~\cite{Lee:2015uza}.  This EFT model is
distributed with the \fs package under the name \modelname{THDMIIMSSMBC}.

\subsection{Application: High-scale MSSM (\HSSUSY)}
\label{sec:application_HSSUSY}

\fbsm has already been applied in
Refs.~\cite{Bagnaschi:2015pwa,Athron:2016fuq,Bagnaschi:2017xid} to perform a
pure EFT calculation of the lightest \CP-even Higgs mass in the MSSM,
assuming that all SUSY particles are integrated out at a heavy SUSY
scale $\MS$.  The \fs spectrum generator constructed for this purpose
is called \HSSUSY and is based on \sarah's SM model file
(\modelname{SM}).  In \fs 1.2.3, \HSSUSY implemented the 1-loop and leading
2-loop threshold corrections of $O(\at\as + \at^2)$ to the quartic Higgs
coupling of the SM from
Refs.~\cite{Bagnaschi:2014rsa,Vega:2015fna} at the SUSY scale.  In
\fstwo, the generalized 2-loop expressions of $O(\at^2)$ for general stop
masses as well as the new 2-loop contributions from Ref.~\cite{Bagnaschi:2017xid},
which involve the bottom and tau Yukawa couplings, are
included.  As a result, the version of \HSSUSY included in \fstwo uses
the 2-loop threshold corrections of
$O(\at\as + \ab\as + (\at+\ab)^2 + \ab\atau + \atau^2)$ for general
SUSY spectra.

In the \HSSUSY model file
(\code{model_files/HSSUSY/FlexibleSUSY.m.in}), these threshold
corrections are implemented as \mathematica expressions in the
high-scale boundary condition:
\begin{lstlisting}
HighScaleInput = {
    {\[Lambda], lambdaTree                     (* tree-level *)
                + lambda1LReg + lambda1LPhi    (* 1-loop *)
                + lambda1LChi1 + lambda1LChi2  (* 1-loop *)
                + lambda1Lbottom + lambda1Ltau (* 1-loop *)
                + ...                          (* 2-loop *)
    }
};

(* arXiv:1407.4081, Eq. (3) *)
lambdaTree = 1/4 (g2^2 + 3/5 g1^2) Cos[2 ArcTan[TanBeta]]^2;

(* arXiv:1407.4081, Eq. (9) *)
lambda1LReg = 1/(4 Pi)^2 (
    - 9/100 g1^4 - 3/10 g1^2 g2^2
    - (3/4 - Cos[2 ArcTan[TanBeta]]^2/6) * g2^4
    );

(* arXiv:1407.4081, Eq. (10) *)
lambda1LPhi = 1/(4 Pi)^2 (
    3 Yu[3,3]^2 (
        Yu[3,3]^2
        + 1/2 (g2^2-g1^2/5) Cos[2 ArcTan[TanBeta]]
      ) Log[msq2[3,3]/SCALE^2]
    + 3 Yu[3,3]^2 (
        Yu[3,3]^2
        + 2/5 g1^2 Cos[2 ArcTan[TanBeta]]
      ) Log[msu2[3,3]/SCALE^2]
    + Cos[2 ArcTan[TanBeta]]^2/300 (
        3 (g1^4 + 25 g2^4) (
            + Log[msq2[1,1]/SCALE^2]
            + Log[msq2[2,2]/SCALE^2]
            + Log[msq2[3,3]/SCALE^2]
        )
        + ...
      )
    + ...
    );

...
\end{lstlisting}
The full expressions for the threshold corrections can be found in the
\HSSUSY model file.  In addition, \HSSUSY makes use of the known
3-loop SM $\beta$ functions, 3-loop corrections to the
running top Yukawa coupling, 3-loop threshold corrections to the
strong coupling and up to 3-loop corrections to the Higgs pole mass:
\begin{lstlisting}
UseSM3LoopRGEs    = True; (* 3-loop RGEs *)
UseYukawa3LoopQCD = True; (* 3-loop thresholds for yt *)
UseSMAlphaS3Loop  = True; (* 3-loop thresholds for alpha_s *)
UseHiggs2LoopSM   = True; (* 2-loop contributions to Mh *)
UseHiggs3LoopSM   = True; (* 3-loop contributions to Mh *)
\end{lstlisting}
With all these corrections enabled, \HSSUSY can be regarded as an
improved variant of \susyhd \cite{Vega:2015fna}, the difference being
that \HSSUSY includes generalized 2-loop threshold corrections also
involving $\ab$ and $\atau$, which are not present in \susyhd.  In
\secref{sec:feft}, \HSSUSY is also compared to the fixed-order
calculation in the full MSSM as well as to \feft.  The following
Examples \ref{ex:HSSUSY_parallel}--\ref{ex:HSSUSY_EFT_uncertainty}
show how \HSSUSY can be run and how different sources of uncertainty
can be estimated.\footnote{The example scripts can be found in the
  \code{doc/examples-2.0/} sub-directory of the \fs package.}
\begin{example}[label=ex:HSSUSY_parallel]
  For illustration, we show in this example the Higgs mass prediction
  in the MSSM with \HSSUSY.  In the following example script, a scan
  over the relative \DRbar stop mixing parameter $X_t/\MS$ is performed for
  $\tan\beta(\MS) = 5$ and three different values of the SUSY scale $\MS$.
  This is done in parallel on all available CPU cores.
\begin{lstlisting}
Get["models/HSSUSY/HSSUSY_librarylink.m"];

CalcMh[TB_, Xt_, MS_] := Module[{handle, spectrum},
    handle = FSHSSUSYOpenHandle[
        fsSettings -> {
            precisionGoal -> 1.*^-5,
            calculateStandardModelMasses -> 1,
            poleMassLoopOrder -> 2,
            ewsbLoopOrder -> 2,
            betaFunctionLoopOrder -> 3,
            thresholdCorrectionsLoopOrder -> 2,
            poleMassScale -> 173.34
        },
        fsModelParameters -> {
            TanBeta -> TB,
            MEWSB -> 173.34,
            MSUSY -> MS,
            M1Input -> MS,
            M2Input -> MS,
            M3Input -> MS,
            MuInput -> MS,
            mAInput -> MS,
            AtInput -> (Xt + 1/TB) MS,
            msq2 -> MS^2 IdentityMatrix[3],
            msu2 -> MS^2 IdentityMatrix[3],
            msd2 -> MS^2 IdentityMatrix[3],
            msl2 -> MS^2 IdentityMatrix[3],
            mse2 -> MS^2 IdentityMatrix[3],
            LambdaLoopOrder -> 2,
            TwoLoopAtAs -> 1,
            TwoLoopAbAs -> 1,
            TwoLoopAtAb -> 1,
            TwoLoopAtauAtau -> 1,
            TwoLoopAtAt -> 1
        }
    ];
    spec = FSHSSUSYCalculateSpectrum[handle];
    FSHSSUSYCloseHandle[handle];
    If[spec =!= $Failed, Pole[M[hh]] /. (HSSUSY /. spec), 0]
];

LaunchKernels[];
DistributeDefinitions[CalcMh];

data = {
    ParallelMap[{#, CalcMh[5, #, 1000 ]}&, Range[-3.5, 3.5, 0.1]],
    ParallelMap[{#, CalcMh[5, #, 2000 ]}&, Range[-3.5, 3.5, 0.1]],
    ParallelMap[{#, CalcMh[5, #, 10000]}&, Range[-3.5, 3.5, 0.1]]
};
\end{lstlisting}
When plotting this data, the following figure results:
\begin{center}
  \includegraphics[width=0.5\textwidth]{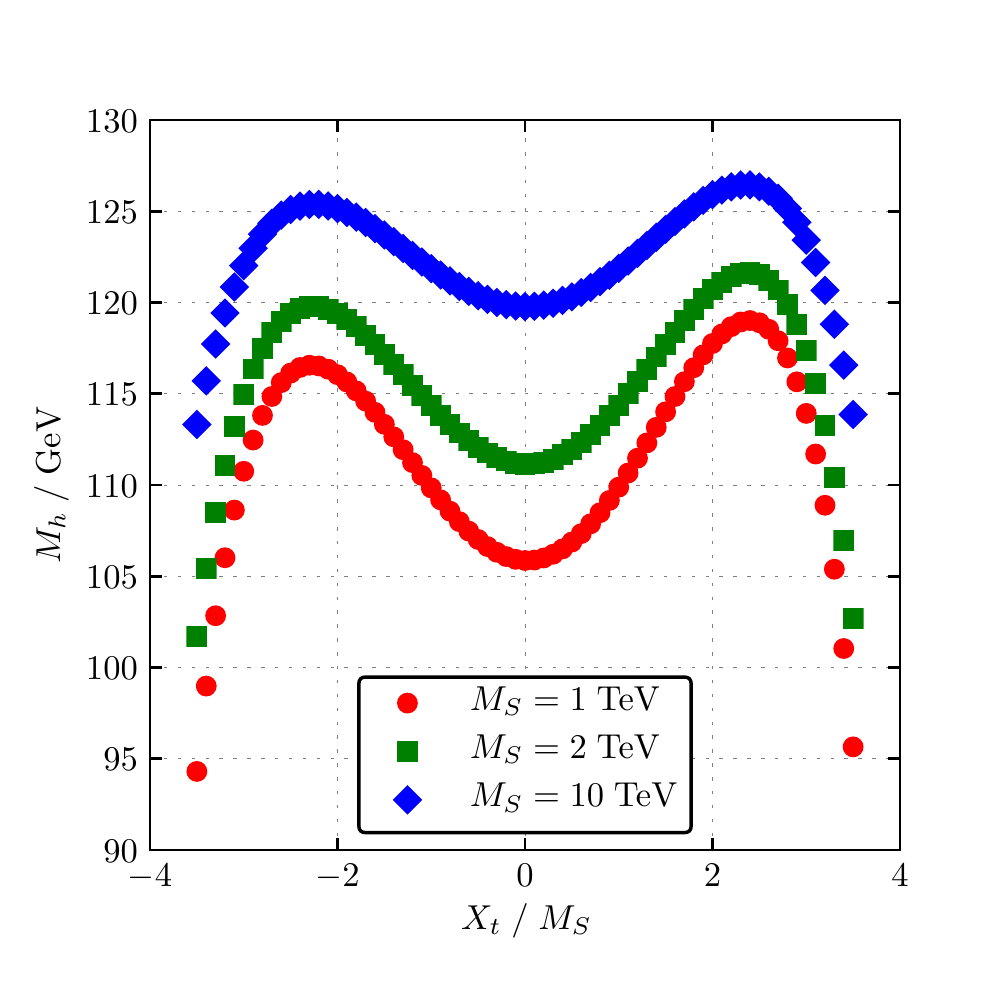}
\end{center}
\end{example}

\begin{example}[label=ex:HSSUSY_uncertainty]
  In this example, we slightly extend \exref{ex:HSSUSY_parallel}
  to make a partial uncertainty estimate of the Higgs pole
  mass predicted by \HSSUSY at the 2-loop level.  We do this by varying the
  threshold correction loop orders which determine the running \MSbar
  top Yukawa coupling
  $y_t$ and the strong coupling in the SM from 2-loop to
  3-loop.  In addition, the renormalization scale,
  at which the Higgs pole mass is calculated,
  is varied by a factor $2$.  The uncertainty
  estimated in this way is referred to as ``Standard Model
  uncertainty'' in the literature \cite{Vega:2015fna,Athron:2016fuq}
  and is one part of the full uncertainty of the EFT calculation
  of \HSSUSY.
  \begin{lstlisting}
Get["models/HSSUSY/HSSUSY_librarylink.m"];

(* generate logarithmically spaced range [start, stop] *)
LogRange[start_, stop_, steps_] :=
    Exp /@ Range[Log[start], Log[stop],
                 (Log[stop] - Log[start])/steps];

(* generate logarithmically spaced range [Q / 2, 2 Q] *)
GenerateScales[Q_] := LogRange[Q/2, 2 Q, 10];

CalcMh[MS_, TB_, Xt_, ytLoops_, asLoops_, Qpole_] :=
    Module[{handle, spec},
    handle = FSHSSUSYOpenHandle[
        fsSettings -> {
            precisionGoal -> 1.*^-5,
            calculateStandardModelMasses -> 1,
            poleMassLoopOrder -> 2,
            ewsbLoopOrder -> 2,
            betaFunctionLoopOrder -> 3,
            thresholdCorrectionsLoopOrder -> 3,
            poleMassScale -> Qpole,
            thresholdCorrections -> 120111021 +
                ytLoops * 10^6 + asLoops * 10^2
        },
        fsModelParameters -> {
            TanBeta -> TB,
            MEWSB -> 173.34,
            MSUSY -> MS,
            M1Input -> MS,
            M2Input -> MS,
            M3Input -> MS,
            MuInput -> MS,
            mAInput -> MS,
            AtInput -> (Xt + 1/TB) * MS,
            msq2 -> MS^2 IdentityMatrix[3],
            msu2 -> MS^2 IdentityMatrix[3],
            msd2 -> MS^2 IdentityMatrix[3],
            msl2 -> MS^2 IdentityMatrix[3],
            mse2 -> MS^2 IdentityMatrix[3],
            LambdaLoopOrder -> 2,
            TwoLoopAtAs -> 1,
            TwoLoopAbAs -> 1,
            TwoLoopAtAb -> 1,
            TwoLoopAtauAtau -> 1,
            TwoLoopAtAt -> 1
        }
    ];
    spec = FSHSSUSYCalculateSpectrum[handle];
    FSHSSUSYCloseHandle[handle];
    If[spec =!= $Failed, Pole[M[hh]] /. (HSSUSY /. spec), 0]
];

(* calculate Higgs mass with uncertainty estimate *)
CalcDMh[MS_, TB_, Xt_] :=
    Module[{Mh, MhYt3L, MhAs3L, varyQpole, DMh},
           Mh     = CalcMh[MS, TB, Xt, 2, 2, 0];
           MhYt3L = CalcMh[MS, TB, Xt, 3, 2, 0];
           MhAs3L = CalcMh[MS, TB, Xt, 2, 3, 0];
           varyQpole = CalcMh[MS, TB, Xt, 2, 2, #]& /@
                       GenerateScales[173.34];
           (* combine uncertainty estimates *)
           DMh = Max[Abs[Max[varyQpole] - Mh],
                     Abs[Min[varyQpole] - Mh]] +
                 Abs[Mh - MhYt3L] + Abs[Mh - MhAs3L];
           { Mh, DMh }
          ];

LaunchKernels[];
DistributeDefinitions[CalcDMh];

data = {
    ParallelMap[{#, CalcDMh[1000 , 5, #]}&, Range[-3.5, 3.5, 0.1]],
    ParallelMap[{#, CalcDMh[2000 , 5, #]}&, Range[-3.5, 3.5, 0.1]],
    ParallelMap[{#, CalcDMh[10000, 5, #]}&, Range[-3.5, 3.5, 0.1]]
};
\end{lstlisting}
In the function \code{CalcDMh[]}, the three sources of uncertainty are
combined linearly.  When drawing the uncertainty $\Delta
M_h^{\SM}$, estimated in this way, around the central value as $M_h
\pm \Delta M_h^{\SM}$, the following figure results:
\begin{center}
  \includegraphics[width=0.5\textwidth]{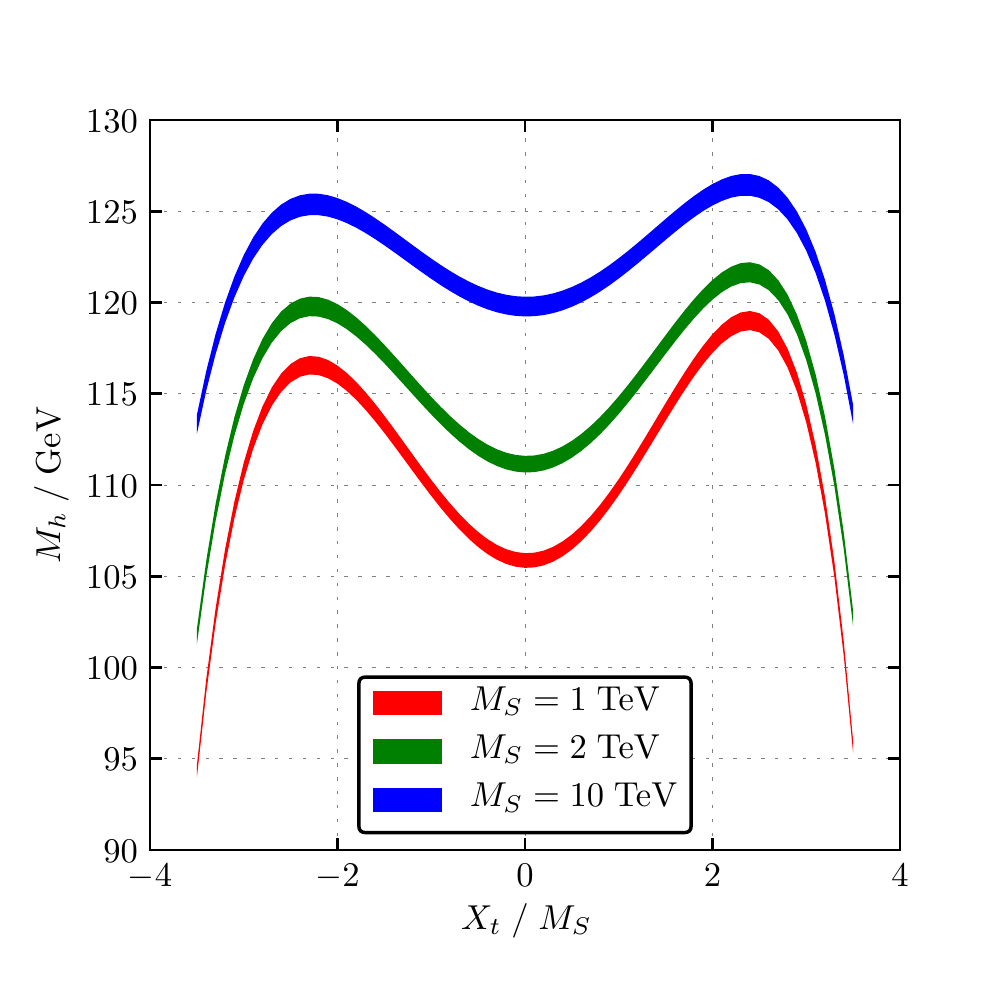}
\end{center}
We find that the uncertainty $\Delta
M_h^{\SM}$ is around or below $500\unit{MeV}$.
\end{example}

\begin{example}[label=ex:HSSUSY_EFT_uncertainty]
  In this example, we make an estimate of the ``EFT uncertainty''
  for the Higgs pole mass predicted by the pure-EFT calculation of
  \HSSUSY in the MSSM\@.  The pure-EFT calculation performed by \HSSUSY
  neglects terms of
  $O(v^2/\MS^2)$.  We estimate these missing terms by multiplying the
  individual 1-loop contributions by the term $(1 +
  \texttt{DeltaEFT}\times
  v^2/\MS^2)$, where \texttt{DeltaEFT} is an input parameter of
  \HSSUSY.  By varying \texttt{DeltaEFT} between $0$ and
  $1$ we obtain an estimate of the effect of these missing terms.
  This method has also been used in
  Refs.~\cite{Vega:2015fna,Athron:2016fuq,Bagnaschi:2017xid}.  The
  following code snippet illustrates how to calculate the Higgs mass
  with \HSSUSY and estimate the ``EFT uncertainty'' as a function of
  the SUSY scale for maximal stop mixing, $X_t =
  \sqrt{6}\MS$ and $\tan\beta = 5$.
  \begin{lstlisting}
Get["models/HSSUSY/HSSUSY_librarylink.m"];

(* generate logarithmically spaced range [start, stop] *)
LogRange[start_, stop_, steps_] :=
    Exp /@ Range[Log[start], Log[stop],
                 (Log[stop] - Log[start])/steps];

CalcMh[MS_, TB_, Xt_, deltaEFT_] :=
    Module[{handle, spec},
    handle = FSHSSUSYOpenHandle[
        fsSettings -> {
            precisionGoal -> 1.*^-5,
            calculateStandardModelMasses -> 1,
            poleMassLoopOrder -> 2,
            ewsbLoopOrder -> 2,
            betaFunctionLoopOrder -> 3,
            thresholdCorrectionsLoopOrder -> 3,
            thresholdCorrections -> 122111221
        },
        fsModelParameters -> {
            TanBeta -> TB,
            MEWSB -> 173.34,
            MSUSY -> MS,
            M1Input -> MS,
            M2Input -> MS,
            M3Input -> MS,
            MuInput -> MS,
            mAInput -> MS,
            AtInput -> (Xt + 1/TB) * MS,
            msq2 -> MS^2 IdentityMatrix[3],
            msu2 -> MS^2 IdentityMatrix[3],
            msd2 -> MS^2 IdentityMatrix[3],
            msl2 -> MS^2 IdentityMatrix[3],
            mse2 -> MS^2 IdentityMatrix[3],
            LambdaLoopOrder -> 2,
            TwoLoopAtAs -> 1,
            TwoLoopAbAs -> 1,
            TwoLoopAtAb -> 1,
            TwoLoopAtauAtau -> 1,
            TwoLoopAtAt -> 1,
            DeltaEFT -> deltaEFT
        }
    ];
    spec = FSHSSUSYCalculateSpectrum[handle];
    FSHSSUSYCloseHandle[handle];
    If[spec =!= $Failed, Pole[M[hh]] /. (HSSUSY /. spec), 0]
];

(* calculate Higgs mass with uncertainty estimate *)
CalcDMh[MS_, TB_, Xt_] :=
    Module[{Mh, MhEFT},
           Mh    = CalcMh[MS, TB, Xt, 0];
           MhEFT = CalcMh[MS, TB, Xt, 1];
           { Mh, Abs[Mh - MhEFT] }
          ];

LaunchKernels[];
DistributeDefinitions[CalcDMh];

data = ParallelMap[{#, Sequence @@ CalcDMh[#, 5, Sqrt[6]]}&,
                   LogRange[173.34, 10^5, 60]];
\end{lstlisting}
In the function \code{CalcDMh[]}, the difference between the Higgs masses
calculated with $\texttt{DeltaEFT} = 0$ and $\texttt{DeltaEFT} =
1$ is used as the uncertainty estimate ($\texttt{DeltaEFT} =
0$ corresponds to the standard \HSSUSY calculation).  When drawing the
uncertainty estimated in this way symmetrically around the central
value, the following figure results:
\begin{center}
  \includegraphics[width=0.5\textwidth]{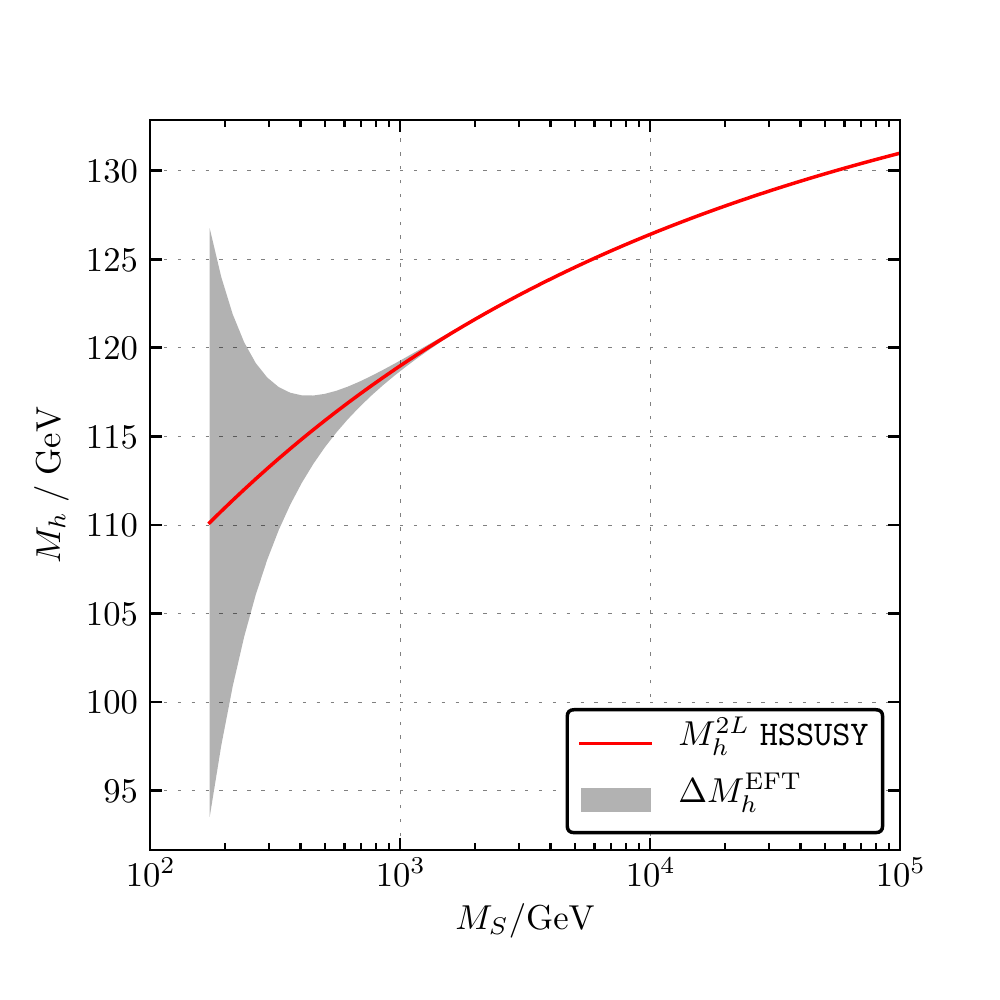}
\end{center}
As expected, we find that the ``EFT uncertainty'' decreases as
$\MS$ increases and falls below $150\unit{MeV}$ for $\MS \gtrsim 2\unit{TeV}$.
\end{example}

\subsection{Tests and comparisons with other spectrum generators}

We have performed various direct and indirect tests of \fbsm and its
components to ensure the correctness of the code:
\begin{itemize}
\item We have performed an analytic comparison of the RGEs generated
  with \sarah 4.5.3 for \fs's split-MSSM model file
  (\modelname{SplitMSSM}) against the RGEs presented in
  Ref.~\cite{Benakli:2013msa} and we found exact agreement.
\item We have performed a detailed numerical comparison of \fs's
  \HSSUSY model against \susyhd 1.0.1 and found excellent agreement
  \cite{Athron:2016fuq}.  The small differences between the two
  programs are of $O(100\unit{MeV})$ and originate from a
  different determination of $y_t$ in the SM at the low-energy scale,
  a different procedure to calculate the Higgs pole mass in the SM and
  the inclusion of additional 2-loop corrections in \HSSUSY which
  involve $\ab$ and $\atau$.
\item The effective \THDM shown in \secref{sec:application_THDM} and
  \appref{app:THDMIIMSSMBC} reproduces the results of
  Ref.~\cite{Lee:2015uza} and \MhEFT 1.0 \cite{MhEFT} for scenarios with
  heavy Higgsinos and gauginos, see for example
  \figref{fig:THDMIIMSSMBC_Mh_MS}.
\item The \fs package contains various EFT scenarios of the MSSM with
  boundary conditions from the literature (\HSSUSY,
  \modelname{SplitMSSM}, \modelname{THDMIIMSSMBC}, \modelname{HTHDMIIMSSMBC},
  \modelname{HGTHDMIIMSSMBC}).  For all these models, we have performed
  various analytic tests of the implemented MSSM boundary conditions,
  checking for example the renormalization scale dependence, relations
  among the parameters and threshold corrections at the
  1-loop level.
\item We have checked the numeric equality of the 3-loop MSSM $\beta$
  functions implemented in \fs and \softsusy 3.7.0.
\item We have checked the correctness of the renormalization scale
  dependent part of the 2-loop QCD corrections to the \MSbar top
  Yukawa coupling in the SM by deriving the 2-loop
  threshold corrections for $y_t$ from the SM to the MSSM
  and checking that no large logarithms appear
  \cite{ThomasKwasnitza:2016yqj}.
\item We have also analytically checked the correctness of the
  renormalization scale dependent part of the 2-loop and 3-loop QCD
  corrections to the \MSbar top Yukawa coupling in the SM
  by proving the renormalization scale invariance of the top quark
  pole mass in the SM at the 3-loop QCD level.
\end{itemize}

\section{\famu extension}
\label{sec:famu}

\begin{figure}[tbh]
  \centering
  \subfloat[Type FFS]{
    \includegraphics{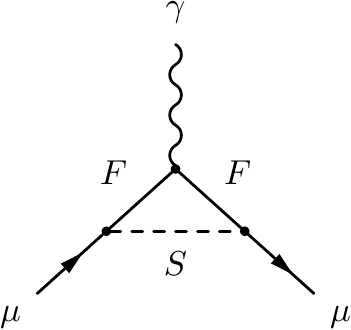}
    \label{fig:amu_diagram_I}
  }
  \qquad
  \subfloat[Type SSF]{
    \includegraphics{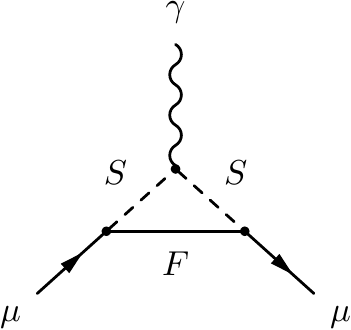}
    \label{fig:amu_diagram_II}
  }
  \caption{Generic diagram types implemented in \fs to calculate the
    1-loop contribution to $\amuBSM$.  The solid line in the loop
    represents any contributing non-SM fermion $F$ and
    the dashed line any contributing scalar particle $S$.}
  \label{fig:amu_diagram_types}
\end{figure}

\fstwo introduces a calculation of the BSM contributions to the
anomalous magnetic moment of the muon, \amuBSM, in the \MSbar/\DRbar
scheme at the 1-loop level in the model under consideration plus the
universal 2-loop QED contributions
\cite{Degrassi:1998es,vonWeitershausen:2010zr}.  The 1-loop
diagram types that are taken into account are shown in
\figref{fig:amu_diagram_types}.
The general result for these two Feynman diagram types reads, in a
notation based on
Refs.~\cite{Stockinger:2006zn,Martin:2001st,Fargnoli:2013zia},
\begin{align}
  \label{eq:amu BSM 1L FFS}
  \amu^{\BSM\text{,1L,FFS}} &=
  \frac{c}{(4\pi)^2} \frac{M_\mu m_\mu}{m_{S}^2}\left(
    \frac{1}{12} A_{SF} F_1^C(x) + \frac{m_F}{3m_\mu} B_{SF} F_2^C(x)
  \right) \,, \\
  \label{eq:amu BSM 1L SSF}
  \amu^{\BSM\text{,1L,SSF}} &=
  - \frac{c}{(4\pi)^2} \frac{M_\mu m_\mu}{m_{S}^2}\left(
    \frac{1}{12} A_{SF} F_1^N(x) + \frac{m_F}{6m_\mu} B_{SF} F_2^N(x)
  \right) \,,
\end{align}
where $x=m_F^2/m_{S}^2$ is the squared running mass ratio between $F$ and $S$, $M_\mu$ is the muon pole mass,
$m_\mu$ is the muon \MSbar/\DRbar mass and $c$ denotes the electric charge of
the particle coupling to the photon divided by the muon charge.  The
$A_{SF}$ and $B_{SF}$ constants are defined as
\begin{align}
  \label{eq:ASF}
  A_{SF} &= z_L z_L^* + z_R z_R^* \,,\\
  \label{eq:BSF}
  B_{SF} &= z_L z_R^* + z_R z_L^* \,,
\end{align}
where $z_L$ and $z_R$ are the left- and right-handed components of the
scalar--fermion--muon vertex, $\Gamma_{SF\mu} =
ie(z_L P_L + z_R P_R)$ with $e$ being the \MSbar/\DRbar electromagnetic coupling constant.
The loop functions $F_i^C(x)$ and $F_i^N(x)$ read
\begin{align}
  F_1^C(x) &= \frac{2}{(1-x)^4} \Big[ 2 + 3x - 6x^2 + x^3 + 6x\log x \Big], \\
  F_2^C(x) &= \frac{3}{2(1-x)^3} \Big[ -3 + 4x - x^2 - 2\log x \Big], \\
  F_1^N(x) &= \frac{2}{(1-x)^4} \Big[ 1 - 6x + 3x^2 + 2x^3 - 6x^2\log x \Big], \\
  F_2^N(x) &= \frac{3}{(1-x)^3} \Big[ 1 - x^2 + 2x\log x \Big].
\end{align}
To calculate $\amu^{\BSM\text{,1L}}$, \fs sums over all possible
instantiations of these two 1-loop diagram types with at least one
non-SM particle in the loop,
\begin{align}
  \label{eq:amu BSM 1L}
  \amu^{\BSM\text{,1L}} =
  \sum_{\substack{\text{FFS}\\\text{diagrams}}} \amu^{\BSM\text{,1L,FFS}}
  + \sum_{\substack{\text{SSF}\\\text{diagrams}}} \amu^{\BSM\text{,1L,SSF}} \,.
\end{align}
The calculation is performed at the renormalization scale $\MLCP$, which
is defined as the running mass of the lightest electrically charged
\BSM particle contributing to $\amu^{\BSM\text{,1L}}$.  In addition to
this 1-loop \BSM contribution, \famu adds the universal leading
logarithmic 2-loop QED contribution \cite{Degrassi:1998es},
\begin{align}
  \amu^{\QED,\text{2L}} = \amu^{\BSM\text{,1L}} \times \Bigg[
    16 \, \frac{\aem^\BSM(\MLCP)}{4\pi}
    \log\frac{M_\mu}{\MLCP}
  \Bigg] \,.
\end{align}
The overall result for \amuBSM is then given by the sum of 1-loop and
the 2-loop QED contribution,
\begin{align}
  \amuBSM = \amu^{\BSM\text{,1L}} + \amu^{\QED,\text{2L}} \,.
\end{align}

\subsection{Choosing \famu in the model file}

In order to calculate \amuBSM with \fs, the
\code{FlexibleSUSYObservable`aMuon} symbol has to be added to an
output block, see \tabref{tab:famu_options} and
\exref{ex:FlexibleAMU_in_model_file}.
\begin{OptionTable}{\famu model file options.}{tab:famu_options}[lllX]
    \multilinecell[t]{\code{FlexibleSUSYObservable`}\\\code{aMuon}}
    & -- & -- & Represents \amuBSM calculated with \famu \\
    \multilinecell[t]{\code{FlexibleSUSYObservable`}\\\code{aMuonGM2Calc}}
    & -- & -- & Represents \amuMSSM calculated with \GMTCalc at 2-loop level \\
    \multilinecell[t]{\code{FlexibleSUSYObservable`}\\\code{aMuonGM2CalcUncertainty}}
    & -- & -- & Represents the uncertainty $\Delta\amuMSSM$ calculated with \GMTCalc \\
    \bottomrule
\end{OptionTable}

\begin{example}[label=ex:FlexibleAMU_in_model_file]
In the following code snippet, the \code{FlexibleSUSYObservable`aMuon} symbol
is added to an output block named \code{FlexibleSUSYLowEnergy}.
The calculated \amuBSM will be written to the entry $21$ of this
block.
\begin{lstlisting}
ExtraSLHAOutputBlocks = {
   {FlexibleSUSYLowEnergy,
           {{21, FlexibleSUSYObservable`aMuon} } }
};
\end{lstlisting}
\end{example}

Note that \amuBSM is treated as an observable in \fs.  The
calculation of observables can be enabled/disabled using the flag $15$
of the \code{FlexibleSUSY} block.

As an alternative to \famu, \GMTCalc \cite{Athron:2015rva} can be used
to calculate \amuMSSM in \fs at the 2-loop level employing results from
Refs.~\cite{Fargnoli:2013zia,Fargnoli:2013zda,vonWeitershausen:2010zr}
in MSSM models without flavor violation. This MSSM-specific option was
first introduced in \fsbreak 1.3.0.  In order to let \GMTCalc calculate the
anomalous magnetic moment of the muon, the symbol
\code{FlexibleSUSYObservable`aMuonGM2Calc} must be added to an output
block.  In addition, the symbol
\code{FlexibleSUSYObservable`aMuonGM2CalcUncertainty} can be used to
calculate the estimated corresponding theory uncertainty for \amuMSSM.

\begin{example}
Calculating \amuMSSM with \GMTCalc in the \modelname{CMSSMNoFV} is
enabled by defining:
\begin{lstlisting}
ExtraSLHAOutputBlocks = {
   {FlexibleSUSYLowEnergy,
           {{1, FlexibleSUSYObservable`aMuonGM2Calc},
            {2, FlexibleSUSYObservable`aMuonGM2CalcUncertainty} } }
};
\end{lstlisting}
\end{example}

\GMTCalc is incorporated into \fs in the form of an addon.
In order to enable the \GMTCalc addon, the
\code{--with-addons=GM2Calc} argument can be passed to the
\code{configure} script during the \fs configuration step.

\subsection{Structure of \famu code}

The C++ interface for \famu is defined in the
\code{<model>_a_muon.hpp} file.  The interface consists of a single
function, which takes a model object as the argument and returns the value
of \amuBSM.

\begin{example}
In the \modelname{CMSSM}, the interface function reads:
\begin{lstlisting}
namespace CMSSM_a_muon {
   double calculate_a_muon(const CMSSM_mass_eigenstates& model);
}
\end{lstlisting}
\end{example}

\subsection{Tests and comparisons with other calculations}

We have tested \famu in the MSSM against \GMTCalc 1.3.3
\cite{Athron:2015rva}, which is currently the most precise code
available to calculate \amuMSSM.  We usually find a $1$--$10\%$
deviation depending on the parameter choice, see
\tabref{tab:test_famu}, except for specific parameter points, where
the deviation is much larger due to a large renormalization scale
dependence, see below.  The $1$--$10\%$ deviation is caused mainly by
the fact that \GMTCalc calculates \amuMSSM in an on-shell scheme and
includes all known 2-loop corrections, while \famu performs the
calculation in the \MSbar/\DRbar scheme at the lightest charged BSM particle scale $\MLCP$ and only
includes the leading logarithmic 2-loop QED correction.  A comparison
of \famu with \spheno 4.0.2 shows differences of up to $10\%$.  These
differences are caused by the fact that \famu includes the leading
logarithmic 2-loop QED contribution and by the different choice of the
renormalization scale: \spheno performs the calculation at the $Z$ pole
mass scale, while \famu performs it at $\MLCP$.

In specific parameter scenarios, where the \MSbar/\DRbar smuon masses
show a high sensitivity to the renormalization scale, the predictions
of $\amuBSM$ in \famu and \spheno are expected to have a large theory
uncertainty, see for example BM4$^\prime$ in \tabref{tab:test_famu}
and see the discussion in Ref.~\cite{Athron:2015rva}.  One
reason for this large uncertainty are the missing 2-loop corrections
in \famu and \spheno, which would (if included) cancel the
renormalization scale dependence at the 2-loop level.  To illustrate
the sensitivity on the renormalization scale, we show in the third
column of \tabref{tab:test_famu} the variation of $\amuBSM$ in \famu
when the scale is varied in the interval $[\MLCP/2,2\MLCP]$.  For BM4$^\prime$,
$\amuBSM$ varies by around $66\%$, which indicates a very imprecise
prediction for this point.  Such a large scale uncertainty is avoided
in \GMTCalc due to the renormalization in an on-shell scheme.  Another
reason for the larger uncertainty in \famu and \spheno is the
renormalization scheme used to renormalize the smuon masses: In the
\MSbar/\DRbar scheme, smuon self energy contributions which are
quadratic in the BSM particle masses can lead to large 1-loop
corrections to the smuon masses.  Such large corrections are avoided in
the on-shell scheme, where they are absorbed into the smuon mass
counter term.
\begin{table}[tbh]
  \centering
  \begin{tabular}{lrrrr}
    \toprule
    Point & \famu & \famu           & \GMTCalc & \spheno \\
          &       & scale variation &          &         \\
    \midrule
    SPS1a  & $29.77$ & $0.46$ & $29.31 \pm 2.36$ & $31.00$ \\
    SPS1b  & $32.46$ & $0.45$ & $32.38 \pm 2.40$ & $32.68$ \\
    SPS3   & $13.80$ & $0.12$ & $13.52 \pm 2.33$ & $14.99$ \\
    SPS4   & $50.02$ & $1.02$ & $52.45 \pm 2.64$ & $45.64$ \\
    BM1$^\prime$  & $42.08$ & $1.58$ & $42.34 \pm 2.33$ & $43.72$ \\
    BM2$^\prime$  & $25.79$ & $0.10$ & $25.67 \pm 2.32$ & $26.16$ \\
    BM3$^\prime$  & $27.81$ & $0.68$ & $27.95 \pm 2.34$ & $27.98$ \\
    BM4$^\prime$  &  $8.11$ & $5.41$ & $33.11 \pm 2.31$ &  $2.19$ \\
    \bottomrule
  \end{tabular}
  \caption{Comparison of $\amuMSSM\cdot 10^{10}$ calculated with \famu,
    \GMTCalc 1.3.3 and \spheno 4.0.2 for the CMSSM benchmark points
    presented in Ref.~\cite{Allanach:2002nj} and the parameter points shown
    in \tabref{tab:bm_amu}.  The third column shows the variation
    of $\amuBSM$ when the renormalization scale is varied between
    $\MLCP/2$ and $2\MLCP$.
    For BM4$^\prime$ the value of \amu calculated by the \DRbar
    programs \famu and \spheno suffers from a high renormalization scale
    sensitivity due to the large values of $M_2$ and $(m_{\tilde{l}})_{ii}$,
    which leads to a very imprecise result and to a huge deviation
    compared to \GMTCalc.}
  \label{tab:test_famu}
\end{table}
\begin{table}[tbh]
  \centering
  \begin{tabular}{lrrrr}
    \toprule
                & BM1$^\prime$  & BM2$^\prime$   & BM3$^\prime$   & BM4$^\prime$ \\
    \midrule
    $\mu$ / GeV & $350$ & $1300$ & $2000$ & $-160$ \\
    $\tan\beta$ & $40$  & $40$   & $40$   & $50$ \\
    $M_1$ / GeV & $150$ & $150$  & $150$  & $140$ \\
    $M_2$ / GeV & $300$ & $300$  & $300$  & $2000$ \\
    $(m_{\tilde{e}})_{ii}$ / GeV & $400$ & $400$  & $400$  & $200$ \\
    $(m_{\tilde{l}})_{ii}$ / GeV & $400$ & $400$  & $400$  & $2000$ \\
    $(m_{\tilde{q}})_{ii}$, $(m_{\tilde{u}})_{ii}$, $(m_{\tilde{d}})_{ii}$ / GeV
    & $400$ & $600$  & $700$ & $2000$ \\
    \bottomrule
  \end{tabular}
  \caption{Definition of the MSSM benchmark points BM1$^\prime$--BM4$^\prime$, inspired
    by the points BM1--BM4 presented in Ref.~\cite{Fargnoli:2013zda}.
    All parameters are defined in the \DRbar scheme at the scale
    $Q = 454.7\unit{GeV}$, except for $\tan\beta$, which is defined at $M_Z$.
    The trilinear couplings and off-diagonal
    elements of the sfermion mass parameters are set to zero and we
    have fixed $m_A = 2\unit{TeV}$.}
  \label{tab:bm_amu}
\end{table}

\section{\fcpv extension}
\label{sec:fcpv}
\subsection{Setting up a \fcpv model}

Since \fs 1.1.0, the model parameters are no longer restricted to be real,
but can be complex.  Whether a parameter is real or complex is
specified in the corresponding \sarah model file.  Parameters can be
forced to be treated as real in \fs by adding them to the
\code{RealParameters} list in the \fs model file, see
\tabref{tab:fcpv_options} and the following examples.  For
compatibility with \fs 1.0, the \code{RealParameters} list is by default
set to \code{\{All\}} meaning that all parameters are assumed to be real.
\begin{OptionTable}{\fcpv model file options.}{tab:fcpv_options}
    \code{RealParameters} & \code{\{ All \}} &
    List of model parameters or \code{\{\}} or \code{\{ All \}} &
    List of parameters to be treated as real\\
    \bottomrule
\end{OptionTable}

\begin{example}
In the MSSM, the $\mu$ parameter, the Yukawa
couplings, the soft-breaking trilinear couplings, the soft-breaking
scalar mass parameters, the soft-breaking gaugino mass parameters and
the $B\mu$ parameter can be complex.  In order to choose all of these
parameters to be complex, except for $B\mu$, one can set
\begin{lstlisting}
RealParameters = { B[\[Mu]] };
\end{lstlisting}
\end{example}

\begin{example}
In order to treat all MSSM parameters defined in
the \sarah model file for the MSSM as complex, set \code{RealParameters} to the
empty list:
\begin{lstlisting}
RealParameters = {};
\end{lstlisting}
\end{example}

\subsection{Application: CMSSM with \CP-violation}

In \sarah's \modelname{MSSM} model, the phase factor between the two Higgs
doublets is set to zero.  Therefore, this model does not allow for
\CP-violation in the Higgs sector.  In order to enable \CP-violation in
the MSSM Higgs sector, \sarah's \modelname{MSSM/CPV} model file can be
used, which allows for a non-zero relative phase factor $e^{i\eta}$
between the Higgs doublets.  In the \modelname{MSSM/CPV}, there are three
linearly independent EWSB equations.  Therefore, in this model three
EWSB output parameters have to be chosen.  In \fs's \modelname{CMSSMCPV}
model file, these are chosen to be $\re B\mu$, $\im B\mu$ and $|\mu|$
by setting the \code{EWSBOutputParameters} variable to
\begin{lstlisting}
EWSBOutputParameters = { Re[B[\[Mu]]], Im[B[\[Mu]]], \[Mu] };
\end{lstlisting}
Since only the magnitude of the $\mu$ parameter is fixed by the EWSB
equations, \fs introduces the phase of $\mu$ as a free parameter,
\code{Phase[\\[Mu]]} = $e^{i\phi_\mu}$.  This phase should be specified in
an SLHA-2 compliant way by reading the real and imaginary parts of
$e^{i\phi_\mu} = \cos\phi_\mu + i \sin\phi_\mu$ from the \code{MINPAR}
and \code{IMMINPAR} block entries $4$ and fixing $e^{i\phi_\mu}$ at the
SUSY scale:
\begin{lstlisting}
MINPAR   = { {4, CosPhiMu} };
IMMINPAR = { {4, SinPhiMu} };

SUSYScaleInput = {
    {Phase[\[Mu]], CosPhiMu + I SinPhiMu}
};
\end{lstlisting}
In the \modelname{CMSSMCPV}, the phase $\eta$ is read from the \code{EXTPAR}
block entry $100$ and also chosen to be input at the SUSY scale:
\begin{lstlisting}
EXTPAR = {
    {100, etaInput}
};

SUSYScaleInput = {
    {eta, etaInput}
};
\end{lstlisting}
The complete \fs \modelname{CMSSMCPV} model file can be found in
\appref{app:CMSSMCPV}.

\subsection{Application: Electric dipole moments of fermions}

\fstwo can calculate $\edmBSM{f}$, new physics contributions to the
electric dipole moment (EDM) of a fermion $f$, in the given model
in the \DRbar scheme at the 1-loop level.
The procedure is very similar to the calculation of \amuBSM
described in \secref{sec:famu}.
This is expected from the following effective Lagrangian,
\begin{equation}
  \Delta \mathcal{L}_\text{eff} =
  - \frac{D_f}{2}\bar{f}_L \sigma_{\mu\nu} f_R F^{\mu\nu} + \mathrm{h.c.},
\end{equation}
where the real and the imaginary parts of the Wilson coefficient
$D_f$ are proportional to the magnetic
(see, e.g., Refs.~\cite{Cho:2001hx,Baek:2002cc})
and the electric (see, e.g., Ref.~\cite{Pokorski:1999hz}) dipole moments
of $f$, respectively.
More precisely,
\begin{equation}
  a_f = - \frac{2 M_f}{e} \re D_f, \qquad
  \edm{f} = \im D_f,
\end{equation}
where $M_f$ is the pole mass of $f$
and $e$ is the running electromagnetic coupling constant in the BSM model.
It is then obvious that
the EDM is given by the sum of all FFS-type and SSF-type diagrams,
\begin{align}
  \edm{f}^{\BSM\text{,1L}} =
  \sum_{\substack{\text{FFS}\\\text{diagrams}}}
  \edm{f}^{\BSM\text{,1L,FFS}}
  + \sum_{\substack{\text{SSF}\\\text{diagrams}}}
  \edm{f}^{\BSM\text{,1L,SSF}}
  \,,
\end{align}
as in Eq.\ \eqref{eq:amu BSM 1L}.  The type of diagram refers to those
shown in \figref{fig:amu_diagram_types}, resulting in the contributions,
\begin{align}
  \frac{1}{e}\,\edm{f}^{\BSM\text{,1L,FFS}} &=
  \frac{c}{(4\pi)^2}
    \frac{m_F}{6{m_{S}^2}} \widetilde{B}_{SF} F_2^C(x)
  \,, \\
  \frac{1}{e}\,\edm{f}^{\BSM\text{,1L,SSF}} &=
  - \frac{c}{(4\pi)^2}
    \frac{m_F}{12{m_{S}^2}} \widetilde{B}_{SF} F_2^N(x)
  \,,
\end{align}
which are essentially the second terms of Eq.\ \eqref{eq:amu BSM 1L FFS} and
Eq.\ \eqref{eq:amu BSM 1L SSF}, respectively, divided by $2 M_f$, except that
the coupling factor is instead
\begin{equation}
  \widetilde{B}_{SF} = 2 \im ( z_L z_R^* ) .
\end{equation}
There are no imaginary parts corresponding to
the first terms of Eq.\ \eqref{eq:amu BSM 1L FFS} and Eq.\ \eqref{eq:amu BSM 1L SSF}
as can be guessed from Eq.\ \eqref{eq:ASF}.
The calculation is performed at the renormalization scale $\MS$ specified
in the model file, which is typically set to the stop mass scale
in supersymmetric models.

One can have the EDM of particle \code{f} calculated
by adding to an output block the form: \code{FlexibleSUSYObservable`EDM[f_]}.
Then $\edm{f}$ is reported in units of $\unit{GeV}^{-1}$.
\begin{example}
The output includes
the EDMs of the electron, muon, and tau if
the following code snippet is inserted into the \fs model file:
\begin{lstlisting}
ExtraSLHAOutputBlocks = {
   {FlexibleSUSYLowEnergy,
           {{23, FlexibleSUSYObservable`EDM[Fe[1]]},
            {24, FlexibleSUSYObservable`EDM[Fe[2]]},
            {25, FlexibleSUSYObservable`EDM[Fe[3]]} } }
};
\end{lstlisting}
\end{example}

\begin{figure}
  \centering
  \includegraphics[width=0.49\textwidth]{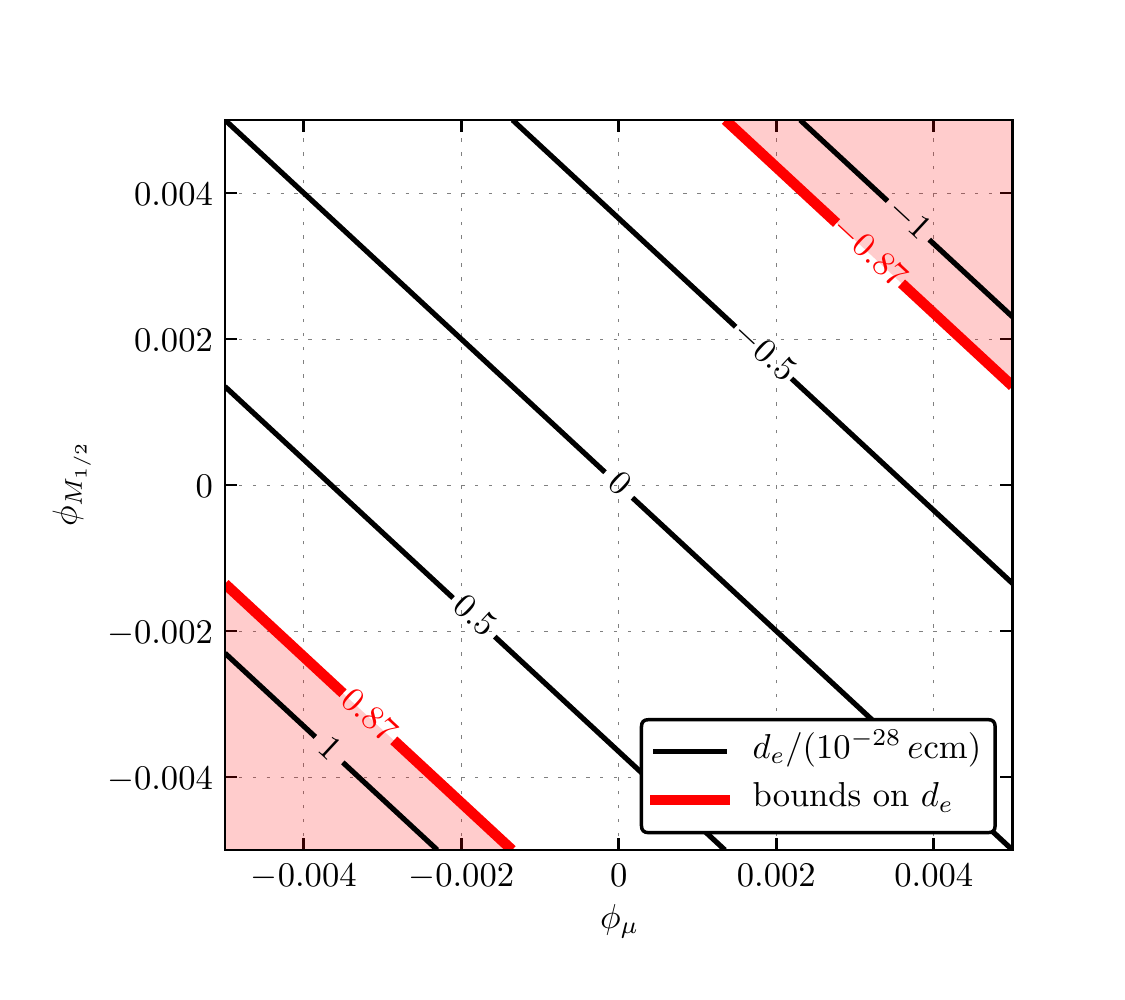}\hfill
  \includegraphics[width=0.49\textwidth]{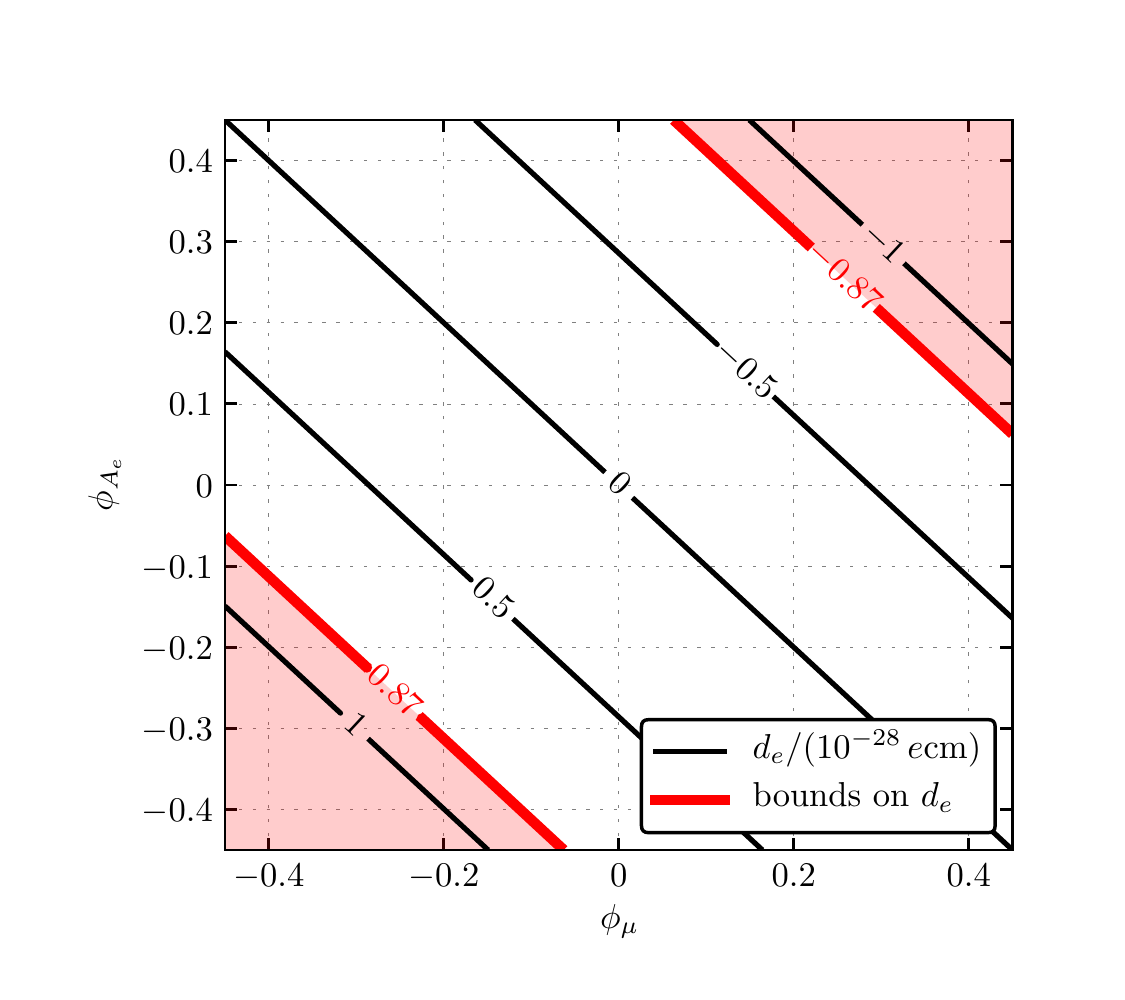}
  \caption{Left panel: contours of the electron EDM in the
    \CP-violating MSSM at the 1-loop level as a function of
    the phases of $\mhalf = M_1 = M_2$ and $\mu$,
    with $\mu A_e$ kept positive real.
    Right panel: same plot except for the vertical axis showing
    the $A_e$ phase and $\mu \mhalf$ kept positive real.
    In both panels, the other relevant parameters are:
    $\tan\beta = 10$, $|\mu| = |\mhalf| = \mzero = |A_e| = 2\unit{TeV}$.
    The thick red lines show the experimental upper bound on $|\edm{e}|$
    \cite{Baron:2013eja}.}
  \label{fig:EDM plots}
\end{figure}
Using a model file thus configured, one can do a quick test of
the electron EDM evaluation in the \CP-violating MSSM for instance.
In this model, only $\arg(\mu M_{1,2,3})$ and $\arg (\mu A_f)$ are
physical among the flavor-conserving phases
apart from those already present in the SM
(see, e.g., Refs.~\cite{Pokorski:1999hz,Stockinger:2006zn}).
For simplicity,
the gaugino masses are assumed to have an equal (complex) value
$\mhalf$, and both soft masses of the left- and right-handed
selectron as well as the approximate tree-level heavy Higgs masses%
\footnote{$\sqrt{\re\!B\mu / (\cos\!\beta \sin\!\beta)}$.}
shall be $\mzero$.
The moduli of $\mu$, $\mhalf$, $\mzero$, and $A_e$
are specified at the SUSY scale,
all of which including the scale are set to $2\unit{TeV}$.
Finally, we set $\tan\beta = 10$ at the $M_Z$ scale.

The resulting electron EDM is displayed as contours in
\figref{fig:EDM plots}.
In the left panel, $\phi_{\mhalf} \equiv \arg \mhalf$ and
$\phi_\mu \equiv \arg \mu$ are varied
while $\arg(\mu A_e)$ is constrained to be zero.
From the directions of the contours it is clear
that \fs correctly reproduces the behaviour of
$\edm{e}$ depending only on the rephasing invariant $\arg(\mu \mhalf)$.
In the right panel, the roles of $\phi_{A_e} \equiv \arg A_e$ and
$\phi_{\mhalf}$ are swapped so that $\arg(\mu \mhalf)$ stays at zero.
Again, the contours verify that
$\edm{e}$ from \fs is determined by the physical phase $\arg(\mu A_e)$.
For reference,
the experimental upper limit on $|\edm{e}|$ at the 90\% confidence level
is shown as the thick red lines \cite{Baron:2013eja}.

\subsection{Tests and comparisons with other spectrum generators}
\begin{figure}[tbh]
  \centering
  \includegraphics[width=0.49\textwidth]{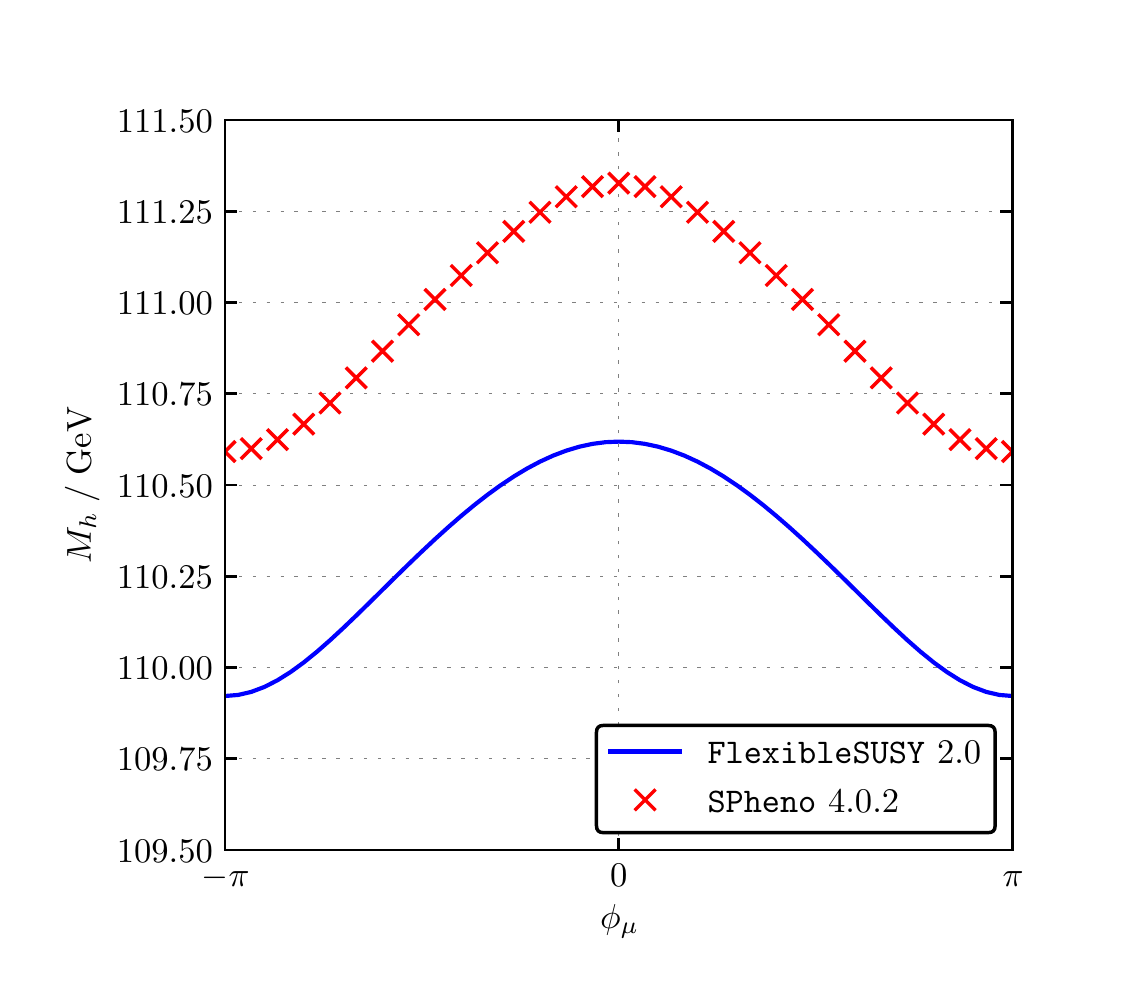}\hfill
  \includegraphics[width=0.49\textwidth]{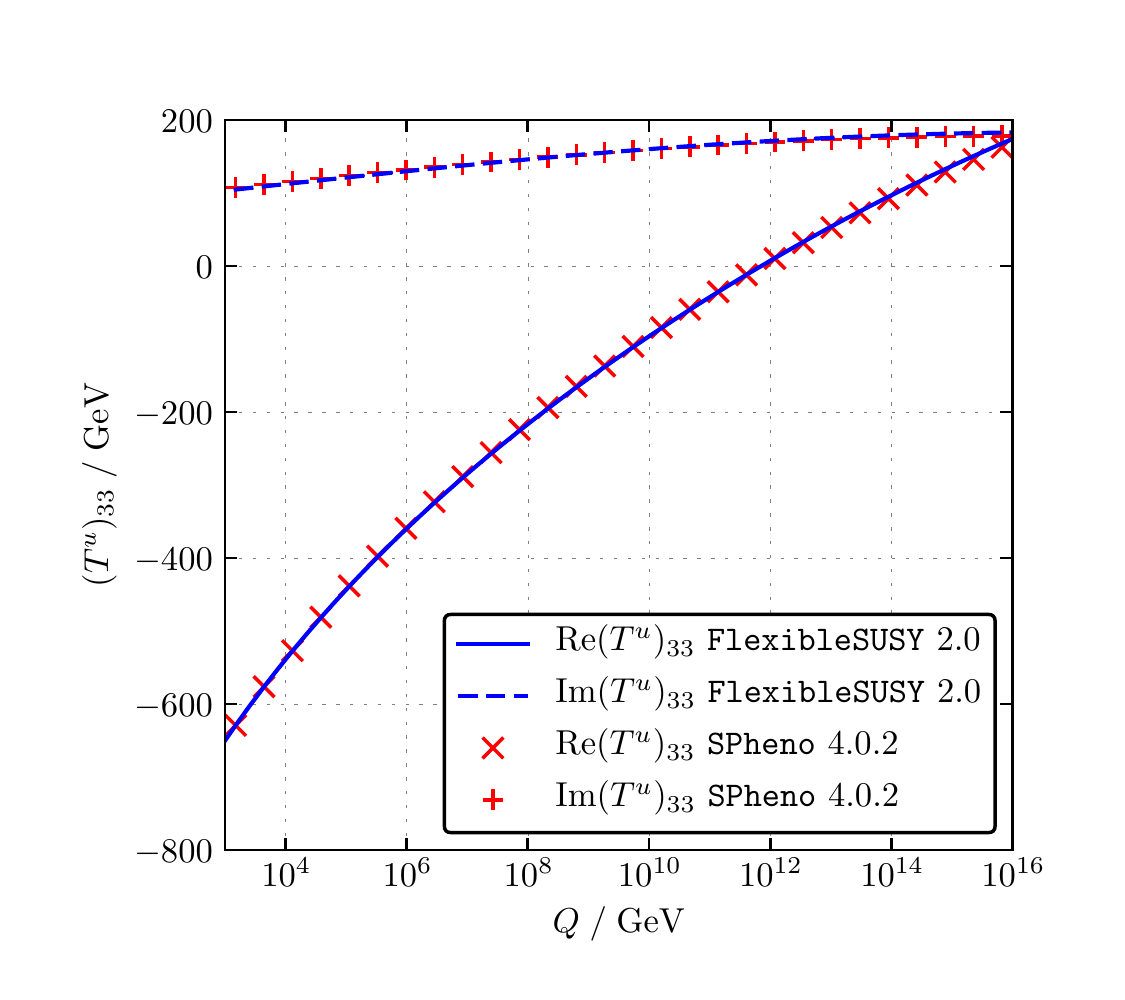}
  \caption{Left panel: lightest Higgs pole mass calculated in the
    \CP-violating CMSSM at the 1-loop level with \fstwo and
    \spheno 4.0.2 as a function of the phase angle of the $\mu$
    parameter for $\mzero = \mhalf = 500\unit{GeV}$, $\tan\beta = 10$
    and $\azero = 0$.  Right panel: renormalization group running of the
    trilinear coupling $(T^u)_{33}$ for $\mzero = \mhalf =
    500\unit{GeV}$, $\tan\beta = 10$, $\phi_\mu = 0$ and $\azero = 500\,
    e^{i \pi/4} \unit{GeV}$.}
  \label{fig:CMSSMCPV_plots}
\end{figure}

In the left panel of \figref{fig:CMSSMCPV_plots}, we show the lightest
Higgs pole mass calculated in the \CP-violating CMSSM at the 1-loop
level with \fstwo and \spheno~4.0.2 as a function of the phase angle
of the complex $\mu$ parameter, $\phi_\mu \equiv \arg \mu$.
We use a low-energy scenario with $\mzero = \mhalf = 500\unit{GeV}$,
$\tan\beta = 10$ and $\azero = 0$.  Even though this scenario is
excluded, the figure illustrates that both \fs and \spheno show the
same behaviour of the Higgs pole mass as a function of $\phi_\mu$.
The $\phi_\mu$-independent shift of around $0.7\unit{GeV}$ between the
Higgs masses calculated by the two programs is mainly caused by the
different treatment of higher-order corrections to
the running \DRbar top Yukawa coupling, which has been discussed in
Refs.~\cite{Staub:2015aea,ThomasKwasnitza:2016yqj}.

The right panel of \figref{fig:CMSSMCPV_plots} shows the
renormalization group running of the $(3,3)$-component of the complex
trilinear coupling $T^u$ for the same CMSSM scenario, except that we
use $\phi_\mu = 0$ and $\azero = 500\, e^{i \pi/4} \unit{GeV}$.  The
lines show the running in \fstwo and the dots the
running in \spheno 4.0.2.  We find very good agreement of
the running of $(T^u)_{33}$ between the two programs over the shown
$14$ orders of magnitude. The maximum deviation between \fs and
\spheno is around $2\%$ for $Q \approx 10^{16} \unit{GeV}$ and is
mainly caused by a different determination of the
dimensionless MSSM parameters at the electroweak scale.

\section{\fmw extension}
\label{sec:fmw}
\fmw is a major new feature which is released in \fstwo.
It allows a more accurate determination of the electroweak gauge
couplings $g_1$ and $g_2$, the prediction of the $W$ pole mass and thus
enables direct comparisons with electroweak precision data.

The running electroweak gauge couplings are related to the running
electromagnetic coupling~$e$, the weak mixing angle~$\theta_W$ and the
normalization factors~$N$ by
\begin{align}
  g_1 &= N_{g_Y}\,\frac{e}{\cos\theta_W}\,,
  & g_2 &= N_{g_L}\,\frac{e}{\sin\theta_W} \,.
\end{align}
The electromagnetic gauge coupling can be obtained from the known
value of $\aem^{\SM(5)}(M_Z)$ as described in
\secref{sec:fbsm} using the known general form of the threshold
correction, Eq.\ \eqref{eq:alpha_em_general_threshold}.
The weak mixing angle, together with the $W$ and $Z$ pole masses and
the muon decay constant $G_F$, form a set of four electroweak
precision quantities. Within the theory, only two of them are
independent. Hence, the running weak
mixing angle can be calculated in different ways:
\begin{enumerate}
\item Using the $W$ and $Z$ pole masses $M_W$ and $M_Z$ as input and
  calculating the cosine of the running weak mixing angle for example
  as:
  \begin{align}
    \cos^2\theta_W = \frac{m_W^2}{m_Z^2} \,,
    \label{eq:cosW_from_MWMZ}
  \end{align}
  where $m_W$ and $m_Z$ are the running $W$ and $Z$ masses which are
  obtained from the corresponding pole masses by
  \begin{align}
    m_V^2(Q) = M_V^2 + \re \Sigma_{V,T}(p^2 = M_V^2, Q) \,, \qquad (V=W,Z)
  \end{align}
  with $\Sigma_{V,T}$ being the transverse part of the vector boson
  self energy. In models with Higgs triplets having the
  vacuum expectation value $v_T$ for example,
  Eq.~\eqref{eq:cosW_from_MWMZ} is adapted accordingly:
  \begin{align}
    \cos^2\theta_W = \frac{m_W^2 - g_2^2 v_T^2}{m_Z^2} \,.
  \end{align}
  This approach was implemented in the first release of \fs \cite{Athron:2014yba},
  as a generalization of the calculation in Ref.\ \cite{Athron:2012pw}.
\item Using the $Z$ pole mass and the measured muon decay constant
  $G_F$ as input and determining $\theta_W$ and the $W$ pole mass
  as described below.
\end{enumerate}
The first approach has the advantage that it can be easily applied to
any \BSM model with a $W$ and $Z$ boson, because the 1-loop calculation of
the running $W$ and $Z$ masses can be fully automatized since only
self energies are necessary.  However, this approach has the
disadvantage that the parametric uncertainty of the calculated
electroweak gauge couplings is then limited by that
of the measured $W$ pole mass of the order $0.02\%$ \cite{Olive:2016xmw}.
This prohibits a meaningful computation of other electroweak precision
observables for which more precise experimental data exist.

The second approach is more complicated to automatize for all \BSM
models, because also 1-loop vertex and box diagrams contributing to
the muon decay have to be taken into account.  However, the approach
has the advantage that the parametric uncertainty of the calculated
electroweak gauge couplings is related to those
of $G_F$ and $M_Z$, which are of the order $0.00005\%$
and $0.002\%$, respectively \cite{Olive:2016xmw}. As a result, the $W$
pole mass is now a meaningful prediction, which can be used to
constrain BSM models.

Before version 2.0, only the first approach could be used in \fs for
all \BSM models.  The second approach was only available in models
which are SM-like or (N)MSSM-like, an option introduced in \fs 1.1.0 (as described in Ref.\ \cite{Staub:2015aea}).
\fstwo is now able to apply the second approach to all \BSM models
which have a $W$ and a $Z$ boson and which have the SM gauge
group as a gauge group factor.

For the implementation of the second approach, \fs uses a
generalization of the procedure presented in Ref.~\cite{Degrassi:1990tu}
for the SM, which has been adapted to the MSSM in Ref.~\cite{Pierce:1996zz}.
The running weak mixing angle is extracted from the relation
\cite{Degrassi:1990tu}
\begin{align}
  \sin^2\theta_W \cos^2\theta_W =
  \frac{\pi\,\aem}
   {\sqrt{2}\,M_Z^2\,G_F\,\hat\rho_\text{tree} \left(1-\Delta\hat{r}\right)}\,
  \label{eq:muon_decay_master}
\end{align}
with the renormalization scale consistently being set to $M_Z$,
where $\aem$ is the electromagnetic coupling of the \BSM model in the
\MSbar/\DRbar scheme and
\begin{align}
  \Delta\hat{r} &= \Delta\hat{r}_\text{1L} + \Delta\hat{r}_\text{2L}^\SM \\
  &\text{with}\;\;
  \Delta\hat{r}_\text{1L} = \frac{1}{1-\Delta\hat\rho}\,\frac{\re\Sigma_{W,T}(0)}{M_W^2}
  - \frac{\re\Sigma_{Z,T}(M_Z^2)}{M_Z^2} + \deltaVB\,,
  \label{eq:Delta_r_hat}\\
  \Delta\hat\rho &= \frac{1}{1 + \frac{\re\Sigma_{Z,T}(M_Z^2)}{M_Z^2}}
  \left[\frac{\re\Sigma_{Z,T}(M_Z^2)}{M_Z^2} - \frac{\re\Sigma_{W,T}(M_W^2)}{M_W^2}
   + \Delta\hat\rho_\text{2L}^\SM\right].
   \label{eq:Delta_rho_hat}
\end{align}
In the occurring self energies $\Sigma$, the top quark mass is chosen
to be the pole mass $M_t$ in order to include partial 2-loop corrections not
contained in $\Delta\hat{r}_\text{2L}^\SM$ and $\Delta\hat\rho_\text{2L}^\SM$
\cite{Fanchiotti:1992tu}.
Since $\Delta\hat{r}$ and $\Delta\hat\rho$ themselves depend
on $\theta_W$, an iteration including these equations has to be
performed to get a self-consistent solution.

The quantity
\begin{align}
  \hat\rho_\text{tree} &= \rho_0\,\frac{m_{Z,\SM}^2}{m_{Z,\text{mix}}^2}
  \label{eq:rho_hat_tree}
\end{align}
introduces two different generalizations. On the one hand, corrections
from higher dimensional Higgs multiplets are included via \cite{Langacker:1991an}
\begin{align}
  \rho_0 &= \frac{\sum_i \left(t_i^2 - t_{3i}^2 + t_i \right)|v_{\varphi_i}|^2}
   {\sum_i 2\,t_{3i}^2\,|v_{\varphi_i}|^2}\,,
\end{align}
where the sums run over all neutral Higgs fields~$\varphi_i$ with vacuum
expectation value~$v_{\varphi_i}$, weak isospin~$t_i$ and its third
component~$t_{3i}$. On the other hand, corrections from extra $U(1)$
gauge groups are included via the ratio of the SM-like tree-level $Z$
boson mass $m_{Z,\SM}$ and the tree-level $Z$ boson mass $m_{Z,\text{mix}}$
including mixing with additional $Z^\prime$ bosons
\cite{Degrassi:1989mu,Leike:1991if}.

The leading SM 2-loop contributions to $\Delta\hat{r}$ and
$\Delta\hat\rho$ \cite{Fanchiotti:1992tu,Pierce:1996zz} are given
by\footnote{The specific numerical values in these formulas depend
on the renormalization scale, which is assumed to be $M_Z$, and on
the choice of the top pole mass in the self energies from
Eqs.~\eqref{eq:Delta_r_hat} and \eqref{eq:Delta_rho_hat}.}
\begin{align}
 \begin{split}
  \Delta\hat{r}_\text{2L}^\SM &= \frac{\aem \,\as^{\SM(5)}}
   {4 \pi^2 \sin^2\theta_W \cos^2\theta_W}
  \left[2.145\,\frac{M_t^2}{M_Z^2} + 0.575 \log\mathopen{}\left(\frac{M_t}{M_Z}
  \right)\mathclose{} - 0.224 - 0.144\,\frac{M_Z^2}{M_t^2}\right] \\
  & \phantom{={}} - \delta_\text{Higgs} \,\frac{1 - \Delta\hat{r}_\text{1L}}
   {1 - \Delta\hat\rho}\,,
   \label{eq:Delta_r_hat_SM_2L}
 \end{split} \\[1em]
  \Delta\hat\rho_\text{2L}^\SM &= \frac{\aem \,\as^{\SM(5)}}
   {4 \pi^2 \sin^2\theta_W}
  \left[-2.145\,\frac{M_t^2}{M_W^2} + 1.262 \log\mathopen{}\left(\frac{M_t}{M_Z}
  \right)\mathclose{} - 2.24 - 0.85\,\frac{M_Z^2}{M_t^2}\right] + \delta_\text{Higgs}\,,
  \label{eq:Delta_rho_hat_SM_2L}
\end{align}
where the SM strong coupling $\as^{\SM(5)}$ is taken at the scale $M_t$ and
we have generalized the Higgs dependent part as
\begin{align}
   \delta_\text{Higgs} &= 3 \left(\frac{G_F \,M_t \,v_\SM}{8 \pi^2 \sqrt{2}}\right)^2
   \,\sum_{i}\left(|a_{\varphi_i tt}|^2 - |b_{\varphi_i tt}|^2\right)
   \rho^{(2)}\mathopen{}\left(\frac{m_{\varphi_i}}{M_t}\right)\mathclose{}
\end{align}
to include corrections from all neutral Higgs fields $\varphi_i$
coupling to the top quark via the vertex
$(\mathbbm{1}\,a_{\varphi_i tt} + \gamma_5\,b_{\varphi_i tt})$.
For this generalization to work properly, the SM-like vacuum
expectation value $v_\SM$ has to be defined and normalized
to the value $\approx 246$ GeV in the corresponding \sarah model file.
The utilized expansions of the function $\rho^{(2)}$ can be found in
Ref.~\cite{Fleischer:1993ub}.

\begin{figure}[tbh]
  \centering
  \subfloat[]{
  \quad\includegraphics{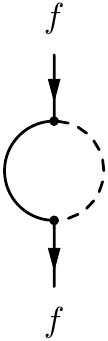}\quad
  \label{fig:delta_VB_wave_diagrams}
  }
  \subfloat[]{
  \qquad\includegraphics{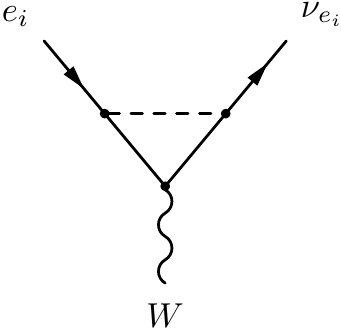}\quad
  \includegraphics{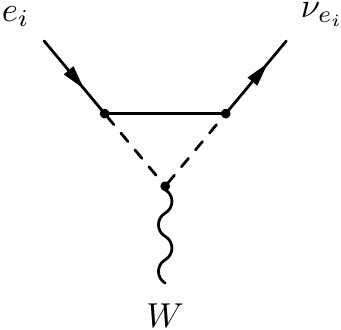}\qquad
  \label{fig:delta_VB_vertex_diagrams}
  }
  \subfloat[]{
  \quad\includegraphics{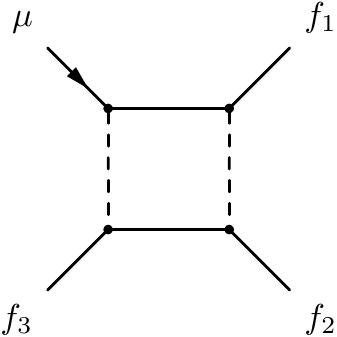}\quad
  \label{fig:delta_VB_box_diagrams}
  }
  \caption{Generic diagram types implemented in \fs to calculate
  the BSM 1-loop contributions to the muon decay, $\deltaVB^\BSM$.
  The solid and dashed lines in the loops represent any valid
  combination of fermions and scalars.
  \protect\subref{fig:delta_VB_wave_diagrams} wave-function
  renormalization diagrams: $f$ stands for $e$, $\mu$, $\nu_e$
  or~$\nu_\mu$.
  \protect\subref{fig:delta_VB_vertex_diagrams} vertex diagrams:
  $e_i$ stands for $e$ or~$\mu$.
  \protect\subref{fig:delta_VB_box_diagrams} box diagrams:
  the triplet $(f_1, f_2, f_3)$ stands for any permutation of
  $e$, $\nu_e$ and $\nu_\mu$.}
  \label{fig:delta_VB_diagram_types}
\end{figure}

The model-specific correction in Eq.~\eqref{eq:Delta_r_hat}
consists of an SM and a BSM part,
\begin{align}
   \deltaVB &= \deltaVB^\SM + \deltaVB^\BSM\,,
\end{align}
where the SM contribution $\deltaVB^\SM$
\cite{Degrassi:1990tu}, originating
from diagrams with additional internal gauge bosons, is given by
Eq.~(C.12) from Ref.~\cite{Pierce:1996zz} with the replacement
$\hat\rho \rightarrow 1/(1-\Delta\hat\rho)$.\footnote{The expression
  for $\deltaVB^\SM$ implemented in \fs has been extended by a term
  expressing the scale dependence so that the calculation can be
  performed consistently at the 1-loop level at scales different from
  $M_Z$.  However, the 2-loop contributions
  $\Delta\hat{r}_\text{2L}^\SM$ and $\Delta\hat\rho_\text{2L}^\SM$ are
  omitted if $Q \neq M_Z$, because their explicit scale dependence is
  currently not taken into account.}
The BSM contribution $\deltaVB^\BSM$ contains corrections from
1-loop external wave-function renormalizations, vertex and
box diagrams, which are put together as in Eq.~(C.13) from the
same reference. For the different corrections, \fs considers the
diagram types shown in \figref{fig:delta_VB_diagram_types} and sums
over all possible instantiations of these by inserting the valid
combinations of particles into the loop.\footnote{ Note that
self energy, vertex and box diagrams with internal vector bosons
are not considered outside of $\deltaVB^\SM$.}

After the weak mixing angle~$\theta_W$ has been calculated via
Eq.~\eqref{eq:muon_decay_master}, the $W$ pole mass can be
computed as
\begin{align}
   M_W = \sqrt{M_Z^2\,\cos^2\theta_W\,\frac{\hat\rho_\text{tree}}{1-\Delta\hat\rho}}\,.
   \label{eq:MW_calculation}
\end{align}

\subsection{Choosing \fmw in the model file}
The method to determine the weak mixing angle can be selected by
setting the following variable in the \fs model file:
\begin{lstlisting}
(* possible values: Automatic, FSFermiConstant or FSMassW *)
FSWeakMixingAngleInput = Automatic;
\end{lstlisting}
By default, \code{FSWeakMixingAngleInput} is set to \code{Automatic},
in which case the weak mixing angle is determined from the muon decay,
if all conditions are fulfilled, otherwise the $W$ mass is used.
Further possible values are \code{FSFermiConstant} and
\code{FSMassW} to explicitly select the muon decay or the $W$ mass method,
respectively.

\subsection{Structure of \fmw code}
The C++ interface for the determination of the running weak mixing
angle~$\theta_W$ and the $W$ pole mass via muon decay is provided by the class
\code{<model>_weinberg_angle} defined in
\code{<model>_weinberg_angle.hpp}.  To construct an object from this
class, two arguments have to be provided: a model object, which
represents the set of running BSM parameters, and a struct of type
\code{Sm_parameters}. The latter is defined within the
\code{<model>_weinberg_angle} class and contains the required SM parameters,
namely the Fermi constant $G_F$ as well as the pole
masses $M_W$, $M_Z$, $M_t$ and $\as^{\SM(5)}(M_t)$.
The values of $\sin\theta_W$ and $M_W$ are calculated and returned by
the class member function \code{calculate()}, which includes all of the
Eqs.~\eqref{eq:muon_decay_master}--\eqref{eq:MW_calculation} and the
necessary iteration.\footnote{The input $W$ pole mass is used as an
initial value while the function \code{calculate()} returns a more fitting
one. By updating the utilized value of $M_W$ during the iteration of the
spectrum generator, a self-consistent solution is ensured.}
\begin{example}
  In the \modelname{MSSM}, the running weak mixing angle~$\theta_W$ and the
  $W$ pole mass for a given \code{model} object can be calculated with
  the following code:
\begin{lstlisting}
MSSM_weinberg_angle::Sm_parameters sm_pars;
sm_pars.fermi_constant = 1.1663787e-05;
sm_pars.mw_pole = 80.385;
sm_pars.mz_pole = 91.1876;
sm_pars.mt_pole = 173.34;
sm_pars.alpha_s = 0.1079;

MSSM_weinberg_angle weinberg(model, sm_pars);

const auto sw_mw     = weinberg.calculate();
const double theta_w = std::asin(sw_mw.first);
const double MW      = sw_mw.second;
\end{lstlisting}
\end{example}

\subsection{Tests and comparisons with other spectrum generators}
The implementation of the muon decay method for the determination
of $\theta_W$ and $M_W$ provided as \fmw has been tested in the SM
and CMSSM by comparing to the results obtained using the algorithm
from \softsusy, which has been added in \fs 1.1.0. We have found
excellent agreement between the two implementations and also added
unit tests performing these comparisons to the \fs test suite.

Furthermore, the automatically generated expression for the
generalized $\hat\rho_\text{tree}$ given in Eq.~\eqref{eq:rho_hat_tree}
has been analytically checked for many models, such as the UMSSM,
MRSSM and \ESSM.

\begin{figure}[tbh]
  \centering
  \includegraphics[width=0.49\textwidth]{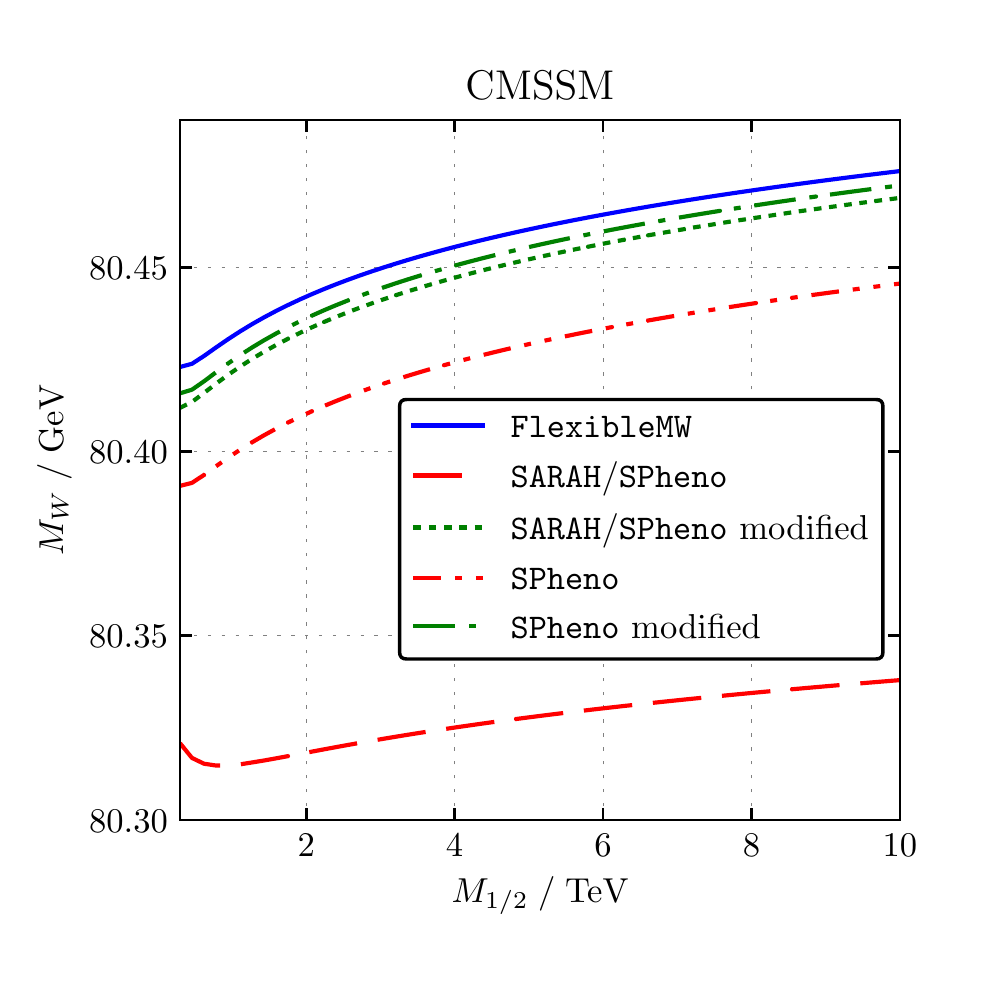}\hfill
  \includegraphics[width=0.49\textwidth]{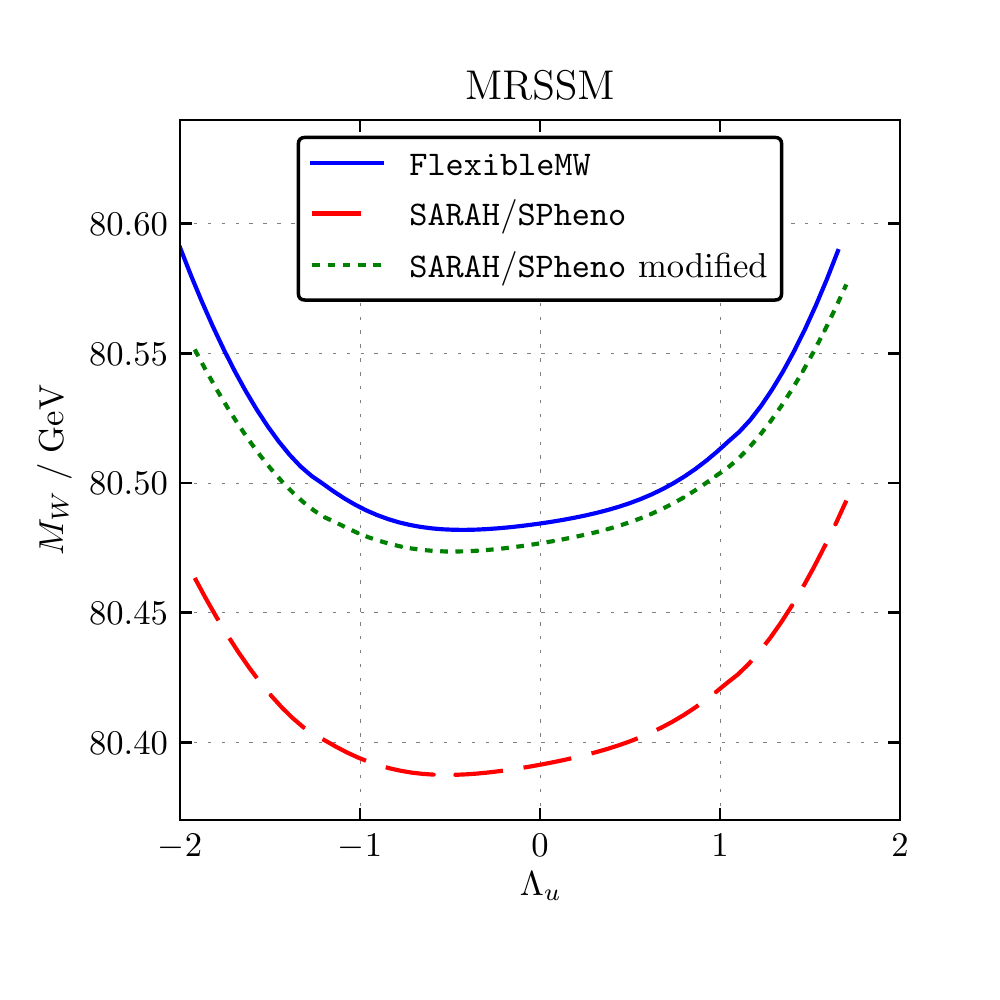}
  \caption{Left panel: $W$ pole mass in the CMSSM as a function of
  $\mhalf$ with $\mzero = 1\unit{TeV}$, $\tan\beta = 10$,
  $\sign\mu = +1$ and $\azero = 0$. Right panel: $W$ pole mass in
  the MRSSM as a function of $\Lambda_u$ with the other parameters
  fixed as for BMP1 in Table~2 from Ref.~\cite{Diessner:2014ksa}.
  For both panels, the differences between the lines are explained
  in the text.}
  \label{fig:FlexibleMW_vs_SPheno}
\end{figure}

Finally, we compared numerical results for the $W$ pole mass in the
CMSSM and MRSSM obtained with \fmw to the ones from \spheno code
generated by \sarah~4.12.2. \figref{fig:FlexibleMW_vs_SPheno} shows $M_W$
as a function of $\mhalf$ (CMSSM, see left panel) or the superpotential
parameter $\Lambda_u$ (MRSSM, see right panel) while all
the other parameters are specified as described in the caption.
In the case of the CMSSM, additionally the results calculated with
\spheno~4.0.3 are plotted. For both models, there is a large
discrepancy between the \fmw values illustrated by the blue solid
line and the results from \sarah/\spheno presented by the red dashed
line. A thorough comparison of the two implementations has revealed
two major differences. On the one hand, \sarah/\spheno partly uses the
\DRbar top mass in the self energies occurring in
Eqs.~\eqref{eq:Delta_r_hat} and \eqref{eq:Delta_rho_hat} as well as
the SM 2-loop corrections Eqs.~\eqref{eq:Delta_r_hat_SM_2L} and
\eqref{eq:Delta_rho_hat_SM_2L} while \fmw always uses the top pole mass,
as suggested by Ref.~\cite{Fanchiotti:1992tu}. Not utilizing the top
pole mass in all of these formulas spoils the correctness of the
included SM 2-loop corrections. On the other hand, the
\sarah/\spheno code contains an inconsistency in the final computation of
$M_W$, which is similar to Eq.~\eqref{eq:MW_calculation} but partly
neglects the SM 2-loop correction to $\Delta\hat\rho$.
This inconsistency is also existent in the original \spheno code that,
for the CMSSM, produces the results depicted by the red dashed-double-dotted
line in the left panel of \figref{fig:FlexibleMW_vs_SPheno}.
After fixing these issues within the \sarah/\spheno and \spheno code,
we get the modified results illustrated by the green dotted and
dashed-dotted line, respectively. These show good agreement with the
values from \fmw for both the CMSSM and MRSSM\@.
The remaining small discrepancies are well
understood and mainly caused by minor differences in the implemented
formulas and the various utilized \DRbar parameters. In addition,
the \sarah/\spheno and \spheno codes use $\as^{\SM(5)}(M_Z)$ while \fmw uses
$\as^{\SM(5)}(M_t)$ as preferred by Ref.~\cite{Fanchiotti:1992tu}.

\section{\fsas extension}
\label{sec:fsas}
In this section, we introduce a new BVP
solver, a major new feature released in \fstwo. The specification of the
running parameters of a BSM model at multiple scales connected by RGEs
constitutes a BVP that must be solved in order to compute the mass spectrum.
\fs was designed with the intention of allowing multiple solvers for this
problem. In \fsone only one supported solver was
distributed, the two-scale solver, which uses a fixed point iteration
with boundary conditions set at both the high scale and the low scale
and is the one used in MSSM spectrum generators. However, additional
solvers are important because convergence of valid points is not
guaranteed with the two-scale solver and viable regions of the
parameter space can be missed if the only option is the two-scale
solver.  This is already true in the CMSSM where it has also been
shown that there can even be multiple solutions to the BVP
\cite{Allanach:2013cda,Allanach:2013yua}, of which the two-scale solver will
only find at most one.  Furthermore, the two-scale
solver cannot find solutions at all for a large class of models, which
includes the fully constrained NMSSM (CNMSSM) \cite{Ellwanger:1993xa,
  Ellwanger:1995ru,Ellwanger:1996gw} and constrained \ESSM
(C\ESSM) \cite{Athron:2009bs}.

A basic problem in these models is that parameters exist that would
naturally be computed at the weak scale by EWSB conditions but
are at the same time constrained by GUT scale conditions.\footnote{In
  principle this could be avoided, e.g., in the CNMSSM and C\ESSM by including
  $\tan \beta$ or $\lambda$ in the set of parameters fixed by minimization
  conditions.  However, $\tan\beta$ affects the top, bottom and $\tau$ Yukawa
  couplings at tree level, and $\lambda$ appears in the 1-loop RGEs for these
  couplings and therefore can have a significant impact on the RG evolution.
  This makes it difficult to obtain a convergent fixed point iteration with
  such a setup.} In the constrained NMSSM, where the soft scalar mass,
$m_S$, is typically an EWSB output, solutions have previously been
found by varying one of the input parameters, $\tan \beta$, and tuning
this until a solution is found where $m_S$ is sufficiently close to
the universal scalar mass, $\mzero$ \cite{Djouadi:2008yj,Djouadi:2008uj}.
For the constrained \ESSM again the soft scalar masses need to be EWSB
outputs.  In that case a new BVP solver using semi-analytic solutions
was invented to find solutions \cite{Athron:2009ue,Athron:2009bs} and
extended and improved in further studies
\cite{Athron:2011wu,Athron:2012sq,Athron:2012pw}, though the precision
was limited due to analytical approximations used in the code.  In
general, the challenges in finding solutions to the mass spectrum in
BSM models can be quite different to those found in the MSSM and
therefore additional solvers can be of great benefit in increasing the
scope of \fs.

In \fstwo, an additional BVP solver has been added that solves the BVP
by using semi-analytic solutions to the RGEs for a subset of the
running parameters in the model.  The semi-analytic algorithm takes
advantage of the hierarchical structure of the RGEs in any model: the parameters
can be split into a sequence of sets such that the running of the parameters
in each set is independent of all of the parameters in the following
sets.  For example, in SUSY models the SUSY preserving parameters run
independently of the soft breaking parameters; in general renormalizable
models the dimensionless parameters run independently of the mass
parameters.\footnote{This partitioning is already reflected to some extent in
  the original C++ class structure for the parameters and
RGEs, which for SUSY (non-SUSY) models is split up so that
the RGEs for the SUSY preserving (dimensionless) parameters can be integrated
separately to those for the soft (dimensionful) parameters; see
\secref{sec:fbsm}.}
It is clear from the general form of the RGEs
\cite{Machacek:1983tz,Machacek:1983fi,Machacek:1984zw,Martin:1993zk,
  Yamada:1994id,Jack:1997eh,Luo:2002ti,Fonseca:2011vn,Goodsell:2012fm,
  Fonseca:2013bua,Sperling:2013eva,Sperling:2013xqa} that further
divisions are possible on the basis of the mass dimension of the
parameters.  The simplest case is the evolution of the mass dimension
one parameters $m^i(Q)$, which is described by a system of linear homogeneous
differential equations that can be solved in the form
\begin{align}
m^i(Q) &= \scoeff{}{}^i_j m^j(Q_0) ,
\end{align}
i.e., in terms of a linear combination of initial values and a set of
dimensionless coefficients that only depend on dimensionless parameters.
The RGEs for higher mass dimension parameters can then be cast as linear,
non-homogeneous systems by substituting in the solutions for the lower
mass dimension parameters, for which the general solutions can easily
be written down.\footnote{See, for example, the derivation given in
  Ref.~\cite{Athron:2009bs} for the case of a SUSY model with real
  soft parameters.}  In the CMSSM, doing so leads to well-known expressions
for the soft masses in terms of the universal parameters $\mzero$, $\mhalf$,
and $\azero$.  For instance, the solutions for the soft scalar masses read
(see also Eq.~\eqref{eq:cmssm-semi-analytic-solns} below)
\begin{equation*}
  m_i^2(Q) = a_i(Q) \mzero^2 + b_i(Q) \mhalf^2
  + c_i(Q) \mhalf \azero + d_i(Q) \azero^2 \, ,
\end{equation*}
where the dimensionless coefficients $a_i$, $b_i$, $c_i$, and $d_i$ are
determined by the running of the gauge and Yukawa couplings, and are computed
numerically.  The basic idea of the semi-analytic solver is to generalize this
approach to other models: first, the analytic forms of the solutions are
determined from the boundary conditions, and then the appearing coefficients
are determined numerically.  Thus the solver obtains a collection of
semi-analytic solutions which express parameters at the scale $Q$ directly in
terms of parameters at the boundary scale $Q_0$.

To make the discussion concrete, consider a general SUSY model with bilinear
and linear superpotential parameters $\mu^{ij}$ and $L^i$, together with a set
of soft breaking gaugino masses $M_a$, scalar trilinear, bilinear and linear
couplings $T^{ijk}$, $b^{ij}$ and $t^i$, and a set of soft scalar masses
$(m^2)_i^j$.  The semi-analytic solutions are obtained from the general 2-loop
RGEs given in Refs.~\cite{Martin:1993zk,Yamada:1994id,Fonseca:2011vn}.
Since the superpotential parameters are only multiplicatively
renormalized, the semi-analytic solutions for the parameters $\mu^{ij}$
and $L^i$ are particularly simple, taking the form
\begin{align}
  \mu^{ij}(Q) &= \scoeff{\mu}{\mu}^{ij}_{kp} \mu^{kp}(Q_0) \, ,
  \label{eq:general-susy-bilinear-solution} \\
  L^i(Q) &= \scoeff{L}{L}^i_p L^p(Q_0) \, ,
  \label{eq:general-susy-linear-solution}
\end{align}
where the coefficients $\scoeff{\mu}{\mu}$ and $\scoeff{L}{L}$ satisfy the
initial conditions $[c^\mu_\mu(Q_0)]^{ij}_{kp} = \delta^i_k
\delta^j_p$, $[c^L_L(Q_0)]^i_p = \delta^i_p$. Again, the point is that
the coefficients $c(Q)$ can be determined once the
running dimensionless SUSY parameters are known, and the equations then express
parameters at $Q$ in  terms of parameters at $Q_0$ with numerically known
coefficients.

The semi-analytic solutions for the dimension one soft breaking parameters can
be written as
\begin{align}
  T^{ijk}(Q) &= \scoeff{T}{T}^{ijk}_{lmn} T^{lmn}(Q_0)
  + \scoeff{T}{M}^{ijkb} M_b(Q_0) \, ,
  \label{eq:general-soft-trilinear-solution} \\
  M_a(Q) &= \scoeff{M}{T}_{almn} T^{lmn}(Q_0)
  + \scoeff{M}{M}_a^b M_b(Q_0) \, .
  \label{eq:general-gaugino-mass-solution}
\end{align}
Upon substituting these solutions into the RGEs for the soft breaking bilinears
$b^{ij}$ and the soft scalar masses $(m^2)_i^j$, one finds the semi-analytic
solutions
\begin{equation} \label{eq:general-soft-bilinear-solution}
  b^{ij}(Q) = \scoeff{b}{b}^{ij}_{kl} b^{kl}(Q_0)
  + \scoeff{b}{\mu T}^{ij}_{abmno} \mu^{ab}(Q_0) T^{mno}(Q_0)
  + \scoeff{b}{\mu M}^{ijc}_{ab} \mu^{ab}(Q_0) M_c(Q_0)
\end{equation}
and
\begin{align}
  (m^2)_i^j(Q) &= \scoeff{m^2}{m^2}^{jl}_{ik} (m^2)^k_l(Q_0)
  + \scoeff{m^2}{MM^*}_i^{jab} M_a(Q_0) M_b^*(Q_0) \nonumber \\
  & \quad {} + \scoeff{m^2}{T M^*}_{iklm}^{ja} T^{klm}(Q_0) M_a^*(Q_0)
  + \scoeff{m^2}{T^* M}_{iklm}^{ja} T^{klm*}(Q_0)
  M_a(Q_0) \nonumber \\
  & \quad {} + \scoeff{m^2}{T^* T}_{iklmnop}^j T^{klm*}(Q_0) T^{nop}(Q_0) \, .
  \label{eq:general-soft-scalar-mass-solution}
\end{align}
Finally, the semi-analytic solutions for the soft breaking linear parameters
$t^i$ are given by
\begin{align}
  t^i(Q) &= \scoeff{t}{t}^i_j t^j(Q_0) + \scoeff{t}{L T}^i_{jklm} L^j(Q_0)
  T^{klm}(Q_0) + \scoeff{t}{LM}^{ia}_j L^j(Q_0) M_a(Q_0) \nonumber \\
  & \quad {} + \scoeff{t}{\mu \mu T}^i_{jklmnop} \mu^{jk}(Q_0) \mu^{lm}(Q_0)
  T^{nop}(Q_0) + \scoeff{t}{\mu \mu M}^{ia}_{jklm} \mu^{jk}(Q_0) \mu^{lm}(Q_0)
  M_a(Q_0)
  \nonumber \\
  & \quad {} + \scoeff{t}{\mu b}^i_{jklm} \mu^{jk}(Q_0) b^{lm}(Q_0)
  + \scoeff{t}{\mu^* M M^*}^{iab}_{jk} \mu^{jk*}(Q_0) M_a(Q_0) M_b^*(Q_0)
  \nonumber \\
  & \quad {} + \scoeff{t}{\mu^* T T^*}^i_{jklmnopq} \mu^{jk*}(Q_0) T^{lmn}(Q_0)
  T^{opq*}(Q_0) \nonumber \\
  & \quad {} + \scoeff{t}{\mu^* T M^*}^{ia}_{jklmn} \mu^{jk*}(Q_0) T^{lmn}(Q_0)
  M_a^*(Q_0) \nonumber \\
  & \quad {} + \scoeff{t}{\mu^* T^* M}^{ia}_{jklmn} \mu^{jk*}(Q_0) T^{lmn*}(Q_0)
  M_a(Q_0) \nonumber \\
  & \quad {} + \scoeff{t}{b^* T}^i_{jklmn} b^{jk*}(Q_0) T^{lmn}(Q_0)
  + \scoeff{t}{b^* M}^{ia}_{jk} b^{jk*}(Q_0) M_a(Q_0) \nonumber \\
  & \quad {} + \scoeff{t}{\mu^* m^2}^{il}_{jkm} \mu^{jk*}(Q_0) (m^2)^m_l(Q_0)
  \, . \label{eq:general-soft-linear-solution}
\end{align}
For models in which Dirac gaugino masses $m_{Da}^i$ are also present, the
solutions for the parameters $t^i$ are modified, $t^i \to t^i + \Delta t^i$,
where $\Delta t^i$ is of the form
\begin{align}
  \Delta t^i(Q) &= \scoeff{t}{m_D m^2}^{iaj}_{kl} m_{Da}^k(Q_0) (m^2)_j^l(Q_0)
  + \scoeff{t}{m_D M M^*}^{iabc}_j m_{Da}^j(Q_0) M_b(Q_0) M_c^*(Q_0)
  \nonumber \\
  & \quad {} + \scoeff{t}{m_D T M^*}^{iab}_{jklm} m_{Da}^j(Q_0) T^{klm}(Q_0)
  M_b^*(Q_0) \nonumber \\
  & \quad {} + \scoeff{t}{m_D T^* M}^{iab}_{jklm} m_{Da}^j(Q_0) T^{klm*}(Q_0)
  M_b(Q_0) \nonumber \\
  & \quad {} + \scoeff{t}{m_D T^* T}^{ia}_{jklmnop} m_{Da}^j(Q_0) T^{klm*}(Q_0)
  T^{nop}(Q_0) \nonumber \\
  & \quad {} + \scoeff{t}{m_D^* m_D^* T}^{iab}_{jklmn} m_{Da}^{j*}(Q_0)
  m_{Db}^{k*}(Q_0) T^{lmn}(Q_0) \nonumber \\
  & \quad {} + \scoeff{t}{m_D^* m_D^* M}^{iabc}_{jk} m_{Da}^{j*}(Q_0)
  m_{Db}^{k*}(Q_0) M_c(Q_0) \nonumber \\
  & \quad {} + \scoeff{t}{m_D m_D \mu}^{iab}_{jklm} m_{Da}^j(Q_0) m_{Db}^k(Q_0)
  \mu^{lm}(Q_0) \, . \label{eq:extra-soft-trilinear-solution}
\end{align}
The semi-analytic solutions for the Dirac gaugino masses themselves follow
from the known general 2-loop RGEs \cite{Goodsell:2012fm}, and can be
written in the form
\begin{equation} \label{eq:general-dirac-gaugino-solutions}
  m_{Da}^i(Q) = \scoeff{m_{Da}}{m_{Da}}^i_j m^j_{Da}(Q_0) \, .
\end{equation}

In non-SUSY models, the semi-analytic solutions follow from the known
results for the 2-loop RGEs in a general gauge theory \cite{Machacek:1983tz,
  Machacek:1983fi,Machacek:1984zw,Luo:2002ti,Fonseca:2013bua}.  For a non-SUSY
model containing a set of real scalar trilinear couplings $h^{ijk}$ and squared
scalar masses $(m^2)^{ij}$, and a set of fermion masses $(M_f)^{ij}$, the
semi-analytic solutions for the mass dimension one parameters read
\begin{align}
  h^{ijk}(Q) &= \scoeff{h}{h}^{ijk}_{lmn} h^{lmn}(Q_0)
  + \scoeff{h}{M_f}^{ijk}_{lm} (M_f)^{lm}(Q_0)
  + \scoeff{h}{M_f^*}^{ijk}_{lm} (M_f)^{lm*}(Q_0) \, ,
  \label{eq:general-dimension-one-trilinear-solution} \\
  (M_f)^{ij}(Q) &= \scoeff{M_f}{h}^{ij}_{lmn} h^{lmn}(Q_0)
  + \scoeff{M_f}{M_f}^{ij}_{lm} (M_f)^{lm}(Q_0)
  + \scoeff{M_f}{M_f^*}^{ij}_{lm} (M_f)^{lm*}(Q_0) \, .
  \label{eq:general-dimension-one-fermion-mass-solution}
\end{align}
In general, all of the dimension one parameters must be considered together,
unlike in SUSY models where they can be separated into superpotential and soft
breaking masses.  After substituting the solutions
Eq.~\eqref{eq:general-dimension-one-trilinear-solution} and
Eq.~\eqref{eq:general-dimension-one-fermion-mass-solution} into the RGEs for
the squared scalar masses, the semi-analytic solutions for the scalar masses
are found to be
\begin{align}
  (m^2)^{ij}(Q) &= \scoeff{m^2}{m^2}^{ij}_{kl} (m^2)^{kl}(Q_0)
  + \scoeff{m^2}{h h}^{ij}_{klmnop} h^{klm}(Q_0) h^{nop}(Q_0) \nonumber \\
  & \quad {} + \scoeff{m^2}{h M_f}^{ij}_{klmpq} h^{klm}(Q_0) (M_f)^{pq}(Q_0)
  + \scoeff{m^2}{h M_f^*}^{ij}_{klmpq} h^{klm}(Q_0) (M_f)^{pq*}(Q_0)
  \nonumber \\
  & \quad {} + \scoeff{m^2}{M_f M_f}^{ij}_{klmn} (M_f)^{kl}(Q_0) (M_f)^{mn}(Q_0)
  \nonumber \\
  & \quad {} + \scoeff{m^2}{M_f M_f^*}^{ij}_{klmn} (M_f)^{kl}(Q_0)
  (M_f)^{mn*}(Q_0) \nonumber \\
  & \quad {} + \scoeff{m^2}{M_f^* M_f^*}^{ij}_{klmn} (M_f)^{kl*}(Q_0)
  (M_f)^{mn*}(Q_0) \, .
  \label{eq:general-scalar-mass-solution}
\end{align}

The solver algorithm implemented by \fsas automatically determines
the above semi-analytic solutions in the model,\footnote{In non-SUSY models,
  \sarah currently does not calculate RGEs for linear scalar couplings
  $L^i$, which have been given in, e.g., Ref.~\cite{Goodsell:2012fm}, and
  therefore the semi-analytic solutions for these parameters are also not used
  in \fs.} given the set of boundary conditions at some scale.  Since the
required coefficients may be determined knowing only the running of the
dimensionless or SUSY preserving parameters, each step of the main fixed point
iteration is split up into two parts.  Firstly, the BVP for the dimensionless
parameters is solved iteratively.  The semi-analytic coefficients at any scale
can then be calculated.  In this second stage, the soft breaking or
dimensionful parameters are expanded in terms of the semi-analytic solutions to
the RGEs.  The low-energy EWSB conditions and masses are thus expressed
explicitly in terms of the boundary values at $Q_0$, allowing, for example,
unknown quantities at one scale to be directly constrained at another.

\subsection{Choosing \fsas in the model file}
\label{sec:chosing_SAS}

The BVP solver algorithms that are applicable in a given model can be
specified in the \fs model file using the list
\code{FSBVPSolvers}.  The elements of this list correspond to
the desired BVP solvers that should be enabled for the model, identified
using the predefined symbols \code{TwoScaleSolver} for the two-scale
algorithm and \code{SemiAnalyticSolver} for the semi-analytic solver.  By
default, if a list of BVP solvers is not specified in the model file, only
the two-scale algorithm is enabled, as summarized in \tabref{tab:fsas_options}.
\begin{OptionTable}{\fsas model file options.}{tab:fsas_options}
  \code{FSBVPSolvers} & \code{\{ TwoScaleSolver \}} & Non-empty list
  containing \code{TwoScaleSolver} or \code{SemiAnalyticSolver} or
  both &
  List of BVP solvers to be used \\
  \code{TwoScaleSolver} & -- & -- & Represents the two-scale BVP solver \\
  \code{SemiAnalyticSolver} & -- & -- & Represents the semi-analytic BVP
  solver \\
  \bottomrule
\end{OptionTable}

In addition to specifying that the semi-analytic solver should be used,
the user should ensure that the boundary conditions for the model are
compatible with its use.  Currently, this requires that those parameters
that will be expanded using semi-analytic solutions of the RGEs
are fixed in the same boundary condition.  So in SUSY models, the
boundary values for all of the soft SUSY breaking parameters should be given
at a single scale, while in non-SUSY models the same should be done for all
of the dimensionful parameters.  The expressions for the boundary values
must also only be polynomials in any dimensionful  parameters, such as
universal scalar masses.

Boundary values for the running parameters of the model can be specified
in terms of input parameters such as those defined in the \code{MINPAR}
and \code{EXTPAR} variables.  In \fstwo, the user can also define extra
parameters by using the list \code{FSAuxiliaryParameterInfo}.   Each entry in
this list should contain the name of the extra parameter being defined and a
list of its properties.  The possible properties for the
new parameters are specified in \tabref{tab:parameter_info}.  In particular,
the mass dimensions of the new parameters may be specified to allow for
simplifying the forms of the semi-analytic solutions, in which dimensionless
parameters can be absorbed into the definitions of the semi-analytic
coefficients.  Input or auxiliary parameters that are not scalars may be
defined by setting the \code{ParameterDimensions} property to a list of the
form \code{\{M,N\}} for an $M \times N$ matrix or \code{\{N\}} for an
$N$-dimensional vector.  A value of \code{\{1\}} corresponds to a scalar.  Note
that in versions of \fs prior to version 2.0, this functionality was available
for input parameters using the variable \code{FSExtraInputParameters}.
However, please note that this variable has been removed in \fstwo; definitions
that were previously given in \code{FSExtraInputParameters} must now be given
in the new list \code{FSAuxiliaryParameterInfo}.
\begin{OptionTable}{Allowed properties for extra parameters}{tab:parameter_info}
  \code{InputParameter} & \code{False} & \code{True} or \code{False}
  & Indicates whether the new parameter is an input parameter \\
  \code{ParameterDimensions} & \code{\{1\}} & A list of the form \code{\{M,N\}}
  or \code{\{N\}} & Specifies the dimensions of the parameter \\
  \code{MassDimension} & -- & A non-negative integer & Specifies the mass
  dimension of the parameter \\
  \code{LesHouches} & -- & A symbol or string, or a list of the form
  \code{\{block, entry\}} & Specifies the SLHA block from which the parameter
  should be read if it is an input parameter \\
  \bottomrule
\end{OptionTable}

\begin{example}
In the MSSM, input parameters giving the values of the
soft SUSY breaking trilinears as $3 \times 3$ matrices can be defined:
\begin{lstlisting}
FSAuxiliaryParameterInfo = {
   {Ae  , { LesHouches -> AEIN,
            ParameterDimensions -> {3,3},
            InputParameter -> True
          } },
   {Ad  , { LesHouches -> ADIN,
            ParameterDimensions -> {3,3},
            InputParameter -> True
          } },
   {Au  , { LesHouches -> AUIN,
            ParameterDimensions -> {3,3},
            InputParameter -> True
          } }
};
\end{lstlisting}
\end{example}

Extra parameters defined in this way can be used in the specification of the
boundary conditions for running model parameters, or indeed themselves be
fixed in the boundary conditions.  A special case is if the new parameters
are to be fixed by the EWSB conditions.  To facilitate this usage, the new
variables \code{EWSBInitialGuess} and \code{EWSBSubstitutions} may be
given in the model file.  The former allows for explicit initial
guesses to be provided for the EWSB output parameters, in the same format
as used for specifying the boundary conditions.  The latter, a list of
two-component lists, can be used to define any substitutions that should
be made in the EWSB equations before attempting to solve them.

\begin{example}
In the so-called VCMSSM \cite{Ellis:2003pz,Ellis:2004qe}, the value of the soft
breaking bilinear $B\mu$ is fixed at the GUT scale, $M_X$, in terms of the
universal soft trilinear $\azero$ and scalar mass $\mzero$ according to
$B\mu(M_X) = \mu(M_X) (\mzero + \azero)$.  Consequently, $B\mu$ can no longer
be fixed to ensure proper EWSB, as is usually done in the CMSSM\@.  Instead,
$|\mu|^2$ and $\tan\beta$ are used as EWSB outputs.  This can be achieved in a
\modelname{VCMSSM} model file through the definitions:
\begin{lstlisting}
FSAuxiliaryParameterInfo = {
   {TanBeta, { ParameterDimensions -> {1},
               MassDimension -> 0 } },
   {MuSq,    { ParameterDimensions -> {1},
               MassDimension -> 2 } },
   {vMSSM,   { ParameterDimensions -> {1},
               MassDimension -> 1 } }
};

EWSBOutputParameters = { TanBeta, MuSq };

EWSBSubstitutions = {
   {vd, vMSSM Cos[ArcTan[TanBeta]]},
   {vu, vMSSM Sin[ArcTan[TanBeta]]},
   {\[Mu], Sign[\[Mu]] Sqrt[MuSq]}
};

EWSBInitialGuess = {
   {TanBeta, vu / vd},
   {MuSq, \[Mu]^2}
};

SUSYScaleInput = {
   {vMSSM, Sqrt[vd^2 + vu^2]},
   FSSolveEWSBFor[EWSBOutputParameters]
};
\end{lstlisting}
\end{example}

Note that \fs automatically substitutes the semi-analytic solutions into
the EWSB conditions for the soft or dimensionful parameters.  Therefore, it
is not necessary for the user to explicitly provide them.  For the purpose of
making use of the semi-analytic algorithm, the only required addition to the
model file is the inclusion of \code{SemiAnalyticSolver} in the list
\code{FSBVPSolvers}.

\begin{example}
The CNMSSM is characterized by universal soft scalar masses at the
high scale.  Typically, this constraint is relaxed somewhat to allow the
soft singlet mass $m_S^2$ to differ from the common scalar mass $\mzero^2$ at
the high scale.  This is done to allow fixing $m_S^2$ using the EWSB conditions
to ensure correct EWSB, and is the approach taken in the \modelname{NMSSM} model
included with \fs, as well as in other public spectrum generators such as
\softsusy, \spheno and \nmspec.  Universality of the soft scalar masses can
alternatively be maintained by using the semi-analytic BVP solver, which is
achieved in the \modelname{CNMSSM} model file by specifying the BVP solvers to
be used,
\begin{lstlisting}
FSBVPSolvers = { SemiAnalyticSolver };
\end{lstlisting}
In this set-up, $\mzero$ ceases to be an input parameter and is fixed by the
EWSB conditions.  This is achieved in the model file by removing $\mzero$ from
the list of input parameters and defining it as an additional parameter, as
follows:
\begin{lstlisting}
(* CNMSSM input parameters *)

MINPAR = {
   {2, m12},
   {3, TanBeta},
   {4, Sign[vS]},
   {5, Azero}
};

EXTPAR = {
   {61, LambdaInput}
};

FSAuxiliaryParameterInfo = {
   {m0Sq,        { ParameterDimensions -> {1},
                   MassDimension -> 2 } },
   {LambdaInput, { ParameterDimensions -> {1},
                   MassDimension -> 0 } }
};
\end{lstlisting}
Note that the definition of \code{LambdaInput} in
\code{FSAuxiliaryParameterInfo} is not compulsory, but allows the semi-analytic
solutions to be simplified by using the fact that it is a dimensionless
parameter.  The value of $\mzero^2$ is then determined from the EWSB conditions
by setting
\begin{lstlisting}
EWSBOutputParameters = { \[Kappa], vS, m0Sq };
\end{lstlisting}
To impose the universality constraint, the condition $m_S^2 = \mzero^2$ at the
GUT scale must be added to the high-scale boundary condition, by defining
\begin{lstlisting}
HighScaleInput={
   {T[Ye], Azero*Ye},
   {T[Yd], Azero*Yd},
   {T[Yu], Azero*Yu},
   {mq2, UNITMATRIX[3] m0Sq},
   {ml2, UNITMATRIX[3] m0Sq},
   {md2, UNITMATRIX[3] m0Sq},
   {mu2, UNITMATRIX[3] m0Sq},
   {me2, UNITMATRIX[3] m0Sq},
   {mHu2, m0Sq},
   {mHd2, m0Sq},
   {ms2, m0Sq},
   {\[Lambda], LambdaInput},
   {T[\[Kappa]], Azero \[Kappa]},
   {T[\[Lambda]], Azero LambdaInput},
   {MassB, m12},
   {MassWB,m12},
   {MassG,m12}
};
\end{lstlisting}
The remainder of the model implementation is otherwise rather similar to that
for the NMSSM solved using the two-scale solver.  The full \modelname{CNMSSM}
model file is given in \appref{app:CNMSSM}.
\end{example}

\subsection{Structure of the generated code}

In keeping with the design goal of \fs to produce generated code that is highly
modular in nature, the implementation of the BVP solvers is separated from
the details of specific physics models.  In \fs, a general BVP solver algorithm
is represented by the templated \code{RGFlow<T>} class.  Particular algorithms
are provided as specializations of this class, with the two-scale and
semi-analytic solvers corresponding to the classes \code{RGFlow<Two_scale>}
and \code{RGFlow<Semi_analytic>}, respectively.  Each realizes an abstract
implementation of the appropriate algorithm, with no dependence on the details
of any particular model.  The required model-dependent information is provided
by separate classes representing the model and boundary and matching conditions,
which are linked to the desired BVP solver class.  New algorithms can easily be
added simply by writing additional specializations of the \code{RGFlow} class.

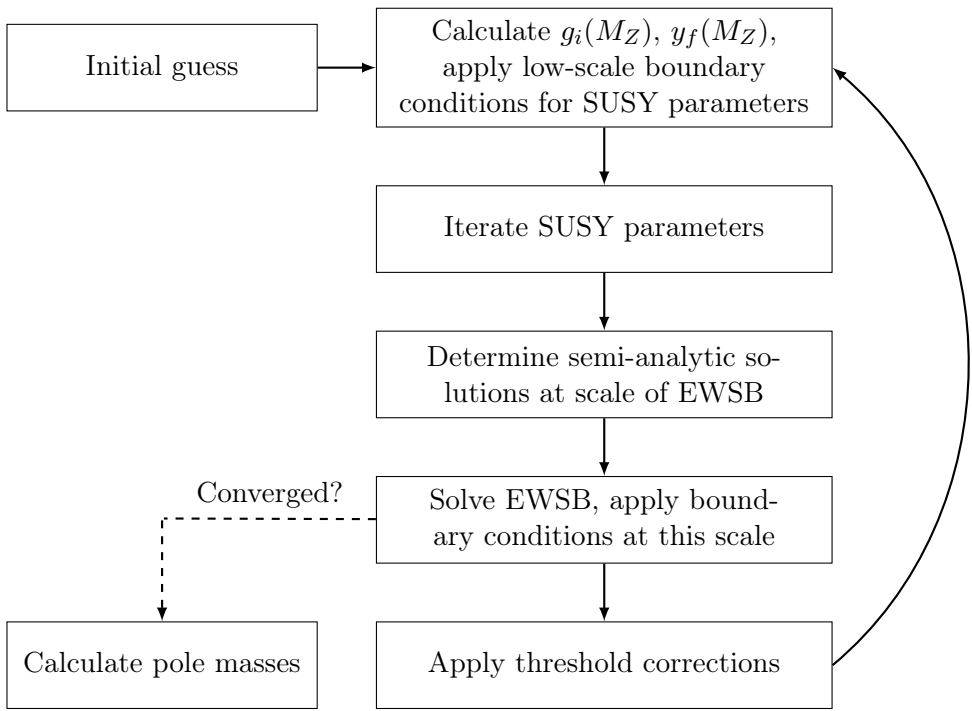
\begin{figure}[tbh]
  \centering
  \begin{tikzpicture}[node distance = 1cm, auto]
    \tikzstyle{block} = [rectangle, draw, text width=15em,
      text centered, minimum height=3em, minimum width=10em]
    \tikzstyle{arrow} = [draw, -latex, thick]
    \node[block] (MZ)
         {Calculate $g_i(M_Z)$, $y_f(M_Z)$, apply low-scale boundary
           conditions for SUSY parameters};
    \node[block, below = 2em of MZ] (susy) {Iterate SUSY parameters};
    \path[arrow] (MZ) --node {} (susy);
    \node[block, below = 2em of susy] (coeffs)
         {Determine semi-analytic solutions at scale of EWSB};
    \path[arrow] (susy) --node {} (coeffs);
    \node[block, below = 2em of coeffs] (MS) {Solve EWSB, apply boundary
      conditions at this scale};
    \path[arrow] (coeffs) --node {} (MS);
    \node[block, below = 2em of MS] (thresh) {Apply threshold corrections};
    \path[arrow] (MS) --node {} (thresh);
    \node[rectangle, draw, text width=10em,
          text centered, minimum height=3em, minimum width=10em,
          left = 2em of MZ] (guess) {Initial guess};
    \path[arrow] (guess) --node {} (MZ);
    \path[-latex, thick] (thresh.east) edge[bend right=50] (MZ.east);
    \node[rectangle, draw, text width=10em,
          text centered, minimum height=3em, minimum width=10em,
          left = 2em of thresh] (pole) {Calculate pole masses};
    \node[inner sep=0pt, above = 3.5em of pole] (convergence) {};
    \path[draw,thick,dashed,above] (MS) --node {Converged?} (convergence);
    \path[arrow,dashed] (convergence) --node {} (pole);
  \end{tikzpicture}
  \caption{Semi-analytic algorithm for calculating the mass spectrum in a
    SUSY model; in a non-SUSY model, the SUSY parameters are replaced by
    the dimensionless model parameters instead.}
  \label{fig:semi-analytic-algorithm}
\end{figure}

The semi-analytic solver algorithm requires two nested iterations.  An
inner iteration, carried out at each step, determines consistent values for
the SUSY preserving (in SUSY models) or dimensionless parameters (in non-SUSY
models) at the low- and high-scale boundaries.  Updated estimates for these
scales are simultaneously calculated during the iteration if necessary, for
example if the high scale $M_X$ is defined in the model file by gauge
unification, $g_1(M_X) = g_2(M_X)$.  Once this has converged, the resulting
estimate for these parameters is used to compute the semi-analytic solutions
for the soft SUSY breaking or dimensionful parameters, at which point the
EWSB conditions may be solved and the \DRbar/\MSbar mass spectrum
calculated.  The new values of the soft or dimensionful parameters are then used
in the inner iteration for computing the required threshold corrections.
This sequence of steps is illustrated in \figref{fig:semi-analytic-algorithm}.
For a single high-scale model such as the CMSSM, the algorithm proceeds as
follows:
\subparagraph{Initial guess:} The initial guess involves a first run of the
inner iteration.  In all of the steps below, threshold corrections are ignored.
\begin{enumerate}
\item The known values of the SM gauge couplings at the scale $M_Z$ are
  used to estimate the values of $g_1$, $g_2$ and $g_3$ at the scale
  $M_t$, ignoring threshold corrections.
\item The user-defined initial guess at the low scale, as given in
  \code{InitialGuessAtLowScale}, is imposed at the scale $M_t$.
\item The SUSY preserving or dimensionless parameters are run to the initial
  guess for $M_X$, given by \code{HighScaleFirstGuess}, and the
  high-scale boundary condition for these parameters, defined in
  \code{HighScaleInput}, is imposed.  The initial guess at the high scale,
  defined in \code{InitialGuessAtHighScale}, is then applied.
\item \label{initial-guess-low-scale} The model is run to the guess for the
  low scale, initially set to the value defined in \code{LowScaleFirstGuess},
  and the low-scale boundary conditions for the SUSY preserving or
  dimensionless parameters defined in \code{LowScaleInput} are applied.
\item The model is run to the current guess for $M_X$.
  \begin{enumerate}
  \item If necessary, the guess for $M_X$ is updated.  For example, in
    the CMSSM with $M_X$ defined to be the scale at which $g_1(M_X) = g_2(M_X)$,
    a new estimate for $M_X$ is calculated according to
    \begin{equation*}
      M_X^\prime = M_X \exp\left( \frac{g_2(M_X)-g_1(M_X)}
      {\beta_{g_1} - \beta_{g_2}} \right ) \, .
    \end{equation*}
  \item The high-scale boundary conditions for the SUSY preserving or
    dimensionless parameters are applied.
  \end{enumerate}
\item If not converged, goto \ref{initial-guess-low-scale}.
\item The model is run to the guess for the low scale.  The semi-analytic
  solutions are calculated at this scale.
\item The EWSB equations are solved at tree level.
\item The \DRbar/\MSbar mass spectrum is calculated.
\end{enumerate}
At this stage, initial guesses for all of the model parameters, boundary
condition scales and the \DRbar/\MSbar mass spectrum are available.  The full
iteration now starts, in which the full set of threshold corrections are
applied.
\subparagraph{Thresholds iteration:}
\begin{enumerate}
\item \label{outer-iteration-step-one} The SUSY preserving or dimensionless
  parameters are determined in an inner iteration analogous to that in the
  initial guess, namely:
  \begin{enumerate}
  \item \label{inner-iteration-step-one} All model parameters are run to the
    low scale (\code{LowScale}) and the \DRbar/\MSbar mass spectrum is
    calculated.
  \item The low scale is recalculated if it is not fixed.
  \item The SM gauge couplings are calculated in the model, including the
    appropriate threshold corrections.
  \item The user-defined constraints for the SUSY preserving or dimensionless
    parameters are applied.
  \item All model parameters are run to the high scale (\code{HighScale}).
  \item The high scale is recalculated if necessary.
  \item The user-defined boundary conditions at this scale for the
    SUSY preserving or dimensionless parameters are applied.
  \item The model parameters are run to the SUSY scale (\code{SUSYScale})
    and the SUSY scale is updated if necessary.
  \item The boundary conditions for the SUSY preserving or dimensionless
    parameters are applied.
  \item If not converged, goto \ref{inner-iteration-step-one}.
  \end{enumerate}
\item All model parameters are run to the scale at which the EWSB equations
  are to be solved.
  \begin{enumerate}
  \item The coefficients in the semi-analytic solutions are determined
    at this scale, using the current estimate for the scale at which the
    relevant boundary conditions are imposed.  For example, in the CMSSM, the
    semi-analytic solutions for the soft gaugino masses, trilinears, scalar
    masses and bilinear take the form
    \begin{equation}
      \begin{aligned}
        M_i(Q) &= p_i(Q) \azero + q_i(Q) \mhalf \, , \\
        T_i(Q) &= e_i(Q) \azero + f_i(Q) \mhalf \, , \\
        m_i^2(Q) &= a_i(Q) \mzero^2 + b_i(Q) \mhalf^2
        + c_i(Q) \mhalf \azero + d_i(Q) \azero^2 \, , \\
        B\mu(Q) &= u(Q) B\mu(M_X) + v(Q) \mu(M_X) \mhalf
        + w(Q) \mu(M_X) \azero \, .
      \end{aligned} \label{eq:cmssm-semi-analytic-solns}
    \end{equation}
    The coefficients are determined numerically by varying the values of
    $\mhalf$, $\azero$, $\mzero$ and $B\mu(M_X)$ and integrating the RGEs from
    $M_X$ to $Q$.  For example, the coefficients $p_i(Q)$, $e_i(Q)$, $d_i(Q)$
    and $w(Q)$ are obtained by keeping only $\azero \neq 0$.  A similar approach
    is followed to successively obtain all of the remaining coefficients.
  \item The calculated semi-analytic solutions are used to set the values of
    the soft SUSY breaking or dimensionful parameters at this scale.
  \item The \DRbar/\MSbar mass spectrum is calculated and the scale at which
    EWSB occurs is updated.
  \item The EWSB conditions are solved at the loop level.
  \end{enumerate}
\item If not converged, goto \ref{outer-iteration-step-one}.  Otherwise
  the iteration finishes.
\end{enumerate}
If the iteration converges, all running parameters in the model are determined
between the low  and high scales.  The remainder of the calculation, that is,
the calculation of the pole mass spectrum and observables, then proceeds in
the same way as in the two-scale algorithm.  Alternatively, the iteration
may fail to converge or may encounter problems that render the parameter point
physically invalid.  As for the two-scale solver, the specific problems that
are encountered for a given parameter space point are stored in the
\code{Problems} class, and may be accessed using the \code{get_problems()}
function of the model class.\footnote{Note that in \fstwo a separate class,
\code{BVP_solver_problems}, is used to store those problems that are associated
only with failures of the BVP solver algorithm, such as a failure to converge,
and which do not necessarily mean the parameter point is ruled out.  A summary
combining all of the problems that arise during a run of the spectrum generator
can then be obtained by calling the \code{get_problems()} method of the spectrum
generator class.}

\subsection{Tests and comparisons with other spectrum generators}

The semi-analytic algorithm provided by \fsas has been tested in the
CMSSM, CNMSSM and the C\ESSM by comparing to the results
obtained using the existing two-scale solver.  We have
carried out consistency checks between the two by confirming that the same
solution can be found in both solvers, provided it is a stable fixed
point in both.  In each of the models, the existing model solved using
the two-scale algorithm has been compared with versions of the model
using alternative boundary conditions.  For example, in the CMSSM,
instead of the traditional approach of fixing $|\mu|^2$ and $B\mu$
using the EWSB conditions, the value of $\mu$ is provided as an input
and $\mzero^2$ and $B\mu(M_X)$ are determined from the EWSB
conditions.\footnote{This alternative approach is useful in scenarios where
  one wishes to have direct control over the Higgsino masses, and therefore
  the composition of the lightest neutralino in the CMSSM, as was done in
  Ref.\ \cite{Athron:2016gor}.}  The benchmark points used as inputs for the
semi-analytic solver in each model are displayed in
\tabref{tab:semi_analytic_bms}.  In all three models, the
running parameters and pole mass spectra are found to differ at or
below the level of $0.1$\%.  Unit tests that perform these
comparisons have also been added to the \fs test suite.
\begin{table}[tbh]
  \centering
  \begin{tabularx}{\textwidth}{lX}
    \toprule
    Model & Unit test benchmark points for the semi-analytic solver\\
    \midrule
    CMSSM & $\mhalf = 500$ GeV, $\tan\beta = 10$, $\azero = 0$ GeV,
    $\mu(\MS) = 623.36$ GeV\\
    CNMSSM & $\mhalf = 133.33$ GeV, $\tan\beta = 10$,
    $\sign \mu_{\text{eff}} = -1$, $\azero = -300$ GeV, $\lambda(M_X) =-0.05$\\
    C\ESSM & $\tan\beta = 10$, $\lambda_3(M_X) = 0.12$, $\kappa(M_X) = 0.2$,
    $\mu^\prime(M_X) = 10$ TeV, $B^\prime\mu^\prime(M_X) = 0$ GeV$^2$,
    $s(\MS) = 4$ TeV, $\lambda_{1,2}(M_X) = 0.1$\\
    \bottomrule
  \end{tabularx}
  \caption{Input parameter values used for the unit tests comparing the
    results of the two-scale and semi-analytic algorithms in the CMSSM,
    CNMSSM and C\ESSM.  The notation for the CNMSSM follows that in
    Refs.~\cite{Allanach:2013kza,Ellwanger:2009dp}, while for the
    C\ESSM we use the notation of Ref.~\cite{Athron:2009bs}.}
  \label{tab:semi_analytic_bms}
\end{table}

In addition to carrying out unit tests on individual benchmark points, we have
also performed extensive scans in the CMSSM to confirm that the semi-analytic
solver produces results in agreement with the two-scale solver.  For most
points, this is found to be the case; however, we have also observed important
exceptions where non-negligible differences are found between the two
solvers.  In these cases, one solver may fail to converge to a stable
solution, or multiple solutions are present \cite{Allanach:2013cda} with
different stability properties in the two solvers.  In this latter case, note
that both solvers always return the first solution to
which they converge,\footnote{When multiple convergent solutions exist,
  the one first obtained will depend, for instance, on the initial guess
  used for the iteration.} with the iteration
stopping immediately once a convergent solution is obtained.  That is, neither
method attempts to find all possible solutions for the given input parameters
or automatically select between multiple fixed points.  Since a given solution
might not be a stable fixed point of both iterations, the two solvers need not
converge to the same solution, leading to the observed differences.  We have
checked that such points nevertheless satisfy the boundary conditions imposed at
each scale and are indeed valid solutions of the BVP.  More generally we have
checked that solutions found in one solver also correspond to (not necessarily
stable) fixed point solutions of the other solver algorithm; that is, they
satisfy all of the boundary conditions so that the parameter values remain
unchanged after applying a single step of the iteration.

\begin{figure}[tbh]
  \centering
  \includegraphics[width=0.6\textwidth]{%
    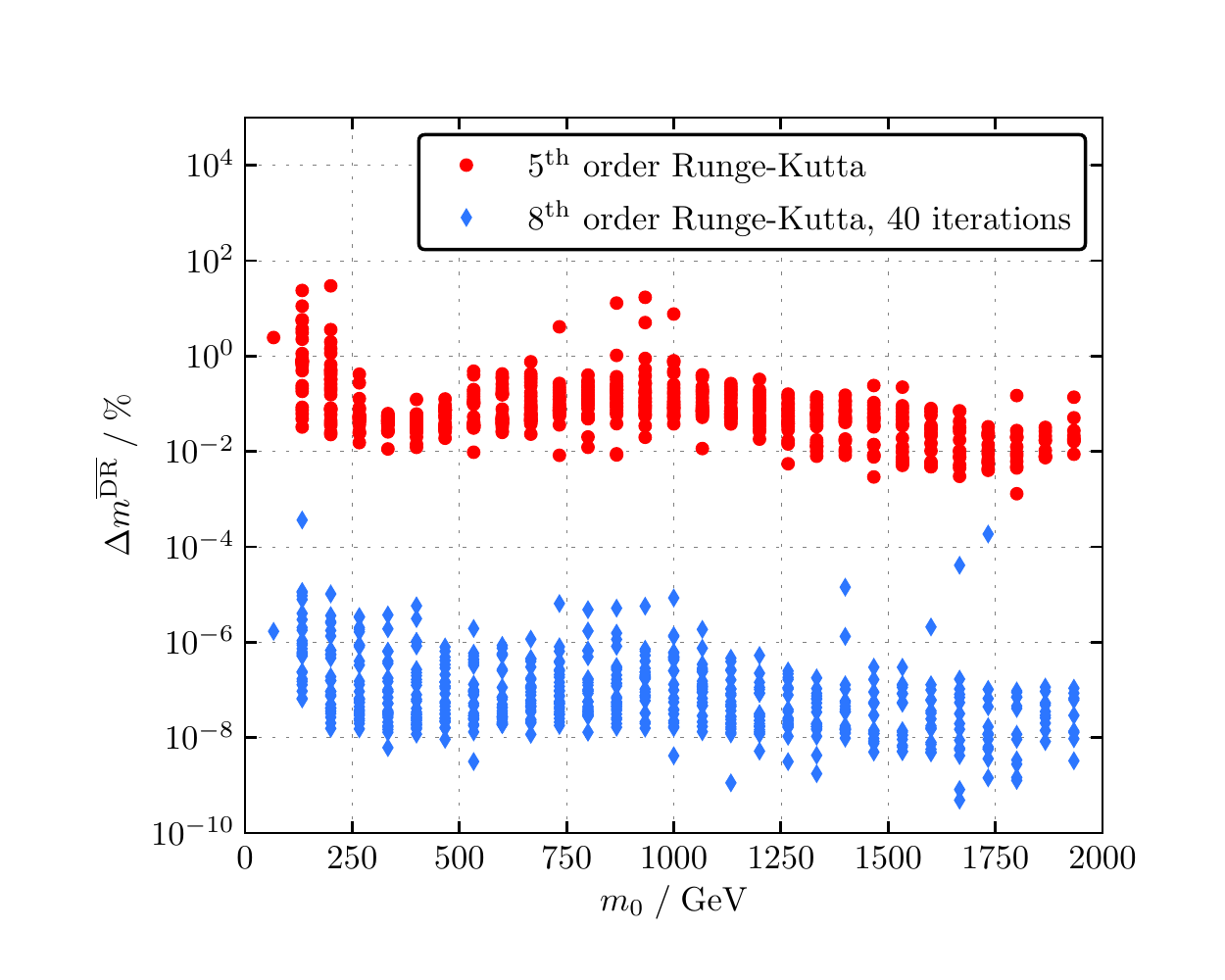}
  \caption{Percentage changes in the CMSSM \DRbar mass spectrum after
    applying a single step of the semi-analytic solver to points obtained
    using the two-scale solver in a linear scan over $\mhalf \in
    [0,300]$ GeV, $\mzero \in [0,2]$ TeV, $\azero = 0$ GeV, $\tan\beta = 40$ and
    $\mu > 0$.  When run using the default Runge-Kutta algorithm provided
    by \fs and allowing the iteration to stop as soon as the precision goal of
    $10^{-4}$ is reached (red circles), changes between $1$\% and $30$\% are
    found for a small number of points.  By using the $8^{\text{th}}$ order
    Runge-Kutta integrator and ensuring convergence is reached in the
    estimate for $M_X$ by forcing 40 iterations in the two-scale algorithm,
    these differences are reduced below the level of $0.001$\% (blue
    diamonds).}
  \label{fig:semi_analytic_differences}
\end{figure}

From these tests, we have found that in some cases the agreement
between the two solvers can depend quite sensitively on small differences
between them.  To illustrate this, in \figref{fig:semi_analytic_differences}
we show the percentage changes in the \DRbar mass spectrum in the CMSSM after
running points obtained using the two-scale solver through a single step of
the semi-analytic solver; if the point is also a fixed point of the latter,
this change should be negligible.  In this
scan, the change after a single iteration can be on the level of several
percent for a small number of points, reaching between $20$\% and $30$\% for
some exceptional points.
That these points initially appear not to be fixed points of the semi-analytic
solver arises primarily from the fact that the semi-analytic coefficients,
and hence the EWSB solution, are sensitive to the estimate for the
high scale $M_X$, as well as numerical errors in the integration of the RGEs.
In particular, for the default convergence criteria imposed by \fs, the
two-scale solver's estimate for $M_X$ is not close enough to convergence,
leading to significant differences in the calculated low-energy soft
parameters.  By requiring convergence in the estimate for $M_X$, together with
using a higher-order Runge-Kutta integration\footnote{By default, \fs makes
  use of an adaptive $5^{\text{th}}$ order algorithm; to perform this test we
  have also implemented into \fstwo an $8^{\text{th}}$ order solver that makes
  use of the Runge-Kutta-Fehlberg method provided by the Boost library
  \texttt{odeint}.  This higher-order solver is available to the user by
  choosing it at the C++ level.} and demanding a higher precision for the
obtained EWSB solution, the change after one iteration is reduced below the
permille level.  Thus, it is important to be aware that differences in the
convergence properties of the two solvers can have an impact on the solutions
found, even if a given point would be a fixed point of both solvers.

\begin{figure}[tbh]
  \centering
  \includegraphics[width=0.5\textwidth]{%
    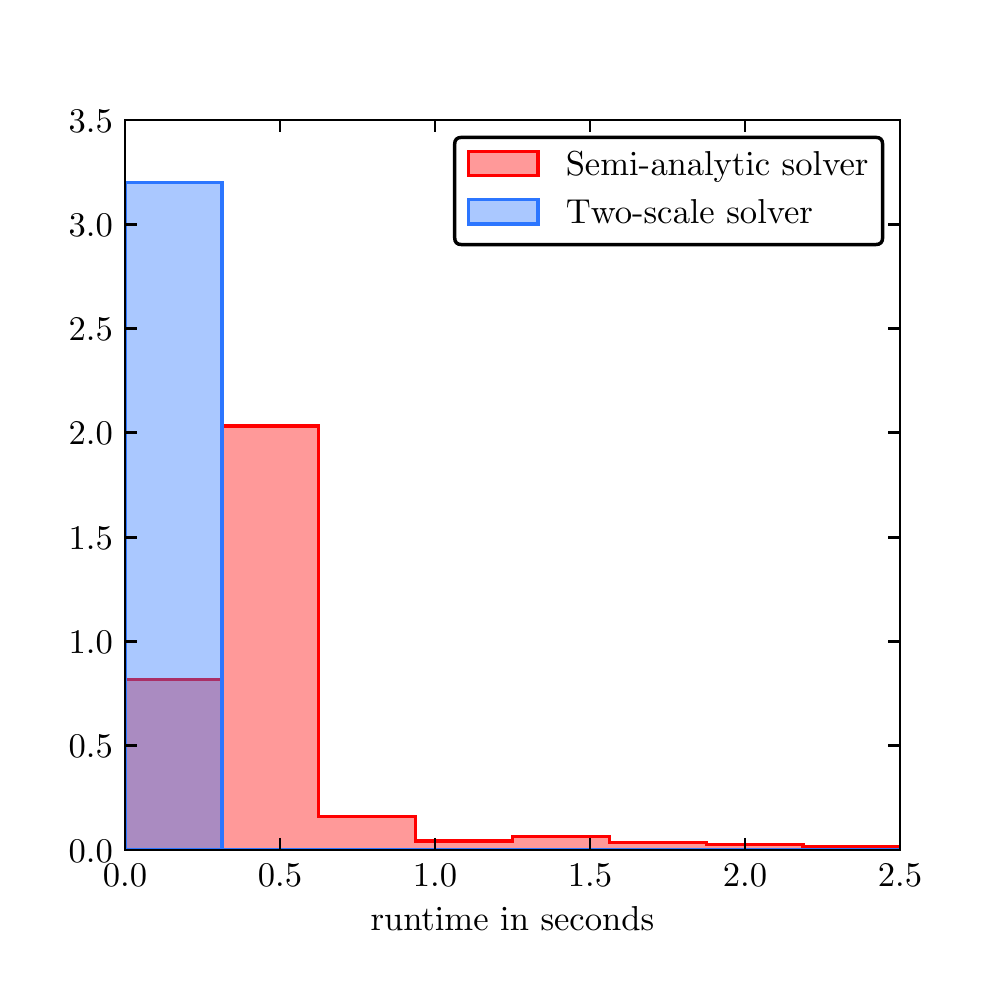}
  \caption{Runtime distributions of the CMSSM spectrum generators created with
    \fstwo using the two-scale and semi-analytic BVP solvers, obtained on
    an Intel i7-4702MQ CPU\@.  The distributions are normalized to have integral
    of unity.}
\label{fig:semi_analytic_benchmark}
\end{figure}

The typical runtimes for the two solvers in the CMSSM are compared in
\figref{fig:semi_analytic_benchmark}.  The distributions are obtained
by randomly sampling the CMSSM input parameters
$\mzero\in [0.2,1]\unit{TeV}$, $\mhalf\in [0.2,1]\unit{TeV}$,
$\tan\beta\in [2,30]$, $\sign\mu\in\{-1,+1\}$ and
$\azero\in [-1,1]\unit{TeV}$.
The runtime of the semi-analytic solver is increased on average by a factor of
$\approx 3$ compared to the two-scale solver.  This increase is
mostly due to the increased number of iterations performed by the semi-analytic
solver.  For each outer iteration of the semi-analytic solver, the inner
iteration typically runs through a similar number of steps as for a full run
of the two-scale solver, with this number decreasing as convergence is
approached on each outer iteration.  Consequently, the total number of
iterations for the semi-analytic solver tends to be larger than that for the
two-scale solver by a similar factor.  There is also an additional cost
associated with running between scales to compute the semi-analytic
coefficients.

\begin{figure}
  \centering
  \includegraphics[width=0.48\textwidth]{%
    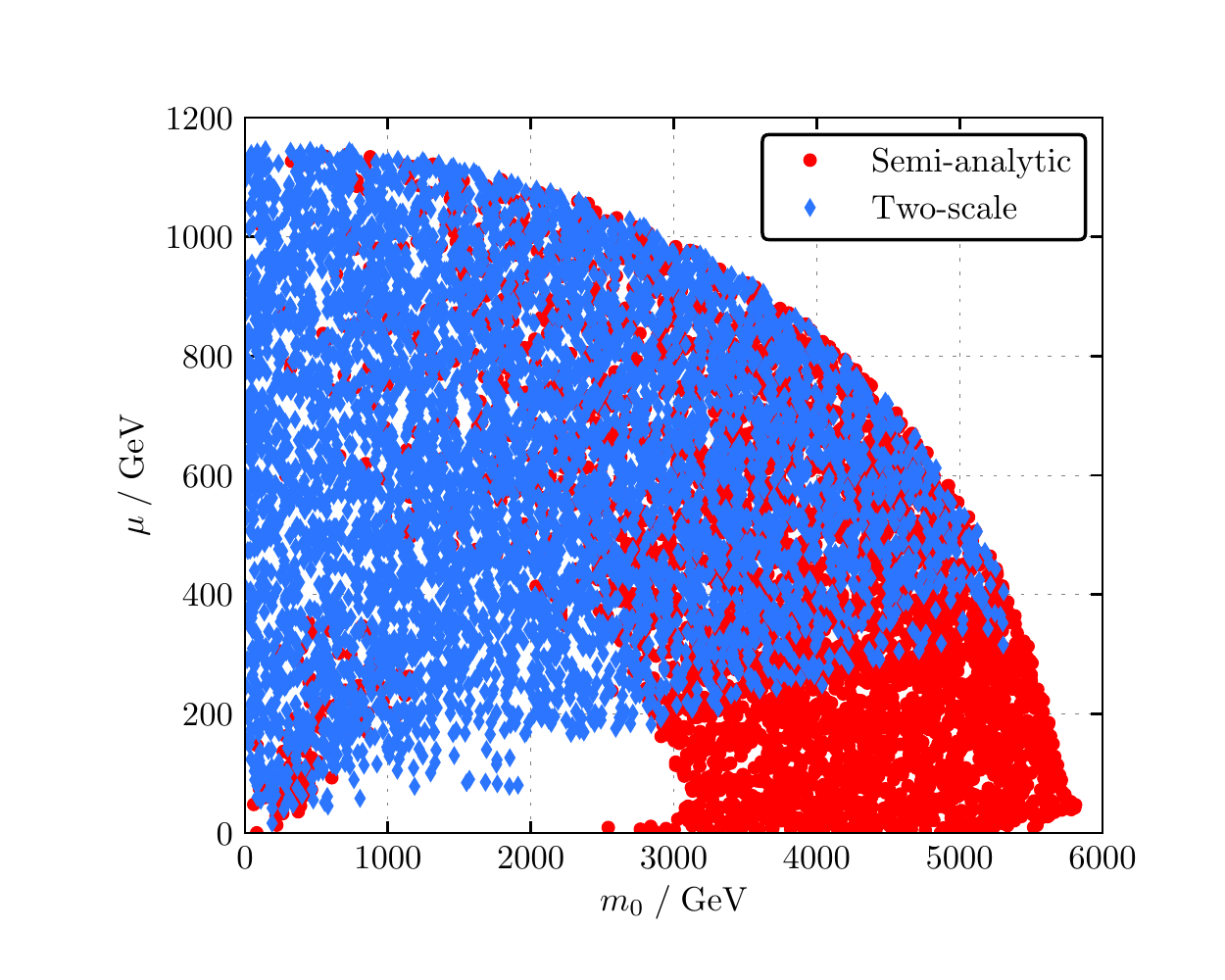}
  \includegraphics[width=0.48\textwidth]{%
    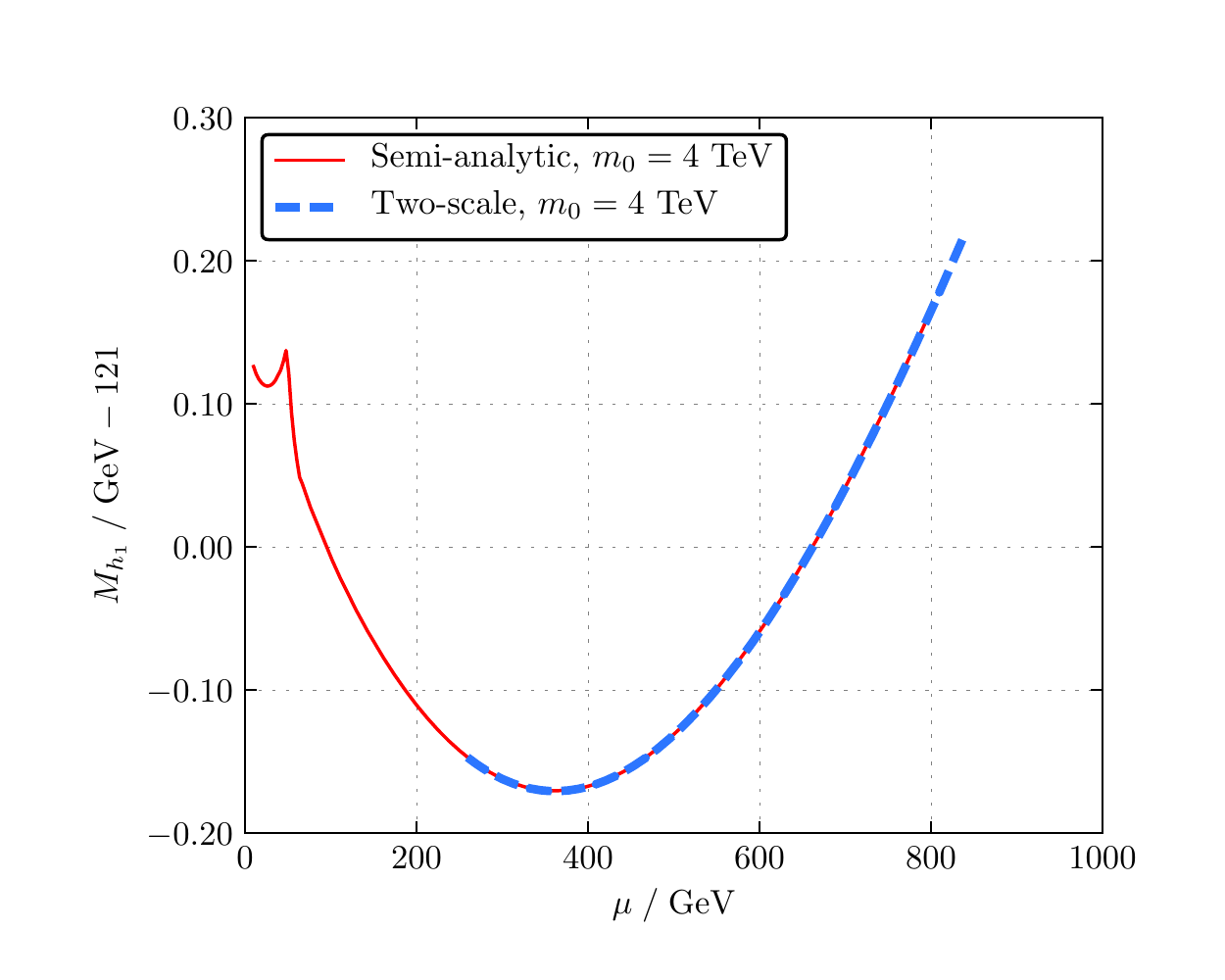}
  \caption{Left panel: CMSSM solutions found by the two-scale solver
    (blue diamonds) and the semi-analytic solver (red circles) in the
    $\mzero - \mu$ plane.  For both solvers, $\mhalf \in [0,1]\unit{TeV}$
    is randomly sampled, while $\tan\beta = 10$ and $\azero = 0$ GeV\@.
    The solutions obtained using the two-scale solver are found by
    randomly sampling $\mzero \in [0, 6]\unit{TeV}$ with $\sign \mu = 1$,
    and those for the semi-analytic solver are found by randomly sampling
    $\mu \in [0,1150] \unit{GeV}$.  Right panel: Change in the calculated
    lightest \CP-even Higgs pole mass $M_{h_1}$ from a reference value of
    $121$ GeV as a function of $\mu$ at fixed $\mzero = 4$ TeV for the two
    solvers.  The solutions shown correspond to a vertical slice at fixed
    $\mzero$ in the left-hand plot; $\mhalf$ is again varied in
    $[0,1]\unit{TeV}$, $\tan\beta = 10$ and $\azero = 0$ GeV.}
  \label{fig:cmssm_solution_regions}
\end{figure}

While the semi-analytic solver suffers from an increased runtime compared to
the two-scale solver, it is also able to provide complementary coverage of the
parameter space to that of the two-scale solver.  This is demonstrated in the
left panel of \figref{fig:cmssm_solution_regions} in the CMSSM, where the
solutions found by each solver are plotted in the $\mzero - \mu$ plane.
In this case, the use of the semi-analytic solver allows for a large number
of solutions to be found in the focus point region \cite{Chan:1997bi,
  Feng:1999hg,Feng:1999mn} at small values of $\mu \ll \mzero$, where the
two-scale solver is unable to find convergent solutions.  Conversely,
the two-scale solver is more effective for finding solutions with small
$\mzero$.  This highlights the fact that, for some parameter points,
the semi-analytic solver generated by \fs might not find a solution
where the two-scale solver is able to, and vice versa.  As noted above, this
can be due to the point in question having different stability properties
under the two different algorithms.  For example, the CMSSM point with
$m_0 = 125\unit{GeV}$, $M_{1/2} = 300\unit{GeV}$, $\tan\beta = 10$,
$\sign \mu = 1$ and $A_0 = 0\unit{GeV}$  is successfully solved using the
two-scale solver, with $\mu \approx 395\unit{GeV}$ required for correct EWSB.
The same point, when run using the semi-analytic solver with the two-scale
solution for $\mu$, fails to converge; for this choice of input parameters
and initial guess, the iteration enters a periodic orbit in which the
approximations to the solution are close to, but do not correspond to, the
fixed point found by the two-scale solver.  To avoid the observed cyclic
behaviour here, it is necessary to either fine-tune the provided input
parameters or otherwise modify the initial guess made by \fs.  Due to the
differing choice of input and output parameters between the two algorithms
in general, certain regions of the parameter space can also be susceptible to
high levels of numerical sensitivity in one approach and not the other,
again necessitating significant fine-tuning.  In the CMSSM, for instance,
large cancellations are typically required in
order to produce a small value of $m_0$ when using the semi-analytic solver.
In such cases, $\mu$ must be carefully fine-tuned to obtain a valid solution.
On the other hand, when using the two-scale solver one has direct control over
the value of $m_0$, and it is not necessary to fine-tune in order to obtain
the desired small value of $m_0$.  The situation is reversed in regions of
parameter space with small values of $\mu$ and large $m_0$, where one now
has direct control over the fine-tuned value of $\mu$ in the semi-analytic
solver but not in the two-scale solver.  In general terms, the regions of
parameter space in which the two solvers are effective need not overlap, and
the use of both in tandem allows for a more complete picture of the parameter
space to be obtained.  Moreover, in the regions in which both solvers do find
solutions, there is excellent agreement between the two algorithms.\footnote{
  Provided the same solution is found if multiple solutions exist.}  This is
illustrated in the right panel of \figref{fig:cmssm_solution_regions}, where
the lightest \CP-even Higgs mass, expressed as the difference from a reference
value of $121$ GeV, is plotted for fixed $\mzero = 4$ TeV\@.  If both solvers
find a solution for this value of $\mzero$, the two values of the Higgs mass
agree very well.

\begin{figure}
  \centering
  \includegraphics[width=0.6\textwidth]{%
    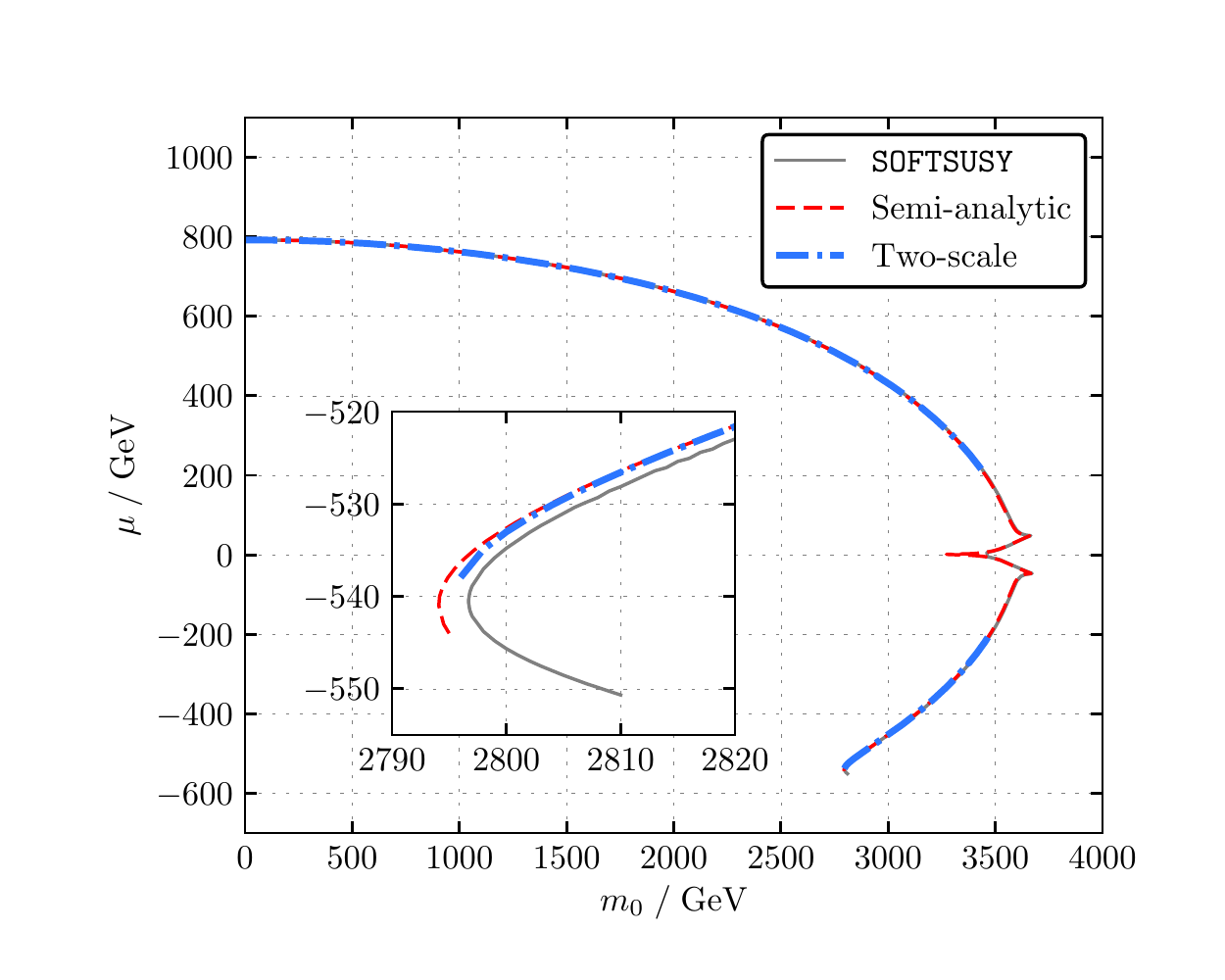}
  \caption{Solutions obtained using the two-scale and semi-analytic solvers
    in the CMSSM for fixed $\mhalf = 660$ GeV, $\azero = 0$ GeV and
    $\tan\beta = 40$.  For comparison, we also reproduce the curve shown in
    Figure~5 of Ref.~\cite{Allanach:2013cda}, where a modified version of
    \softsusy 3.3.7 was used to allow multiple solutions to be found to the
    CMSSM boundary conditions.  The inset plot shows the region $2790\unit{GeV}
    \leq \mzero \leq 2820\unit{GeV}$, $-555\unit{GeV} \leq \mu
    \leq -520\unit{GeV}$ in more detail.}
  \label{fig:cmssm_multiple_solutions}
\end{figure}

It is also evident from \figref{fig:cmssm_solution_regions} that the use of
both solvers allows features in the parameter space to be picked up that would
be missed by either solver alone.  In this case, in
\figref{fig:cmssm_solution_regions} the two-scale solver only finds a single
solution for each value of $\mhalf$, for fixed $\sign\mu = 1$, while the
semi-analytic solver in some cases finds multiple solutions.  These
solutions have different values of $|\mu|$ for the same value of $\mhalf$,
leading to the sharp feature at low values of $\mu$ evident in the right panel
of \figref{fig:cmssm_solution_regions}.  The existence of multiple solutions to
the CMSSM BVP, and the inability of the ordinary two-scale fixed point iteration
to find all such solutions, are well-known and have previously been studied in
Refs.~\cite{Allanach:2013cda,Allanach:2013yua}.  In
\figref{fig:cmssm_multiple_solutions} we compare the results obtained using
the two BVP solvers in \fstwo to those found using the modified version of
\softsusy employed in Ref.~\cite{Allanach:2013cda}.
As expected, in this region of the parameter space the two-scale solver
produced by \fs finds at most only two solutions, corresponding to the two
possible signs of $\mu$.  The automatically generated semi-analytic solver
enables additional solutions to be found, which are in good agreement with
those found using the modified version of \softsusy.  Slight differences in
the values of $\mu$ and $\mzero$, and where solutions are found, arise from
small differences in the incorporated corrections between the two codes.  In
particular, the semi-analytic solver does not find valid solutions for
$\mu < -545$ GeV due to the tree-level mass of the \CP-odd Higgs becoming
tachyonic.  In general, multiple solutions at a given parameter point
can have significantly different phenomenological properties
\cite{Allanach:2013yua}, so that finding them is important to completely
characterize a model.  The availability of multiple solvers improves the
ability of \fs to locate additional solutions, without requiring extensive
modifications to the generated code.

The longer runtime of the semi-analytic solver can also be an
acceptable tradeoff if the model of interest cannot easily be
handled using the two-scale solver.  Constrained models such as the
CNMSSM and C\ESSM are examples of this.  To demonstrate the applicability of
the semi-analytic solver to these models, we have performed
scans over the parameter spaces of these models.

First we have performed scans in the CNMSSM to demonstrate that with the
semi-analytic solver we can sample the parameter space effectively, making
\fstwo the first public spectrum generator that can do this ``out of the box''.
We performed a four dimensional scan of the CNMSSM, using the model file
provided in \appref{app:CNMSSM} to produce a CNMSSM spectrum generator that
was then linked to \multinest 3.10 for efficient sampling of the parameter
space.  The input parameters were varied over the ranges
\begin{gather}
  2 \leq \tan \beta \leq 50, \\
  0\unit{TeV} \leq \mhalf \leq 5\unit{TeV}, \\
  -5\unit{TeV} \leq \azero \leq 0\unit{TeV}, \\
  -0.3 \, \leq \lambda(M_X) \leq 0.3, \\
  \sign \mu_{\textrm{eff}}  = +1,
\end{gather}
and the log likelihood function provided to \multinest was defined to be
\begin{equation*}
  \log L = -\frac{1}{5}({M_h^{\SM}}/{\unit{GeV}} - 125.09)^2 ,
\end{equation*}
where $M_h^{\SM}$ is the SM-like Higgs pole mass.  In this way the
scan is directed towards solutions with the observed Higgs mass of
$125.09$ GeV\@.

\begin{figure}[tbh]
  \centering
  \includegraphics[width=0.48\textwidth]{%
     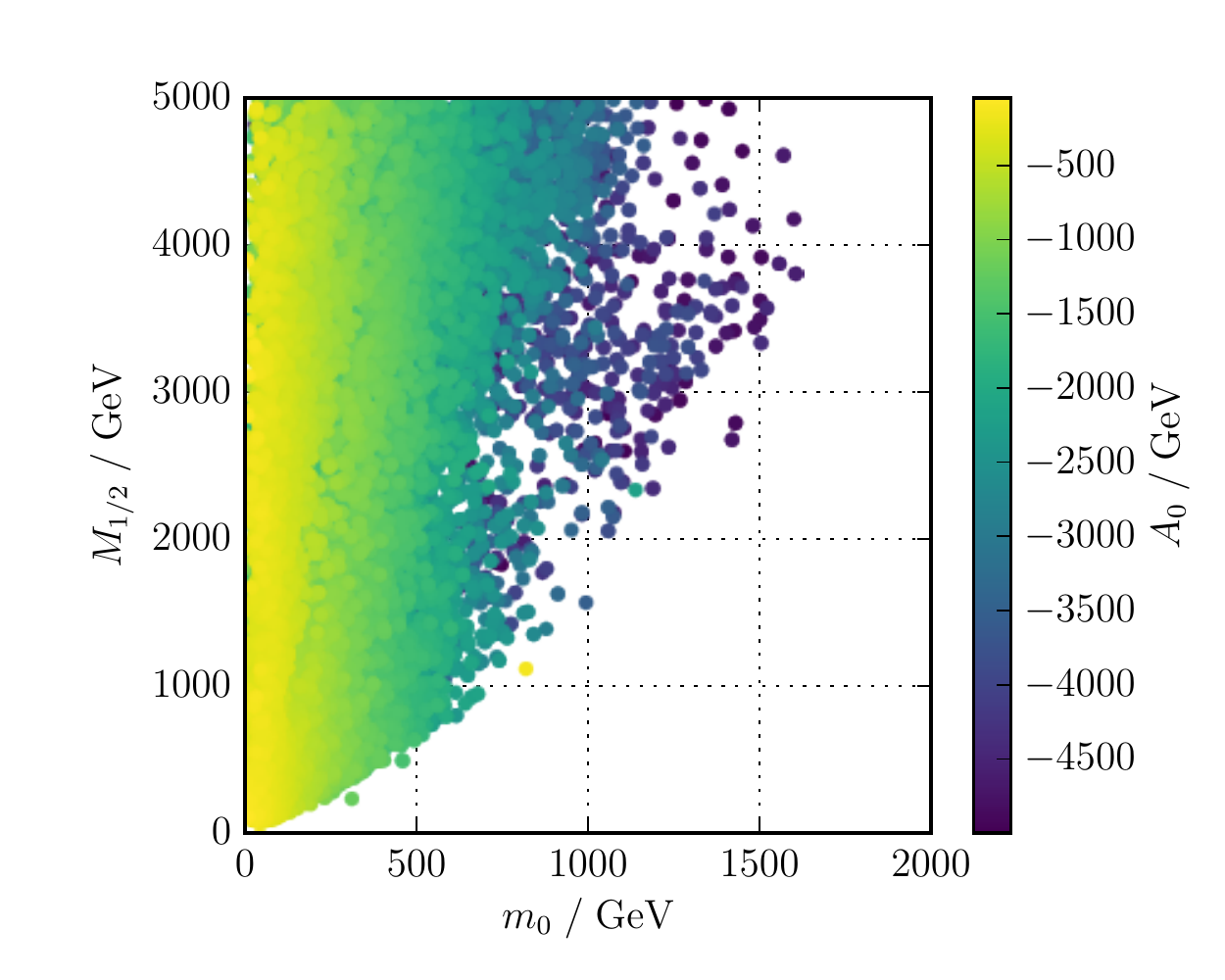}
   \includegraphics[width=0.48\textwidth]{%
     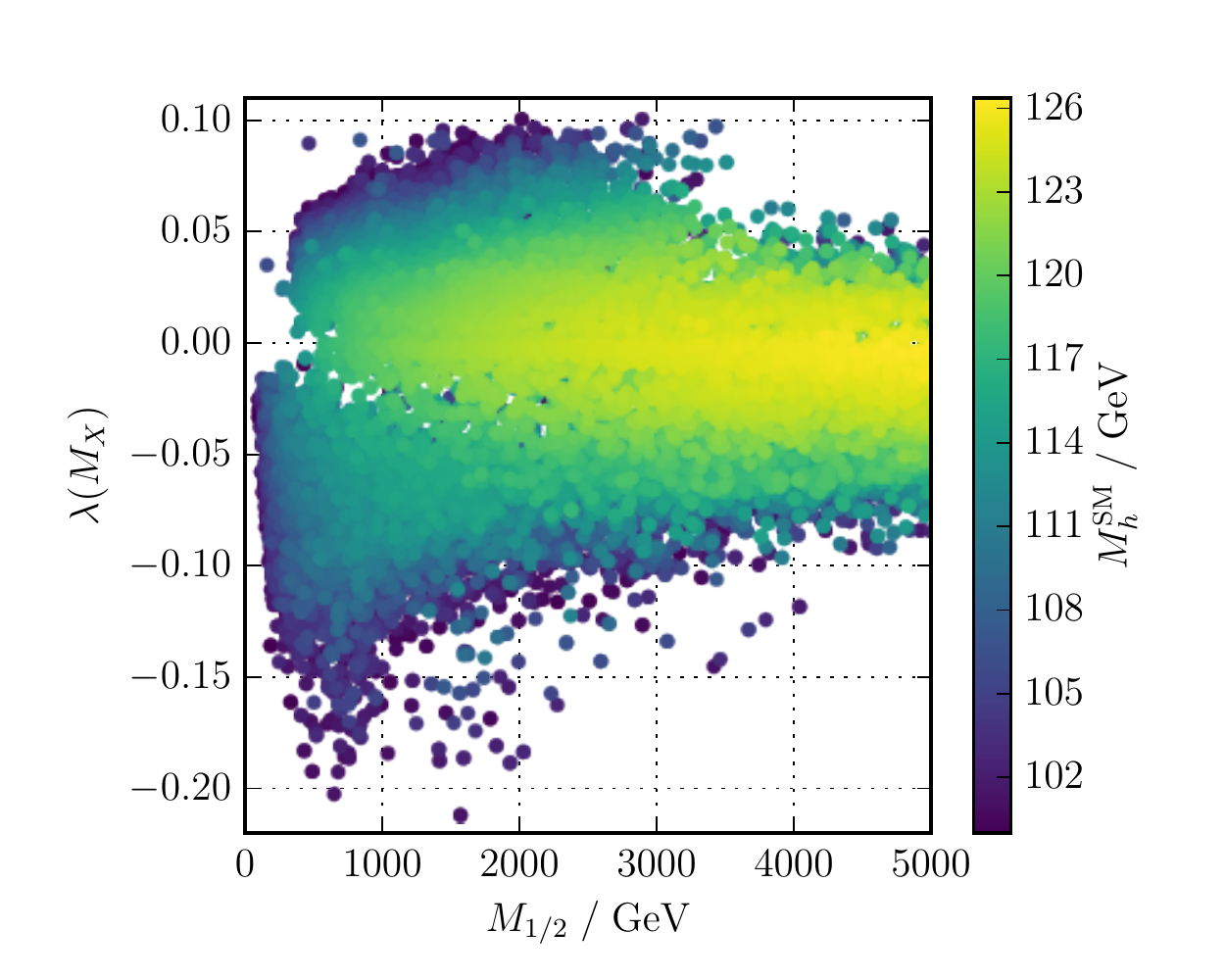} \\
  \includegraphics[width=0.48\textwidth]{%
    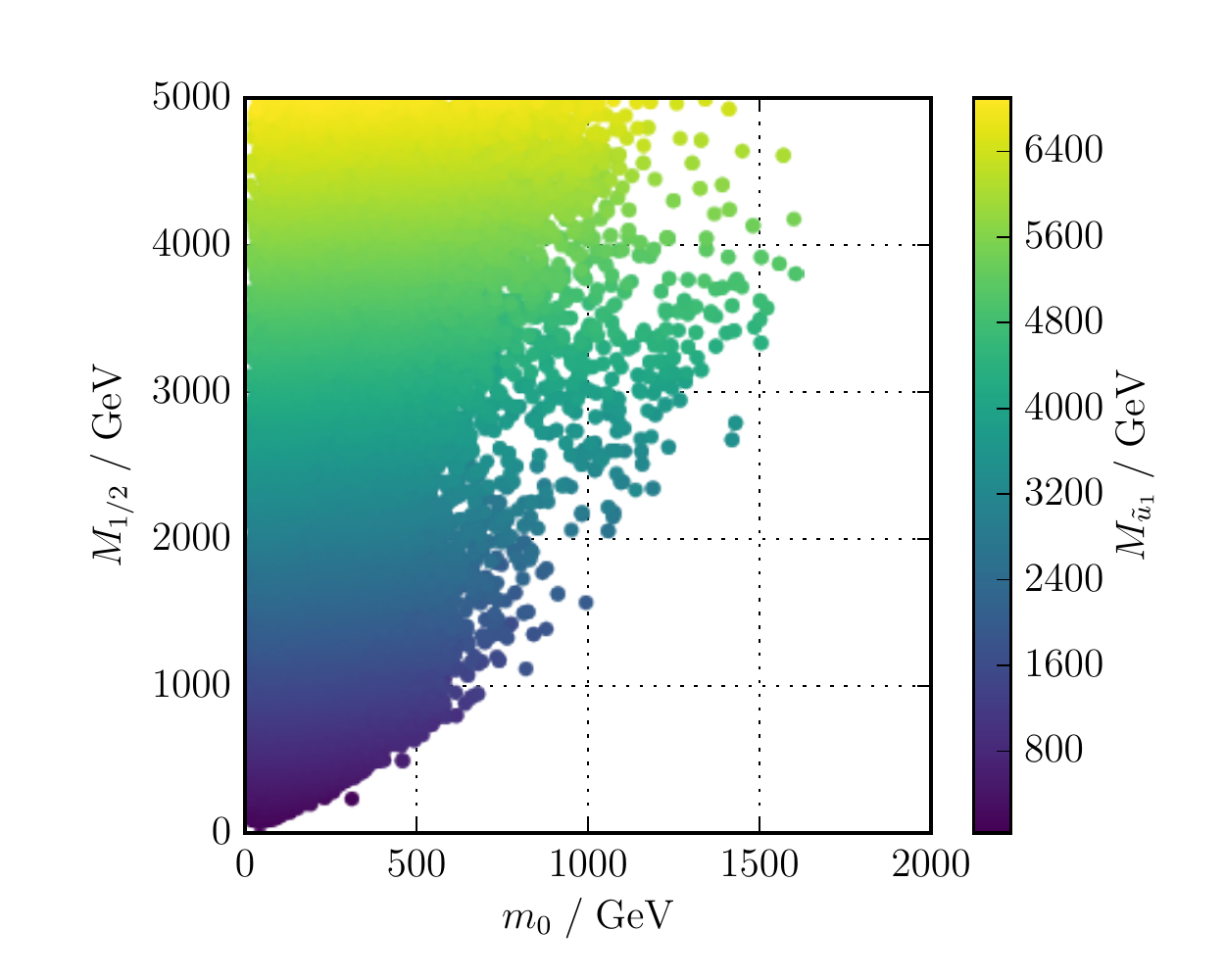}
   \includegraphics[width=0.48\textwidth]{%
    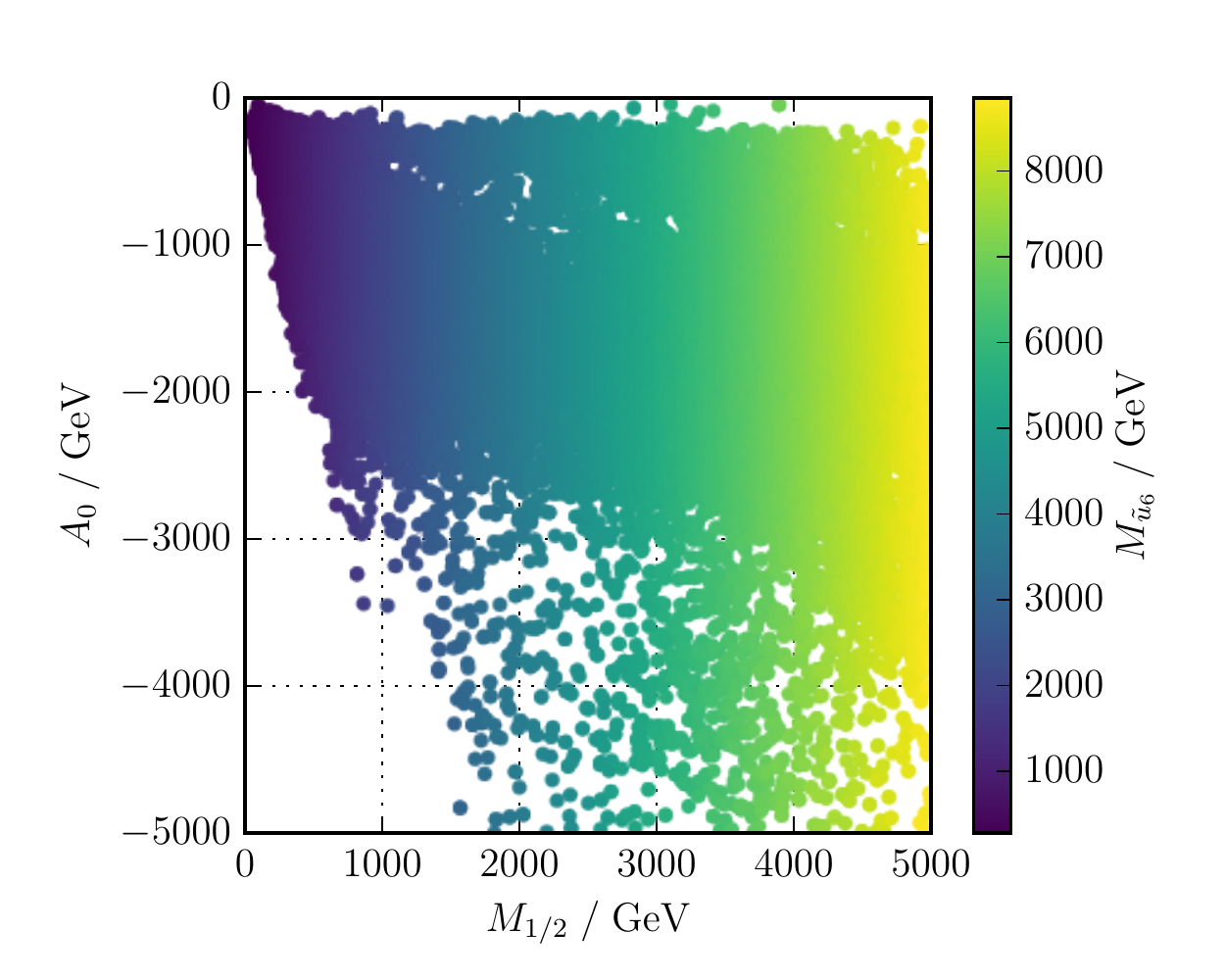}
   \caption{CNMSSM solutions obtained using \fstwo and \multinest 3.10
     shown in the $\mzero-\mhalf$ plane with $\azero$ as a color
     contour (top left panel); $\mhalf-\lambda$ plane showing the
     SM-like Higgs pole mass $M_h^{\SM} $, as a color contour, with
     the range restricted to $M_h^{\SM} > 100$ GeV for clarity (top
     right panel); $\mzero-\mhalf$ plane with the mass of the lightest
     up-type squark, which is predominantly stop (bottom left panel),
     and the $\mhalf-\azero$ plane with the heaviest up-type squark
     mass (bottom right panel).}
   \label{fig:cnmssm_solutions}
\end{figure}

The results are shown in \figref{fig:cnmssm_solutions}, where all points are
required to have $\mzero^2 > 0$.  As can be seen in the top left panel, the range
of $\mzero$ values found is considerably smaller than the range allowed for the
universal soft masses, $\mhalf$ and $\azero$, which are inputs.  The reason for
this is that in the CNMSSM a non-zero singlet VEV must be generated by
developing the correct shape of the scalar potential.  This could be achieved
in the usual way through a negative quadratic term if the soft breaking singlet
mass squared, $m_S^2$, is negative.  In fact, since the NMSSM scalar potential
also contains the cubic singlet terms $\kappa A_\kappa S^3 / 3 + \mathrm{h.c.}$,
the requirement $m_S^2 < 0$ can be relaxed so that an approximate condition for
generating the correct shape of the potential reads $A_\kappa^2 \gtrsim
9 m_S^2$ \cite{Ellwanger:1996gw,Derendinger:1983bz}.  If $\mzero \gg
\azero / 3$, this condition is satisfied only if $m_S^2$ is driven to be
sufficiently small during the RG evolution from the GUT scale to the SUSY
scale.  This in turn can be achieved for large enough values of the
superpotential cubic singlet coupling $\kappa$ and the singlet-Higgs coupling
$\lambda$.

However, in this constrained model, where the soft trilinears are not free
input parameters at the SUSY scale, it is also the case that large values of $\lambda$
always generate substantial singlet mixing that can reduce the lightest
\CP-even Higgs mass.  As a result, a $125$ GeV Higgs mass is obtained with small
values of $\lambda$, to avoid this mixing, as well as large $\mhalf$, as can be
seen in the top right panel of \figref{fig:cnmssm_solutions}.  For such small
singlet Yukawa couplings, the RG flow between the GUT and SUSY scales leaves
$m_S^2(\MS)$ and $A_\kappa(\MS)$ close to their GUT scale values, namely
$\mzero^2$ and $\azero$.  As a result, $\mzero$ is heavily constrained if
the condition $A_\kappa^2 \gtrsim 9 m_S^2$ is to be satisfied in the absence of
significant RG evolution for $m_S$, as the top left panel of
\figref{fig:cnmssm_solutions} demonstrates.

The consequences of such small $\mzero$ values can be seen in the bottom left
and bottom right panels of \figref{fig:cnmssm_solutions}, where we plot the
lightest and heaviest up-type squark mass, respectively, to illustrate that the
squark masses are now predominantly set by the universal gaugino mass, with
little influence from $\mzero$ or $\azero$.  Furthermore, the squarks are always
lighter than the gluino, which has a mass $\approx 2\, \mhalf$.

\begin{figure}[tbh]
  \centering
  \includegraphics[width=0.48\textwidth]{%
     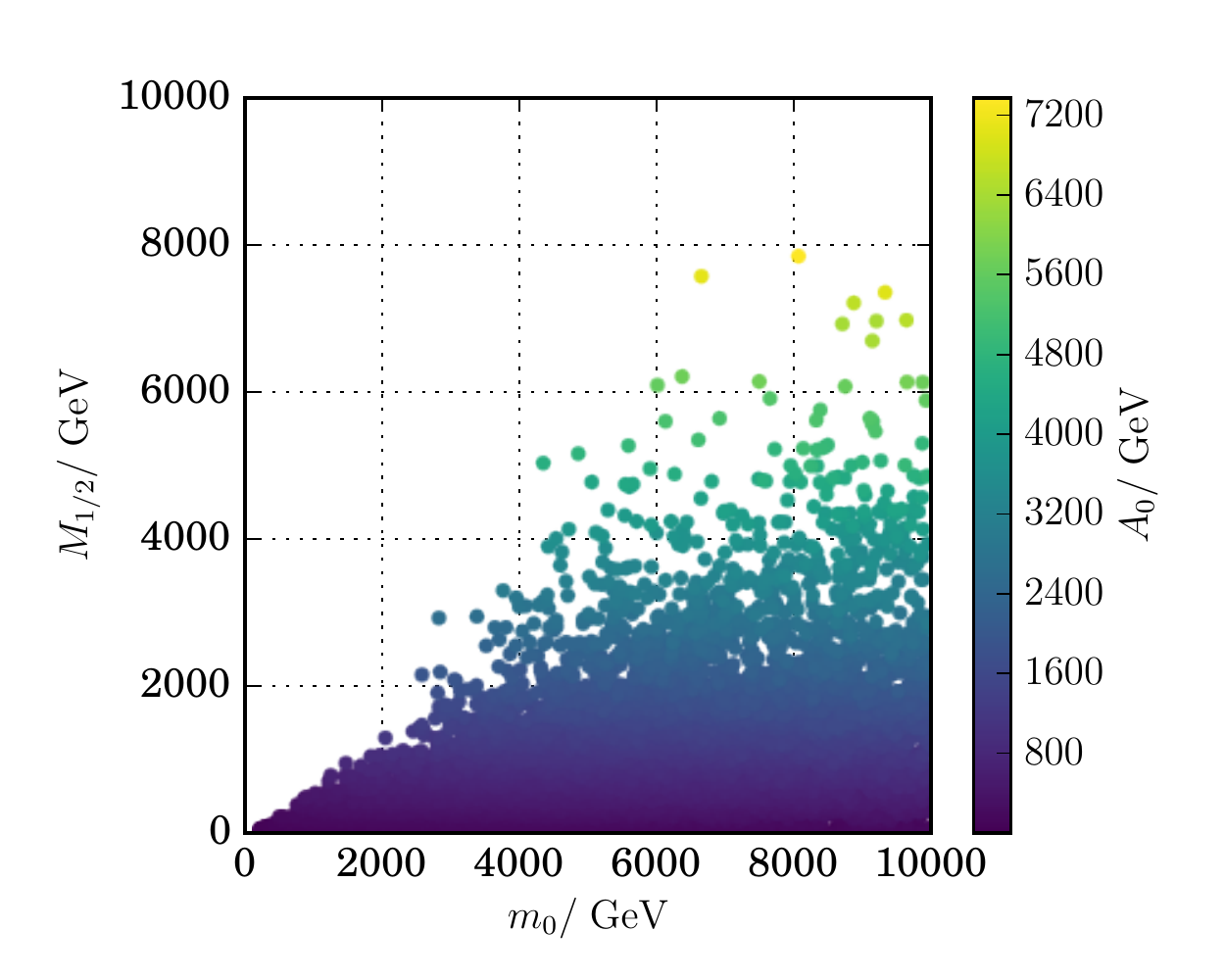}
   \includegraphics[width=0.48\textwidth]{%
     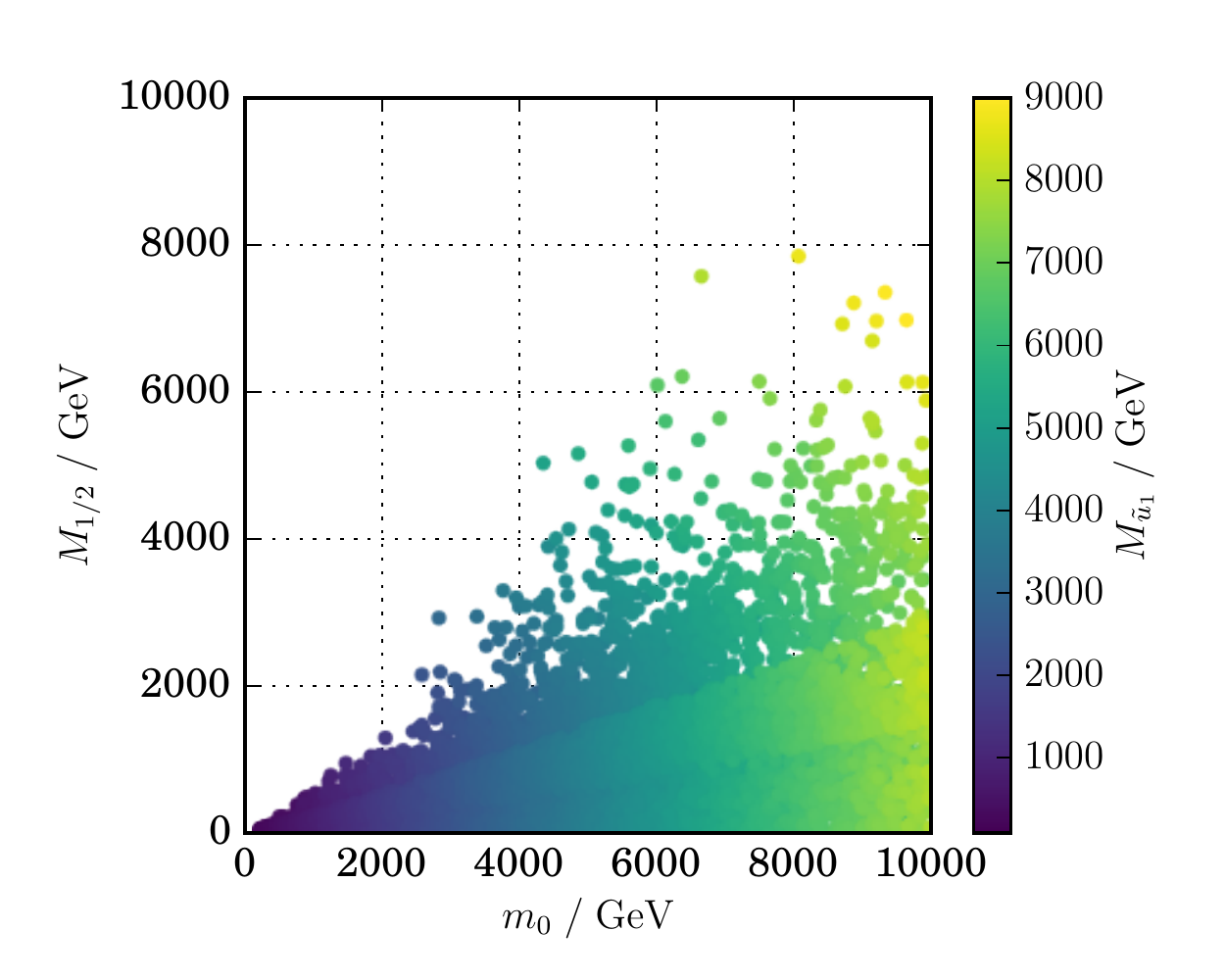} \\
  \includegraphics[width=0.48\textwidth]{%
    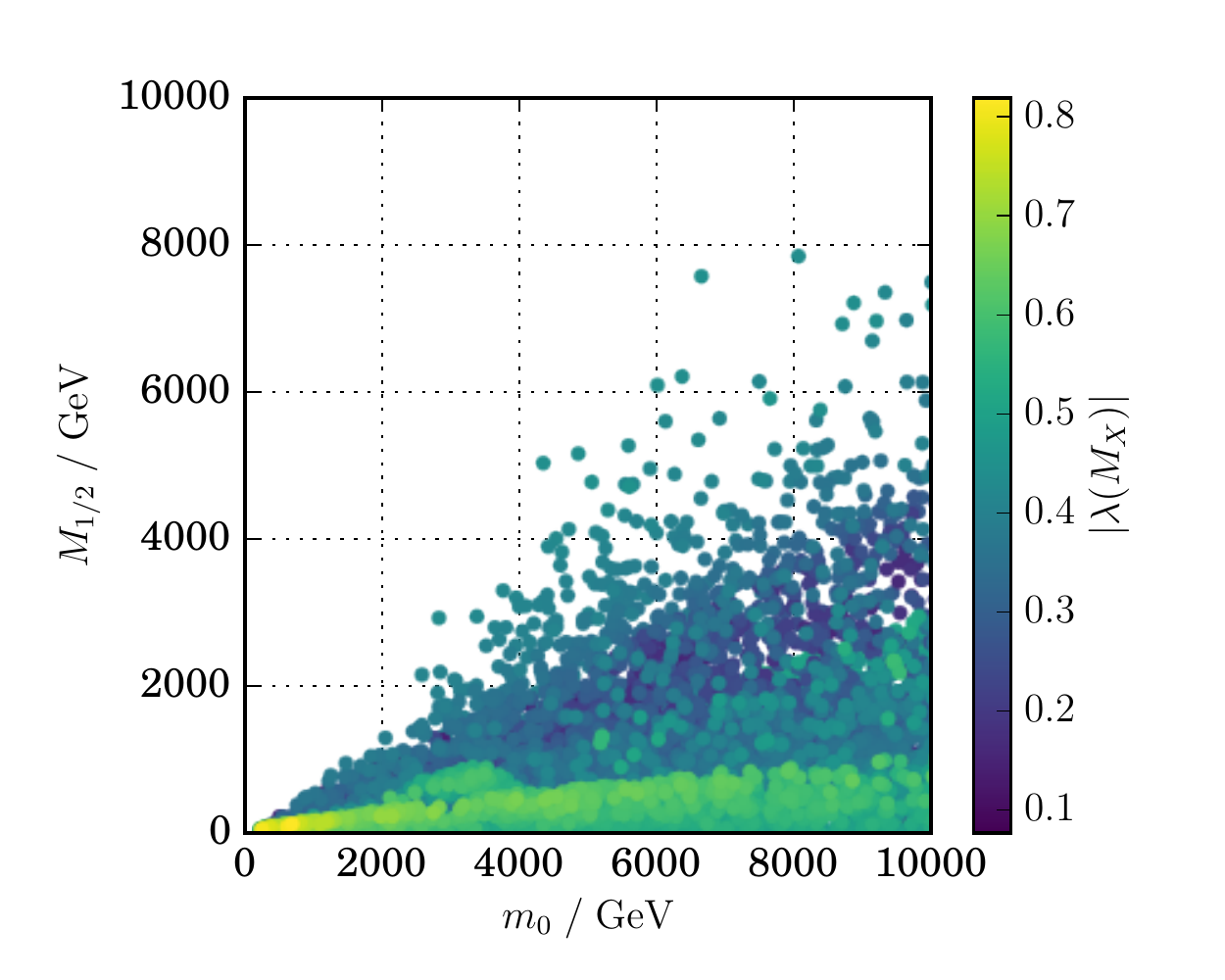}
   \includegraphics[width=0.48\textwidth]{%
    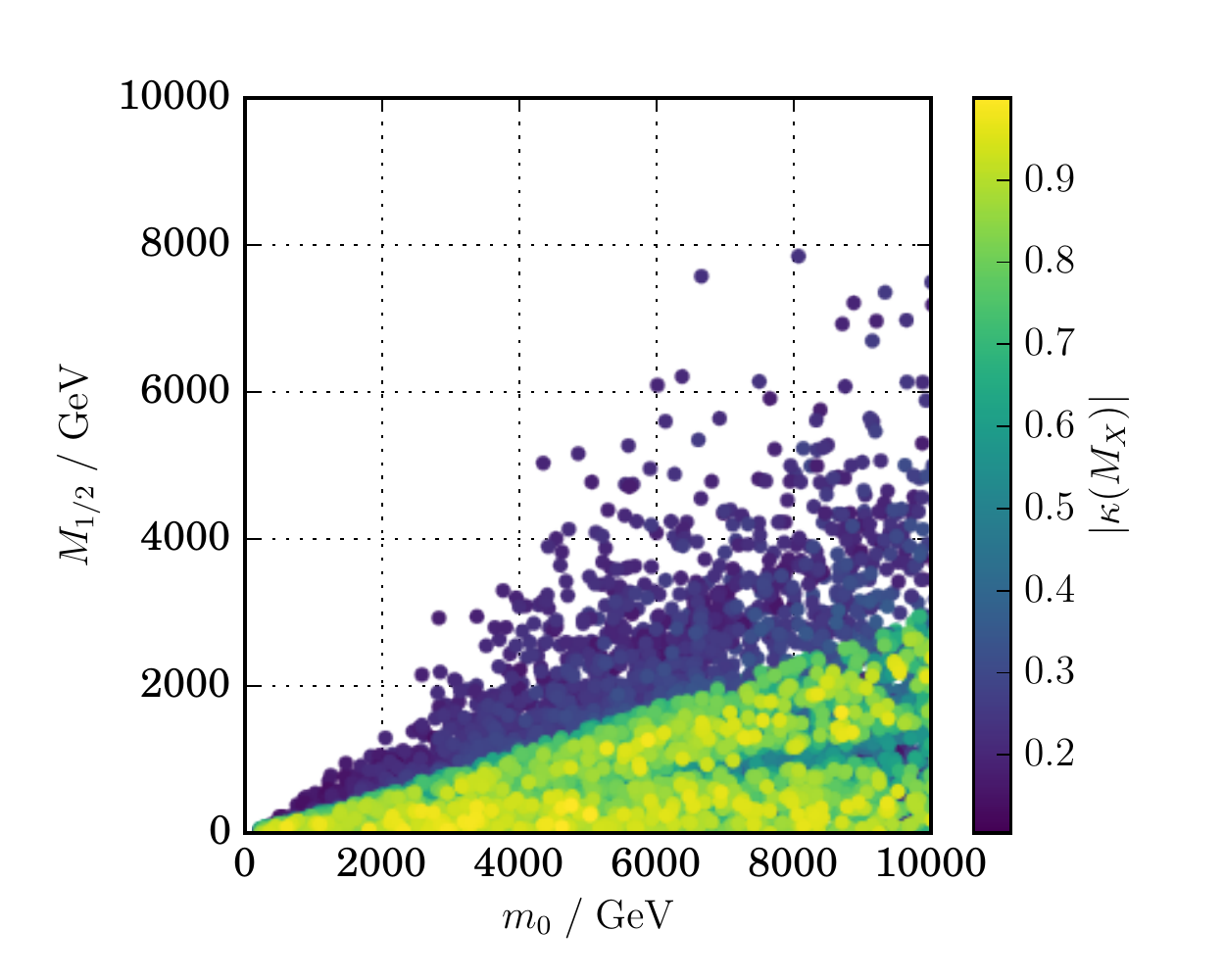}
   \caption{C\ESSM solutions obtained using \fstwo and \multinest 3.10 shown in
     the $\mzero-\mhalf$ plane with $\azero$, the lightest up-type squark pole mass
     $M_{\tilde{u}_1}$,  $|\lambda|$ and $|\kappa|$ as color contours.}
  \label{fig:ce6ssm_solutions}
\end{figure}

Compared to the CNMSSM, we find that the situation in the C\ESSM is rather
different.  The C\ESSM is an alternative to the CNMSSM in which an elementary
$\mu$ term is forbidden by a $U(1)_N$ gauge symmetry, and complete $E_6$ matter
supermultiplets are included to ensure anomaly cancellation.  As with the
CNMSSM, the two-scale solver is ineffective for finding solutions because EWSB
needs to have a soft mass as an output, but one can make a spectrum generator
with the semi-analytic solver where the universal GUT scale masses $\mzero$,
$\mhalf$ and $\azero$ are EWSB outputs.\footnote{The definitions of the
  new Lagrangian parameters and GUT scale constraints in the C\ESSM that we
  use may be found in Ref.~\cite{Athron:2009bs}.}  Using the \modelname{C\ESSM}
model file provided with \fstwo, we carried out a scan in which the
singlet-Higgs Yukawa coupling, $\lambda(M_X) \equiv \lambda_3(M_X)$, and the
exotic Yukawa coupling, $\kappa(M_X)$, were varied over $[-1,1]$ and the singlet
VEV, $s$, was varied over $[1,500] \unit{TeV}$.  The results of this scan are
shown in \figref{fig:ce6ssm_solutions}.

In contrast to the CNMSSM, in the C\ESSM it is very easy to have large
$\mzero$ values as there is an exotic Yukawa coupling between the singlet and
extra colored matter introduced to avoid gauge anomalies that drives the soft
singlet mass negative, providing a radiative symmetry breaking mechanism.
Additionally, whereas $\mzero \lesssim \mhalf$ in the CNMSSM, in this model
it is typically the case that $\mzero$ is larger than $\mhalf$.  This is
in qualitative agreement with the literature \cite{Athron:2009ue,
  Athron:2009bs,Athron:2012sq,Athron:2012pw}, and arises because the new
colored matter results in heavily modified RGEs in which the 1-loop
$\beta$ function of the strong coupling now vanishes.  The squark masses are
mostly set by $\mzero$ as a result, as is shown in the top right panel of
\figref{fig:ce6ssm_solutions}, while the bottom panels show that the range for
$\lambda$ and $\kappa$ at the GUT scale is very wide, despite strong
constraints on $\lambda$ at the electroweak scale coming from requirements for correct
EWSB\@.   We do not make a detailed quantitative comparison to previous work in
the literature here, but note that this is the first time that the C\ESSM
results have been presented with the same level of precision (full 2-loop RGEs,
1-loop pole masses) as is standard in the CMSSM and significant quantitative
differences are to be expected.

\begin{figure}[tbh]
  \centering
  \includegraphics[width=0.48\textwidth]{%
    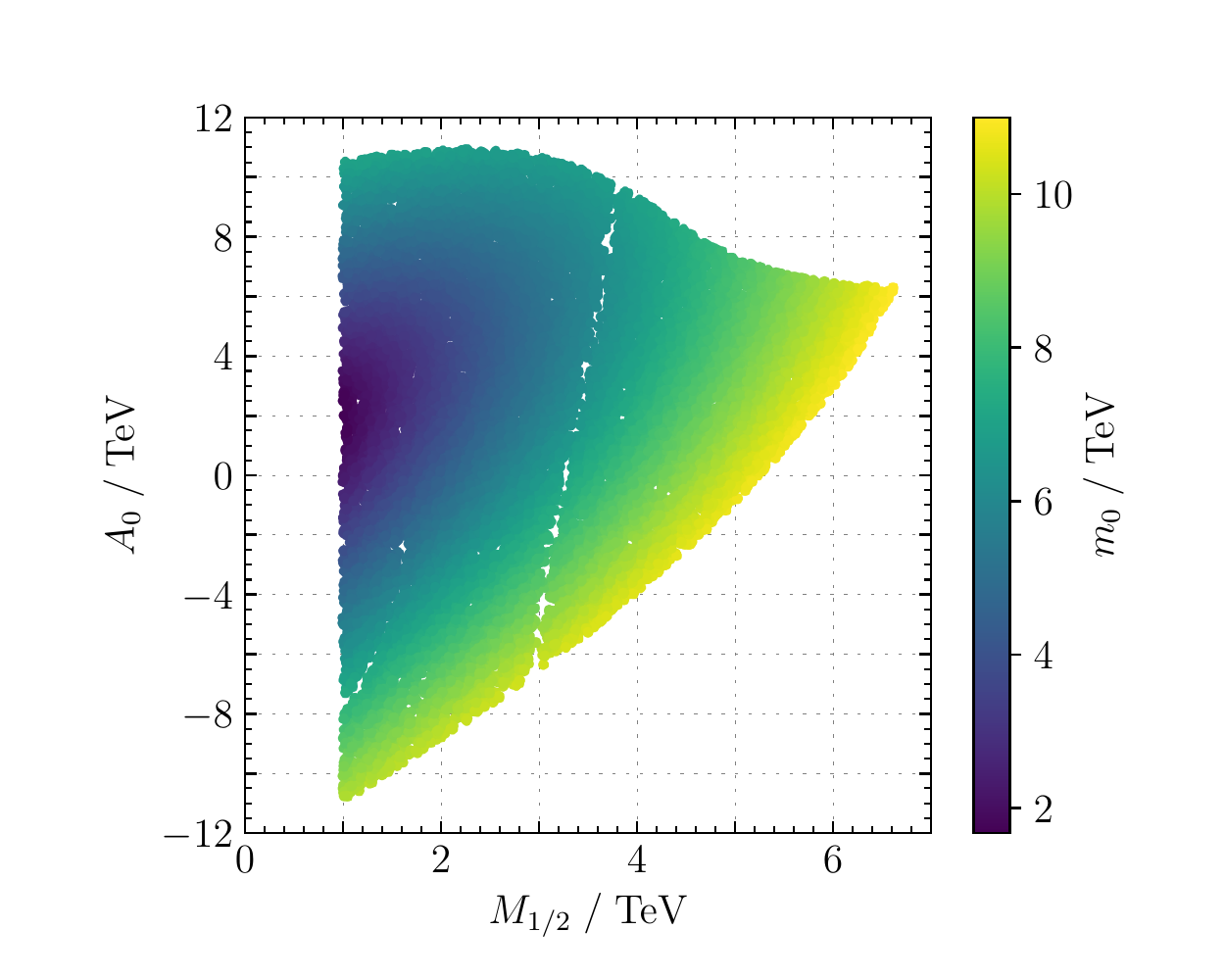}
  \includegraphics[width=0.48\textwidth]{%
    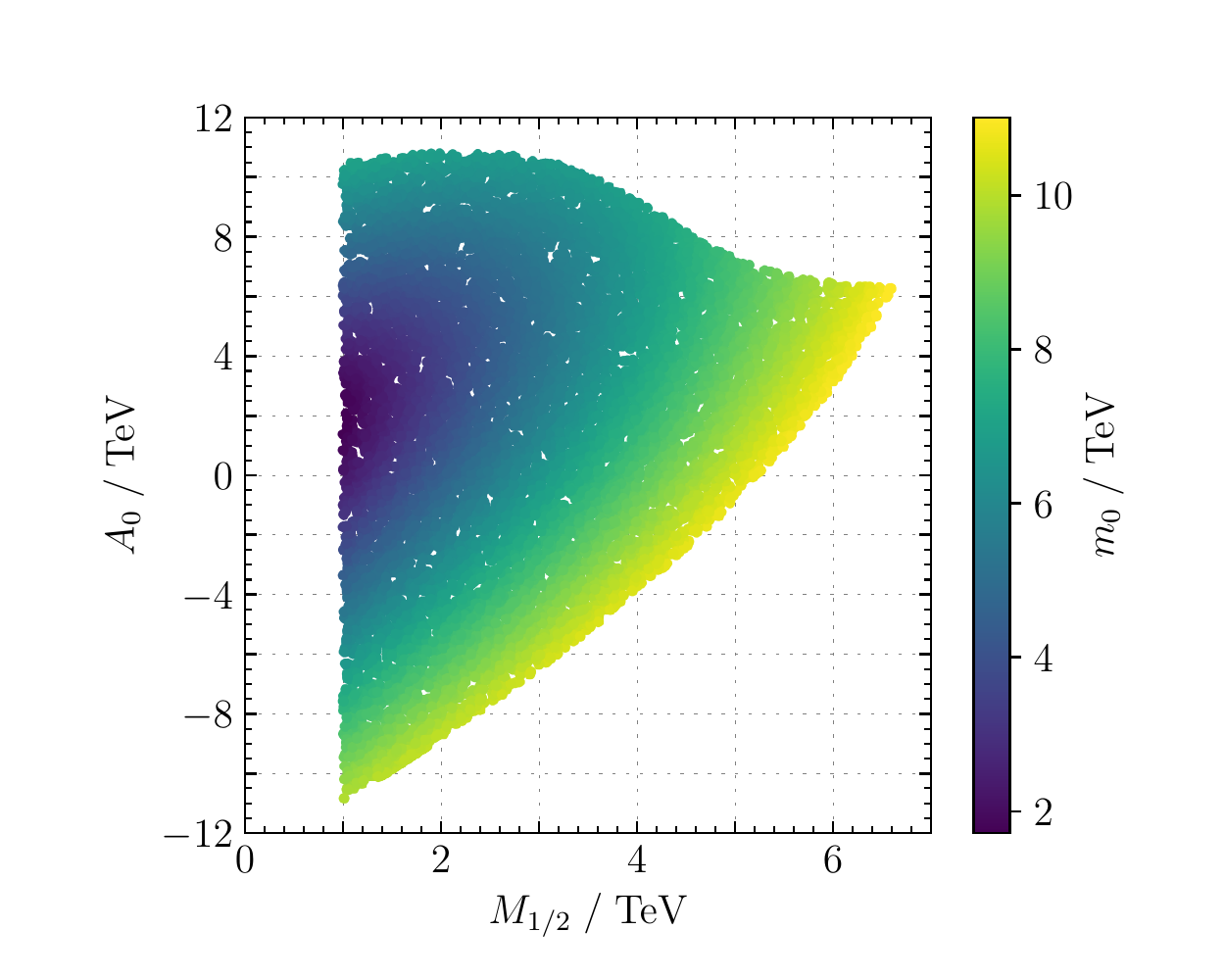}
  \caption{CS\ESSM solutions obtained using the spectrum generator
    automatically generated by \fstwo (left panel) and using the prototype
    spectrum generator used in the numerical analysis of
    Ref.~\cite{Athron:2016gor} (right panel), showing good agreement between
    the two codes.  Note that here we have not applied the limits on the Higgs
    mass or dark matter relic density that lead to additional restrictions on
    the parameter space, as discussed in Ref.~\cite{Athron:2016gor}.}
  \label{fig:cse6ssm_m12_Azero_plane}
\end{figure}

For a more precise comparison between calculations performed at the
same level of precision, we have also performed scans in a recently proposed
variant of the \ESSM, the so-called CS\ESSM \cite{Athron:2015vxg,
  Athron:2016gor}. Here the results obtained using \fstwo have been checked for
agreement with those obtained from a hand-written prototype of the
semi-analytic solver that was implemented for the studies in
Refs.~\cite{Athron:2015vxg,Athron:2016gor}.  The solutions found using the
generated CS\ESSM spectrum generator are compared with those found in
Ref.~\cite{Athron:2016gor} in \figref{fig:cse6ssm_m12_Azero_plane}.  The viable
solution regions and values of the model parameters are found to be in very
good agreement with the results obtained using the earlier code.

\section{\feft}
\label{sec:feft}

 \feft is a method to predict the lightest Higgs pole mass in any \BSM
 model accurately for both high and low new physics scales $\MS$ and
 was presented first in Ref.~\cite{Athron:2016fuq}.  An implementation
 of this method was first released in \fs 1.7.0.  Here we present an
 upgrade at the next-to-leading order and next-to-leading logarithmic
 (NLO+NLL) accuracy, which we release in \fstwo.

\feft combines an EFT approach with a diagrammatic
calculation, allowing for an all-order resummation of large logarithms
of the ratio $\MS/m_t$, together with the inclusion of all
non-logarithmic 1-loop contributions.  In particular, all
power-suppressed 1-loop contributions of $O(v^2/\MS^2)$ are
included in \feft, which would otherwise be neglected in a pure EFT
calculation.  Thanks to these properties, \feft maintains the accuracy
at all scales: For low scales the prediction agrees with a
fixed-order calculation; for large scales it agrees with a pure EFT
calculation. In the intermediate region, where the $O(v^2/\MS^2)$
terms are small, but still non-negligible, \feft gives the correct
fixed-order result plus higher-order logarithms, thus resolving the
ambiguity between the fixed-order and the pure EFT approach.

In Ref.~\cite{Athron:2016fuq} already several versions of the \feft
approach have been extensively discussed and compared with existing
calculations of the lightest Higgs boson mass in the MSSM and other
supersymmetric models. The approach has also been implemented recently
in \sarah/\spheno \cite{Staub:2017jnp}, including 2-loop corrections in
the matching. The version implemented in  \fstwo contains
additional improvements resulting in a higher accuracy. In the
following, we briefly summarize the main idea of \feft, then explain
the details of the implemented version and how to use it. For further
details of the approach and a detailed comparison of theoretical
uncertainties, we refer to Ref.~\cite{Athron:2016fuq}.

\subsection{Basic matching condition}

\feft performs a matching of the \BSM model to the SM,
thereby determining the quartic Higgs coupling $\lambda$ of the
SM\@.  The basic ingredient of \feft to fix $\lambda$ at
the matching scale is a Higgs pole mass matching condition
\begin{align}
  (M_h^\SM)^2 = (M_h^\BSM)^2 \,,
  \label{eq:Higgs_pole_mass_matching}
\end{align}
where $M_h^\SM$ is the Higgs pole mass calculated in the SM at the 1-loop level and $M_h^\BSM$ is the corresponding SM-like
Higgs pole mass in the \BSM model, also at the 1-loop
level.
Generally, the Higgs pole mass is computed in any \BSM model
by solving the following
equation (or a suitable matrix generalization):
\begin{align}
  (M_h^\BSM)^2 = (m_h^\BSM)^2 - \re\Sigma_h^{\BSM}(p^2) + \frac{t_h^\BSM}{v} \,,
  \label{eq:Mh_mom_iteration}
\end{align}
where $m_h^\BSM$ is the respective tree-level mass and
$\Sigma_h^{\BSM}$ and $t_h^\BSM$ are the Higgs self
energy and tadpole in the \MSbar/\DRbar scheme.
In principle, in an all-order calculation,  the
Higgs self energy has to be evaluated at the momentum $p^2 =
(M_h^\SM)^2=(M_h^\BSM)^2$.
  From this condition the quartic Higgs coupling of the SM can be extracted as
\begin{align}
  \lambda = \frac{1}{v^2} \left[
     (M_h^\BSM)^2 + \re\Sigma_h^{\SM}((M_h^\SM)^2) - \frac{t_h^\SM}{v}
  \right] \,.
  \label{eq:lambda_from_Mh_matching}
\end{align}

This matching is equivalent
to the one of pure EFT calculations \cite{Bagnaschi:2014rsa,Vega:2015fna} at the
1-loop level, up to power-suppressed terms. Correspondingly, the resulting
Higgs boson mass is exact at the 1-loop level and takes into account
all leading logarithms \cite{Athron:2016fuq}.

\subsection{New matching procedure in \fstwo}
\label{sec:FS-2_matching_procedure}

\fstwo has an improved implementation of the approach, which is
still exact at the 1-loop level but also
correctly resums next-to-leading logarithms.
This improvement originates from an amended
matching procedure. As mentioned before, the matching procedure
in Eq.~\eqref{eq:lambda_from_Mh_matching}
is equivalent to a pure EFT matching at the 1-loop level. However,
depending on implementation details, it can differ by terms of
2-loop or higher order. If these spurious 2-loop terms contain
 (next-to-leading) large logarithms, they spoil the correct
resummation of (next-to-leading) logarithms by RGE running.

By construction, all versions of \feft discussed in
Ref.~\cite{Athron:2016fuq} and Ref.~\cite{Staub:2017jnp} are correct at the
leading logarithmic level, however not all subleading logarithms are correctly
included.

In the following, we discuss the two potential origins of these subleading
logarithms and how they are avoided by the improved  implementation in
\fstwo.

\paragraph{Insertion of 1-loop parameters into 1-loop \BSM self energies or
  tadpoles}

The first potential source of large 2-loop logarithms in the matching
procedure is the insertion of parameters, which have been obtained from the SM
via a 1-loop matching, into the 1-loop self energies or tadpoles of the \BSM
model.  We illustrate this effect with the most important parameter, the top
Yukawa coupling:
The running top Yukawa coupling of the \BSM model  $y_t^\BSM$ is
determined by a matching as\footnote{We ignore potential tree-level
  factors here for brevity.}
\begin{align}
  y_t^\BSM = y_t^\SM + \Delta y_t \,, \label{eq:yt_SM_BSM}
\end{align}
where $\Delta y_t$ is of 1-loop order (but without large logarithms).
At the same time, the Higgs pole mass calculations on the left-hand side and
right-hand side of Eq.~\eqref{eq:Higgs_pole_mass_matching} are of the form
\begin{align}
  (M_h^\SM)^2 &= (m_h^\SM)^2 + \propto \frac{(v^\SM)^2(y_t^\SM)^4}{(4\pi)^2} \log\frac{m_t^\SM}{Q} + \cdots \,,
  \label{eq:Mh_top_SM}\\
  (M_h^\BSM)^2 &= (m_h^\BSM)^2 + \propto \frac{(v^\BSM)^2(y_t^\BSM)^4}{(4\pi)^2} \log\frac{m_t^\BSM}{Q} + \cdots \,,
  \label{eq:Mh_top_BSM}
\end{align}
where the matching scale $Q$ is of the order $\MS$ and we have also
introduced $m_h^\SM$, $m_t^\SM$ and $m_t^\BSM$ for the running SM
Higgs mass, running SM top mass and running BSM top mass,
keeping our convention of using an upper case `$M$' for pole masses
and lower case `$m$' for running tree-level masses.  If
Eq.\ \eqref{eq:Mh_top_SM} and Eq.\ \eqref{eq:Mh_top_BSM} are set equal
and the relation Eq.\ \eqref{eq:yt_SM_BSM} is inserted, potentially
large 2-loop terms for example of the form
\begin{align}
  \frac{4(v^\SM)^2(y_t^\SM)^3 \Delta y_t}{(4\pi)^2} \log\frac{m_t^\SM}{Q} + \cdots
\label{eq:twolooplogsyt}
\end{align}
remain. Such terms effectively shift the quartic Higgs coupling of the
SM by  next-to-leading logarithmic 2-loop terms.%
\footnote{If the self energies are evaluated at the
  2-loop level, the problem repeats itself one order higher, i.e., the
  term in Eq.~\eqref{eq:twolooplogsyt} is cancelled but similar terms of
  next-to-leading logarithmic 3-loop order remain.}

In order to avoid large higher-order logarithms originating from the
insertion of 1-loop parameters into 1-loop \BSM self energies and
tadpoles, \fstwo maintains two different sets of running \BSM
parameters: One parameter set which has been obtained from the
SM using a tree-level matching, and another set
from the SM using  1-loop matching.
The tree-level parameter set is used to evaluate the 1-loop self
energies and tadpoles on the right-hand side of
Eq.~\eqref{eq:Mh_mom_iteration}.  In this way, no terms like the ones
in Eq.~\eqref{eq:twolooplogsyt} are generated.
The 1-loop parameter set is used to evaluate the tree-level Higgs
mass (matrix) of the \BSM model.  In this way, the desired 1-loop
corrections to the quartic Higgs coupling $\lambda$ are generated.

\paragraph{Momentum iteration}

The second source of large 2-loop logarithms in the matching has to
do with the momentum argument of the self energies entering
Eqs.~\eqref{eq:Higgs_pole_mass_matching} and \eqref{eq:Mh_mom_iteration}.
Writing $p^2=(m_h^\BSM)^2+\Delta p^2$, we see
that the momentum argument of Eq.~\eqref{eq:Mh_mom_iteration}
contains the 1-loop term $\Delta p^2$ (which
also involves large logarithms).
The difference between the left-hand side and the right-hand side of the
matching condition, Eq.~\eqref{eq:Higgs_pole_mass_matching},
then contains 2-loop terms, which can be expanded as
\begin{align}
\left.\left(
\frac{\partial}{\partial
        p^2}\re\Sigma_h^{\BSM}(p^2)
-
\frac{\partial}{\partial
        p^2}\re\Sigma_h^{\SM}(p^2)
\right)
\right|_{p^2 = (m_h^\BSM)^2}
\,\Delta p^2
\,.
\end{align}
If the self energies
are evaluated at the 1-loop level and $p^2$ is determined as
described above, these terms do not cancel against anything.
Like the terms discussed in Eq.~\eqref{eq:twolooplogsyt}, these terms
would then lead to large 2-loop next-to-leading logarithms in the
determination of $\lambda$.%

To avoid large higher-order logarithmic contributions coming from the
momentum argument, \fstwo does not  perform the usual momentum
iteration when the Higgs pole masses in the SM and in the
\BSM model are calculated at the matching scale for
Eq.~\eqref{eq:Higgs_pole_mass_matching}.  Instead, the SM Higgs pole mass at
the matching scale is now calculated as
\begin{align}
  (M_h^\SM)^2 = (m_h^\SM)^2 - \re\Sigma_h^{\SM}((m_h^\BSM)^2) + \frac{t_h^\SM}{v} \,,
  \label{eq:Higgs_pole_mass_matching_new}
\end{align}
where the self energy momentum is set to the tree-level
\MSbar/\DRbar Higgs mass $m_h^\BSM$ in the \BSM model at the matching
scale, which is calculated in terms of running \BSM parameters which
have been obtained by a tree level matching.
A similar expression is used to calculate the Higgs pole mass in the
\BSM model $M_h^\BSM$, where we also insert $p^2 = (m_h^\BSM)^2$ as the
self energy momentum in order to enable cancellation of
momentum-dependent terms.  For example, if the \BSM Higgs is a
singlet, the \BSM Higgs pole mass is calculated as
\begin{align}
  (M_h^\BSM)^2 = (m_h^\BSM)^2 - \re\Sigma_h^{\BSM}((m_h^\BSM)^2) + \frac{t_h^\BSM}{v} \,.
  \label{eq:Higgs_pole_mass_matching_new_BSM_singlet}
\end{align}
On the other hand, if the \BSM Higgs is a multiplet and the $k$-th
element is the SM-like Higgs, then the SM-like \BSM Higgs pole mass is
the $k$-th eigenvalue of the loop-corrected mass matrix $\mathsf{M}_h$
in the interaction eigenstate basis,
\begin{align}
  (\mathsf{M}_h)_{ij} =
  (m_h^\BSM)^2_{ij} - \re\Sigma_{h,ij}^{\BSM}((m_{h_k}^\BSM)^2) + \frac{t_{h_i}^\BSM}{v_i}\delta_{ij}  \,.
  \label{eq:Higgs_pole_mass_matching_new_BSM_multiplet}
\end{align}

By employing this new matching procedure, \feft consistently avoids
large higher-order logarithms and thereby
resums the leading and next-to-leading logarithms and includes all
non-logarithmic 1-loop contributions.

\subsection{Comparison of old and improved \feft implementations}

\begin{figure}[tbh]
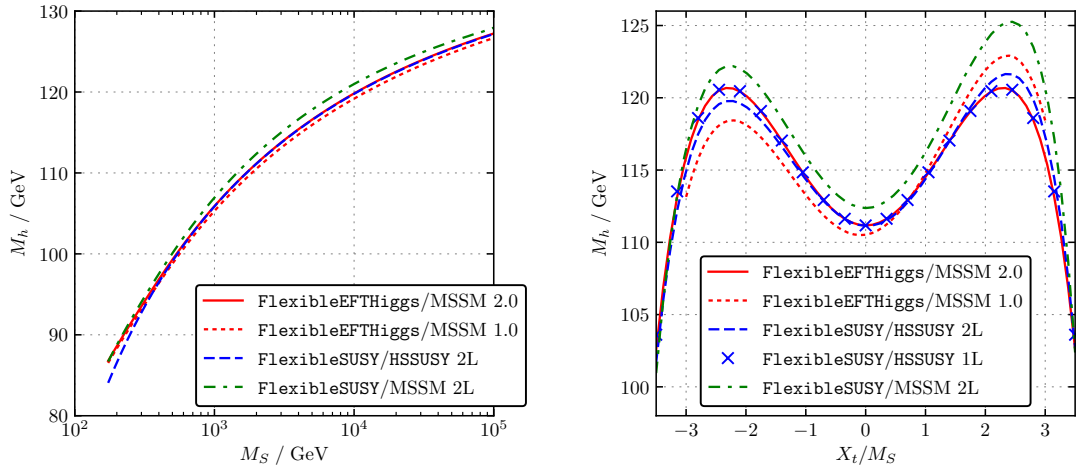

  \centering
  \includegraphics[width=0.49\textwidth]{{{plots/FlexibleEFTHiggs/scan_Mh_MS_TB-5_Xt-0}}} \hfill
  \includegraphics[width=0.49\textwidth]{{{plots/FlexibleEFTHiggs/scan_Mh_Xt_TB-5_MS-2000}}}
  \caption{Comparison of the predicted lightest \CP-even Higgs pole mass in the
    MSSM using the \feft implementations of \fs 1.7.4 and 2.0 for
    $\tan\beta = 5$.  In the left panel we use $X_t = 0$ and in the
    right panel $\MS = 2\unit{TeV}$.}
  \label{fig:feft_comparison}
\end{figure}

In \figref{fig:feft_comparison} we show a comparison of the predicted
lightest \CP-even Higgs mass in the MSSM between the old \feft
implementation of \fs 1.7.4 (red dotted line) and the improved version
in \fs 2.0 (red solid line).  For small SUSY scales of $\MS <
300\unit{GeV}$, we find that the improved version still reproduces the
fixed-order calculation.
As can be seen in the left panel of \figref{fig:feft_comparison} for
vanishing stop mixing, $X_t = 0$, both the old and the improved
version closely reproduce the 2-loop pure EFT calculation with
\HSSUSY: For SUSY scales above $10\unit{TeV}$, the old version
deviates from \HSSUSY-2L by around $600\unit{MeV}$ while the
improved one deviates by around $10\unit{MeV}$.  This is due to the
fact that for $X_t = 0$, the 2-loop threshold correction to the
quartic Higgs coupling at the SUSY scale is negligible.
However, for maximal stop mixing, $X_t/\MS=\sqrt{6}$, which is the
region where the old implementation showed the largest theoretical
uncertainty, we find up to $3\unit{GeV}$ difference between the old
and the improved implementation, see the right panel of
\figref{fig:feft_comparison}.  This difference manifests the consequences of the
different treatment of higher-order terms in the two versions,
especially the inclusion of large 2-loop logarithms in the old
implementation.

The figure shows furthermore that the improved version (which
performs a 1-loop calculation) is now able to perfectly reproduce the
1-loop pure EFT calculation of \HSSUSY (blue crosses) for
\emph{arbitrary} stop mixing and SUSY scales above $\approx
1\unit{TeV}$.  This is in contrast to the old version, which shows a
stronger deviation of around $2\unit{GeV}$ from the 1-loop pure EFT
calculation for large stop mixing, see the right panel of
\figref{fig:feft_comparison}.  Compared to the 2-loop pure EFT
calculation of \HSSUSY (blue dashed line), both the old and the
improved version deviate by around $1$--$2\unit{GeV}$ for non-zero
stop mixing.  This deviation can be attributed to genuine 2-loop
contributions.  Note that \HSSUSY does not reproduce the Higgs mass
prediction of the fixed-order calculation for
$\MS \lesssim 400\unit{GeV}$ in the shown scenario with $X_t = 0$,
because of the neglected terms of $O(v^2/\MS^2)$.  In other scenarios
the $O(v^2/\MS^2)$ terms may be important up to
$\MS \approx 1\unit{TeV}$.

\subsection{Choosing \feft in the model file}

In order to build a \feft spectrum generator, the
\code{FlexibleEFTHiggs} flag can be set to \code{True} in the model
file, see \tabref{tab:feft_info}:
\begin{lstlisting}
FlexibleEFTHiggs = True;
\end{lstlisting}
In \feft spectrum generators, the low-energy boundary condition cannot
be modified, because it is fixed internally to perform a matching of
the $\SM(5)$ to the full SM\@.  The considered \BSM model is
matched to the SM at the \BSM matching scale, $\Qmatch$,
which is set to the \code{SUSYScale} by default.  In this matching,
the running normalized gauge couplings $g_i^\BSM(\Qmatch)$
($i=1,2,3$), the Yukawa coupling matrices $Y_f^\BSM(\Qmatch)$
($f=u,d,e$) and the SM-like vacuum expectation value $v^\BSM(\Qmatch)$
of the \BSM model are determined automatically from the following matching
conditions on pole masses and running couplings
\begin{subequations}
\begin{align}
  (M_V^\BSM)^2 &= (M_V^\SM)^2, \qquad V = W, Z,
  \label{eq:WZ-matching} \\
  M_f^\BSM &= M_f^\SM, \qquad f = e, \mu, \tau, u, d, c, s, t, b,
  \label{eq:f-matching} \\
  \aem^\BSM(\Qmatch) &= \aem^\SM(\Qmatch)\times (1 + \Delta\aem),
  \label{eq:aem-matching} \\
  \as^\BSM(\Qmatch) &= \as^\SM(\Qmatch)\times (1 + \Delta\as),
  \label{eq:as-matching}
\end{align}%
\label{eq:FSEFTHiggs_matching}%
\end{subequations}%
where $\Delta\aem$ and $\Delta\as$ are the known 1-loop
threshold corrections \cite{Hall:1980kf}, including potential \MSbar
to \DRbar conversion terms \cite{Martin:1993yx}.
\fs imposes the individual matching conditions in
Eqs.~\eqref{eq:FSEFTHiggs_matching} at the appropriate loop orders
such that no large 2-loop logarithms are generated, as
described in \secref{sec:FS-2_matching_procedure}.  For example, to
obtain the correct Higgs mass in the MSSM at the 1-loop level,
Eqs.~\eqref{eq:WZ-matching} and \eqref{eq:aem-matching} are imposed at
the 1-loop level while Eqs.~\eqref{eq:f-matching} and
\eqref{eq:as-matching} are imposed at the tree level.  The SM-like
vacuum expectation value $v^\BSM(\Qmatch)$ of the \BSM model is
defined as
\begin{align}
  v^\BSM(\Qmatch) = \frac{2 m_Z^\BSM(\Qmatch)}{\sqrt{\big(g_Y^\BSM(\Qmatch)\big)^2 + \big(g_2^\BSM(\Qmatch)\big)^2}},
  \label{eq:feft_vev_definition}
\end{align}
where $m_Z^\BSM(\Qmatch)$ is the running $Z$ boson mass and
$g_Y^\BSM(\Qmatch)$ and $g_2^\BSM(\Qmatch)$ are the running
electroweak gauge couplings in the \BSM model at the matching scale.

The running \BSM model parameters can be given as input at either the
\code{SUSYScale} or at the \code{HighScale}.  The following example
demonstrates how to fix the \DRbar parameters of the MSSM at the SUSY
scale.
\begin{example}[label=ex:MSSMEFTHiggs]
In the \feft/MSSM model (\code{MSSMEFTHiggs})
the soft-breaking MSSM parameters, the $\mu$ parameter and $\tan\beta$
are input at the SUSY scale, \MS.  Thus, the boundary condition at the SUSY
scale has the form
\begin{lstlisting}
SUSYScaleInput = {
    {vu, Sqrt[vu^2 + vd^2] Sin[ArcTan[TanBeta]]},
    {vd, Sqrt[vu^2 + vd^2] Cos[ArcTan[TanBeta]]},
    {MassB, M1Input},
    {MassWB, M2Input},
    {MassG, M3Input},
    {mq2, mq2Input},
    {mu2, mu2Input},
    {md2, md2Input},
    {ml2, ml2Input},
    {me2, me2Input},
    {\[Mu], MuInput},
    {B[\[Mu]], mAInput^2/(TanBeta + 1/TanBeta)},
    {T[Yu], AuInput Yu},
    {T[Yd], AdInput Yd},
    {T[Ye], AeInput Ye}
};
\end{lstlisting}
The symbols \code{TanBeta}, \code{M1Input}, \code{M2Input},
\code{M3Input}, \code{mq2Input}, \code{mu2Input}, \code{md2Input},
\code{ml2Input}, \code{me2Input}, \code{MuInput}, \code{mAInput},
\code{AuInput}, \code{AdInput}, \code{AeInput} describe the MSSM input
parameters $\tan\beta^{\DRbar}(\MS)$, $M_i(\MS)$, $m_{\tilde{f}}^2(\MS)$
($f=q,u,d,l,e$), $\mu(\MS)$, $m_A(\MS)$ and $A_f(\MS)$
($f=u,d,e$) in the \DRbar scheme at the SUSY scale.

Note that no explicit SUSY scale boundary condition for the gauge
couplings, $g_1$, $g_2$ and $g_3$, and Yukawa couplings, $Y_u$, $Y_d$
and $Y_e$, of the MSSM has to be specified, because they are all fixed
automatically at \Qmatch\ using the \feft matching conditions, Eqs.\
\eqref{eq:FSEFTHiggs_matching}.

However, there is a subtlety with the vacuum expectation values: In the
above boundary condition, the input value $\tan\beta^{\DRbar}(\MS)$ is
used to fix the ratio of $v_u(\MS)$ and $v_d(\MS)$.  However, their
magnitude $\sqrt{v_u^2 + v_d^2}$ is unfixed so far.  To fix
it, we can use the value of $v^\MSSM(\Qmatch)$, which is
automatically determined by \fs at the matching scale, see
Eq.~\eqref{eq:feft_vev_definition}.  Therefore, we want to set
\begin{subequations}
\begin{align}
  v_u(\Qmatch) &= v^\MSSM(\Qmatch) \sin\beta^{\DRbar}(\Qmatch) \,, \\
  v_d(\Qmatch) &= v^\MSSM(\Qmatch) \cos\beta^{\DRbar}(\Qmatch) \,.
\end{align}%
\label{eq:VEV_matching}%
\end{subequations}%
Such a matching is not done automatically by \feft.  The user must
specify how the VEVs of any Higgs fields that have electroweak
interactions are related to the electroweak VEV, $v^{\BSM}(\Qmatch)$,
which is given above for the MSSM\@.  To do this, the model file has an
additional constraint list: \code{MatchingScaleInput}.  Conditions
to relate or fix model parameters at the matching scale
$\Qmatch$ can be expressed using the \code{MatchingScaleInput} list.  This can actually be used to set any BSM parameter or to override the automatic \feft matching conditions if the user wishes.  However, it is only required that the user specifies the matching for the VEVs here.
To express the relations of Eqs.~\eqref{eq:VEV_matching}, we set
\begin{lstlisting}
MatchingScaleInput = {
    {vu, VEV Sin[ArcTan[vu/vd]]},
    {vd, VEV Cos[ArcTan[vu/vd]]}
};
\end{lstlisting}
The symbol \code{VEV} is reserved by \fs and represents the running
SM-like vacuum expectation value $v^\MSSM(\Qmatch)$ at the matching
scale, as defined in Eq.~\eqref{eq:feft_vev_definition}.
\end{example}
\begin{OptionTable}{\feft model file options}{tab:feft_info}[lllX]
  \code{FlexibleEFTHiggs} & \code{False} & \code{True} or \code{False} &
  Flag to enable/ disable \feft \\
  \code{VEV} & -- & -- & SM-like VEV in the \BSM model, $v^\BSM(\Qmatch)$ \\
  \code{MatchingScaleInput} & \code{\{\}} & list of 2-tuples &
  boundary conditions for \BSM parameters at the matching scale $\Qmatch$\\
  \bottomrule
\end{OptionTable}

Once the model file is written, a spectrum generator can be created and
run in the usual way.  For example, to build the model described in
\exref{ex:MSSMEFTHiggs}, one may run:
\begin{lstlisting}[language=bash]
$ ./createmodel --name=MSSMEFTHiggs
$ ./configure --with-models=MSSMEFTHiggs
$ make
\end{lstlisting}
These commands create the \fs spectrum generator for the
\modelname{MSSMEFTHiggs} model.  The generated spectrum generator can then
be run from the command line as
\begin{lstlisting}[language=bash]
$ cd models/MSSMEFTHiggs
$ ./run_MSSMEFTHiggs.x --slha-input-file=LesHouches.in.MSSMEFTHiggs
\end{lstlisting}

The only difference with the SLHA interface is that there are new
\feft-specific options in the SLHA file. In \feft, the pole masses of
the BSM particles are calculated at the scale $\QpoleBSM$, which is
set to the \code{SUSYScale} by default.  The scale $\QpoleBSM$ can be
changed by setting the entry \code{FlexibleSUSY[17]} to a non-zero
value in the SLHA input file.  Similarly, in the \mathematica interface
$\QpoleBSM$ can be changed by setting \code{poleMassScale} to a
non-zero value.
The pole masses of the SM particles are calculated at the scale
$\QpoleSM$, which is set to the top pole mass $M_t$ by default.  The
scale $\QpoleSM$ can be changed by setting the entry
\code{FlexibleSUSY[18]} to a non-zero value in the SLHA input file.
In the \mathematica interface, $\QpoleSM$ can be changed by setting
\code{eftPoleMassScale} to a non-zero value.
The matching scale $\Qmatch$ is set to the \code{SUSYScale} by
default.  It can be changed by setting the entry \code{FlexibleSUSY[19]} to a
non-zero value in the SLHA input file.  In the \mathematica interface,
$\Qmatch$ can be changed by setting \code{eftMatchingScale} to a
non-zero value.

\begin{example}[label=ex:MSSMEFTHiggs_uncertainty_estimate]
  This example demonstrates how a partial uncertainty estimate of the
  lightest Higgs pole mass can be made with \feft.  The
  uncertainty is estimated by varying the matching scale $\Qmatch$ and
  the scale $\QpoleSM$, at which the lightest Higgs pole mass is
  calculated, both by a factor 2.
  \begin{lstlisting}
Get["models/MSSMEFTHiggs/MSSMEFTHiggs_librarylink.m"];

Mtpole = 173.34;

(* generate logarithmically spaced range [start, stop] *)
LogRange[start_, stop_, steps_] :=
    Exp /@ Range[Log[start], Log[stop],
                 (Log[stop] - Log[start])/steps];

(* generate logarithmically spaced range [Q / 2, 2 Q] *)
GenerateScales[Q_] := LogRange[Q/2, 2 Q, 10];

(* run MSSMEFTHiggs spectrum generator *)
RunMSSMEFTHiggs[MS_, TB_, Xt_, Qpole_, Qmatch_] :=
    Module[{handle, spectrum},
           handle = FSMSSMEFTHiggsOpenHandle[
               fsSettings -> {
                   precisionGoal -> 1.*^-5,
                   maxIterations -> 10000,
                   poleMassLoopOrder -> 2,
                   ewsbLoopOrder -> 2,
                   betaFunctionLoopOrder -> 3,
                   thresholdCorrectionsLoopOrder -> 2,
                   poleMassScale -> 0,
                   eftPoleMassScale -> Qpole,
                   eftMatchingScale -> Qmatch,
                   eftMatchingLoopOrderUp -> 1,
                   eftMatchingLoopOrderDown -> 1,
                   calculateBSMMasses -> 0
               },
               fsSMParameters -> {
                   Mt -> Mtpole
               },
               fsModelParameters -> {
                   MSUSY   -> MS,
                   M1Input -> MS,
                   M2Input -> MS,
                   M3Input -> MS,
                   MuInput -> MS,
                   mAInput -> MS,
                   TanBeta -> TB,
                   mq2Input -> MS^2 IdentityMatrix[3],
                   mu2Input -> MS^2 IdentityMatrix[3],
                   md2Input -> MS^2 IdentityMatrix[3],
                   ml2Input -> MS^2 IdentityMatrix[3],
                   me2Input -> MS^2 IdentityMatrix[3],
                   AuInput -> {{MS/TB, 0    , 0},
                               {0    , MS/TB, 0},
                               {0    , 0    , MS/TB + Xt MS}},
                   AdInput -> MS TB IdentityMatrix[3],
                   AeInput -> MS TB IdentityMatrix[3]
               }
           ];
           spectrum = FSMSSMEFTHiggsCalculateSpectrum[handle];
           FSMSSMEFTHiggsCloseHandle[handle];
           spectrum
          ];

(* extract lightest Higgs pole mass Pole[M[hh]] from spectrum *)
RunMSSMEFTHiggsMh[pars__] :=
    (Pole[M[hh]] /. (MSSMEFTHiggs /. RunMSSMEFTHiggs[pars]))[[1]];

(* calculate Higgs mass and perform scale variation *)
RunMSSMEFTHiggsUncertainty[MS_, TB_, Xt_] :=
    Module[{MhMean, DMh, varyQpole, varyQmatch},
           MhMean = RunMSSMEFTHiggsMh[MS, TB, Xt, 0, 0];
           varyQpole  = RunMSSMEFTHiggsMh[MS, TB, Xt, #, 0]& /@
                        GenerateScales[Mtpole];
           varyQmatch = RunMSSMEFTHiggsMh[MS, TB, Xt, 0, #]& /@
                        GenerateScales[MS];
           (* combine uncertainty estimates *)
           DMh = Max[Abs[Max[varyQpole] - MhMean],
                     Abs[Min[varyQpole] - MhMean]] +
                 Max[Abs[Max[varyQmatch] - MhMean],
                     Abs[Min[varyQmatch] - MhMean]];
           { MhMean, DMh }
          ];

{Mh, DMh} = RunMSSMEFTHiggsUncertainty[2500, 20, Sqrt[6]];

Print["Mh = (", Mh, " +- ", DMh, ") GeV"];
  \end{lstlisting}
  The output of the script could read
  \begin{lstlisting}
Mh = (125.047 +- 1.52741) GeV
  \end{lstlisting}
\end{example}

\section{Current limitations and workarounds}
\label{sec:limitations}

Currently, the models and scenarios which can be constructed with
\fstwo are limited to the following cases:
\begin{itemize}
\item The couplings of the model(s) must remain perturbative at all
  scales between the highest and lowest boundary condition.
\item The considered models are required to have a gauge symmetry
  that has the SM gauge group
  $\GSM = SU(3)_C\times SU(2)_L\times U(1)_Y$ as a factor.  This
  restriction is currently necessary to perform an unambiguous
  matching of the model to the $\SM(5)$ at the low-energy scale.  If a
  model that does not have $\GSM$ as a gauge group
  factor is considered, then the gauge couplings of the model must be fixed
  by hand in one of the boundary conditions.  See
  \secref{sec:determination_of_g_Y} for an example of a model with a
  left-right symmetry.
\item Tensor-like Lagrangian parameters of rank 3, which would arise
  in $R$-parity violating SUSY models for example, are currently not
  supported.  As a workaround, the rank 3 tensor-like couplings could
  be decomposed into a sum of terms with rank 2 matrix-like couplings.
\item The extraction of the running Yukawa couplings in models where
  a 4$^\text{th}$ generation of fermions mixes with the SM fermions is currently
  not supported.  As a workaround, the running Yukawa couplings can
  manually be fixed by using the running SM fermion masses, which can
  be accessed in the model file via the \code{upQuarksDRbar} =
  $\diag(m_u, m_c, m_t)$, \code{downQuarksDRbar} =
  $\diag(m_d, m_s, m_b)$ and \code{downLeptonsDRbar} =
  $\diag(m_e, \allowbreak m_\mu, \allowbreak m_\tau)$ symbols.  In the
  $\mu\nu$SSM \cite{LopezFogliani:2005yw} (\sarah/\fs model name:
  \modelname{munuSSM}), for example, the Yukawa coupling matrix $Y_e$ of
  the down-type leptons can approximately be fixed at the low-energy
  scale as
  \begin{align}
    Y_e(Q) = \frac{\sqrt{2}}{v_d} \diag(m_e, m_\mu, m_\tau) \,,
  \end{align}
  which is expressed in the \fs model file as
  \begin{lstlisting}
LowScaleInput = {
   {Ye, Sqrt[2] downLeptonsDRbar / vd},
   ...
};
  \end{lstlisting}
\end{itemize}
Due to the modular nature of the generated code, adaptation and extension
to overcome restrictions in scope are quite straightforward.

\section{Conclusions}
\label{sec:conclusions}

In order to study the vast zoo of models beyond the SM,
tools for each model are necessary to calculate the mass spectrum and observables.
\fs is a meta-tool for automatized generation of such tools which reliably operate at high precision and speed
for a broad class of BSM models.

In this paper, we have presented all of the substantial updates to \fs
available in version 2.0.  These include many model-specific
higher-order corrections, as well as extensions to support non-SUSY
models (\fbsm), models with complex parameters (\fcpv) and a new
solver which allows the EWSB
outputs to be defined at the high scale (\fsas).  Furthermore, \fs can
now calculate in any given
model: the anomalous magnetic moment of the muon (\famu) as well as the muon decay and the $W$ mass (\fmw),
including partial 2-loop contributions.  \fstwo also comes with an
update of the hybrid EFT/fixed-order calculation of the Higgs mass
(\feft) with a higher-order log resummation.

Altogether, these represent a significant extension to the
calculations that can be performed in BSM models at high precision. To
illustrate the variety of potential applications of \fstwo, we have
presented many physics examples.  These include large-scale parameter
scans performed efficiently on multiple CPU cores (see
\secref{sec:application_HSSUSY}), and the construction of low-energy
effective field theories of SUSY models with complicated boundary
conditions at the matching scale (see Sections
\ref{sec:application_THDM} and \ref{sec:application_HSSUSY}).  Indeed,
\fs is already being used extensively for such cases, including global
fits by the \GAMBIT collaboration and major studies of precision Higgs
mass predictions.  Furthermore, the modularity of the generated
spectrum generators allows easy implementation of model-specific
higher-order corrections, which has been done in the past to include
3-loop Higgs mass contributions from the \Himalaya library (see
\secref{sec:MSSM-specific_ho_corrections}) and to add power-suppressed
terms of $O(v^2/\MS^2)$ to \HSSUSY.
Furthermore, we have illustrated how to calculate the anomalous
magnetic moment of the muon and electric dipole moments with \fstwo in
Sections \ref{sec:famu}--\ref{sec:fcpv}.  Various physics applications
for \fsas have been presented in \secref{sec:fsas}, which include the
study of multiple solutions to the boundary value problem of the CMSSM
as well as parameter scans in the CNMSSM and C\ESSM.

The \fs system has been extensively tested for correctness against
results from the literature and other spectrum generators.  In
addition, speed tests have been carried out, with results proving its
effectiveness in large-scale scans.  The auto-generated C++ code is
designed in such a way that users can easily read and reuse
its components to develop their own analysis tools.

\section*{Acknowledgments}

We especially thank Florian Staub and Mark Goodsell for many helpful
discussions and technical help with \sarah.  Furthermore, we kindly
thank Ben Allanach for providing components of \softsusy, which are
included in \fs.  We also thank Pietro Slavich and Emanuele Bagnaschi
for providing the 2-loop contributions to the Higgs mass(es) in the
full MSSM and in \HSSUSY.
Furthermore, we thank Ben
Allanach and Alexander Bednyakov for providing the 2-loop \SQCD
contributions to the \DRbar top, bottom and $\tau$ Yukawa couplings as
well as to the strong coupling in the MSSM and both Alexander
Bednyakov and the authors of \susyhd for providing the 3-loop SM $\beta$
functions.  We also thank the authors of the \Himalaya
package for providing 3-loop contributions to the light \CP-even Higgs mass
in the MSSM\@.
We kindly thank and acknowledge the role of Tom Steudtner for writing
the first version of \feft.
A.V.\ is very grateful to the KUTS working group, especially Emanuele
Bagnaschi, Henning Bahl, Johannes Braathen, Mark Goodsell, Robert
Harlander, Luminita Mihaila, Pietro Slavich, Florian Staub and Georg
Weiglein, for countless insights and helpful discussions about Higgs
mass precision predictions in supersymmetry over the last years.
P.A.\ is very grateful to his \GAMBIT collaborators, especially Ben
Farmer and James McKay, for helpful suggestions and feedback on
\fs. P.A., A.V.\ and D.H.\ thank their collaborators on the
``Precision tools and models to narrow in on the $750\unit{GeV}$
diphoton resonance'' project for helpful comments and advice that have
influenced the development of \fs.  Similarly, P.A.\ and A.V.\ also
thank their collaborators in the ``Higgs mass predictions of public
NMSSM spectrum generators'' project for the helpful discussions during
that project.  P.A.\ and D.H.\ warmly thank Roman Nevzorov for many
helpful discussions about the semi-analytic solver approach and
collaboration with him which had a lot of impact on the development of
this code.  The work of P.A.\ and D.H.\ was supported by the Australian
Research Council through the ARC Centre of Excellence for Particle Physics
at the Terascale (CoEPP) (grant CE110001104).  The work of P.A. is also
supported by the Australian Research Council Future Fellowship grant
FT160100274.  The work of D.H.\ was supported by the University of Adelaide,
and through an Australian Government Research Training Program Scholarship.
D.H.\ also acknowledges financial support from the Grant Agency of the Czech
Republic (GACR), contract 17-04902S\@.  This work has also been supported by the
German Research Foundation DFG through Grant No.~STO876/2.

\appendix
\renewcommand*{\thesection}{\Alph{section}}

\section{SM input parameters}
\label{app:sm_input_parameters}
The list of all SM input parameters is given
in \tabref{tab:sm_mma_options}.  In the SLHA interface of \fs, the
SM parameters can be given in four different input blocks:
\begin{itemize}
\item The block \texttt{SMINPUTS} contains the electromagnetic and
  strong coupling, the Fermi constant and the masses of SM
  particles, as defined in the SLHA-2 standard \cite{Allanach:2008qq}.
  The definitions of the individual block entries are shown in the
  first part of \tabref{tab:sm_mma_options}.  If a parameter is
  omitted from the \texttt{SMINPUTS} block, then it is set to the
  default value defined in \tabref{tab:sm_mma_options}.
\item The CKM matrix is given as input in the block \texttt{VCKMIN} in
  the Wolfenstein parametrization, as defined in
  Ref.~\cite{Allanach:2008qq}.  If the \texttt{VCKMIN} block or an entry is
  missing, the corresponding parameter is set to zero.
  The given CKM matrix elements can be accessed in the \fs model file
  via the \texttt{CKM} symbol to set the Yukawa matrices at the
  low-energy scale.  See \fs's \modelname{CMSSMCKM} model for an example.
\item The PMNS matrix is given as input in the block
  \texttt{UPMNSIN}, as defined in Ref.~\cite{Allanach:2008qq}.  If the
  \texttt{UPMNSIN} block or an entry is missing, the corresponding
  parameter is set to zero.
  The given PMNS matrix elements can be accessed in the \fs model file
  via the \texttt{PMNS} symbol to fix potential neutrino mass
  parameters at the low-energy scale.
\item For special applications such as the calculation of
  $\amu$ with \GMTCalc, further input parameters are
  needed.  These can be given in the \texttt{FlexibleSUSYInput} block.
  The block entries are defined in \tabref{tab:sm_mma_options}.  If
  the block or a block entry is missing, the input parameters are set
  to their default values as defined in the table.
\end{itemize}
In SLHA format the blocks with their respective default values read:
\begin{lstlisting}
Block SMINPUTS               # Standard Model inputs
    1   1.279160000e+02      # alpha^(-1) SM MSbar(MZ)
    2   1.166378700e-05      # G_Fermi
    3   1.184000000e-01      # alpha_s(MZ) SM MSbar
    4   9.118760000e+01      # MZ(pole)
    5   4.180000000e+00      # mb(mb) SM MSbar
    6   1.733400000e+02      # mtop(pole)
    7   1.776990000e+00      # mtau(pole)
    8   0.000000000e+00      # mnu3(pole)
    9   8.038500000e+01      # MW pole
   11   5.109989020e-04      # melectron(pole)
   12   0.000000000e+00      # mnu1(pole)
   13   1.056583715e-01      # mmuon(pole)
   14   0.000000000e+00      # mnu2(pole)
   21   4.750000000e-03      # md(2 GeV) MS-bar
   22   2.400000000e-03      # mu(2 GeV) MS-bar
   23   1.040000000e-01      # ms(2 GeV) MS-bar
   24   1.270000000e+00      # mc(mc) MS-bar
Block VCKMIN                 # CKM matrix input (Wolfenstein parameters)
    1                 0      # lambda(MZ) SM DR-bar
    2                 0      # A(MZ) SM DR-bar
    3                 0      # rhobar(MZ) SM DR-bar
    4                 0      # etabar(MZ) SM DR-bar
Block UPMNSIN                # PMNS matrix input
    1                 0      # theta_12
    2                 0      # theta_23
    3                 0      # theta_13
    4                 0      # delta
    5                 0      # alpha_1
    6                 0      # alpha_2
Block FlexibleSUSYInput
    0   0.00729735           # alpha_em(0)
    1   125.09               # Mh pole
\end{lstlisting}

In \fs's \mathematica interface, the SM parameters must be
passed to the \code{FS<model>OpenHandle[fsSMParameters -> \{...\}]}
function in the form of replacement rules.  The symbols associated to the
SM input parameters are given in
\tabref{tab:sm_mma_options}.\footnote{Note that in \fs's \mathematica
  interface, the fine structure constant $\aem^{\SM(5)}(M_Z)$ is input,
  not its inverse as in the SLHA standard.}
Unset parameters are set to their
default values defined in the table.  Note that in the \mathematica
interface of \fs, the CKM matrix parameters are given in the
exact parametrization in terms of the angles $\theta_{12}$,
$\theta_{23}$, $\theta_{13}$ and $\delta$ \cite{Olive:2016xmw}.  A
call of \code{FS<model>OpenHandle[]} with all parameters set to their
respective default values would read:
\begin{lstlisting}{Mathematica}
handle = FS<model>OpenHandle[
    fsSMParameters -> {
        alphaEmMZ -> 1/127.916,      (* SMINPUTS[1] *)
        GF -> 1.166378700*^-5,       (* SMINPUTS[2] *)
        alphaSMZ -> 0.1184,          (* SMINPUTS[3] *)
        MZ -> 91.1876,               (* SMINPUTS[4] *)
        mbmb -> 4.18,                (* SMINPUTS[5] *)
        Mt -> 173.34,                (* SMINPUTS[6] *)
        Mtau -> 1.77699,             (* SMINPUTS[7] *)
        Mv3 -> 0,                    (* SMINPUTS[8] *)
        MW -> 80.385,                (* SMINPUTS[9] *)
        Me -> 0.000510998902,        (* SMINPUTS[11] *)
        Mv1 -> 0,                    (* SMINPUTS[12] *)
        Mm -> 0.1056583715,          (* SMINPUTS[13] *)
        Mv2 -> 0,                    (* SMINPUTS[14] *)
        md2GeV -> 0.00475,           (* SMINPUTS[21] *)
        mu2GeV -> 0.0024,            (* SMINPUTS[22] *)
        ms2GeV -> 0.104,             (* SMINPUTS[23] *)
        mcmc -> 1.27,                (* SMINPUTS[24] *)
        CKMTheta12 -> 0,
        CKMTheta23 -> 0,
        CKMTheta13 -> 0,
        CKMDelta -> 0,
        PMNSTheta12 -> 0,            (* UPMNSIN[1] *)
        PMNSTheta23 -> 0,            (* UPMNSIN[2] *)
        PMNSTheta13 -> 0,            (* UPMNSIN[3] *)
        PMNSDelta -> 0,              (* UPMNSIN[4] *)
        PMNSAlpha1 -> 0,             (* UPMNSIN[5] *)
        PMNSAlpha2 -> 0,             (* UPMNSIN[6] *)
        alphaEm0 -> 1/137.035999074, (* FlexibleSUSYInput[0] *)
        Mh -> 125.09                 (* FlexibleSUSYInput[1] *)
    }
]
\end{lstlisting}
\begin{table}[tbh!]
  \centering
  \resizebox{\textwidth}{!}{%
  \begin{tabular}{llll}
    \toprule
    Index & \mathematica symbol & Default & Description \\
    \midrule
    \multicolumn{4}{c}{\texttt{Block SMINPUTS}}\\
    \midrule
     1 & \code{alphaEmMZ} & $1/127.916$ & electromagnetic coupling, $\aem^{\SM(5)}(M_Z)$ \\
     2 & \code{GF} & $1.1663787\cdot 10^{-5}$ & Fermi coupling constant, $G_F\times\text{GeV}^2$ \\
     3 & \code{alphaSMZ} & $0.1184$ & strong coupling, $\as^{\SM(5)}(M_Z)$ \\
     4 & \code{MZ} & $91.1876$ & $Z$ pole mass, $M_Z/\text{GeV}$ \\
     5 & \code{mbmb} & $4.18$ & running bottom mass, $m_b^{\SM(5)}(m_b)/\text{GeV}$ \\
     6 & \code{Mt} & $173.34$ & top pole mass, $M_t/\text{GeV}$ \\
     7 & \code{Mtau} & $1.77699$ & $\tau$ pole mass, $M_\tau/\text{GeV}$ \\
     8 & \code{Mv3} & $0$ & heaviest neutrino pole mass, $M_{\nu_3}/\text{GeV}$ \\
     9 & \code{MW} & $80.385$ & $W$ pole mass, $M_W/\text{GeV}$ \\
    11 & \code{Me} & $0.000510998902$ & electron pole mass, $M_e/\text{GeV}$ \\
    12 & \code{Mv1} & $0$ & lightest neutrino pole mass, $M_{\nu_1}/\text{GeV}$ \\
    13 & \code{Mm} & $0.1056583715$ & muon pole mass, $M_\mu/\text{GeV}$ \\
    14 & \code{Mv2} & $0$ & 2$^\text{nd}$ lightest neutrino pole mass, $M_{\nu_2}/\text{GeV}$ \\
    21 & \code{md2GeV} & $0.00475$ & running down mass, $m_d(2\unit{GeV})/\text{GeV}$ \\
    22 & \code{mu2GeV} & $0.0024$ & running up mass, $m_u(2\unit{GeV})/\text{GeV}$ \\
    23 & \code{ms2GeV} & $0.104$ & running strange mass, $m_s(2\unit{GeV})/\text{GeV}$ \\
    24 & \code{mcmc} & $1.27$ & running charm mass, $m_c^{\SM(4)}(m_c)/\text{GeV}$ \\
    \midrule
    \multicolumn{4}{c}{\texttt{Block VCKMIN}}\\
    \midrule
     1 & & $0$ & CKM Wolfenstein parameter $\lambda$ \\
     2 & & $0$ & CKM Wolfenstein parameter $A$ \\
     3 & & $0$ & CKM Wolfenstein parameter $\bar{\rho}$ \\
     4 & & $0$ & CKM Wolfenstein parameter $\eta$ \\
    \midrule
       & \code{CKMTheta12} & $0$ & CKM matrix parameter $\theta_{12}$ \\
       & \code{CKMTheta23} & $0$ & CKM matrix parameter $\theta_{23}$ \\
       & \code{CKMTheta13} & $0$ & CKM matrix parameter $\theta_{13}$ \\
       & \code{CKMDelta} & $0$ & CKM matrix parameter $\delta$ \\
    \midrule
    \multicolumn{4}{c}{\texttt{Block UPMNSIN}}\\
    \midrule
     1 & \code{PMNSTheta12} & $0$ & PMNS solar angle $\theta_{12}$ \\
     2 & \code{PMNSTheta23} & $0$ & PMNS atmospheric angle $\theta_{23}$ \\
     3 & \code{PMNSTheta13} & $0$ & PMNS matrix parameter $\theta_{13}$ \\
     4 & \code{PMNSDelta}   & $0$ & PMNS Dirac phase $\delta$ \\
     5 & \code{PMNSAlpha1}  & $0$ & PMNS 1$^\text{st}$ Majorana phase $\alpha_1$ \\
     6 & \code{PMNSAlpha2}  & $0$ & PMNS 2$^\text{nd}$ Majorana phase $\alpha_2$ \\
    \midrule
    \multicolumn{4}{c}{\texttt{Block FlexibleSUSYInput}}\\
    \midrule
     0 & \code{alphaEm0} & $1/137.035999074$ & $\aem$ in the Thomson limit \\
     1 & \code{Mh} & $125.09$ & SM Higgs pole mass $M_h/\text{GeV}$ \\
    \bottomrule
  \end{tabular}}
  \caption{SLHA input block entries and \mathematica symbols to
    specify the SM input parameters.  The first column
    represents the index in the corresponding SLHA input block and the
    second column the symbol used in the \mathematica interface.}
  \label{tab:sm_mma_options}
\end{table}

\section{FlexibleSUSY configuration options}
\label{app:fs_config_options}

\fs provides many configuration options to switch on/off contributions
and choose/fine-tune the solver algorithm(s).  All runtime
configuration options are listed in \tabref{tab:sg_mma_options}.
In the SLHA interface of \fs, the configuration options are read from
the \fs block.  In addition, some information is also read
from the \texttt{MODSEL} block, see below.  In the SLHA format all \fs
configuration entries with their respective default values read:
\begin{lstlisting}
Block MODSEL
   12   0            # output scale of running parameters (0 = SUSY scale)
Block FlexibleSUSY
    0   1e-04        # precision goal
    1   0            # max. iterations (0 = automatic)
    2   0            # solver (0 = all, 1 = two-scale, 2 = semi-analytic)
    3   0            # calculate SM pole masses
    4   2            # pole mass loop order
    5   2            # EWSB loop order
    6   3            # beta-functions loop order
    7   2            # threshold corrections loop order
    8   1            # Higgs 2L corrections O(alpha_t alpha_s)
    9   1            # Higgs 2L corrections O(alpha_b alpha_s)
   10   1            # Higgs 2L corrections O((alpha_t + alpha_b)^2)
   11   1            # Higgs 2L corrections O(alpha_tau^2)
   12   0            # force output
   13   1            # Top quark 2L corrections QCD
   14   1e-11        # beta-function zero threshold
   15   0            # calculate observables (a_muon, ...)
   16   0            # force positive majorana masses
   17   0            # pole mass renormalization scale (0 = SUSY scale)
   18   0            # pole mass renormalization scale in the EFT
                     # (0 = min(SUSY scale, Mt))
   19   0            # EFT matching scale (0 = SUSY scale)
   20   2            # EFT loop order for upwards matching
   21   1            # EFT loop order for downwards matching
   22   0            # EFT index of SM-like Higgs in the BSM model
   23   1            # calculate BSM pole masses
   24   123111321    # individual threshold correction loop orders
   25   0            # ren. scheme for Higgs 3L corrections
                     # (0 = DR, 1 = MDR)
   26   1            # Higgs 3L corrections O(alpha_t alpha_s^2)
   27   1            # Higgs 3L corrections O(alpha_b alpha_s^2)
   28   1            # Higgs 3L corrections O(alpha_t^2 alpha_s)
   29   1            # Higgs 3L corrections O(alpha_t^3)
\end{lstlisting}
In the \mathematica interface of \fs, the configuration options are
passed to the function \code{FS<model>OpenHandle[fsSettings ->
  \{...\}]} in form of replacement rules.  The symbols associated to
the configuration options are given in \tabref{tab:sg_mma_options}.
Unset options are set to their default values defined in the table.
A call of \code{FS<model>OpenHandle[]} with all configuration options
set to their default values would read:
\begin{lstlisting}
handle = FS<model>OpenHandle[
    fsSettings -> {
        precisionGoal -> 1.*^-4,           (* FlexibleSUSY[0] *)
        maxIterations -> 0,                (* FlexibleSUSY[1] *)
        solver -> 0,                       (* FlexibleSUSY[2] *)
        calculateStandardModelMasses -> 0, (* FlexibleSUSY[3] *)
        poleMassLoopOrder -> 2,            (* FlexibleSUSY[4] *)
        ewsbLoopOrder -> 2,                (* FlexibleSUSY[5] *)
        betaFunctionLoopOrder -> 3,        (* FlexibleSUSY[6] *)
        thresholdCorrectionsLoopOrder -> 2,(* FlexibleSUSY[7] *)
        higgs2loopCorrectionAtAs -> 1,     (* FlexibleSUSY[8] *)
        higgs2loopCorrectionAbAs -> 1,     (* FlexibleSUSY[9] *)
        higgs2loopCorrectionAtAt -> 1,     (* FlexibleSUSY[10] *)
        higgs2loopCorrectionAtauAtau -> 1, (* FlexibleSUSY[11] *)
        forceOutput -> 0,                  (* FlexibleSUSY[12] *)
        topPoleQCDCorrections -> 1,        (* FlexibleSUSY[13] *)
        betaZeroThreshold -> 1.*^-11,      (* FlexibleSUSY[14] *)
        forcePositiveMasses -> 0,          (* FlexibleSUSY[16] *)
        poleMassScale -> 0,                (* FlexibleSUSY[17] *)
        eftPoleMassScale -> 0,             (* FlexibleSUSY[18] *)
        eftMatchingScale -> 0,             (* FlexibleSUSY[19] *)
        eftMatchingLoopOrderUp -> 2,       (* FlexibleSUSY[20] *)
        eftMatchingLoopOrderDown -> 1,     (* FlexibleSUSY[21] *)
        eftHiggsIndex -> 0,                (* FlexibleSUSY[22] *)
        calculateBSMMasses -> 1,           (* FlexibleSUSY[23] *)
        thresholdCorrections -> 123111321, (* FlexibleSUSY[24] *)
        higgs3loopCorrectionRenScheme -> 0,(* FlexibleSUSY[25] *)
        higgs3loopCorrectionAtAsAs -> 1,   (* FlexibleSUSY[26] *)
        higgs3loopCorrectionAbAsAs -> 1,   (* FlexibleSUSY[27] *)
        higgs3loopCorrectionAtAtAs -> 1,   (* FlexibleSUSY[28] *)
        higgs3loopCorrectionAtAtAt -> 1,   (* FlexibleSUSY[29] *)
        parameterOutputScale -> 0          (* MODSEL[12] *)
    }
]
\end{lstlisting}
\begin{table}
  \centering
  \resizebox{\textwidth}{!}{%
  \begin{tabularx}{1.2\textwidth}{lllX}
    \toprule
    Index & \mathematica symbol & Default & Description \\
    \midrule
    \multicolumn{4}{c}{\texttt{Block FlexibleSUSY}}\\
    \midrule
     0 & \code{precisionGoal}                 & \code{1.*^-4} &
    precision goal of RG running and mass spectrum\\
     1 & \code{maxIterations}                 & \code{0} &
    maximum number of iterations for the running between the scales ($0$ = automatic)\\
     2 & \code{solver}                        & \code{0} &
    BVP solver ($0$ = all, $1$ = two-scale solver, $2$ = semi-analytic solver)\\
     3 & \code{calculateStandardModelMasses}  & \code{0} &
    switch to enable/disable calculation of pole masses of SM particles ($0$ = disabled)\\
     4 & \code{poleMassLoopOrder}             & \code{2} &
    pole mass loop order\\
     5 & \code{ewsbLoopOrder}                 & \code{2} &
    EWSB loop order (should be set equal to the pole mass loop order)\\
     6 & \code{betaFunctionLoopOrder}         & \code{3} &
    loop order for renormalization group running\\
     7 & \code{thresholdCorrectionsLoopOrder} & \code{2} &
    global switch for loop order of threshold corrections when
    converting the $\SM(5)$ parameters to the BSM parameters\\
     8 & \code{higgs2loopCorrectionAtAs}      & \code{1} &
    enable/disable 2-loop corrections $O(\at\as)$ to $M_{h,H,A}$\\
     9 & \code{higgs2loopCorrectionAbAs}      & \code{1} &
    enable/disable 2-loop corrections $O(\ab\as)$ to $M_{h,H,A}$\\
    10 & \code{higgs2loopCorrectionAtAt}      & \code{1} &
    enable/disable 2-loop corrections $O(\at^2)$ to $M_{h,H,A}$\\
    11 & \code{higgs2loopCorrectionAtauAtau}  & \code{1} &
    enable/disable 2-loop corrections $O(\atau^2)$ to $M_{h,H,A}$\\
    12 & \code{forceOutput}                   & \code{0} &
    force output, even if problems occurred\\
    13 & \code{topPoleQCDCorrections}         & \code{1} &
    QCD corrections to calculate $M_t$ (0 = 1-loop, 1 = 2-loop, 2 = 3-loop)\\
    14 & \code{betaZeroThreshold}             & \code{1.*^-11} &
    below this threshold $\beta$ functions are treated as zero\\
    15 & & \code{0} & enable/disable calculation of observables\\
    16 & \code{forcePositiveMasses}           & \code{0} &
    make Majorana masses positive (violates SLHA)\\
    17 & \code{poleMassScale}                 & \code{0} &
    scale at which pole masses are calculated (0 = SUSY scale)\\
    18 & \code{eftPoleMassScale}              & \code{0} &
    scale at which SM pole masses are calculated in \feft (0 = $M_t$)\\
    19 & \code{eftMatchingScale}              & \code{0} &
    matching scale in \feft (0 = SUSY scale)\\
    20 & \code{eftMatchingLoopOrderUp}        & -- &
    ignored\\
    21 & \code{eftMatchingLoopOrderDown}      & \code{1} &
    loop order for $\Delta\lambda$ in \feft\\
    22 & \code{eftHiggsIndex}                 & \code{0} &
    index of SM-like Higgs in BSM Higgs multiplet in \feft\\
    23 & \code{calculateBSMMasses}            & \code{1} &
    enable/disable calculation of BSM pole masses\\
    24 & \code{thresholdCorrections}          & \code{123111321} &
    individual threshold correction loop orders, see \tabref{tab:sg_mma_tc}\\
    25 & \code{higgs3loopCorrectionRenScheme} & \code{0} &
    renormalization scheme for 3-loop MSSM Higgs corrections
    (0 = \DRbar, 1 = \MDRbar)\\
    26 & \code{higgs3loopCorrectionAtAsAs}    & \code{1} &
    enable/disable 3-loop corrections $O(\at\as^2)$ to $M_{h,H,A}$\\
    27 & \code{higgs3loopCorrectionAbAsAs}    & \code{1} &
    enable/disable 3-loop corrections $O(\ab\as^2)$ to $M_{h,H,A}$\\
    28 & \code{higgs3loopCorrectionAtAtAs}    & \code{1} &
    enable/disable 3-loop corrections $O(\at^2\as)$ to $M_{h}$\\
    29 & \code{higgs3loopCorrectionAtAtAt}    & \code{1} &
    enable/disable 3-loop corrections $O(\at^3)$ to $M_{h}$\\
    \midrule
    \multicolumn{4}{c}{\texttt{Block MODSEL}}\\
    \midrule
    12 & \code{parameterOutputScale}          & \code{0} &
    output scale for running parameters (0 = SUSY scale)\\
    \bottomrule
  \end{tabularx}}
  \caption{SLHA input block entries and corresponding \mathematica
    symbols to specify the configuration options for \fs's spectrum
    generators.  The symbols $M_{h,H,A}$ and $M_t$ denote the Higgs
    and top quark pole masses, respectively.}
  \label{tab:sg_mma_options}
\end{table}
\begin{table}[tbh]
  \centering
  \begin{tabular}{lll}
    \toprule
    digit position $n$    & default value & parameter \\
    (from the right) & (prefactor of $10^n$) & \\
    \midrule
    $0$ & $1$ (1-loop) & $\alpha^\BSM_{\text{em}}$ \\
    $1$ & $2$ (2-loop) & $\sin^\BSM(\theta_W)$ \\
    $2$ & $3$ (3-loop) & $\as^\BSM$ \\
    $3$ & $1$ (1-loop) & $m^\BSM_Z$ \\
    $4$ & $1$ (1-loop) & $m^\BSM_W$ \\
    $5$ & $1$ (1-loop) & $m^\BSM_h$ \\
    $6$ & $3$ (3-loop) & $m^\BSM_t$ \\
    $7$ & $2$ (2-loop) & $m^\BSM_b$ \\
    $8$ & $1$ (1-loop) & $m^\BSM_\tau$ \\
    \bottomrule
  \end{tabular}
  \caption{Specification of individual loop orders of threshold
    corrections for extracting the running masses, couplings and
    Weinberg angle in the BSM model at the low-energy scale using the
    field \texttt{FlexibleSUSY[24]} in the SLHA interface or the
    symbol \texttt{thresholdCorrections} in the \mathematica
    interface, respectively.  The digit position is counted from the
    right, starting at $0$.  Setting all loop orders to their default
    values results in the integer $123111321$.}
  \label{tab:sg_mma_tc}
\end{table}
The individual configuration options have the following meaning:
\begin{itemize}
\setlength{\itemindent}{1cm}
\item[\texttt{FlexibleSUSY[0]},] \texttt{precisionGoal}: This option
  describes the numeric precision of the renormalization group
  running, the mass spectrum calculation, the electroweak symmetry
  breaking and the calculation of the observables.  For most models a
  precision of $10^{-4}$ is sufficient.  For models with various
  3-loop corrections, like \HSSUSY or MSSM-like models, a
  precision of $10^{-5}$ might be better.

\item[\texttt{FlexibleSUSY[1]},] \texttt{maxIterations}: This option
  describes the maximum number of iterations for the renormalization
  group running between the various scales.  If it is set to $0$, then
  the maximum number of iterations $N_{\text{max,it}}$ is chosen
  according to the precision goal $p$ (see above) as
  \begin{align}
    N_{\text{max,it}} = - 10\log_{10} p \,.
  \end{align}

\item[\texttt{FlexibleSUSY[2]},] \texttt{solver}: This option chooses
  the BVP solver to be used.  If set to $0$, all
  solvers that have been enabled in the model file (see
  \secref{sec:chosing_SAS}) are used.  In this case, each of the
  enabled solvers will be tried in turn, in the same order as given in
  \code{FSBVPSolvers}, until a solution is found, at which point \fs
  will return this solution and no further solvers are tried.  In the
  event that no solver obtains a valid solution, \fs reports the status of the
  last solver tried.  Non-zero values of \texttt{FlexibleSUSY[2]}
  select a single solver to be used.  If set to $1$, the two-scale
  solver is used if it has been enabled in the model file.  If set
  to $2$, then the semi-analytic solver is used if it has been enabled
  in the model file.  If a solver that has not been enabled in the model
  file is chosen, \fs stops with an error.

\item[\texttt{FlexibleSUSY[3]},]
  \texttt{calculateStandardModelMasses}: This option allows the user
  to enable/disable the calculation of the pole masses of the SM
  particles.  Note that this switch must be set to $1$ in
  \HSSUSY to calculate the Higgs pole mass, because in \HSSUSY the
  Higgs pole mass is calculated in the SM\@.

\item[\texttt{FlexibleSUSY[4]},] \texttt{poleMassLoopOrder}: This
  option allows the user to select the loop order at which the pole
  masses are calculated.  If set to $0$, the running tree-level masses
  are output.  If set to $1$, the pole masses are calculated at the
  1-loop level.  If set to $2$ or $3$, then model-specific 2-loop or
  3-loop corrections are taken into account, respectively, if they have been
  enabled in the model file (see
  \secref{sec:model_specific_contributions}).  \textbf{Important
    note:} In order to obtain a consistent pole mass spectrum, the
  loop order of the electroweak symmetry breaking (see
  \texttt{FlexibleSUSY[5]}, \texttt{ewsbLoopOrder}) must be set to the
  \emph{same} value as the pole mass loop order!

\item[\texttt{FlexibleSUSY[5]},] \texttt{ewsbLoopOrder}: This option
  allows the user to select the loop order at which the electroweak
  symmetry breaking (EWSB) equations are solved.  If set to $0$, the
  EWSB equations are solved at the tree level.  If set to $1$, the
  EWSB equations are solved at the 1-loop level.  If set to $2$ or
  $3$, then model-specific 2-loop or 3-loop corrections are taken into
  account, respectively, if they have been enabled in the model file (see
  \secref{sec:model_specific_contributions}).  \textbf{Important
    note:} In order to obtain a consistent pole mass spectrum, the
  loop order of the electroweak symmetry breaking must be set to the
  \emph{same} value as the pole mass loop order (see
  \texttt{FlexibleSUSY[4]}, \texttt{poleMassLoopOrder})!

\item[\texttt{FlexibleSUSY[6]},] \texttt{betaFunctionLoopOrder}: With
  this option the user can select the loop level of the $\beta$
  functions used to integrate the RGEs.  If set to $1$, 1-loop $\beta$
  functions are used.  If set to $2$, 2-loop $\beta$ functions are used.  If
  set to $3$, then model-specific 3-loop $\beta$ functions are used (see
  \secref{sec:model_specific_contributions}).  Note that \sarah can
  generate 2-loop $\beta$ functions for all model parameters (except
  scalar tadpole terms), so 2-loop running can always be used.

\item[\texttt{FlexibleSUSY[7]},]
  \texttt{thresholdCorrectionsLoopOrder}: With this option the user
  can choose the \emph{maximum} loop level of the threshold corrections
  used to determine the running gauge couplings $g_1$, $g_2$, $g_3$
  and the running Yukawa coupling matrices $Y_u$, $Y_d$, $Y_e$ of the
  \BSM model at the low-energy scale (\code{LowScale}) from the given
  SM input parameters ($\aem^{\SM(5)}(M_Z)$,
  $\as^{\SM(5)}(M_Z)$, $G_F$, $M_Z$, $M_t$, $m_b^{\SM(5)}(m_b)$,
  $m_c^{\SM(4)}(m_c)$, \ldots).  See \secref{sec:determination_of_g_Y}
  and \ref{sec:model_specific_contributions} for a description on how
  the running gauge and Yukawa couplings are calculated and how
  model-specific higher-order corrections can be included.  If the
  threshold corrections loop order is set to $1$, then no 2-loop
  threshold corrections or higher are taken into account.  If set to
  $2$, then no 3-loop threshold corrections or higher are taken into
  account.  If set to $3$, then no 4-loop threshold corrections are
  taken into account.
  Note that threshold corrections for individual parameters can be
  disabled by using \texttt{FlexibleSUSY[24]} or
  \texttt{thresholdCorrections}, respectively.

\item[\texttt{FlexibleSUSY[8]},] \texttt{higgs2loopCorrectionAtAs}:
  With this option the 2-loop contributions to the Higgs pole mass(es)
  of $O(\at\as)$ can be enabled/disabled.  Note that this
  option has an effect only if 2-loop contributions have been
  activated in the \fs model file.  See
  \secref{sec:model_specific_contributions} for details on how to
  activate 2-loop contributions to the Higgs pole mass(es) of
  $O(\at\as)$ in the SM, (N)MSSM or split-MSSM\@.

\item[\texttt{FlexibleSUSY[9]},] \texttt{higgs2loopCorrectionAbAs}:
  With this option the 2-loop contributions to the Higgs pole mass(es)
  of $O(\ab\as)$ can be enabled/disabled.  Note that this
  option has an effect only if 2-loop contributions have been
  activated in the \fs model file.  See
  \secref{sec:model_specific_contributions} for details on how to
  activate 2-loop contributions of $O(\ab\as)$ to the Higgs pole
  mass(es) in the (N)MSSM\@.

\item[\texttt{FlexibleSUSY[10]},] \texttt{higgs2loopCorrectionAtAt}:
  With this option the 2-loop contributions to the Higgs pole mass(es)
  of $O(\at^2)$ or $O((\at+\ab)^2)$ can be enabled/disabled.
  Note that this option has an effect only if 2-loop contributions
  have been activated in the \fs model file.  See
  \secref{sec:model_specific_contributions} for details on how to
  activate 2-loop contributions of these orders to the Higgs pole
  mass(es) in the SM or (N)MSSM\@.

\item[\texttt{FlexibleSUSY[11]},]
  \texttt{higgs2loopCorrectionAtauAtau}: With this option the 2-loop
  contributions to the Higgs pole mass(es) of $O(\atau^2)$
  can be enabled/disabled.  Note that this option has an effect only
  if 2-loop contributions have been activated in the \fs model file.
  See \secref{sec:model_specific_contributions} for details on how to
  activate 2-loop contributions of $O(\atau^2)$ to the Higgs pole
  mass(es) in the (N)MSSM\@.

\item[\texttt{FlexibleSUSY[12]},] \texttt{forceOutput}: This option
  allows the user to force an output of \fs, even if a physical
  problem has occurred (tachyon, non-perturbative parameter, no EWSB,
  \ldots).  If set to $0$, \fs does not give an output if a problem
  has occurred.  If set to $1$, an output is always given, even if a
  problem has occurred.  Please be very careful and check for
  potential warnings/problems when forcing the output!

\item[\texttt{FlexibleSUSY[13]},] \texttt{topPoleQCDCorrections}: With
  this option the user can enable additional loop contributions when
  the top quark \emph{pole mass} is re-calculated.  Note that the top pole mass
  is only re-calculated if \texttt{FlexibleSUSY[3]} or
  \texttt{calculateStandardModelMasses} is set to $1$.  If set to
  $0$, then 1-loop (SUSY-)QCD contributions are taken into account (but
  only if \texttt{FlexibleSUSY[4]} or \texttt{poleMassLoopOrder} is
  set to $1$).  If set to $1$, then 2-loop (SUSY-)QCD contributions are
  taken into account (but only if \texttt{FlexibleSUSY[4]} or
  \texttt{poleMassLoopOrder} is set to $2$).  If set to $2$, then
  3-loop (SUSY-)QCD contributions are taken into account (but only if
  \texttt{FlexibleSUSY[4]} or \texttt{poleMassLoopOrder} is set to
  $3$).

\item[\texttt{FlexibleSUSY[14]},] \texttt{betaZeroThreshold}: With
  this option a numerical threshold can be defined below which a
  $\beta$ function is treated as being exactly zero.  A small but
  non-zero threshold can avoid numerical problems when integrating the
  RGEs.

\item[\texttt{FlexibleSUSY[15]}:] With this option the calculation of
  the observables ($\amu$, EDMs, effective couplings of
  $h\rightarrow\gamma\gamma$ and $h\rightarrow gg$) can be
  enabled/disabled.  Note that in the \mathematica interface, the
  observables are calculated by the function
  \code{FS<model>CalculateObservables[]}, see
  \secref{sec:Mathematica_interface}.

\item[\texttt{FlexibleSUSY[16]},] \texttt{forcePositiveMasses}: With
  this option the masses of Majorana fermions can be forced to be
  positive in the SLHA output of \fs.  If set to $1$, then Majorana
  masses are always positive, but the corresponding mixing matrices
  are in general complex (note that this violates the SLHA
  convention).  If set to $0$, then the Majorana masses can be
  positive or negative, but the corresponding mixing matrices are
  guaranteed to be real (SLHA convention).

\item[\texttt{FlexibleSUSY[17]},] \texttt{poleMassScale}: With this
  option the user can choose the scale (in GeV) at which the pole mass
  spectrum is calculated.  If set to $0$, then the value assigned to
  the \code{SUSYScale} variable in the model file is used.  In \feft,
  the pole mass scale is defined to be the scale at which the pole
  masses in the \emph{full} \BSM model are calculated.  To vary the
  scale at which the pole masses in the effective theory (the
  SM) are calculated, use \texttt{FlexibleSUSY[18]} or
  \texttt{eftPoleMassScale} in the SLHA or \mathematica interface,
  respectively.

\item[\texttt{FlexibleSUSY[18]},] \texttt{eftPoleMassScale}: This
  option applies only to \feft models.  With this option the user can
  choose the scale (in GeV) at which the pole masses in the effective
  field theory (i.e., in the SM) are calculated.  If the
  scale is set to $0$, then $Q = M_t$ is used.  This option can be
  used to estimate a partial uncertainty of the Higgs mass
  prediction in \feft by varying the pole mass scale around $Q = M_t$,
  see \exref{ex:MSSMEFTHiggs_uncertainty_estimate}.

\item[\texttt{FlexibleSUSY[19]},] \texttt{eftMatchingScale}: This
  option applies only to \feft models.  With this option the user can
  specify the scale at which the matching of the \BSM model to the
  effective theory (i.e., the SM) is performed.  If the
  scale is set to $0$, then the value assigned to the \code{SUSYScale}
  variable in the \fs model file is used.  This option can be used to
  estimate a partial uncertainty of the Higgs mass prediction
  in \feft by varying the matching scale around $Q = \MS$, see
  \exref{ex:MSSMEFTHiggs_uncertainty_estimate}.

\item[\texttt{FlexibleSUSY[20]},] \texttt{eftMatchingLoopOrderUp}:
  This option is ignored in \fstwo.

\item[\texttt{FlexibleSUSY[21]},] \texttt{eftMatchingLoopOrderDown}:
  This option applies only to \feft models.  With this option the user
  can select the loop order at which the quartic Higgs coupling
  $\lambda$ of the SM is fixed when matching the \BSM
  model to the SM in \feft.  If set to $0$, then $\lambda$
  is fixed using only tree-level matching.  If set to $1$
  (recommended), then $\lambda$ is fixed by a 1-loop matching
  condition.

\item[\texttt{FlexibleSUSY[22]},] \texttt{eftHiggsIndex}: This option
  applies only to \feft models.  With this option the user can choose
  which field in the Higgs multiplet of the \BSM model corresponds to
  the SM-like Higgs.  If set to $0$, the lightest field in the Higgs
  multiplet is interpreted as SM-like Higgs.  If set to $1$, the
  2$^{\text{nd}}$ lightest field is interpreted as SM-like Higgs, etc.
  The chosen field is then used in the matching condition
  $M_h^{\SM} = M_{h_i}^{\BSM}$, where $i$ is the index of the chosen
  field in the Higgs multiplet ($i=0,1,2,\ldots$).

\item[\texttt{FlexibleSUSY[23]},] \texttt{calculateBSMMasses}: This
  option allows the user to enable/disable the calculation of the pole
  masses of the \BSM particles.  If set to $0$, then the \BSM pole
  masses are not calculated.  If set to $1$, then the \BSM pole masses
  are calculated.  This option is useful in \feft for example: If one
  is only interested in the prediction of the SM-like Higgs pole mass,
  then this option can be set to $0$ to suppress the calculation of
  the masses of the heavy Higgs bosons, the charginos, neutralinos and sfermions.

\item[\texttt{FlexibleSUSY[24]},] \texttt{thresholdCorrections}: With
  this option the user has a finer control over the threshold
  corrections to the individual model parameters.  The value assigned
  to this option is an integer number, where each digit (with respect to base
  $10$) represents the threshold correction loop order for a
  particular running \BSM parameter.  The association between the
  digits and the parameters as well as the default loop orders are
  shown in \tabref{tab:sg_mma_tc}.
  \begin{example}
    The following table shows example values for the integer number
    which specifies the individual threshold correction loop orders,
    together with the list of included loop corrections.
    \begin{center}
    \begin{tabular}{rl}
      \toprule
      integer & used threshold corrections \\
      \midrule
      0       & no threshold corrections, everything at tree level \\
      1       & only $\Delta\aem^{1L}$, everything else at tree level \\
      100     & only $\Delta\as^{1L}$, everything else at tree level \\
      101     & only $\Delta\aem^{1L}$ and  $\Delta\as^{1L}$, everything else at tree level \\
      3000101 & only $\Delta\aem^{1L}$ and  $\Delta\as^{1L}$ and  $\Delta y_t^{3L}$, everything else at tree level \\
      \bottomrule
    \end{tabular}
  \end{center}
  \end{example}
  If the field \texttt{FlexibleSUSY[24]} or the symbol
  \texttt{thresholdCorrections} is omitted, then the whole option is
  set to the default value given in \tabref{tab:sg_mma_tc}.  If the
  field is not omitted, then all loop orders must be given.  Note
  that setting the loop orders larger than the value set in
  \texttt{FlexibleSUSY[7]} or \texttt{thresholdCorrectionsLoopOrder}
  has no effect, see above.
  \begin{example}[label=ex:HSSUSY_thresholds]
    In the model file of \HSSUSY, 3-loop QCD corrections to the running
    top Yukawa coupling $y_t$ are enabled (\code{UseYukawa3LoopQCD =
      True}).  Switching between the 2-loop and 3-loop QCD corrections
    to $y_t$ can be used to estimate a partial uncertainty of
    the 2-loop Higgs pole mass.  In order to do this,
    \code{FSHSSUSYOpenHandle[]} must be called twice, setting
    \begin{align*}
      \texttt{thresholdCorrections} \rightarrow 123111321 \\
      \texttt{thresholdCorrections} \rightarrow 122111321
    \end{align*}
    respectively, and setting each time
    \code{thresholdCorrectionsLoopOrder -> 3} to enable the 3-loop
    corrections globally.  Note that the digit at the 6$^\text{th}$ position
    from the right (the prefactor of $10^6$) has been changed from $3$
    to $2$ to change the threshold correction loop order of $y_t$ from
    3-loop to 2-loop.  \exref{ex:HSSUSY_uncertainty} makes use
    of this method to estimate a partial uncertainty of
    \HSSUSY based on changing the threshold correction loop orders for
    $y_t$ and $\as$ in the SM\@.
  \end{example}

\item[\texttt{FlexibleSUSY[25]},]
  \texttt{higgs3loopCorrectionRenScheme}: This option applies only to
  MSSM models in which the 3-loop Higgs pole mass contributions from
  \Himalaya are enabled (the flag \code{UseHiggs3LoopMSSM = True} is
  set in the model file), see
  \secref{sec:model_specific_contributions}.  With this option the
  user can choose between the \DRbar and \MDRbar renormalization
  scheme.  If this option is set to $0$, then the \DRbar scheme is
  used.  If set to $1$, the \MDRbar scheme is used.

\item[\texttt{FlexibleSUSY[26]},] \texttt{higgs3loopCorrectionAtAsAs}:
  With this option the user can enable/disable 3-loop contributions to
  the Higgs pole mass(es) of $O(\at\as^2)$.  Note that this option
  has an effect only if model-specific 3-loop contributions to the
  Higgs pole mass(es) of this order have been enabled in the \fs model file.  See
  \secref{sec:model_specific_contributions} on how to enable 3-loop
  contributions of this order in the SM and in the MSSM\@.

\item[\texttt{FlexibleSUSY[27]},] \texttt{higgs3loopCorrectionAbAsAs}:
  With this option the user can enable/disable 3-loop contributions to
  the Higgs pole mass(es) of $O(\ab\as^2)$.  Note that this option
  has an effect only if model-specific 3-loop contributions to the
  Higgs pole mass(es) of this order have been enabled in the \fs model file.
  See \secref{sec:model_specific_contributions} on how to enable
  3-loop contributions of this order in the SM and in the
  MSSM\@.

\item[\texttt{FlexibleSUSY[28]},] \texttt{higgs3loopCorrectionAtAtAs}:
  With this option the user can enable/disable 3-loop contributions to
  the Higgs pole mass(es) of $O(\at^2\as)$.  Note that this option
  has an effect only if model-specific 3-loop contributions to the
  Higgs pole mass(es) of this order have been enabled in the \fs model file.
  See \secref{sec:model_specific_contributions} on how to enable
  3-loop contributions of this order in the SM\@.

\item[\texttt{FlexibleSUSY[29]},] \texttt{higgs3loopCorrectionAtAtAt}:
  With this option the user can enable/disable 3-loop contributions to
  the Higgs pole mass(es) of $O(\at^3)$.  Note that this option has
  an effect only if model-specific 3-loop contributions to the Higgs
  pole mass(es) of this order have been enabled in the \fs model file.  See
  \secref{sec:model_specific_contributions} on how to enable 3-loop
  contributions of this order in the SM\@.

\item[\texttt{MODSEL[12]},] \texttt{parameterOutputScale}: With this
  option the scale (in GeV) can be specified, at which the running
  \MSbar/\DRbar model parameters are output.  If set to $0$, then the
  running parameters are output at the scale assigned to the
  \code{SUSYScale} variable in the \fs model file.
\end{itemize}

\section{CMSSMCPV model file}
\label{app:CMSSMCPV}
\begin{lstlisting}[language=Mathematica]
FSModelName = "@CLASSNAME@";
FSEigenstates = SARAH`EWSB;
FSDefaultSARAHModel = MSSM/CPV;

MINPAR = { {1, m0},
           {2, m12},
           {3, TanBeta},
           {4, CosPhiMu},
           {5, Azero},
           {100, Phase[\[Mu]]} };

IMMINPAR = { {2, Imm12},
             {4, SinPhiMu},
             {5, ImAzero} };

EXTPAR = {
    {100, etaInput}
};

RealParameters = {};

EWSBOutputParameters = { Re[B[\[Mu]]], Im[B[\[Mu]]], \[Mu] };

SUSYScale = Sqrt[Product[M[Su[i]]^(Abs[ZU[i,3]]^2 + Abs[ZU[i,6]]^2), {i,6}]];

SUSYScaleFirstGuess = Sqrt[m0^2 + 4 m12^2];

SUSYScaleInput = {
    {eta, etaInput},
    {Phase[\[Mu]], CosPhiMu + I SinPhiMu}
};

HighScale = g1 == g2;

HighScaleFirstGuess = 2.0 10^16;

HighScaleInput = {
   {T[Ye], (Azero + I ImAzero) Ye},
   {T[Yd], (Azero + I ImAzero) Yd},
   {T[Yu], (Azero + I ImAzero) Yu},
   {mq2, UNITMATRIX[3] m0^2},
   {ml2, UNITMATRIX[3] m0^2},
   {md2, UNITMATRIX[3] m0^2},
   {mu2, UNITMATRIX[3] m0^2},
   {me2, UNITMATRIX[3] m0^2},
   {mHu2, m0^2},
   {mHd2, m0^2},
   {MassB, m12 + I Imm12},
   {MassWB,m12 + I Imm12},
   {MassG, m12 + I Imm12}
};

LowScale = LowEnergyConstant[MZ];

LowScaleFirstGuess = LowEnergyConstant[MZ];

LowScaleInput = {
   {Yu, Automatic},
   {Yd, Automatic},
   {Ye, Automatic},
   {vd, 2 MZDRbar / Sqrt[GUTNormalization[g1]^2 g1^2 + g2^2] \
        Cos[ArcTan[TanBeta]]},
   {vu, 2 MZDRbar / Sqrt[GUTNormalization[g1]^2 g1^2 + g2^2] \
        Sin[ArcTan[TanBeta]]}
};

InitialGuessAtLowScale = {
   {vd, LowEnergyConstant[vev] Cos[ArcTan[TanBeta]]},
   {vu, LowEnergyConstant[vev] Sin[ArcTan[TanBeta]]},
   {\[Mu]   , LowEnergyConstant[MZ]},
   {B[\[Mu]], LowEnergyConstant[MZ]^2},
   {Yu, Automatic},
   {Yd, Automatic},
   {Ye, Automatic}
};

InitialGuessAtHighScale = {};

UseHiggs2LoopMSSM = False;

ExtraSLHAOutputBlocks = {
   {FlexibleSUSYOutput, NoScale,
           {{0, Hold[HighScale]},
            {1, Hold[SUSYScale]},
            {2, Hold[LowScale]} } },
   {FlexibleSUSYLowEnergy,
           {{21, FlexibleSUSYObservable`aMuon},
            {23, FlexibleSUSYObservable`EDM[Fe[1]]},
            {24, FlexibleSUSYObservable`EDM[Fe[2]]},
            {25, FlexibleSUSYObservable`EDM[Fe[3]]} } },
   {EFFHIGGSCOUPLINGS, NoScale,
           {{1, FlexibleSUSYObservable`CpHiggsPhotonPhoton},
            {2, FlexibleSUSYObservable`CpHiggsGluonGluon},
            {3, FlexibleSUSYObservable`CpPseudoScalarPhotonPhoton},
            {4, FlexibleSUSYObservable`CpPseudoScalarGluonGluon} } },
   {ALPHA, NoScale,
           {{ArcSin[Pole[ZH[2,2]]]}}},
   {HMIX , {{1, Re[\[Mu]]},
            {2, vu / vd},
            {3, Sqrt[vu^2 + vd^2]},
            {101, Re[B[\[Mu]]]},
            {102, vd},
            {103, vu} } },
   {ImHMIX,{{1, Im[\[Mu]]},
            {101, Im[B[\[Mu]]]} } },
   {Au,    {{1, 1, Re[T[Yu][1,1] / Yu[1,1]]},
            {2, 2, Re[T[Yu][2,2] / Yu[2,2]]},
            {3, 3, Re[T[Yu][3,3] / Yu[3,3]]} } },
   {Ad,    {{1, 1, Re[T[Yd][1,1] / Yd[1,1]]},
            {2, 2, Re[T[Yd][2,2] / Yd[2,2]]},
            {3, 3, Re[T[Yd][3,3] / Yd[3,3]]} } },
   {Ae,    {{1, 1, Re[T[Ye][1,1] / Ye[1,1]]},
            {2, 2, Re[T[Ye][2,2] / Ye[2,2]]},
            {3, 3, Re[T[Ye][3,3] / Ye[3,3]]} } },
   {ImAu,  {{1, 1, Im[T[Yu][1,1] / Yu[1,1]]},
            {2, 2, Im[T[Yu][2,2] / Yu[2,2]]},
            {3, 3, Im[T[Yu][3,3] / Yu[3,3]]} } },
   {ImAd,  {{1, 1, Im[T[Yd][1,1] / Yd[1,1]]},
            {2, 2, Im[T[Yd][2,2] / Yd[2,2]]},
            {3, 3, Im[T[Yd][3,3] / Yd[3,3]]} } },
   {ImAe,  {{1, 1, Im[T[Ye][1,1] / Ye[1,1]]},
            {2, 2, Im[T[Ye][2,2] / Ye[2,2]]},
            {3, 3, Im[T[Ye][3,3] / Ye[3,3]]} } },
   {MSOFT, {{1, Re[MassB]},
            {2, Re[MassWB]},
            {3, Re[MassG]},
            {21, mHd2},
            {22, mHu2},
            {31, SignedAbsSqrt[Re[ml2[1,1]]]},
            {32, SignedAbsSqrt[Re[ml2[2,2]]]},
            {33, SignedAbsSqrt[Re[ml2[3,3]]]},
            {34, SignedAbsSqrt[Re[me2[1,1]]]},
            {35, SignedAbsSqrt[Re[me2[2,2]]]},
            {36, SignedAbsSqrt[Re[me2[3,3]]]},
            {41, SignedAbsSqrt[Re[mq2[1,1]]]},
            {42, SignedAbsSqrt[Re[mq2[2,2]]]},
            {43, SignedAbsSqrt[Re[mq2[3,3]]]},
            {44, SignedAbsSqrt[Re[mu2[1,1]]]},
            {45, SignedAbsSqrt[Re[mu2[2,2]]]},
            {46, SignedAbsSqrt[Re[mu2[3,3]]]},
            {47, SignedAbsSqrt[Re[md2[1,1]]]},
            {48, SignedAbsSqrt[Re[md2[2,2]]]},
            {49, SignedAbsSqrt[Re[md2[3,3]]]} } },
   {ImMSOFT,
           {{1, Im[MassB]},
            {2, Im[MassWB]},
            {3, Im[MassG]},
            {31, SignedAbsSqrt[Im[ml2[1,1]]]},
            {32, SignedAbsSqrt[Im[ml2[2,2]]]},
            {33, SignedAbsSqrt[Im[ml2[3,3]]]},
            {34, SignedAbsSqrt[Im[me2[1,1]]]},
            {35, SignedAbsSqrt[Im[me2[2,2]]]},
            {36, SignedAbsSqrt[Im[me2[3,3]]]},
            {41, SignedAbsSqrt[Im[mq2[1,1]]]},
            {42, SignedAbsSqrt[Im[mq2[2,2]]]},
            {43, SignedAbsSqrt[Im[mq2[3,3]]]},
            {44, SignedAbsSqrt[Im[mu2[1,1]]]},
            {45, SignedAbsSqrt[Im[mu2[2,2]]]},
            {46, SignedAbsSqrt[Im[mu2[3,3]]]},
            {47, SignedAbsSqrt[Im[md2[1,1]]]},
            {48, SignedAbsSqrt[Im[md2[2,2]]]},
            {49, SignedAbsSqrt[Im[md2[3,3]]]} } }
};
\end{lstlisting}

\section{\THDM model file}
\label{app:THDMIIMSSMBC}
\begin{lstlisting}[language=Mathematica,numbers=left,numberstyle=\tiny\sffamily]
FSModelName = "@CLASSNAME@";
FSEigenstates = SARAH`EWSB;
AutomaticInputAtMSUSY = False;
FSDefaultSARAHModel = "THDM-II";

MINPAR = {
    {3, TanBeta}
};

EXTPAR = {
    {0, MSUSY},
    {1, MEWSB},
    {2, MuInput},
    {6, MAInput},
    {7, AtInput},
    {8, AbInput},
    {9, AtauInput},
    {100, LambdaLoopOrder}
};

EWSBOutputParameters = { M112, M222 };

(* The high scale where we match to the MSSM *)
HighScale = MSUSY;

HighScaleFirstGuess = MSUSY;

HighScaleInput = {
    {Lambda1, 1/2 (1/4 ( (GUTNormalization[g1] g1)^2 + g2^2)
                   + UnitStep[LambdaLoopOrder-1] (
                        deltaLambda1th1L + deltaLambda1Phi1L)
                   + UnitStep[LambdaLoopOrder-2] deltaLambda1th2L)},
    {Lambda2, 1/2 (1/4 ( (GUTNormalization[g1] g1)^2 + g2^2)
                   + UnitStep[LambdaLoopOrder-1] (
                        deltaLambda2th1L + deltaLambda2Phi1L)
                   + UnitStep[LambdaLoopOrder-2] deltaLambda2th2L)},
    {Lambda3, 1/4 (-(GUTNormalization[g1] g1)^2 + g2^2)
              + UnitStep[LambdaLoopOrder-1] (
                   deltaLambda3th1L + deltaLambda3Phi1L)
              + UnitStep[LambdaLoopOrder-2] deltaLambda3th2L},
    {Lambda4, -1/2 g2^2
              + UnitStep[LambdaLoopOrder-1] (
                   deltaLambda4th1L + deltaLambda4Phi1L)
              + UnitStep[LambdaLoopOrder-2] deltaLambda4th2L},
    {Lambda5, 0
              + UnitStep[LambdaLoopOrder-1] (
                   deltaLambda5th1L + deltaLambda5Phi1L)
              + UnitStep[LambdaLoopOrder-2] deltaLambda5th2L},
    {Lambda6, 0
              + UnitStep[LambdaLoopOrder-1] (
                   deltaLambda6th1L + deltaLambda6Phi1L)
              + UnitStep[LambdaLoopOrder-2] deltaLambda6th2L},
    {Lambda7, 0
              + UnitStep[LambdaLoopOrder-1] (
                   deltaLambda7th1L + deltaLambda7Phi1L)
              + UnitStep[LambdaLoopOrder-2] deltaLambda7th2L}
};

(* The scale where we impose the EWSB conditions
   and calculate the spectrum *)
SUSYScale = MEWSB;

SUSYScaleFirstGuess = MEWSB;

SUSYScaleInput = {
    {M122  , MAInput^2 Sin[ArcTan[v2/v1]] Cos[ArcTan[v2/v1]]}
};

LowScale = LowEnergyConstant[MT];

LowScaleFirstGuess = LowEnergyConstant[MT];

LowScaleInput = {
   {Yu, Automatic},
   {Yd, Automatic},
   {Ye, Automatic},
   {v1, 2 MZMSbar / Sqrt[GUTNormalization[g1]^2 g1^2 + g2^2] *
        Cos[ArcTan[TanBeta]]},
   {v2, 2 MZMSbar / Sqrt[GUTNormalization[g1]^2 g1^2 + g2^2] *
        Sin[ArcTan[TanBeta]]}
};

InitialGuessAtLowScale = {
   {v1, LowEnergyConstant[vev] Cos[ArcTan[TanBeta]]},
   {v2, LowEnergyConstant[vev] Sin[ArcTan[TanBeta]]},
   {Yu, Automatic},
   {Yd, Automatic},
   {Ye, Automatic},
   {M122, MAInput^2 Sin[ArcTan[TanBeta]] Cos[ArcTan[TanBeta]]}
};

DefaultPoleMassPrecision = MediumPrecision;
HighPoleMassPrecision    = {hh};
MediumPoleMassPrecision  = {};
LowPoleMassPrecision     = {};

SMParticles = {
    Electron, TopQuark, BottomQuark,
    VectorP, VectorZ, VectorG, VectorW, Neutrino,
    gP, gG, gZ, gWm, gWmC
};

(* abbreviations *)
At   = AtInput;
Ab   = AbInput;
Atau = AtauInput;
Lambda1WagnerLee = 2 Lambda1;
Lambda2WagnerLee = 2 Lambda2;

(* arxiv:1508.00576, Eq. (45) *)
deltaLambda1th1L = With[{
    kappa = 1/(4 Pi)^2,
    ht = Yu[3,3],
    hb = Yd[3,3],
    htau = Ye[3,3],
    gY = GUTNormalization[g1] g1,
    muMS = MuInput / MSUSY,
    AbMS = Ab / MSUSY,
    AtauMS = Atau / MSUSY
    },
    (
        - kappa/2 ht^4 muMS^4
        + 6 kappa hb^4 AbMS^2 (1 - AbMS^2/12)
        + 2 kappa htau^4 AtauMS^2 (1 - AtauMS^2/12)
        + kappa (g2^2 + gY^2)/4 (3 ht^2 muMS^2 - 3 hb^2 AbMS^2
                                 - htau^2 AtauMS^2)
    )
];

(* arxiv:1508.00576, Eq. (46) *)
deltaLambda2th1L = With[{
    kappa = 1/(4 Pi)^2,
    ht = Yu[3,3],
    hb = Yd[3,3],
    htau = Ye[3,3],
    gY = GUTNormalization[g1] g1,
    muMS = MuInput / MSUSY,
    AbMS = Ab / MSUSY,
    AtauMS = Atau / MSUSY,
    AtMS = At / MSUSY
    },
    (
        6 kappa ht^4 AtMS^2 (1 - AtMS^2/12)
        - kappa/2 hb^4 muMS^4
        - kappa/6 htau^4 muMS^4
        - kappa (g2^2 + gY^2)/4 (3 ht^2 AtMS^2 - 3 hb^2 muMS^2
                                 - htau^2 muMS^2)
    )
];

(* arxiv:1508.00576, Eq. (47)-(48) *)
deltaLambda3th1L = With[{
    kappa = 1/(4 Pi)^2,
    ht = Yu[3,3],
    hb = Yd[3,3],
    htau = Ye[3,3],
    gY = GUTNormalization[g1] g1,
    muMS = MuInput / MSUSY,
    AbMS = Ab / MSUSY,
    AtauMS = Atau / MSUSY,
    AtMS = At / MSUSY
    },
    (
        kappa/6 muMS^2 (3 ht^4 (3 - AtMS^2)
                        + 3 hb^4 (3 - AbMS^2)
                        + htau^4 (3 - AtauMS^2))
        + kappa/2 ht^2 hb^2 (3 (AtMS + AbMS)^2
                             - (muMS^2 - AtMS AbMS)^2
                             - 6 muMS^2)
        - kappa/2 (g2^2 - gY^2)/4 (3 ht^2 (AtMS^2 - muMS^2)
                                   + 3 hb^2 (AbMS^2 - muMS^2)
                                   + htau^2 (AtauMS^2 - muMS^2))
    )
];

(* arxiv:1508.00576, Eq. (49) *)
deltaLambda4th1L = With[{
    kappa = 1/(4 Pi)^2,
    ht = Yu[3,3],
    hb = Yd[3,3],
    htau = Ye[3,3],
    gY = GUTNormalization[g1] g1,
    muMS = MuInput / MSUSY,
    AbMS = Ab / MSUSY,
    AtauMS = Atau / MSUSY,
    AtMS = At / MSUSY
    },
    (
        kappa/6 muMS^2 (3 ht^4 (3 - AtMS^2)
                        + 3 hb^4 (3 - AbMS^2)
                        + htau^4 (3 - AtauMS^2))
        - kappa/2 ht^2 hb^2 (3 (AtMS + AbMS)^2
                             - (muMS^2 - AtMS AbMS)^2
                             - 6 muMS^2)
        + kappa/2 g2^2/2 (3 ht^2 (AtMS^2 - muMS^2)
                          + 3 hb^2 (AbMS^2 - muMS^2)
                          + htau^2 (AtauMS^2 - muMS^2))
    )
];

(* arxiv:1508.00576, Eq. (50) *)
deltaLambda5th1L = With[{
    kappa = 1/(4 Pi)^2,
    ht = Yu[3,3],
    hb = Yd[3,3],
    htau = Ye[3,3],
    muMS = MuInput / MSUSY,
    AbMS = Ab / MSUSY,
    AtauMS = Atau / MSUSY,
    AtMS = At / MSUSY
    },
    (
        - kappa/6 muMS^2 (3 ht^4 AtMS^2 + 3 hb^4 AbMS^2 + htau^4 AtauMS^2)
    )
];

(* arxiv:1508.00576, Eq. (51) *)
deltaLambda6th1L = With[{
    kappa = 1/(4 Pi)^2,
    ht = Yu[3,3],
    hb = Yd[3,3],
    htau = Ye[3,3],
    muMS = MuInput / MSUSY,
    AbMS = Ab / MSUSY,
    AtauMS = Atau / MSUSY,
    AtMS = At / MSUSY,
    gbar = ((GUTNormalization[g1] g1)^2 + g2^2) / 4
    },
    (
        kappa/6 muMS (+ 3 ht^4 muMS^2 AtMS
                      + 3 hb^4 AbMS (AbMS^2 - 6)
                      + htau^4 AtauMS (AtauMS^2 - 6))
        (* arxiv:hep-ph/9307201, Eq. (6.13)-(6.14) *)
        + gbar/2 kappa muMS (+ 3 AbMS hb^2
                             - 3 AtMS ht^2
                             + AtauMS htau^2)
    )
];

(* arxiv:1508.00576, Eq. (52) *)
deltaLambda7th1L = With[{
    kappa = 1/(4 Pi)^2,
    ht = Yu[3,3],
    hb = Yd[3,3],
    htau = Ye[3,3],
    muMS = MuInput / MSUSY,
    AbMS = Ab / MSUSY,
    AtauMS = Atau / MSUSY,
    AtMS = At / MSUSY,
    gbar = ((GUTNormalization[g1] g1)^2 + g2^2) / 4
    },
    (
        kappa/6 muMS (+ 3 ht^4 AtMS (AtMS^2 - 6)
                      + 3 hb^4 muMS^2 AbMS
                      + htau^4 muMS^2 AtauMS)
        (* arxiv:hep-ph/9307201, Eq. (6.13)-(6.14) *)
        - gbar/2 kappa muMS (+ 3 AbMS hb^2
                             - 3 AtMS ht^2
                             + AtauMS htau^2)
    )
];

(* arxiv:1508.00576, Eq. (53) *)
deltaLambda1Phi1L = With[{
    kappa = 1/(4 Pi)^2,
    ht = Yu[3,3],
    hb = Yd[3,3],
    htau = Ye[3,3],
    gY = GUTNormalization[g1] g1,
    muMS = MuInput / MSUSY,
    AbMS = Ab / MSUSY,
    AtauMS = Atau / MSUSY
    },
    (
        -kappa/6 (g2^2 + gY^2)/2 (3 ht^2 muMS^2 + 3hb^2 AbMS^2
                                  + htau^2 AtauMS^2)
    )
];

(* arxiv:1508.00576, Eq. (54) *)
deltaLambda2Phi1L = With[{
    kappa = 1/(4 Pi)^2,
    ht = Yu[3,3],
    hb = Yd[3,3],
    htau = Ye[3,3],
    gY = GUTNormalization[g1] g1,
    muMS = MuInput / MSUSY,
    AtMS = At / MSUSY
    },
    (
        -kappa/6 (g2^2 + gY^2)/2 (3 ht^2 AtMS^2 + 3 hb^2 muMS^2
                                  + htau^2 muMS^2)
    )
];

(* arxiv:1508.00576, Eq. (55) *)
deltaLambda3Phi1L = With[{
    kappa = 1/(4 Pi)^2,
    ht = Yu[3,3],
    hb = Yd[3,3],
    htau = Ye[3,3],
    gY = GUTNormalization[g1] g1,
    muMS = MuInput / MSUSY,
    AbMS = Ab / MSUSY,
    AtauMS = Atau / MSUSY,
    AtMS = At / MSUSY
    },
    (
        -kappa/6 (g2^2 - gY^2)/4 (3 ht^2 (AtMS^2 + muMS^2)
                                  + 3 hb^2 (AbMS^2 + muMS^2)
                                  + htau^2 (AtauMS^2 + muMS^2))
    )
];

(* arxiv:1508.00576, Eq. (56) *)
deltaLambda4Phi1L = With[{
    kappa = 1/(4 Pi)^2,
    ht = Yu[3,3],
    hb = Yd[3,3],
    htau = Ye[3,3],
    muMS = MuInput / MSUSY,
    AbMS = Ab / MSUSY,
    AtauMS = Atau / MSUSY,
    AtMS = At / MSUSY
    },
    (
        kappa/6 g2^2/2 (3 ht^2 (AtMS^2 + muMS^2)
                        + 3 hb^2 (AbMS^2 + muMS^2)
                        + htau^2 (AtauMS^2 + muMS^2))
    )
];

(* arxiv:1508.00576, Eq. (57) *)
deltaLambda5Phi1L = 0;
deltaLambda6Phi1L = 0; (* wrong in arxiv:hep-ph/9307201, Eq. (6.17) *)
deltaLambda7Phi1L = 0; (* wrong in arxiv:hep-ph/9307201, Eq. (6.17) *)

(* arxiv:1508.00576, Eq. (59) *)
deltaLambda1th2L = With[{
    kappa = 1/(4 Pi)^2,
    ht = Yu[3,3],
    muMS = MuInput / MSUSY
    },
    (
        -4/3 kappa^2 ht^4 g3^2 muMS^4
    )
];

(* arxiv:1508.00576, Eq. (60) *)
deltaLambda2th2L = With[{
    kappa = 1/(4 Pi)^2,
    ht = Yu[3,3],
    muMS = MuInput / MSUSY,
    AtMS = At / MSUSY
    },
    (
        16 kappa^2 ht^4 g3^2 (-2 AtMS + 1/3 AtMS^3 - 1/12 AtMS^4)
    )
];

(* arxiv:1508.00576, Eq. (61) *)
deltaLambda3th2L = With[{
    kappa = 1/(4 Pi)^2,
    ht = Yu[3,3],
    muMS = MuInput / MSUSY,
    AtMS = At / MSUSY
    },
    (
        4 kappa^2 ht^4 g3^2 AtMS muMS^2 (1 - 1/2 AtMS)
    )
];

deltaLambda4th2L = deltaLambda3th2L;
deltaLambda5th2L = deltaLambda3th2L;

(* arxiv:1508.00576, Eq. (62) *)
deltaLambda6th2L = With[{
    kappa = 1/(4 Pi)^2,
    ht = Yu[3,3],
    muMS = MuInput / MSUSY,
    AtMS = At / MSUSY
    },
    (
        4/3 kappa^2 ht^4 g3^2 muMS^3 (-1 + AtMS)
    )
];

(* arxiv:1508.00576, Eq. (63) *)
deltaLambda7th2L = With[{
    kappa = 1/(4 Pi)^2,
    ht = Yu[3,3],
    muMS = MuInput / MSUSY,
    AtMS = At / MSUSY
    },
    (
        4 kappa^2 ht^4 g3^2 muMS (2 - AtMS^2 + 1/3 AtMS^3)
    )
];
\end{lstlisting}

\section{CNMSSM model file}
\label{app:CNMSSM}
\begin{lstlisting}[language=Mathematica]

FSModelName = "@CLASSNAME@";
FSEigenstates = SARAH`EWSB;
FSDefaultSARAHModel = NMSSM;

FSBVPSolvers = { SemiAnalyticSolver };

(* CNMSSM input parameters *)

MINPAR = {
   {2, m12},
   {3, TanBeta},
   {4, Sign[vS]},
   {5, Azero}
};

EXTPAR = {
   {61, LambdaInput}
};

FSAuxiliaryParameterInfo = {
   {m0Sq,        { ParameterDimensions -> {1},
                   MassDimension -> 2 } },
   {LambdaInput, { ParameterDimensions -> {1},
                   MassDimension -> 0 } }
};

EWSBOutputParameters = { \[Kappa], vS, m0Sq };

SUSYScale = Sqrt[Product[M[Su[i]]^(Abs[ZU[i,3]]^2 + Abs[ZU[i,6]]^2), {i,6}]];

SUSYScaleFirstGuess = Sqrt[14 m12^2 - 3 m12 Azero + Azero^2];

SUSYScaleInput = {};

HighScale = g1 == g2;

HighScaleFirstGuess = 2.0 10^16;

HighScaleInput={
   {T[Ye], Azero*Ye},
   {T[Yd], Azero*Yd},
   {T[Yu], Azero*Yu},
   {mq2, UNITMATRIX[3] m0Sq},
   {ml2, UNITMATRIX[3] m0Sq},
   {md2, UNITMATRIX[3] m0Sq},
   {mu2, UNITMATRIX[3] m0Sq},
   {me2, UNITMATRIX[3] m0Sq},
   {mHu2, m0Sq},
   {mHd2, m0Sq},
   {ms2, m0Sq},
   {\[Lambda], LambdaInput},
   {T[\[Kappa]], Azero \[Kappa]},
   {T[\[Lambda]], Azero LambdaInput},
   {MassB, m12},
   {MassWB,m12},
   {MassG,m12}
};

LowScale = LowEnergyConstant[MZ];

LowScaleFirstGuess = LowEnergyConstant[MZ];

LowScaleInput = {
   {Yu, Automatic},
   {Yd, Automatic},
   {Ye, Automatic},
   {vd, 2 MZDRbar / Sqrt[GUTNormalization[g1]^2 g1^2 + g2^2] Cos[ArcTan[TanBeta]]},
   {vu, 2 MZDRbar / Sqrt[GUTNormalization[g1]^2 g1^2 + g2^2] Sin[ArcTan[TanBeta]]}
};

InitialGuessAtLowScale = {
   {vd, LowEnergyConstant[vev] Cos[ArcTan[TanBeta]]},
   {vu, LowEnergyConstant[vev] Sin[ArcTan[TanBeta]]},
   {\[Lambda], LambdaInput},
   {\[Kappa], 0.1},
   {vS, 1000},
   {m0Sq, LowEnergyConstant[MZ]^2},
   {Yu, Automatic},
   {Yd, Automatic},
   {Ye, Automatic}
};

InitialGuessAtHighScale = {};

UseHiggs2LoopNMSSM = True;
EffectiveMu = \[Lambda] vS / Sqrt[2];
EffectiveMASqr = (T[\[Lambda]] vS / Sqrt[2] + 0.5 \[Lambda] \[Kappa] vS^2) (vu^2 + vd^2) / (vu vd);

PotentialLSPParticles = { Chi, Sv, Su, Sd, Se, Cha, Glu };

DefaultPoleMassPrecision = MediumPrecision;
HighPoleMassPrecision    = {hh, Ah, Hpm};
MediumPoleMassPrecision  = {};
LowPoleMassPrecision     = {};

ExtraSLHAOutputBlocks = {
   {FlexibleSUSYOutput, NoScale,
           {{0, Hold[HighScale]},
            {1, Hold[SUSYScale]},
            {2, Hold[LowScale]} } },
   {EWSBOutputs, NoScale,
           {{1, \[Kappa]},
            {2, vS},
            {3, m0Sq} } },
   {FlexibleSUSYLowEnergy,
           {{0, FlexibleSUSYObservable`aMuon} } },
   {EFFHIGGSCOUPLINGS, NoScale,
           {{1, FlexibleSUSYObservable`CpHiggsPhotonPhoton},
            {2, FlexibleSUSYObservable`CpHiggsGluonGluon},
            {3, FlexibleSUSYObservable`CpPseudoScalarPhotonPhoton},
            {4, FlexibleSUSYObservable`CpPseudoScalarGluonGluon} } },
   {NMSSMRUN,
           {{1, \[Lambda]},
            {2, \[Kappa]},
            {3, T[\[Lambda]] / \[Lambda]},
            {4, T[\[Kappa]] / \[Kappa]},
            {5, \[Lambda] vS / Sqrt[2]},
            {10, ms2} } }
};
\end{lstlisting}

\clearpage

\section*{References}
\addcontentsline{toc}{section}{References}

\bibliographystyle{JHEP}
\bibliography{flexiblesusy-2}

\end{document}